\RequirePackage{ifpdf}
\documentclass[hyper,12pt,A4paper]{article}
\usepackage{jheppub}

\input xy
\xyoption{all}
\input xy
\xyoption{all}

\usepackage{graphicx}
\usepackage{latexsym,amsmath,amsfonts,amssymb}
\usepackage{mathrsfs}
\usepackage[makeroom]{cancel}
\usepackage{bbm}
\usepackage{bm}
\usepackage{subfigure}
\usepackage{paralist}
\usepackage{url}

\makeatletter
\def\@fpheader{\relax}
\makeatother

\newcommand{\vol}{\mathrm{vol}}

\numberwithin{equation}{section}

\newcommand{\nn}{\nonumber}
\newcommand{\mat}[1]{\begin{pmatrix} #1 \end{pmatrix}}

\newcommand{\be}{\begin{equation}} 
\newcommand{\ee}{\end{equation}}
\newcommand{\bea}{\begin{equation} \begin{aligned}} \newcommand{\eea}{\end{aligned} \end{equation}}

\newcommand{\bit}{\begin{itemize}} 
\newcommand{\eit}{\end{itemize}}

\newcommand{\Z}{\mathbb{Z}}
\newcommand{\C}{\mathbb{C}}
\newcommand{\R}{\mathbb{R}}

\renewcommand{\t}{\widetilde }
\newcommand{\wh}{\widehat }
\renewcommand{\d}{\partial }

\renewcommand{\b}{\bar }

\newcommand{\half}{{1\over 2}}

\newcommand{\CC}{\mathcal{C}}

\newcommand{\CF}{\mathcal{F}}

\newcommand{\CH}{\mathcal{H}}
\newcommand{\CI}{\mathcal{I}}

\newcommand{\CK}{\mathcal{K}}

\newcommand{\CM}{\mathcal{M}}
\newcommand{\CN}{\mathcal{N}}
\newcommand{\CO}{\mathcal{O}}

\newcommand{\CT}{\mathcal{T}}

\newcommand{\CV}{\mathcal{V}}

\newcommand{\FR}{\mathfrak{R}}
\newcommand{\Fg}{\mathfrak{g}}

\newcommand{\GG}{\mathbf{G}}

\newcommand{\GH}{\mathbf{H}}

\newcommand{\rk}{{{\rm rk}(\GG)}}

\newcommand{\m}{\mathfrak{m}}

\newcommand{\h}{\hat}

\newcommand{\oneloop}{\text{1-loop}}
\DeclareMathOperator{\Tr}{Tr}

\DeclareMathOperator{\sign}{sign}

\newcommand{\SL}{{\mathscr L}}

\newcommand{\ov}{\over}

\newcommand{\trilog}{{\text{Li}_3}}

\newcommand{\bE}{{\bf E}}
\renewcommand{\P}{{\mathbb{P}}}

\newcommand{\MG}{{\mathbf X}} %M-theory singualrity
\newcommand{\FT}{{\mathcal{T}_\MG}} %5d THEORY
\newcommand{\COD}{\mathbf{E}} %compact divisor
\title{Five-dimensional SCFTs and gauge theory phases: an M-theory/type IIA perspective
}

\author[\flat]{Cyril Closset,}
\author[\sharp]{Michele Del Zotto,}
\author[\natural]{Vivek Saxena}

\affiliation[\flat]{ Mathematical Institute, University of Oxford\\ Woodstock Road, Oxford, OX2 6GG, United Kingdom}
\affiliation[\sharp]{Department of Mathematical Sciences and Centre for Particle Theory\\ Durham University, Durham DH1 3LE, UK}
\affiliation[\natural]{C.N. Yang Institute for Theoretical Physics, Stony Brook University\\Stony Brook, NY 11794-3840, USA}

\preprint{YITP-SB-18-41}

\abstract{We revisit the correspondence between Calabi-Yau (CY) threefold isolated singularities $\mathbf{X}$ and five-dimensional superconformal field theories (SCFTs), which arise at low energy in M-theory on the space-time transverse to  $\mathbf{X}$. 
Focussing on the case of toric CY singularities, we analyze the ``gauge-theory phases'' of the SCFT by exploiting fiberwise M-theory/type IIA duality. In this setup, the low-energy gauge group simply arises on stacks of coincident D6-branes wrapping 2-cycles in some ALE space of type $A_{M-1}$ fibered over a real line, and the map between the K\"ahler parameters of $\mathbf{X}$ and the Coulomb branch parameters of the field theory (masses and VEVs) can be read off systematically. Different type IIA ``reductions'' give rise to different gauge theory phases, whose existence depends on the particular (partial) resolutions of the isolated singularity $\mathbf{X}$. We also comment on the case of non-isolated toric singularities. Incidentally, we propose a slightly modified expression for the Coulomb-branch prepotential of 5d $\mathcal{N}=1$ gauge theories.}

\begin{document}

\maketitle

\section{Introduction}

Five-dimensional gauge theories are infrared-free and become strongly coupled at scales of the order of the inverse gauge coupling, $1/g^{2}$. Being non-renormalizable, they are not, by themselves, well-defined quantum field theories. In the mid-1990's, it was realized that many supersymmetric 5d gauge theories could be viewed as the low-energy description of certain deformations of superconformal field theories (SCFTs) in five and six  dimensions \cite{Seiberg:1996bd, Morrison:1996xf, Douglas:1996xp, Intriligator:1997pq,Aharony:1997bh}.\footnote{For instance, five-dimensional maximal SYM is believed to describe the 6d $\CN=(2,0)$ theory compactified on a  circle with radius $R=g^2$---in the $SU(N)$ case, this is directly implied by the M-theory/type IIA duality \protect\cite{Witten:1995ex,Strominger:1995ac}, since a stack of $N$ M5-branes wrapping a circle is dual to a stack of $N$ D4-branes in flat space.} A key feature of these RG flows is that a given SCFT can admit several deformations of this kind, leading to apparently inequivalent gauge theory phases in the infrared (IR) \cite{Aharony:1997bh}, a phenomenon sometimes called an ``ultraviolet (UV) duality'' \cite{Bao:2011rc, Bergman:2013aca, Taki:2014pba, Bergman:2014kza, Zafrir:2014ywa, Mitev:2014jza}.

Since renormalization group (RG) flows are unidirectional, one should really start with the 5d SCFT and consider its deformations to whatever IR phases are possible. The five-dimensional SCFT itself is necessarily strongly coupled \cite{Chang:2018xmx}, therefore a direct field-theory analysis would be very challenging to perform. Instead, suitable string-theory embeddings of the SCFT allow us to study its deformation to massive phases rather systematically. The purpose of this paper is to provide a detailed exploration of this aspect of the physics of these systems, addressing the infrared phases of selected 5d SCFTs and their deformations. Along the way, we obtain several other new results, a short account of which can be found below. In particular, we obtain a new expression for the well-known 5d Coulomb branch prepotential that correctly accounts for the 5d parity anomaly.

The ideal laboratory for our study is provided by five-dimensional field theories obtained by geometric engineering in M-theory \cite{Morrison:1996xf, Douglas:1996xp, Intriligator:1997pq} --- see also \cite{Katz:1996fh, Katz:1997eq, Hollowood:2003cv} for more on geometric engineering and \cite{Cherkis:2014vfa, DelZotto:2017pti,Xie:2017pfl,Alexandrov:2017mgi,Jefferson:2017ahm,Jefferson:2018irk,Bhardwaj:2018yhy,Bhardwaj:2018vuu,Apruzzi:2018nre} for more recent studies. The conjectured correspondence between local Calabi-Yau (CY) threefold isolated singularities $\MG$ and 5d SCFTs
\be\label{Mtheory X3 SCFT intro}
\text{M-theory on}\:\; \R^{1,4} \times \MG \qquad \longleftrightarrow \qquad \FT \equiv {\rm SCFT}(\MG)~,
\ee
provides geometric tools to address the structure of the Coulomb phase of $\FT$ and of its deformations. Another well-known construction of 5d SCFTs is in terms of $(p,q)$-brane webs in type IIB \cite{Aharony:1997bh,DeWolfe:1999hj,Benini:2009gi,Bergman:2013aca,Zafrir:2014ywa,Hayashi:2014hfa}. Whenever $\MG$ is a toric singularity, the $(p,q)$-fivebrane web and the M-theory setup are dual to each other \cite{Leung:1997tw}. Further evidence for the existence of these 5d fixed points, in some appropriate large $N$ limit, is provided by the AdS/CFT duality \cite{Brandhuber:1999np,Bergman:2012kr,Passias:2012vp,Lozano:2012au,Bergman:2013koa,Apruzzi:2014qva,Bergman:2015dpa,Kim:2015hya,DHoker:2016ysh, Gutperle:2018vdd,Bergman:2018hin}. 

In the present paper, we introduce a complementary description to the M-theory and $(p,q)$-webs, focussing on the case when $\MG$ is toric.~\footnote{Many of the recent results obtained in the context of $(p,q)$-brane webs in IIB with orientifolds \cite{Zafrir:2015ftn,Hayashi:2016jak,Hayashi:2017btw,Hayashi:2018bkd, Hayashi:2018lyv} have non-toric M-theory duals. We leave the natural generalization of our methods to these setups (consisting of including orientifold planes and D8-branes in our analysis) for future work.} 
The M-theory description is purely geometric, while the IIB picture is purely in terms of branes in flat space. 
Our setup gives an intermediate dual configuration in type IIA string theory, in terms of D6-branes and geometry. The arbitrary-looking separation between ``branes'' and ``geometry'' in the type IIA setup will have a very neat interpretation in terms of the ``gauge theory phases'' of the SCFT $\FT$, as we will explain momentarily.

\medskip

In the rest of this introduction, we first review some general features of $\FT$. Then, we summarize our type IIA approach, and state our main results. Finally, we discuss some more subtle point regarding the so-called parity anomaly in five-dimensional field theories.

\subsection{Extended parameter space of a 5d SCFT}

\begin{figure}
\begin{center}
\subfigure[\small $\CM_\FT^C$, with $\mu_j=0$.]{
\includegraphics[height=7cm]{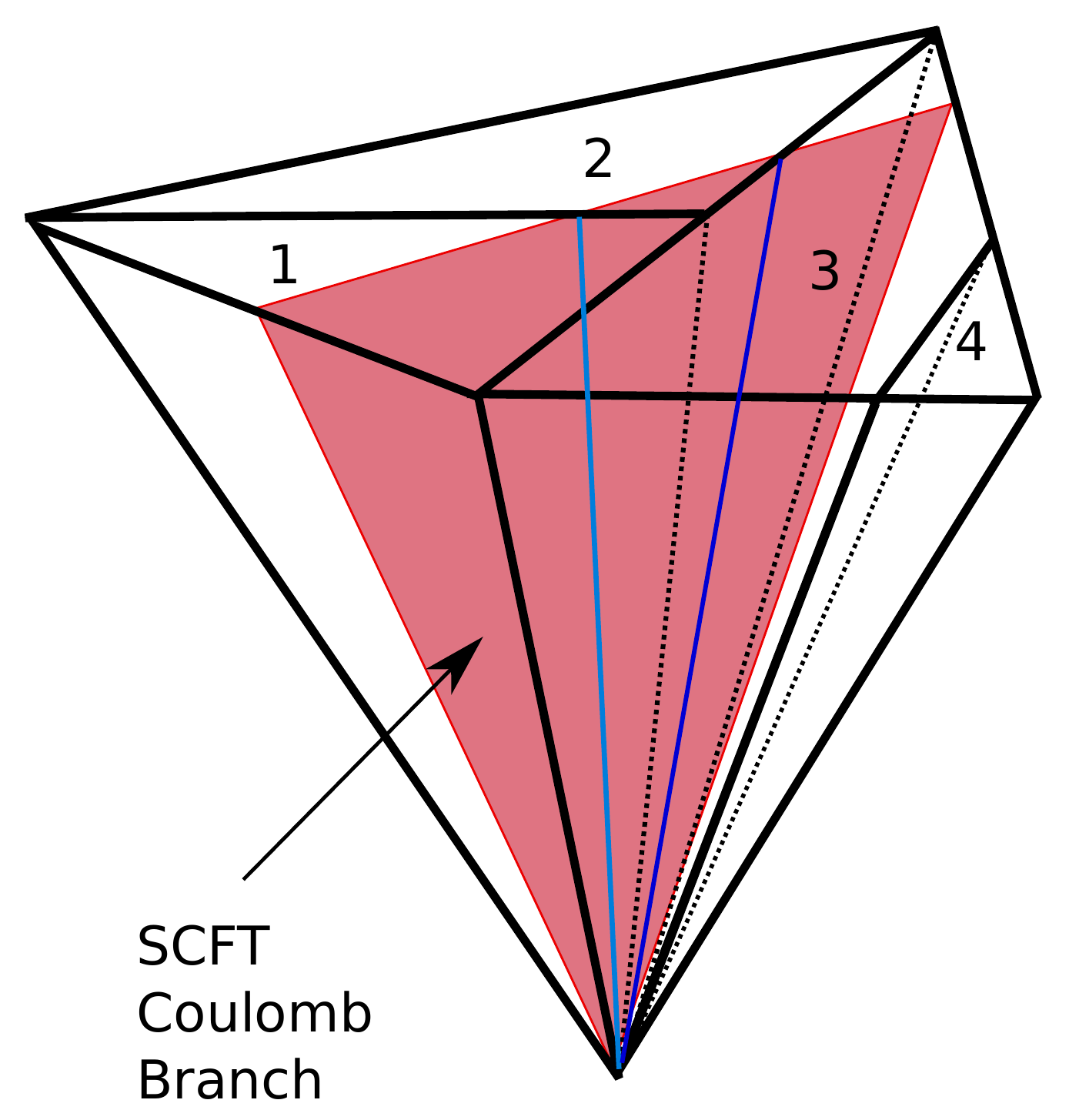}\label{puah1}}\qquad\qquad
\subfigure[\small  $\CM_\FT^C(\mu)$, with $\mu_j=a_j \neq 0$.]{
\includegraphics[height=7cm]{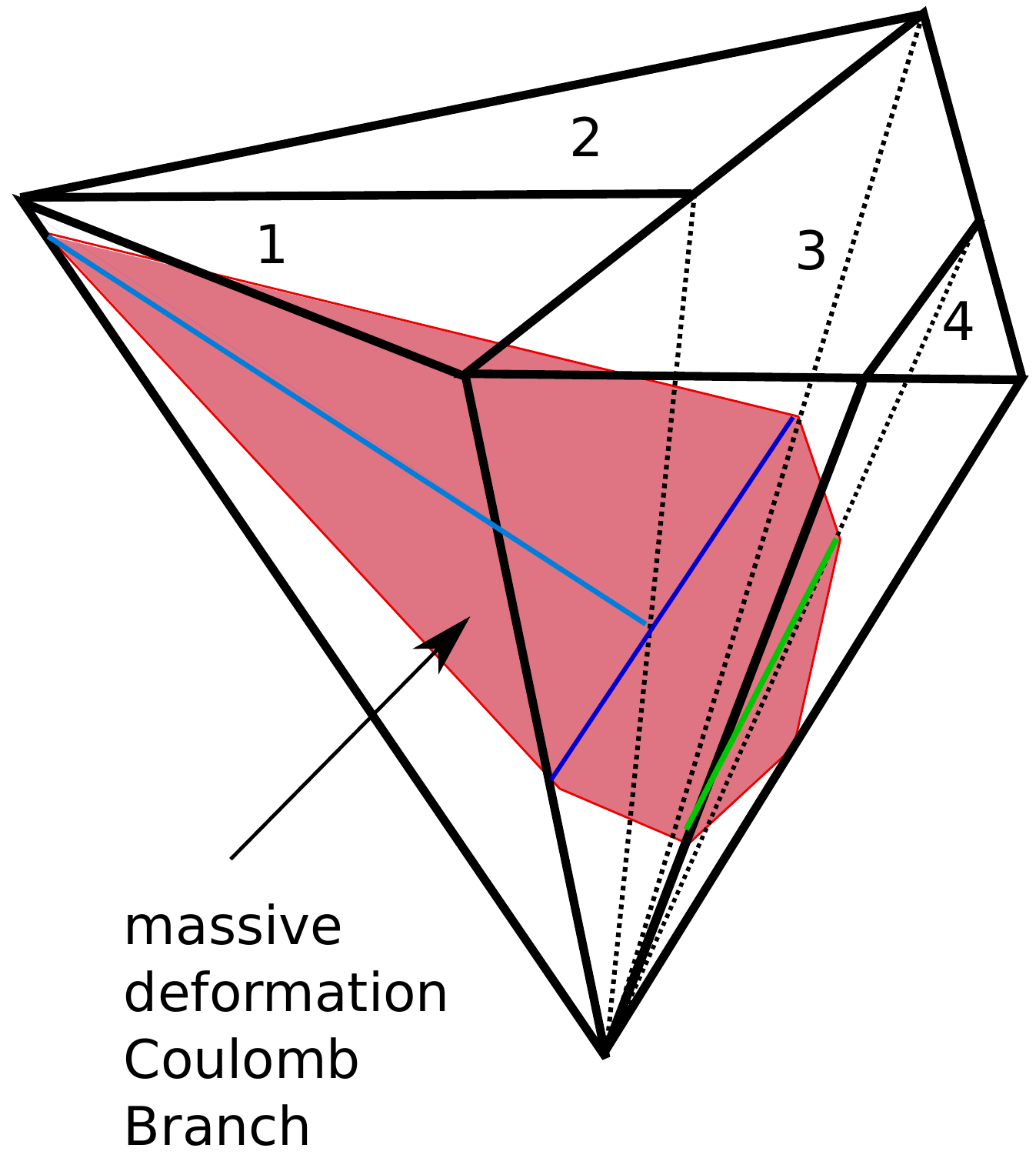}\label{puah2}}
\caption{Structure of the extended Coulomb parameter space of a 5d SCFT. On the left-hand-side, we represent the codimension-$f$ hypersurface $\mu_j = 0$ (in violet), which corresponds to the Coulomb branch, $\CM_\FT^C$, of the SCFT $\FT$ itself.  Any other slicing of the parameter space with $\mu_j =a_j \neq 0$, as depicted on the right-hand-side, corresponds to the Coulomb branch of a theory obtained from the 5d SCFT by a mass deformation.}\label{puah}
\end{center}
\end{figure}

In this paper, we are interested in studying the Coulomb phase of 5d SCFTs. From the geometric engineering \eqref{Mtheory X3 SCFT intro}, the picture that emerges is as follows. The space of all possible Coulomb-branch vacuum expectation values (VEVs) and massive deformations of the field theory $\FT$ is identified with the extended K\"ahler cone of the singularity \cite{Witten:1996qb}, which we refer to as the extended (Coulomb) parameter space of the 5d SCFT---see figure \ref{puah} for a schematic view.~\footnote{ The 5d SCFT has a larger parameter space which includes the VEVs of Higgs branch operators. The latter are identified via the correspondence \eqref{Mtheory X3 SCFT intro} with the space of complex structure deformations of the singularity $\MG$ \cite{Morrison:1996xf, Intriligator:1997pq, Aharony:1997bh}. The 5d SCFT Higgs branch has been studied recently {\it e.g.} in \cite{Cremonesi:2015lsa, Ferlito:2017xdq, Cabrera:2018jxt}. In this work, we focus on the Coulomb branch.} This space is obtained as the union of the K\"ahler cones of all possible crepant resolutions of the singularity $\MG$. The latter are smooth local CY threefolds, that we denote $\widehat \MG_\ell$. Each such resolution gives a chamber on the extended Coulomb parameter space, corresponding to a given phase that we denote by $\mathscr C_\ell$. In Figure~\ref{puah}, we depict a case with four possible phases $\ell = 1,2,3,4$. The VEVs of Coulomb branch operators in the 5d SCFT correspond to the extended K\"ahler moduli dual to compact divisors, while the BPS massive deformations of the 5d SCFT correspond to the extended K\"ahler parameters that are dual to non-compact divisors.~\footnote{In this sentence, we distinguished between the ``K\"ahler moduli'' $\nu$, which are dynamical fields in the low-energy M-theory setup, and the ``K\"ahler parameters'' $\mu$, which are non-dynamical. In the following, by abuse of notation, we will use the two terms interchangeably.}  We denote the former by $\nu_a$, where $a=1,...,r$ and $r$ is the rank of the corresponding SCFT, and the latter by $\mu_j$ where $j=1,...,f$ and $f$ is the rank of the global symmetry group of the SCFT. These real variables $(\nu, \mu)\in \R^{r+f}$ can be thought of as coordinates for an ambient space in which the extended K\"ahler cone is embedded, with the $\nu_a$'s being the Coulomb branch scalars. At any fixed value of the mass parameters $\mu_j$, the Coulomb branch metric (and the rest of the low-energy effective action) is captured by a prepotential that is computed geometrically as a function of triple-intersection numbers of a crepant resolution of $\MG$, corresponding to the values of the extended K\"ahler moduli $\nu_a$ and $\mu_j$:
\be
\mathcal F = \mathcal F(\nu_a, \mu_j) = \vol(\widehat \MG_\ell) \qquad\text{for}\qquad (\nu_a~,\, \mu_j) \in \mathscr C_\ell~,
\ee 
for some properly regularized K\"ahler volume of the non-compact threefold.
This prepotential is non-smooth at codimension-one walls in the interior of the extended K\"ahler cone, corresponding to flop transitions among birationally-equivalent resolutions $\widehat \MG_{\ell}$ and $\widehat \MG_{\ell'}$---these are codimension-one walls along which $\mathscr C_{\ell}$ and $\mathscr C_{\ell'}$ intersect.~\footnote{For instance, in Figure~\protect\ref{puah}, the chambers $\mathscr C_{3}$ and $\mathscr C_4$ are connected by a flop transition, while $\mathscr C_{1}$ and $\mathscr C_4$ are not.}  The external boundaries of the extended K\"ahler cone are characterized by divisors shrinking either to a point or a curve. The former case corresponds to tensionless string, the latter to a gauge enhancement. 
The Coulomb branch of the SCFT is canonically identified as the codimension-$f$ hypersurface:
\be
\mathcal M_\FT ^C \equiv \{\mu_j = 0\} \cap \widehat {\mathcal K} (\MG)~.
\ee
This identification entails that we can compute the SCFT Coulomb branch prepotential as:
\be
\mathcal F_{\rm SCFT} = \mathcal F_{\rm SCFT}(\nu) \equiv  \mathcal F(\nu, \mu) \Big|_{\mu_j = 0,\, \forall j}~.
\ee
This space has been referred to, in the recent literature, as the ``physical Coulomb branch'' \cite{Jefferson:2017ahm}. We stress here that, since the gauge coupling is a specific deformation of the SCFT, this cannot be a gauge theory phase; but, as we shall see below, it is compatible with one such phase whenever the corresponding chamber admits a gauge theory interpretation and survives the $\mu_j \to 0$ limit. The structure of phases along the SCFT Coulomb branch is given by the interesections of $\mathcal M_\FT ^C$ with the chambers $\mathscr{C}_\ell$. For instance in figure \ref{puah} the SCFT Coulomb branch is not compatible with the chamber $\mathscr{C}_4$; yet, it is possible to reach a theory in phase $\mathscr{C}_4$ by deforming the SCFT (see figure \ref{puah2}).

In this sense, the geometry of the resolutions of $\MG$ gives the analogue of the Seiberg-Witten solution of 4d $\CN=2$ theories for this class of 5d $\CN=1$ SCFTs. Exploiting geometry, one can compute the prepotential for any value of the parameters and understand the structure of the fibration of the Coulomb moduli over the space of deformation parameters of the SCFT.

\subsection{Type IIA perspective and 5d $\CN=1$ supersymmetric gauge theories}

In the following, we revisit the problem of understanding which deformations of $\FT$ have a low-energy description as gauge theories. Our main tool is the M-theory/IIA duality. In this section, as in most of the paper, we assume that $\MG$ is toric. As we have reviewed above, any relevant deformation of $\FT$ corresponds to performing a crepant resolution of the singularity:
\be
\pi_\ell \; : \; \widehat \MG_\ell \rightarrow \MG~,
\ee
with $\widehat \MG_\ell$ a smooth (or, at least, less singular) local CY$_3$. We then choose an abelian sugroup $U(1)_M$ which is part of the $T^3$ toric action on the threefold $\widehat \MG_\ell$, and we consider the ``reduction'' to type IIA string theory along that $U(1)_M$, viewed as the ``M-theory circle'' \cite{Aganagic:2009zk, Benini:2009qs, Jafferis:2009th, Benini:2011cma}. On general ground, we have a duality:
\be
\text{M-theory on}\:\; \R^{1,4} \times \widehat \MG_\ell \qquad \leftrightarrow \qquad \text{Type IIA string theory on} \; \;  \R^{1,4} \times \CM_5~,
\ee
where the transverse five-manifold is the quotient of the local CY$_3$ by $U(1)_M$:
\be 
\CM_5 \cong  \widehat \MG_\ell /U(1)_M~.
\ee
The resulting five-dimensional type IIA background is a fibration of the ALE space corresponding to the resolution of the singularity $\mathbb C^2 / \mathbb Z_K$ over a real line. Under certain conditions, which we will spell out explicitly, the type-IIA description is under perturbative control and contains D6-branes.~\footnote{Here, we mean ``perturbative in the open-string sector.'' We do not keep track of the details of the metric of $\CM_5$ beyond the relevant K\"ahler moduli inherited from $\wh\MG$; in particular, the curvatures need not be small.} In particular, there can be $N$ D6-branes wrapped over a $\P^1$ inside the ALE space, which leads to a non-abelian 5d $\CN=1$ supersymmetric $SU(N)$ gauge theory with an inverse gauge coupling ${1\ov g^2}= {\rm vol}(\P^1)$. More generally, as we will see, deformations of toric singularities can only give rise to linear quivers with special unitary gauge groups.~\footnote{To obtain more general gauge groups and/or representations, we should leave the toric realm. We leave that generalization for future work.} Whenever the threefold has chambers that admit such a perturbative map to IIA, we identify a gauge theory chamber. If the gauge theory chamber is compatible with the SCFT Coulomb phase, meaning that it survives in the limit $\mu_j\to 0$ on the K\"ahler parameters, then we can consider the gauge-theory prepotential as valid even in the strong-coupling limit $\mu_j\to 0$, because by construction it is going to match with the geometry prepotential in the chamber considered. This gives a surprising result: it is possible to trust a gauge theory computation in the limit of infinite coupling, precisely when the effective field theory approximation breaks down.   We stress that this is not going to be valid for all gauge theory phases, but only for those that are compatible with the SCFT Coulomb branch. For instance, in Figure~\ref{puah}, if chamber $\mathscr C_4$ is a gauge theory phase, it is not going to be connected to the origin of the moduli space.

\begin{figure}
\begin{center}
\subfigure[\small Triangulated toric diagram.]{
\includegraphics[height=4cm]{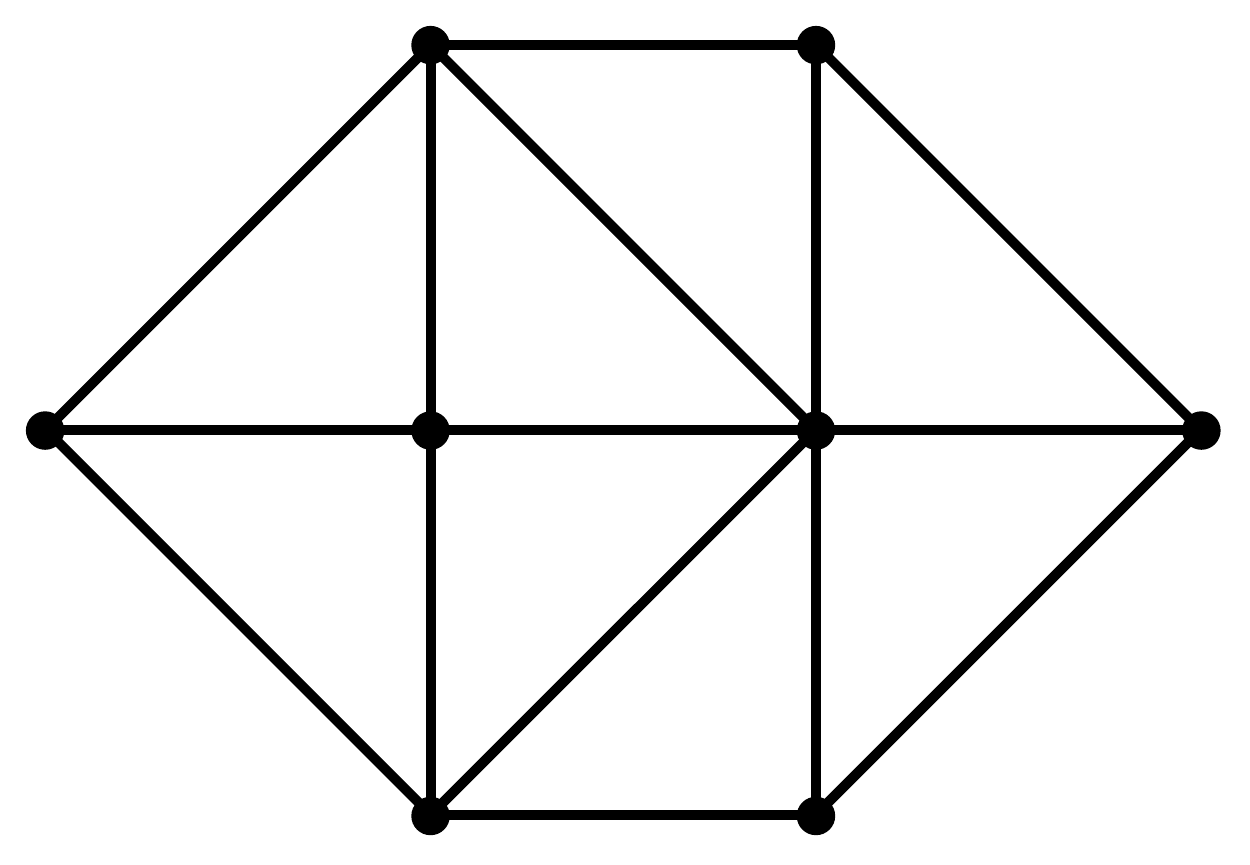}\label{res beetle intro}}\qquad\qquad
\subfigure[\small  Dual $(p,q)$-web.]{
\includegraphics[height=4cm]{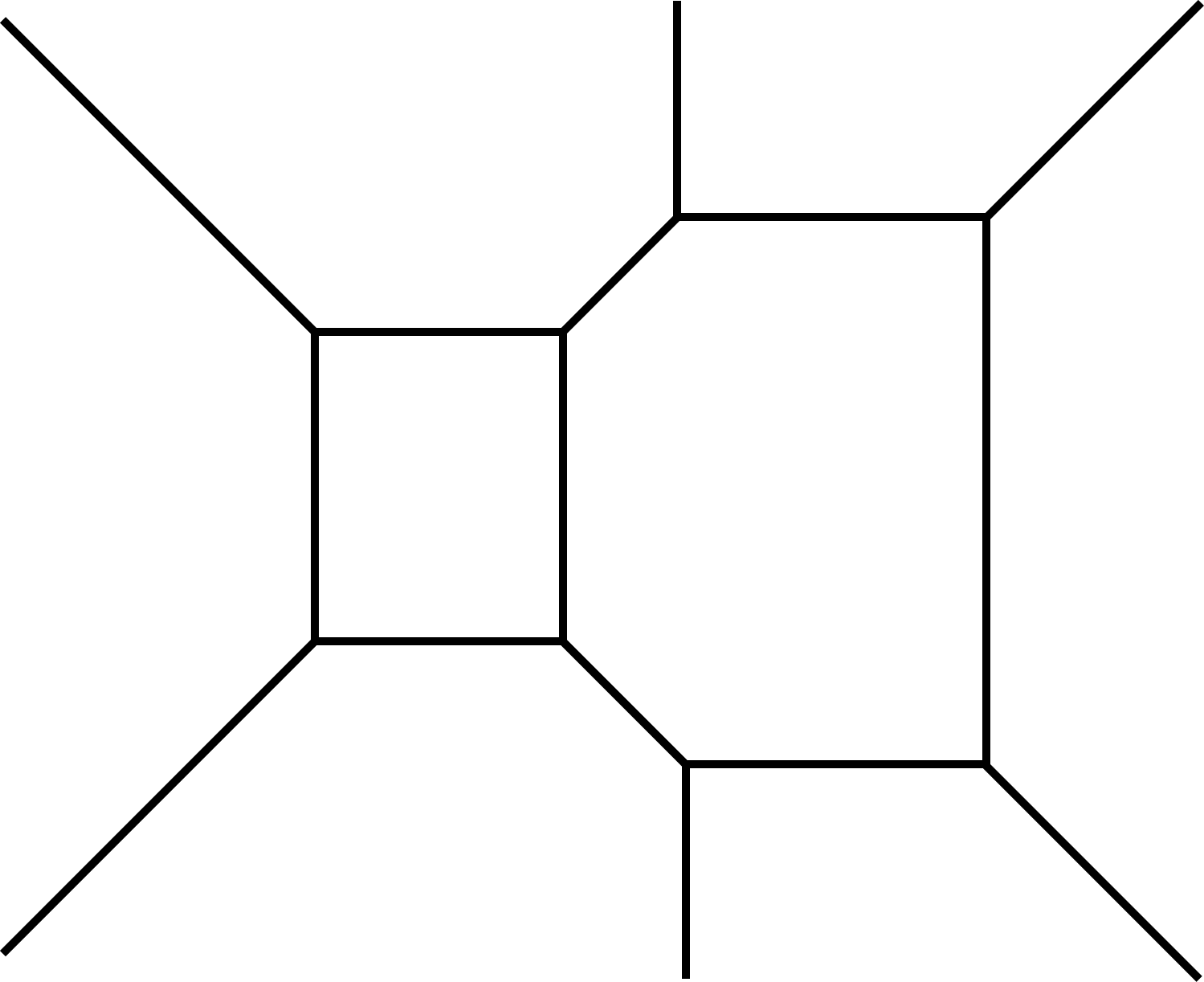}\label{beetle pq intro}}
\caption{Triangulated toric diagram and dual $(p,q)$ web for a particular resolution of the ``beetle'' singularity. This resolution admits both an $SU(3)$ $N_f=2$ and an $SU(2) \times SU(2)$ quiver gauge-theory description. \label{fig: beetle intro}}
\end{center}
\end{figure}

For {\it isolated} toric singularities, we find a perfect matching between the allowed ({\it i.e.} ``perturbative'') IIA reductions and the Coulomb-branch chambers of the gauge theories in question. Our main example will be what we call the ``beetle SCFT,'' a famous rank two SCFT which has gauge theory phases corresponding to $SU(3)$ $N_f=2$ and $SU(2)\times SU(2)$ with a bifundamental hyper. The name ``beetle'' is suggested by the toric diagram and dual $(p,q)$-web of the CY threefold, as shown in Figure~\ref{fig: beetle intro}. 
 The beetle geometry has 24 resolutions, out of which there are 16 resolutions that admit a IIA description with an $SU(3)$ $N_f=2$ gauge-theory interpretation, and 6 resolutions that admit a IIA descriptions with an $SU(2) \times SU(2)$ quiver interpretation. (Moreover, we identify 6 chambers that do not admit any gauge theory interpretation at all.) These numbers match exactly with the number of inequivalent Coulomb chambers one finds in the respective gauge theories. Only 4 of these gauge-theory chambers overlap, meaning that they corresponds to resolutions that have both an $SU(2)\times SU(2)$ interpretation and an $SU(3)$ interpretation. Perhaps not surprisingly, these resolutions are precisely the ones that survive the map to the SCFT Coulomb branch, while the other chambers are not compatible with the $\mu_j\to 0$ limit. 

For {\it non-isolated toric singularities}, this beautiful feature is lost. The number of perturbative type-IIA reductions does not match the number of Coulomb-branch phases of the corresponding gauge theories.
We interpret this as an indication that non-isolated toric singularities do not fully characterize a 5d SCFT. 
In the literature, however, such singularities are often used to described 5d fixed points, since they are particularly convenient to deal with---see {\it e.g.} \cite{Bao:2013pwa, Bergman:2013aca, Hayashi:2014hfa}.  An important class of such examples is provided by the 5d $T_N$ SCFTs, which can seemingly be described as an orbifold toric singularity $\C^3/(\Z_N \times \Z_N)$ in M-theory \cite{Benini:2009gi}.
We conjecture that, in any such case, there exists some non-toric {\it isolated} CY singularity that describe the 5d fixed point and whose extended K\"ahler cone encodes the fully extended parameter space of the SCFT, unlike the toric model. A well-known example is given by the $T_3$ theory, which is the 5d fixed point with  $E_6$ exceptional symmetry first found by Seiberg \cite{Seiberg:1996bd}. In that case, the isolated (non-toric) singularity is the complex cone over the del Pezzo surface $dP_6$, to be compared with the $\mathbb C^3/(\mathbb Z_3 \times Z_3)$ toric singularity.

\subsection{The 5d parity anomaly and the gauge-theory prepotential}
An interesting feature of 5d field theories, as in any odd number of space-time dimension, is the possible presence of parity-odd terms in the effective action.~\footnote{Here, ``parity'' is a slight misnomer, but this terminology is standard. See {\it e.g.} \cite{Seiberg:2016rsg, Witten:2016cio} for a detailed discussion in the 3d case, which is completely similar. Some important work related to this aspect of 5d $\CN=1$ gauge theories was recently carried out in \protect\cite{Chang:2017cdx}.}  In particular, the five-dimensional Chern-Simons term for a 5d (background) gauge field breaks parity explicitly. Moreover, massless Dirac fermions coupled to gauge fields, such as the ones that appear on the walls of Coulomb branch chambers, suffer from the so-called {\it parity anomaly}, which is really a mixed gauge-parity anomaly \cite{AlvarezGaume:1984nf}. In this work, we will use an (implicit) regularization that preserve gauge invariance for both dynamical and background gauge fields (coupling to conserved currents for gauged and global symmetries, respectively). This leads us to a slightly different form of the one-loop-exact prepotential of 5d $\CN=1$ supersymmetric gauge theories, compared to the well-known Intriligator-Morrison-Seiberg result \cite{Intriligator:1997pq} generally quoted in the literature. This is explained in detail in section~\ref{sec:5dgaga} and appendix~\ref{app prepot}, which can be read independently from the rest of this work.

%%%%%%%%%%%%%%
\medskip
\medskip
\noindent
This paper is organized as follows. In section \ref{sec:5dgaga}, we revisit the basic properties of five dimensional gauge theories; in particular we give a new formula for the 5d prepotential from gauge theory. In section \ref{sec: IIA perspective}, we review the M-theory engineering of 5d SCFTs, and we formulate our IIA perspective on gauge theory phases. In sections~\ref{sec:SU2 expls} and~\ref{sec: SUNk},  we revisit some known examples from our perspective. In section~\ref{sec:SDUL}, we comment on the so-called ``UV dualities'' and we  give a detailed study of the phases of the beetle SCFT.  Section \ref{sec:nonTO} is devoted to the analysis of non-isolated toric singularities. Various further computational details are collected in appendices.

%%%%%%%%%%%%%%%%%%%%%%%%%%%%%%%%%%%%%%%
\section{5d $\CN=1$ gauge theories, parity anomaly and prepotential}\label{sec:5dgaga}
In this section, we review and revisit some basic properties of five-dimensional $\CN=1$ supersymmetric gauge theories. We discuss some aspects of the five-dimensional parity anomaly, we revisit the derivation of the 5d prepotential, and we conclude with some remarks about the five-dimensional gauge theory Coulomb phases.

\subsection{Five-dimensional gauge theory basics}
Consider a five-dimensional $\CN=1$ supersymmetric gauge theory with gauge group $\GG$ and Lie algebra $\Fg= {\rm Lie}(\GG)$. Unless otherwise stated, we take $\GG$ compact and connected. We then have:
\be
\GG= \prod_s \GG_s~, 
\ee
where each factor $\GG_s$ is a simple Lie group.
The vector multiplet $\CV$ contains a 5d gauge field $A_\mu$, a real scalar $\varphi$, and the gaugini $\lambda, \t \lambda$, all in the adjoint representation of $\Fg$.
It is coupled to matter fields in hypermultiplets $\CH$, in some (generally reducible) representation $\FR$ of the gauge group. 

These gauge theories have a non-trivial global symmetry: 
\be
\GG_F \times SU(2)_R~,
\ee
 with $SU(2)_R$ the $R$-symmetry. The ``flavor'' symmetry group $\GG_F$ contains a factor $\GG_\CH$ which acts only on the hypermultiplets. It also contains an abelian factor $U(1)_{T_s}$ for each topological symmetry:
 \be
\GG_F= \GG_\CH \times \prod_s U(1)_{T_s}~.
\ee
Recall that, for each simple Lie group $\GG_s$ with connection $A_\mu^{(s)}$, we have a topological symmetry whose conserved current reads:
\be
j^\mu_{T_s} ={1\ov 32 \pi^2} \epsilon^{\mu \nu\rho\sigma\kappa} \,\Tr\big(F^{(s)}_{\nu\rho} F^{(s)}_{\sigma\kappa}\big)~,
\ee
with $F^{(s)}= dA^{(s)}- i A^{(s)}\wedge A^{(s)}$. This is automatically conserved by virtue of the Bianchi identity. The particles charged under $U(1)_{T_s}$ are the 5d uplift of the 4d Yang-Mills $\GG_s$ instantons, therefore the topological symmetry is also called the instanton symmetry.

To keep track of the flavor symmetry, we introduce background vector multiplets $\CV_F$, including an abelian vector multiplet $\CV_{T_s}$ for each topological current. In flat space, a non-trivial supersymmetric background for $\CV_F$ is obtained by turning on constant VEVs for the real scalar $\varphi_F$. We denote these five-dimensional ``real masses'' by $\mu$:
\be\label{real masses mu}
\langle \varphi_F \rangle \equiv \mu =\big(m_\alpha~,\, h_{0,s}\big)~.
\ee
Here,  the ``flavor masses'' for the hypermultiplet flavor group are simply denoted by $m=(m_\alpha)$, where $\alpha= 1, \cdots, {\rm rank}(\GG_\CH)$ runs over a maximal torus of $\GG_\CH$. The masses for the topological symmetries are denoted by $h_0= (h_{0,s})$. The $U(1)_{T_s}$ mass term is also the Yang-Mills Lagrangian for the $\GG_s$ vector multiplet, with:
\be
h_0 = {8\pi^2 \ov g^2}
\ee 
the 5d inverse gauge coupling \cite{Seiberg:1996bd}.

\subsection{Parity anomaly, Chern-Simons terms and real masses}\label{subsec: parity anomaly}
Five-dimensional parity acts by inverting the sign of a single coordinate on $\R^5$, say $x^5 \rightarrow - x^5$. Consider a single Dirac fermion $\psi$, transforming in the ${\bf 4}$ of ${\rm Spin}(5) \cong Sp(2)$, and coupled to a $U(1)$ gauge field $A_\mu$ with charge $1$. It is well-known that $\psi$ suffers from a parity anomaly---one cannot quantize $\psi$ while preserving both parity and gauge invariance.

One can always choose a gauge-invariant quantization, at the expense of violating parity \cite{AlvarezGaume:1984nf}.  The presence of the parity-violating term in the effective action is probed by a quantity denoted by:
\be\label{kappa mod 1}
\kappa \quad \;({\rm mod}\;\; 1)~.
\ee
The coefficient $\kappa$, sometimes called the ``CS contact term,'' appears as a parity-odd term in the three-point function of the $U(1)$ current \cite{Chang:2017cdx, Closset:2012vp}. The mod $1$ ambiguity in \eqref{kappa mod 1} corresponds to the possibility of adding a Chern-Simons term to the 5d action, with integer-quantized level. We will come back to this point below.

\paragraph{Chern-Simons terms.}  Given a five-dimensional $U(1)$ gauge field $A_\mu$, we can write down the 5d Chern-Simons term:
\be\label{5dcs}
S_{\rm CS} = {i k \ov 24 \pi^2} \int_{\CN_5} A \wedge F \wedge F~,
\ee
on some Euclidean five-manifold $\CN_5$.
This is well-defined only if the CS level $k$ is integer quantized, $k\in \Z$.~\footnote{In this normalization, $k=1$ is the minimal CS term when $\CN_5$ is a spin five-manifold with $p_1=0$, as required by the M-theory engineering of these gauge theories \protect\cite{Witten:1996qb}. On the other hand, $k=6$ gives the minimal $U(1)$ CS term on an arbitrary $\CN_5$.} 
The Chern-Simons term \eqref{5dcs} breaks parity explicitly.  It can be completed to the $\CN=1$ supersymmetric action, which takes the schematic form \cite{Seiberg:1996bd, Kim:2012gu}:
\be\label{CS Neq1}
{k \ov 24 \pi^2} \int_{\CN_5} \left( i A \wedge F \wedge F + 6 i \b\lambda\big(D+ {i\ov 2} \gamma^{\mu\nu} F_{\mu\nu}\big)\lambda- {1\ov 6}\varphi  F_{\mu\nu} F^{\mu\nu}+ 12 \varphi D^2+ \cdots \right)~,
\ee
where $D$ denotes an $SU(2)_R$ triplet of auxiliary fields, and we omitted terms involving derivatives of $\lambda$ and $\varphi$. Note that the real field $\varphi$ is mapped to $-\varphi$ under parity.

More generally, we may consider the mixed $U(1)_a-U(1)_b-U(1)_c$ CS terms:
\be
 {i k^{abc} \ov 24 \pi^2} \int_{\CN_5} A_a \wedge F_b \wedge F_c~,
\ee
with quantized levels $k^{abc}\in \Z$. We may also have a non-abelian Chern-Simons term for any simple Lie group with non-trivial cubic index---that is, for $\Fg_s= {\rm Lie}(\GG_s)= {\frak su(N)}$ with $N>2$ \cite{Intriligator:1997pq}. All these CS terms have an $\CN=1$ completion similar to \eqref{CS Neq1}.

\paragraph{Chern-Simons contact term and parity anomaly.}
Consider again a (background) $U(1)$ gauge field coupled to a free fermion $\psi$, and let $j^\mu$ be the conserved current for the $U(1)$ symmetry that acts as $\psi \rightarrow e^{i \alpha}\psi$. 
The three-point function of $j^\mu$ contains the parity-odd contact term \cite{Chang:2017cdx}:
\be\label{kappa cc}
{i \kappa \ov 24\pi^2}  \epsilon_{\mu\nu\rho\sigma\kappa} p^\sigma q^\kappa\; \subset\; \langle j_\mu(p) j_\nu(q) j_\rho(-p-q) \rangle~.
\ee
The Chern-Simons term \eqref{5dcs} obviously contributes an integer to the coefficient $\kappa$---that is, adding the ``local term''  \eqref{5dcs} to the effective action for a free fermion has the effect of shifting $\kappa$ to $\kappa +k$. This explains the ambiguity \eqref{kappa mod 1}. Since $k$ is integer-quantized, the non-integer part of the ``CS contact term'' $\kappa$ is physical \cite{Closset:2012vp}.

For a single {\it massless} Dirac fermion $\psi$, one finds that $\kappa = -\half$ (mod $1$). We fix the integer-valued ambiguity by choosing the ``$U(1)_{-\half}$ quantization,'' such that:
\be\label{kappa half}
\kappa_\psi=-\half
\ee
for a massless free fermion. Any other quantization scheme is related to this one by a shift of $\kappa$ by an integer. In most of the literature on 5d $\CN=1$ gauge theories, this is called a ``CS level $\kappa=-\half$.'' Here, we would like to emphasize that $\kappa$ can be extracted from a gauge-invariant (non-local) effective action, while the CS level $k$ must be an integer by gauge invariance. Therefore, we should distinguish between the CS contact term $\kappa \in \R$ and the properly-quantized CS level $k\in \Z$.~\footnote{We refer to Appendix~A of \cite{Closset:2018ghr} for a pedagogical discussion of this point, in the (completely analogous) three-dimensional case.} 

Since the hypermultiplet (coupled to a $U(1)$ gauge field) contains a Dirac fermion, we have the same parity anomaly in $\CN=1$ supersymmetric theories. By defaults, we will choose the $U(1)_{-\half}$ quantization for every hypermultiplet.

\paragraph{Massive fermions.} 
Another parity-odd term is the real mass:
\be
\delta \SL_m =i m\, \b\psi \psi~,\qquad\quad m\in \R~,
\ee
for any Dirac fermion $\psi$. By sending $|m|\rightarrow \infty$, one can  integrate out $\psi$. A one-loop computation \cite{Witten:1996qb, Bonetti:2013ela} shows that this shifts the parity-odd contact term \eqref{kappa cc} by:
\be\label{kappa shift}
\delta \kappa=  -\half \sign(m)~.
\ee
Let $\kappa_\psi(m)$ denote the CS contact term for the free fermion $\psi$, as a function of the real mass. In the $U(1)_{-\half}$ quantization, we then have:
\be\label{limit m large}
\lim_{m \rightarrow -\infty}  \kappa_\psi(m) = 0~,\qquad\qquad
\lim_{m \rightarrow +\infty}  \kappa_\psi(m) = -1~,
\ee
by adding \eqref{kappa shift} to \eqref{kappa half}.
This is interpreted as follows: for $m$ large and negative, the effective field theory is simply a CS term at level $k=-1$ for the background gauge field $A_\mu$; for $m$ large and positive, we have a completely empty theory in the infrared. The limit \eqref{limit m large} can also serve as the definition of what is meant by ``$U(1)_{-\half}$ quantization.''

\paragraph{General charge.} The above discussion was for a Dirac fermion of unit charge. More generally, for a massless hypermultiplet of $U(1)$ charge $Q\in \Z$, we have
\be\label{kappa Q3}
\kappa_\psi=- {Q^3\ov 2}~,
\ee
and turning on a real mass leads to a shift 
$\delta \kappa=  - {Q^3\ov 2} \sign(m)$ in the IR.

\paragraph{CS contact terms in general theories.} 
More generally, we might consider a more complicated theory whose spectrum might be gapped. In a gapped phase, all the IR contact terms $\kappa$ must be properly quantized, since they should come entirely from Chern-Simons terms in the effective action  \cite{Closset:2012vp}---this can be understood as a 5d version of the Coleman-Hill theorem \cite{COLEMAN1985184}. At a gapless point, on the other hand, the observable $\kappa$ mod $1$ is generally non-trivial; the simplest example being the free fermion, as in \eqref{kappa Q3}.

\subsection{The Coulomb branch prepotential}
By giving expectation values to the adjoint scalar $\varphi$:
\be
\langle\varphi\rangle = {\rm diag}(\varphi_a) = (\varphi_1~,\, \cdots~,\, \varphi_\rk)~,
\ee
we break the gauge group to a maximal torus $\GH$ times the Weyl group:
\be
\GG \rightarrow \GH  \rtimes W_\GG~, \qquad \qquad \GH \cong \prod_{a=1}^\rk U(1)_a~.
\ee
The $\CN=1$ supersymmetric low-energy effective field theory on the Coulomb branch is fully determined by a prepotential $\CF(\varphi, \mu)$.
Here and below, $\varphi= (\varphi_a)$ denotes the low-energy Coulomb branch scalars, which sit in abelian vector multiplet $\CV_a$, and $\mu= (m, h_0)$ denotes the real masses \eqref{real masses mu}---that is, the flavor masses $m_a$ and the inverse gauge couplings $h_{0,I}$.

The five-dimensional prepotential is a cubic polynomial in the real variables $\varphi$ and $\mu$. It is one-loop exact, with the one-loop contribution coming from integrating out the W-bosons and massive hypermultiplets at a generic point on the Coulomb branch  \cite{Seiberg:1996bd, Nekrasov:1996cz, Intriligator:1997pq}. We have: 
\bea\label{F 5d full}
&\CF(\varphi, \mu)\;=\;  \half h_{0,s} K_s^{ab}  \varphi_a \varphi_b +{k^{abc}\ov 6} \varphi_a \varphi_b \varphi_c   \,+{1\ov 6} \sum_{\alpha\in \Delta} \Theta\big(\alpha(\varphi)\big) \big(\alpha(\varphi)\big)^3 \cr
&\qquad\qquad\;\; \; -{1\ov 6}\sum_\omega \sum_{\rho \in \FR} \Theta\big(\rho(\varphi)+ \omega(m)\big) \big(\rho(\varphi)+ \omega(m)\big)^3~,
\eea
where the sum over repeated indices ($s$ and $a,b,c$) is understood. 
Here, $\Theta(x)$ is the Heaviside step function:
\be
\Theta(x)= \begin{cases} 1 \quad & {\rm if} \;  x \geq 0~,\\ 
0 \quad & {\rm if} \;  x < 0.\end{cases}
\ee 
The first term in \eqref{F 5d full} is the classical contribution from the Yang-Mills terms, with $h_{0,s}= {8\pi^2 \ov g^2_s}$ the inverse gauge couplings and $K^{ab}_s$ the Killing forms of the simple factors $\Fg_s$. The  second term  in \eqref{F 5d full}  gives the classical Chern-Simons action, with CS levels $k^{abc}$ in the abelianized theory. For $\GG= \prod_s \GG_s$ with $\GG_s$ a simple gauge group, we only have CS contributions from the $SU(N)$ factors (with $N>2$), with $k^{abc} =k_s \, d_s^{abc}$ for each $s$, in terms of the cubic Casimir $d^{abc}_s$ of $\Fg_s$. The last term on the first line of \eqref{F 5d full} is the one-loop contribution from the W-bosons, with $\Delta$ the set of non-zero roots $\alpha= (\alpha^a)$ of $\Fg$, and $\alpha(\varphi)= \alpha^a \varphi_a$ the natural pairing. The second line of \eqref{F 5d full} is the one-loop contribution from the hypermultiplets. The sum is over the flavor and gauge weights, $\omega$ and $\rho$, respectively, with $\omega(m)= \omega^\alpha m_\alpha$ and $\rho(\varphi)= \rho^a\varphi_a$.

This 5d prepotential is the correct result for the hypermultiplet in the ``$U(1)_{-\half}$ quantization,'' as explained above. That is, for a single hypermultiplet coupled to a $U(1)$ vector multiplet with scalar $\varphi$, we have the continuous function:
\be\label{CF CH}
\CF_\CH(\varphi) = -{1\ov 6} \Theta(\varphi) \varphi^3~.
\ee
Since a $U(1)$ CS term at level $k$ contributes:
\be\label{U1k}
\CF_{U(1)_k}(\varphi) = {k\ov 6} \varphi^3~,
\ee
to the prepotential, the hypermultiplet contribution \eqref{CF CH} precisely reproduces the decoupling limits \eqref{limit m large}, for either sign of the real mass $\varphi$.

The prepotential \eqref{F 5d full} is derived in Appendix~\ref{app prepot}. It should be compared to the result given by Intriligator, Morrison and Seiberg (IMS) in \cite{Intriligator:1997pq}. As we explain in Appendix, the terms of order $\varphi^3$ in \eqref{F 5d full} are the same as in the IMS prepotential (once we correctly map the CS levels), but our prescription gives a slightly different result for the lower-order terms. This is explained by our different treatment of the parity anomaly.

At a generic point on the Coulomb branch, the theory is gapped and therefore, as discussed above, the Chern-Simons contact terms $\kappa$ should all be integer-quantized. This is true not only for the gauge CS levels, but also for the mixed gauge-flavor and purely flavor CS levels:
\be
\kappa^{abc} = \d_{\varphi_a} \d_{\varphi_b}  \d_{\varphi_c}\CF~, \quad
\kappa^{ab \alpha}=  \d_{\varphi_a} \d_{\varphi_b}  \d_{m_\alpha}\CF~, \quad
\kappa^{a \alpha\beta}=  \d_{\varphi_a} \d_{m_\alpha}  \d_{m_\beta} \CF \quad \in \;\;\Z~, 
\ee
and so on (including by taking derivatives with respect to the gauge couplings $h_0$). The effective CS levels that follow from \eqref{F 5d full} are indeed integer quantized, by construction, which is not always the case if we use the IMS prepotential.~\footnote{Using the IMS prepotential together with the usual treatment of the ``parity anomaly,'' $\kappa^{abc}$ is always integer-quantized and agrees with our result, but some of the mixed flavor-gauge effective CS levels can be half-integer. In the usual (and somewhat misleading) language, our prescription for the prepotential can be understood as a correction to the IMS prepotential that adds some explicit ``half-integer CS levels'' on the Coulomb branch to ``cancel the parity anomalies'' for the flavor group.}

\subsection{BPS particles and strings}
Five dimensional gauge theories on their Coulomb branch contain half-BPS particle states and half-BPS string states.

\paragraph{BPS particles.} The VEV of $\varphi$ and the real masses $\mu$ determine the mass $M$ of BPS particle states, according to:
\be
M = |Q^a \varphi_a + Q_F^\alpha m_\alpha + Q_F^s h_{0,s}|~,
\ee
where $Q^a$, $Q_F^\alpha$ and $Q_F^s$ denote the integer-quantized gauge charges, the $\GG_\CH$ flavor charges, and the $U(1)_{T_s}$ instanton charges, respectively. The RHS of the formula above can be interpreted as the absolute value of the real central charge that enters in the 5d $\CN=1$ Poincar\'e superalgebra. The elementary states that carry gauge charge are the W-bosons $W_\alpha$, associated to the roots $\alpha\in \Fg$, with:
\be
M(W_\alpha) = \alpha(\varphi)~.
\ee
The hypermultiplet states, with gauge charges $Q^a=\rho^a$ and flavor charges $Q_F^\alpha= \omega^\alpha$, have masses:
\be
M(\CH_{\rho, \omega}) = \rho(\varphi) + \omega(m)~.
\ee
The prepotential \eqref{F 5d full} is non-smooth across loci where any particle become massless. We will discuss these Coulomb-branch ``walls'' further in subsection~\ref{subsec: CB walls} below.

There are also BPS particles charged under the topological symmetries (that is, with $Q^s_F \neq 0$). These ``instantonic particles'' are solitonic particles in the gauge-theory language, corresponding to $\GG$-instantons in the four-dimensional sense. They are more subtle to understand from the low-energy point of view, but play a crucial role in the UV completion of the non-abelian gauge theories at strong coupling---see {\it e.g.} \cite{Lambert:2014jna, Rodriguez-Gomez:2015xwa, Tachikawa:2015mha, Zafrir:2015uaa, Yonekura:2015ksa, Cremonesi:2015lsa}. We will discuss them further in subsection~\ref{subsec: instanton} below.

\paragraph{BPS strings.} The 5d $\CN=1$ gauge theory also contains BPS strings, corresponding to the five-dimensional uplift of the four-dimensional $\CN=2$ monopoles. They are labelled by GNO-quantized magnetic fluxes through any $S^2$ surrounding them in $\R^5$. In particular, for every $U(1)_a$ in the maximal torus of $\GG$, there is a string with $U(1)_a$ magnetic flux $\m_a=1$, whose tension on the Coulomb branch is given by the first derivative of the prepotential with respect to $\varphi_a$ \cite{Seiberg:1996bd}:
\be
T_a(\varphi, \mu) = {\d\CF\ov \d{\varphi_a} }~.
\ee
In summary, while the second and third derivatives of $\CF$ with respect to $\varphi$ determine the Coulomb-branch low-energy effective action, the first derivatives determine the string tensions.

\subsection{Instanton particles in $SU(N)$ quivers}\label{subsec: instanton}
All of our examples below will be $SU(N)$ quivers---that is, we will have a gauge group:
\be
\GG= \prod_s SU(N_s)
\ee
coupled to bifundamental and/or fundamental hypermultiplets. In that simple case, we can compute the masses of the instanton particles by the following trick \cite{Assel:2018rcw}. Let us first replace each $SU(N)$ gauge group by $U(N)$. The Coulomb branch scalars of $U(N)$ are denoted by:
\be
\phi= (\phi_a)~, \qquad a=1, \cdots, N~.
\ee
The prepotential for a $U(N)$ theory at CS level $k\in \Z$ reads:~\footnote{In our string-theory construction, we will restrict ourselves to $|k|< N$. For $N=2$, the $U(2)$ CS level $k=0$ or $\pm1$ will correspond to the $\Z_2$-valued $SU(2)$ $\theta$-angle $\theta=0$ or $\pm\pi$ \protect\cite{Intriligator:1997pq}.}
\be\label{F UN}
\CF^{U(N)} = {1\ov 2} h_0 \sum_{a=1}^N \phi_a^2 + {k\ov 6} \sum_{a=1}^N \phi_a^3 + {1\ov 6} \sum_{\substack{a,b =1\\a <b}}^N(\phi_a-\phi_b)^3 + \cdots~.
\ee
Here, we choose a Weyl chamber of $U(N)$ such that $\phi_1 > \phi_2 > \cdots >\phi_N$, and the ellipsis is the one-loop contribution from hypermultiplets. The generalization to any $U(N)$ quiver is straightforward. Obviously, we recover the $SU(N)$ result by imposing the traceless condition, $\sum_a \phi_a=0$. We always take:
\be\label{U to SU}
\phi_a = \varphi_a - \varphi_{a-1}~, \quad a=1, \cdots, N~, \qquad \text{with} \qquad \varphi_0=\varphi_N=0~.
\ee
In the M-theory construction, we will see that, for each $SU(N)$ gauge group, there are $N$ ``elementary'' instanton particles $\CI_a$ ($a=1, \cdots, N$) with instanton charge $1$. Their masses are conveniently given by the second derivatives of \eqref{F UN}, that is:
\be\label{I mass heuristic}
M(\CI_a) = {\d^2 \CF^{U(N)} \ov \d \phi_a^2} \Bigg|_{\phi= \phi(\varphi)} = h_0 + \cdots~.
\ee
Thus, we first compute the ``effective $U(1)_a$ gauge coupling'' on the Coulomb branch of the $U(N)$-quiver theory, and then impose the $SU(N)$ tracelessness conditions \eqref{U to SU}. This heuristic prescription will be natural from the point of view of our string-theory  construction. It would be interesting to give a fully gauge-theoretic derivation of the  instanton particle masses. 

For each $SU(N_s)$ gauge group in a given quiver theory, the instanton particle $\CI_s =\CI_{s, a}$ with the lowest mass can be viewed as ``the'' elementary instanton, while any other instanton particle $\CI_{s, b}$, $b\neq a$, can be interpreted as a marginal bound state of $\CI_s$ with other (W-boson and hypermultiplet) BPS particles \cite{Aharony:1997bh}.

\subsection{Physical chambers on the Coulomb branch}\label{subsec: CB walls}
Consider the Coulomb branch $\CM^C(\mu)$ of the gauge theory at some fixed value of the masses $\mu=(h_0, m)$. There can be various real codimension-one ``walls'' on the Coulomb branch, at which locations particles become massless:
\be
M(\varphi; \mu) \rightarrow 0~.
\ee
 These walls divide the moduli space into different chambers, characterized by distinct BPS particle spectra. Note that the central charge is real in 5d, hence there can be no marginal-stability walls---only threshold bound-states are allowed among BPS particles. Since the particle masses are linear in $\varphi$, the walls are hyperplanes in $\R^r \cong \{ \varphi_a\}$.

From the gauge-theory perspective, we have two kinds of ``perturbative'' walls, depending on the state that becomes massless:
\begin{itemize}
\item[(i)] {\bf An hypermultiplet mode $\CH$ becomes massless.} Such a wall is ``traversable''---the theory on the other side simply contains an hypermultiplet with the opposite sign of the real mass $M(\CH)$, with central charge $Z= |M(\CH)|$. The prepotential is non-smooth across such a wall, as is clear from \eqref{F 5d full}.
\item[(ii)] {\bf A W-boson becomes massless.} Such a ``hard wall'' is at the boundary of the Weyl chamber and it is not traversable.~\footnote{Indeed, ``going through'' a Weyl chamber one obtains a gauge-equivalent description, therefore we cannot cross such a wall anymore than we can cross through a mirror. We should fix the gauge once and for all and consider a single fundamental Weyl chamber.}
\end{itemize}
Thus, the ``naive'' field-theory Coulomb branch consists of a fundamental Weyl chamber of the gauge group $\GG$, which is further subdivided into ``field theory chambers'' separated by walls where hypermultiplets go massless.

On the other hand, the low-energy effective field theory on the Coulomb branch is clearly valid only if {\it all} BPS states, whether perturbative or not, are safely massive. Thus, we should also worry about two other kinds of walls, where:
\begin{itemize}
\item[(iii)] {\bf A BPS instanton particle becomes massless.} Such a wall is harder to describe in purely field-theory terms. It might be traversable or not, depending on the theory. We will see examples of this in our study of classic examples. 
\item[(iv)] {\bf A magnetic string become tensionless, $T(\varphi; \mu) \rightarrow 0$.} Note that $T(\varphi; \mu)=0$ is a quadratic equation in $\varphi$, therefore such ``magnetic walls'' are rather more peculiar.
\end{itemize}
The existence of these non-perturbative walls was pointed out in \cite{Jefferson:2017ahm}, where the notion of a {\it physical Coulomb branch} was introduced. The physical Coulomb branch of a 5d $\CN=1$ gauge theory is the subspace of the ``naive'' Coulomb branch within the bounds of both hard instanton walls and magnetic walls.~\footnote{The main focus of  \cite{Jefferson:2017ahm} was on the strong-coupling limit $\mu \rightarrow 0$, while in this work we are interested in the full parameter space. }

%%%%%%%%%%%%%%%%%%%%%%%%%%%%%%%%%%%%%%%%%%
%%%%%%%%%%%%%%%%%%%%%%%%%%%%%%%%%%%%%%%%%%

\section{Type IIA perspective and gauge-theory phases}\label{sec: IIA perspective}
In this section, we review the M-theory approach to 5d SCFTs, we give a pedagogical introduction to some useful toric geometry tools, and we discuss in detail a new and complementary type-IIA perspective.

\subsection{5d SCFTs from M-theory on a CY$_3$ singularity}\label{subsec: Mtheory geom}
The starting point of our analysis is the geometric engineering of the 5d field theory from M-theory on a local Calabi-Yau three-fold (CY$_3$) $\MG$, an isolated canonical singularity. It is believed that $\MG$ defines an SCFT $\FT$ on the transverse space:
\be\label{eq:MtheoryBG}
\text{M-theory on}\:\; \R^{1,4} \times \MG \qquad \leftrightarrow \qquad  \FT \quad \text{SCFT on } \; \R^{1,4 + \varepsilon}.
\ee
If the singularity $\MG$ is elliptic, then $\varepsilon = 1$ and $\FT$ is a 6d SCFT;~\footnote{Shrinking the elliptic fiber to zero size uplifts M-theory to F-theory and gives rise to a 6d compactification \protect\cite{Vafa:1996xn} --- see e.g. \cite{Heckman:2013pva,DelZotto:2014hpa,Heckman:2015bfa,DelZotto:2015isa,DelZotto:2015rca,Morrison:2016nrt,Mekareeya:2017sqh} for work in the context of characterizing the singular geometries for 6d SCFTs.}
 otherwise, $\varepsilon=0$ and $\FT$ is a 5d SCFT. In this paper, we focus on that latter case.

A generic CY$_3$ singularity $\MG$ can be smoothed out by a crepant resolution:~\footnote{One can also consider complex deformations of the singularity, which characterize the Higgs branch of the SCFT \cite{Morrison:1996xf, Intriligator:1997pq}. In this paper, we focus on crepant resolutions and the corresponding Coulomb-branch physics.} 
\be
\pi : \widehat \MG \to \MG~, \qquad\qquad \pi^*K_{\MG} = K_{\widehat \MG}~,
\ee 
giving us a smooth local CY threefold $\widehat \MG$. The exceptional set $\pi^{-1}(0)$ (with $0\in \MG$ the isolated singularity) contains $n_4 \equiv r \geq 0$ compact divisors. The non-negative integer $r$ is called the ``rank'' of $\MG$. It corresponds to the real dimension of the SCFT Coulomb branch:
\be
r= \dim \CM_{\FT}^C~.
\ee
 The resolved space $\wh\MG$ also contains compact curves, denoted by $\CC$, which may intersect the exceptional divisors non-trivially.

The dictionary between such a smooth geometry and a 5d $\CN=1$ field theory can be established, in principle, as a decoupling limit, starting from a M-theory compactification on compact CY threefold $Y$ (see {\it e.g.}  \cite{Cadavid:1995bk,Ferrara:1996hh,Ferrara:1996wv, Alexandrov:2017mgi}),
and scaling the volume of $Y$ to infinity while the volumes of a collection of holomorphic 2-cycles and of holomorphic 4-cycles that are \textit{intersecting} within $Y$ are kept finite; this has the effect of sending the five-dimensional Planck mass to infinity, thus decoupling gravity. We require the collection of 2- and 4- cycles to be intersecting because we are interested in obtaining an interacting SCFT. This gives rise to the local model $\wh\MG$.

Given the local CY threefold $\wh\MG$, we pick a basis $\CC^{\bf a}$ of compact holomorphic 2-cycles in $H_2(\wh\MG, \Z)$. Let $n_2 = r+f$ denote the dimension of $H_2(\wh\MG, \Z)$, with $r$ the rank and $f\geq 0$ some non-negative integer. The two-cycles $\CC^{\bf a}$ are dual to either compact divisors (in the exceptional set) or non-compact divisors. Let us denote by $D_k$ the divisors, compact or not, and let us choose some basis of $n_2$ divisors $\{D_k\}_{k=1}^{n_2}$ such that:
\be
\CC^{\bf a} \cdot D_k = {{\bf Q}^{\bf a}}_k~, \qquad \det {\bf Q} \neq 0~.
\ee
Let $J$ denote the K\"ahler form and let $S$ denote the (Poincar\'e dual) K\"ahler class, which is a particular linear combination of divisors over the real numbers:
\be\label{S def sec 3}
S= \sum_{k=1}^{n_2}  \lambda^k D_k = \sum_{j=1}^f \mu^j D_j +  \sum_{a=1}^r \nu^a \bE_a~.
\ee 
Here, we split the set $\{D_k\}$ into $r$ compact divisors, denoted by $\bE_a$, and $f$ non-compact divisors, denoted by $D_j$. The K\"ahler volumes of the compact curves in $\wh\MG$ are given by:
\be\label{K vol xi gen}
\xi^{\bf a}(\mu, \nu) \equiv \int_{\CC^{\bf a}} J = \CC^{\bf a} \cdot S= {{\bf Q}^{\bf a}}_k \lambda^k~ = {{\bf Q}^{\bf a}}_j\mu^j + {{\bf Q}^{\bf a}}_a \nu^a > 0~.
\ee
Therefore, the parameters $\mu^k \in \R$ and $\nu^a \in \R$ in \eqref{S def sec 3} are essentially the K\"ahler moduli of two-cycles dual to non-compact and compact four-cycles, respectively.

Let us briefly review the correspondence between this geometric structure and the low-energy physics of the SCFT $\FT$ and of its massive deformations \cite{Morrison:1996xf, Intriligator:1997pq}. For generic values of the K\"ahler parameters, the low-energy $\CN=1$ field theory is an abelian theory with gauge group $U(1)^r \cong H^2(\wh\MG, \R)/H^2(\wh\MG, \Z)$. The $U(1)$ gauge fields arise from the periods of the M-theory three-form over the curves $\CC_a$ dual to compact divisors, in the obvious way. Moreover, the exact prepotential for this abelian theory can be computed from the geometry, as:
\be\label{F from geom}
\CF(\mu, \nu) = -{1\ov 6} \int_{\wh\MG} J \wedge J \wedge J = -{1\ov 6} S\cdot S\cdot S~.
\ee
This prepotential is fully determined by the triple-intersection numbers of $\wh \MG$, up to some regularization that is needed to compute the triple-intersection of three non-compact divisors; this introduces some ambiguity in \eqref{F from geom} which, however, only affects the $\nu$-independent part of $\CF(\mu, \nu)$. Since those ``constant terms'' are non-physical for the 5d field theory in flat space, we can mostly ignore them---see Appendix~\ref{app: intersect} for further discussion.
The K\"ahler parameters $\mu$ and $\nu$ in \eqref{S def sec 3} are the mass parameters and the dynamical fields, respectively. The mass parameters $\mu_i$ correspond to deformations of the SCFT by dimension-four operators:
\be
S_{\FT} \rightarrow S_{\FT}+ \mu^j \int d^5 x \, \CO_j~,
\ee
 The operators $\CO_j$ sit at level two in short multiplets $C_1[0,0]^{(2)}_3$ of the $\mathfrak{f}(4)$ superconformal algebra \cite{Cordova:2016xhm}. 
 
 The dynamical fields $\nu$, on the other hand, are the Coulomb branch parameters. At $\mu=0$, they correspond to real VEVs of SCFT operators, thus spanning the intrinsic Coulomb branch $\CM_\FT^C$ of $\CT_{\MG}$. 
 The Coulomb phase of a 5d SCFT can be defined as the branch of its vacuum moduli space where the $SU(2)_R$ symmetry is unbroken. In particular, we expect to flow to a Coulomb phase by giving VEVs to scalars that are singlets with respect to the $SU(2)_R$ symmetry. However, since the Coulomb branch is a real manifold, one do not expect to have an underlying Coulomb branch chiral ring (as opposed, for instance, to the case of 4d $\CN=2$ SCFTs). From the classification of protected unitary representations of $\mathfrak{f}(4)$ in \cite{Cordova:2016emh}, one can see that few (if any) BPS multiplets have the desired features to be Coulomb branch operators, while many non-BPS multiplets do. Hence, one expects that the Coulomb branch corresponds to VEVs of SCFT operators that sit in long multiplets, which are not protected.~\footnote{We are grateful to Thomas Dumitrescu for illuminating correspondence on this point.}

More generally, $\mu$ and $\nu$ can be both non-zero, and the Coulomb branch $\CM_\FT^C(\mu)$ is fibered over the parameter space $\{\mu\}$, 
\be\label{CB fibration}
\CM_\FT^C(\mu) \rightarrow \mathcal P_{\FT} \rightarrow \{\mu\}~,
\ee
as discussed in the introduction.

 The 5d theory corresponding to $\widehat \MG$ admits BPS excitations consisting of electrically charged BPS particles and (dual)  BPS magnetic strings. In geometry, these are realized by M2-branes wrapping the holomorphic curves $\CC^{\bf a}$, and by M5-branes wrapping holomorphic surfaces $\bE_a$, respectively. 
 The masses of the particles are given by the K\"ahler volumes \eqref{K vol xi gen}. The tensions of the magnetic strings are given by the K\"ahler volumes of the compact divisors, which coincide with the first derivatives of the prepotential:
 \be
 T_a(\mu, \nu) \equiv - \d_{\nu^a} \CF(\mu, \nu) = \half \int_{\bE_a} J \wedge J= \text{vol}(\bE_a)~.
 \ee
Finally, the metric of the Coulomb branch is given by:
\be
\tau_{ab}(\mu, \nu) = \partial_{\nu^a} \partial_{\nu^b} \CF(\mu, \nu) = \text{vol}(\COD_a \cdot \COD_b)~,
\ee
which is the volume of a curve $\CC_{ab}=\COD_a \cdot \COD_b$ at the intersection of two exceptional divisors.

\paragraph{The extended parameter space.}
The above discussion focussed on a particular resolution $\wh \MG$ of the singularity $\MG$. In general, there can be many distinct, birationally equivalent local threefolds $\wh \MG_\ell$, which all have the same singular limit $\MG$.
For a given $\wh\MG$, the K\"ahler cone of the singularity $\MG$ is the set of all positive K\"ahler forms:
\begin{align}
\mathcal K(\widehat \MG \backslash \MG) = \{ J  ~|~ \CC \cdot S > 0 ~~\text{for all holomorphic curves $\CC \subset \widehat \MG$} \}~. 
	\end{align}
The extended K\"ahler cone is the closure of the union of all compatible K\"ahler cones,
\be\label{EKC}
\widehat {\mathcal K}(\MG) = \left\{ \bigcup_\ell \mathcal K(\widehat \MG_\ell \backslash \MG)\right\}^c~.
\ee
The extended K\"ahler cone is a fan, with pairs of K\"ahler cones glued along common faces in the interior of $\widehat {\mathcal K}(\MG)$. The boundaries of $\mathcal K(\wh\MG_\ell \backslash \MG)$ correspond to loci where the 3-fold $\wh\MG_\ell$ develops a singularity. The interior boundaries are regions where a holomorphic curve collapses to zero volume and formally develops negative volume in the adjacent K\"ahler cone, signaling a flop transition. This corresponds to a BPS particle becoming massless, which triggers a jump in the third derivatives of the prepotential \eqref{F from geom}. By contrast, the exterior boundaries of $\widehat {\mathcal K}(\MG)$ are loci where one of the 4-cycles $\bE_a$ can collapse to a 2-cycle or a point. The SCFT point is the origin of $\widehat {\mathcal K}(\MG)$, and corresponds to the singularity $\MG$, which is characterized by the connected union of 4-cycles shrinking to a point. All the Coulomb branch VEVs and the massive deformations of the 5d SCFTs obtained from M-theory are encoded in the extended K{\"a}hler moduli space of $\MG$ \cite{Witten:1996qb}:
\be
\mathcal P_{\FT} = \widehat{\mathcal  K}(\MG)~.
\ee
where $\mathcal P_{\FT}$ is the ``extended parameter space'' of the theory $\FT$, including all the VEVs and mass parameters. In a sense, the collection of all crepant resolutions of $\MG$ provides the 5d analogue of the Seiberg-Witten geometry in 4d $\CN=2$ theory, namely, the smooth M-theory geometry $\wh\MG$ corresponding to one such resolution gives a solution to the 5d theory in its Coulomb phase, as captured by the exact prepotential \eqref{F from geom}.

\paragraph{Gauge-theory phases.} 
At scales much smaller than the scale set by $|\mu|$, we often have useful field-theory descriptions of $\CM_\FT^C(\mu)$ in \eqref{CB fibration} as the Coulomb branch of a 5d $\CN=1$ gauge theory; in that case, some deformation parameters $\mu_s=h_{0,s}$ correspond to super-Yang-Mills terms in the low-energy description.

When that happens, one can check the correspondence between geometry and field theory by matching the geometric prepotential  \eqref{F from geom} to the one-loop-exact gauge-theory prepotential \eqref{F 5d full}, namely:
\be\label{geom to FT check}
\CF(\mu, \nu) = \CF(h_0, m, \varphi)~,
\ee
for some linear map between the K\"ahler parameters and the field-theory parameters.
 Surprisingly, in the literature so far, the matching \eqref{geom to FT check} has mostly been checked (in numerous examples) in the limit $\mu^j=0$ (in which case we simply have $\nu_a= - \varphi_a$, in our conventions). This corresponds to a strong-coupling limit in the gauge theory, where the low-energy approximation breaks down. It is also a somewhat degenerate subspace of the extended K\"ahler cone \eqref{EKC}. Remarkably, one finds a perfect match nonetheless \cite{Morrison:1996xf, Intriligator:1997pq}, which probably signals some kind of non-renormalization theorem at work.
 
One of the aims of this paper is to check \eqref{geom to FT check} with generic parameters turned on. Another objective is to explore when, if at all, we can have a {\it non-abelian} gauge-theory description of the low-energy physics (what we call a ``gauge-theory phase''). The standard approach \cite{Intriligator:1997pq, Jefferson:2018irk} is to look for a ``ruling'' of the exceptional set $\pi^{-1}(0) \subset \wh \MG$---that is, we look for a set of surfaces $\bE$ which can be written as a fibration $\mathbb{P}^1 \rightarrow \bE \rightarrow \CC$ over a curve $\CC$ with $\mathbb{P}^1$ fibers. The M2-branes wrapped over the fibers are identified with the W-bosons, which become massless in a limit to a boundary of the K\"ahler cone where the rule surfaces shrink to zero size. Instead of following this purely geometric approach, we will propose a closely related description of the gauge-theory phases by exploiting the M-theory/type IIA duality.

In what follows, we restrict ourselves to the case of $\MG$ a {\it toric} singularity, which simplifies the analysis significantly. 

\subsection{CY$_3$ singularity, toric diagram and GLSM}
\begin{figure}[t]
\begin{center}
\subfigure[\small $\C^3/\Z_3$.]{
\includegraphics[height=3.8cm]{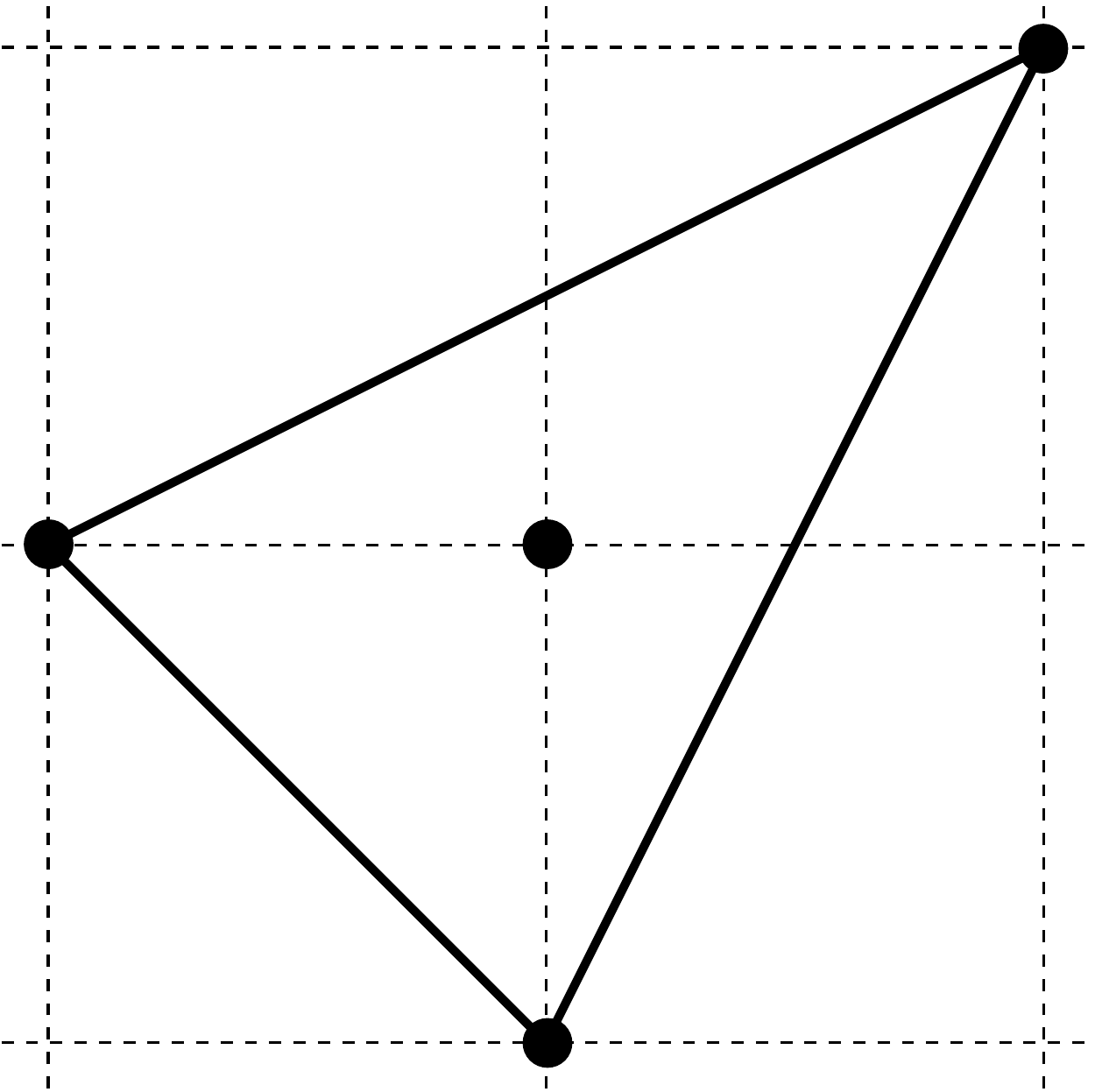}\label{fig:C3Z3 sing}}\qquad\qquad\qquad
\subfigure[\small $\CO(-3) \rightarrow \mathbb{P}^2$.]{
\includegraphics[height=3.8cm]{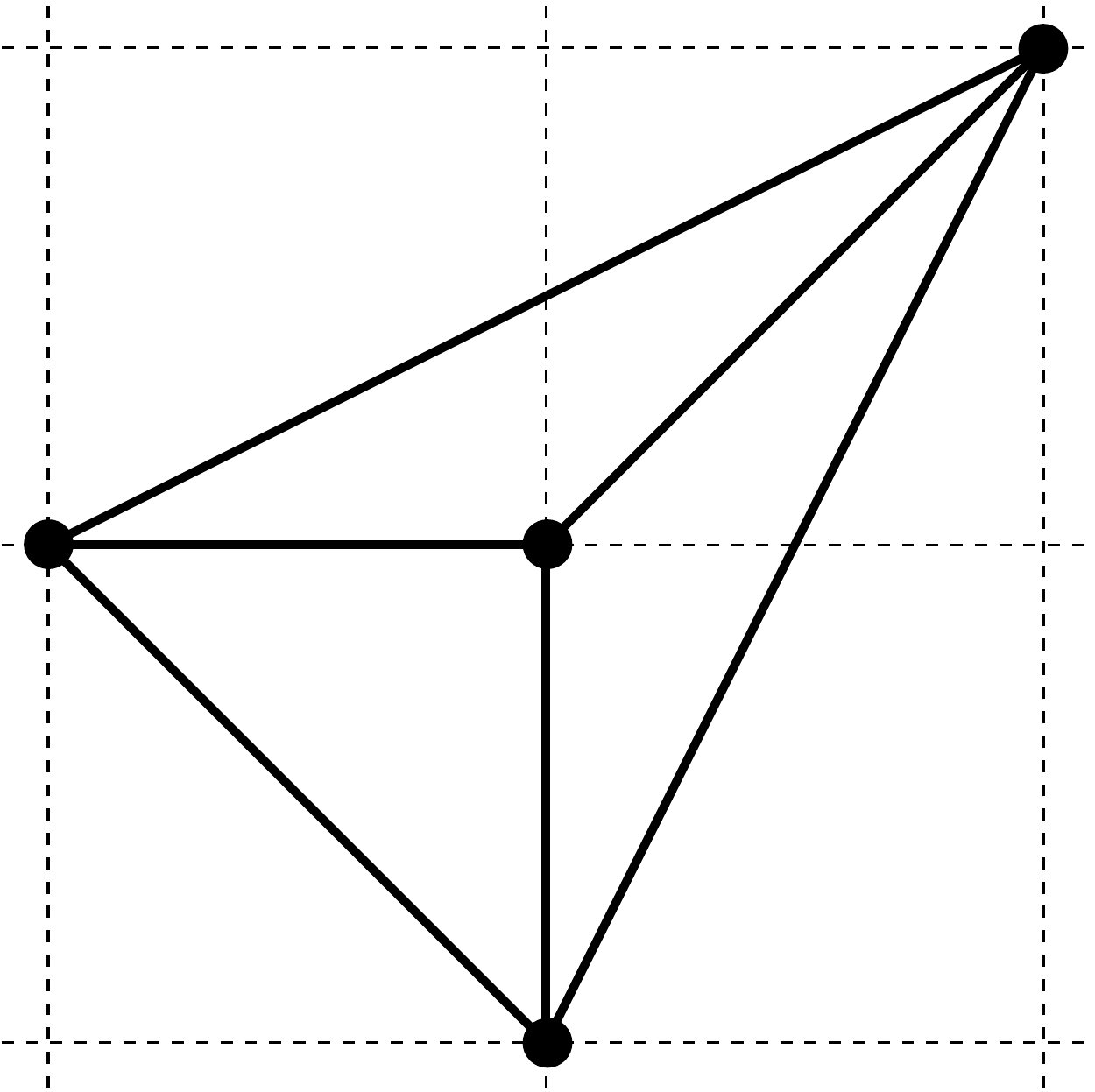}\label{fig:C3Z3res sing}}
\caption{Toric diagram for the $\C^3/\Z_3$ singularity, and its crepant resolution. In this example, there are three external points $w_1=(-1,0)$, $w_2= (0,-1)$, $w_3=(1,1)$ ($n_E=3)$, one internal point $w_0=(0,0)$ ($r=1$), and a unique triangulation of the toric diagram. \label{fig:examples toric dig}}
 \end{center}
 \end{figure}

Let us now review some toric geometry tools that we will use extensively below. The toric CY$_3$ singularity $\MG$ is an affine toric variety. It is determined by its toric  cone $\Delta_0$, a set of $n_E$ vectors $v_i$ in $\Z^3$ which generate a strongly convex rational polyhedral cone. The Calabi-Yau condition implies that all the $v_i$'s are co-planar. In that case, we can perform an $SL(3, \Z)$ transformation to bring the toric vectors onto the plane $v^z=1$, namely:
\be\label{vi ext}
v_i= (w_i, 1)~, \quad i=1, \cdots, n_E~,\quad \qquad\quad w_i= (w^x_i, w^y_i) \in \Z^2~.
\ee
The vectors $w_i$ give us the toric diagram of the singularity, $\Gamma \in \Z^2$. A simple example is shown in Figure~\ref{fig:examples toric dig}. The toric diagram also has a number $r \geq 0$ of internal points.
%%%%%%%%%%

The toric singularity itself can be pictured as a $T^3\cong U(1)^3$ fibration over the dual cone $\Delta_0^\vee$. The toric vectors $v_i$ are outward-pointing orthogonal to the external facets $F_i \subset \Delta_0^\vee$, corresponding to which $U(1)_i \subset U(1)^3$ degenerates at each facet, with $U(1)_i \times U(1)_j$ degenerating at the edge $E_{ij}=  F_i \cap F_j$. Note that each facet corresponds to non-compact four-cycle (the $T^2$ fibration over $F_i$), which is a toric divisor denoted by $D_i$. Each edge $E_{ij}\subset \Delta_0^\vee$ corresponds to a non-compact two-cycle (the $U(1)$ fibration over $E_{ij}$) at the intersection of two toric divisors, denoted by $\CC_{ij}\cong D_i \cdot D_j$.

The  resolved CY$_3$ $\wh\MG$ is obtained by subdividing the toric cone $\Delta_0$ into a more general toric fan $\Delta$. At the level of the toric diagram $\Gamma$, this correspond to including the internal points:
\be
v_a = (w_a, 1)~, \quad a=1, \cdots, r~,
\ee
and to choosing a triangulation of $\Gamma$. A maximal resolution corresponds to a particular complete triangulation of $\Gamma$. Any two maximal resolutions are related by a sequence of flops.

\subsubsection{GLSM description}
We may also describe the toric singularity and its crepant resolutions as a K{\"a}hler quotient:
\be\label{Kq def}
\wh\MG \cong \C^{n}/\!/_{\!\xi}\, U(1)^{n-3}~, \qquad n\equiv n_E +r~.
\ee
The advantage of this perspective is that it keeps track of the K{\"a}hler parameters, $\xi$, which determine the K\"ahler volumes of the exceptional curves. 
This construction can be nicely presented, in the physics language, as a ``gauged linear sigma-model'' (GLSM) \cite{Witten:1993yc}, with homogeneous coordinates $\{z_i\} \in \C^n$ (with index $i=1, \cdots, n$) and K{\"a}hler parameters $\xi_{\bf a}$ (with index ${\bf a}=1, \cdots, n-3$) for the $U(1)_{\bf a}$ actions by which we quotient. The parameters $\xi_{\bf a}$ are also known as Fayet-Iliopololous (FI) parameters for the auxiliary gauge groups $U(1)_{\bf a}$.  
The GLSM data is conveniently summarized by a table:
\be\label{CY3 glsm}
\begin{tabular}{l|c|c}
 & $z_i$ & FI \\
 \hline
    $U(1)_{\bf a}$  &$Q^{\bf a}_i$ &  $\xi_{\bf a}$\\
\end{tabular}~, \qquad\qquad
{\rm span}(Q_i^{\bf a})={\rm ker}(v_1, \cdots, v_{n})~. 
\ee
Here, the charges $Q_i^{\bf a}\in \Z$ are the $U(1)_{\bf a}$ charges of the homogenous coordinates $z_i$. 
The K{\"a}hler quotient \eqref{Kq def} is given explicitly by:
\be
\wh\MG \cong  \Big\{ z_i\, \Big| \,  \sum_i Q_i^{\bf a} |z_i|^2 = \xi_{\bf a} \Big\}/U(1)^{n-3}~.
\ee
Each $z_i$ is associated to a point $w_i\in \Gamma$ in the toric diagram. The corresponding toric divisor 
is defined by:
\be
D_i \cong \{ z_i =0\} \cap \wh\MG~.
\ee
We will also often use the following notation, as in subsection~\ref{subsec: Mtheory geom}, which distinguishes between non-compact and compact toric divisors, corresponding to the external and internal points in the toric diagram, respectively:
\be
D_j~, \quad j=1, \cdots, n_E~, \qquad \qquad \bE_a~, \quad a=1, \cdots, r~.
\ee
The charge vectors $Q^{\bf a}$ are related to the toric vectors $v_i\in \Delta$ (including all the internal points in the toric diagram) by:
\be
\sum_{i=1}^n v_i Q_i^{\bf a}=0~.
\ee
In particular, $\sum_i Q_i^{\bf a}=0$ is the condition for the GLSM target space to be Calabi-Yau. 
These relations also imply three linear relations amongst toric divisors, including the compact ones:
\be
\sum_{i=1}^n w^x_i D_i \cong 0~, \qquad \quad
\sum_{i=1}^n w^y_i D_i \cong 0~, \qquad \quad
\sum_{i=1}^n  D_i \cong 0~.
\ee

\subsubsection{Triangulation, curves and intersection numbers}\label{subsec: triang and inter numb}
Given a fully triangulated toric diagram $\Gamma$, there is a convenient way to write down a (redudant) GLSM and to compute all the triple intersections numbers amongs divisors.
Let us consider the compact curve:
\be\label{CasDD}
\CC_{ij} \cong D_i \cdot D_j \cong  \{ z_i =0, z_j=0\} \cap \wh\MG~.
\ee
It corresponds to an internal line in the toric diagram, denoted by $E_{ij}$. This $E_{ij}$ is at the intersection of two triangles with vertices $w_i, w_j, w_k$ and $w_i, w_j, w_l$, respectively. We can then write a row of the GLSM as:
\be\label{glsm inter construct}
\begin{tabular}{l|ccccc|c}
 & $z_i$ & $z_j$ & $z_k$ & $z_l$ & $z_{m\neq i, j, k, l}$ &  \\
 \hline
    $\CC_{ij}$  &$q_i$ &$q_j$ &$1$ &$1$ &$0$ &  $\xi_{\CC_{ij}}$\\
\end{tabular}~, \qquad
v_i q_i + v_j q_j+v_k + v_l=0~. 
\ee
Here and henceforth, it will be convenient to label the GLSM fields $z_i$ by the corresponding toric divisors, and the rows by the corresponding curves.
By construction, the FI term $\xi_{\CC_{ij}}$ is the K{\"a}hler volume of the curve in $\wh\MG$, which is positive:
\be
\xi_{ij} = \int_{\CC_{ij}} J >0~.
\ee
Repeating this operation for every internal line $E_{ij}$ in $\Gamma$, we obtain a redundant GLSM that capture all the exceptional curves. We can reduce that redundant description by choosing a set of $n-3$ linearly independent curves $\CC_{\bf a}$, giving us a proper GLSM:
\be\label{CY3 glsm bis}
\begin{tabular}{l|c|c}
 & $D_i$ & FI \\
 \hline
    $\CC^{\bf a}$  &$Q^{\bf a}_i$ &  $\xi_{\bf a}$\\
\end{tabular}~.
\ee
This is obviously equivalent to \eqref{CY3 glsm}, but here we have chosen a particular basis for the $U(1)^{n-3}$ gauge group, adapted to a given triangulated toric diagram, such that $\xi_{\bf a}>0$, $\forall {\bf a}$. We loosely refer to $\{\CC^{\bf a} \}$ as the ``generators of the Mori cone.'' The intersections numbers between divisors and curves are simply given by:
\be
D_i \cdot \CC^{\bf a}= Q^{\bf a}_i~.
\ee
Combining this with \eqref{CasDD} and the linear relations \eqref{glsm inter construct}, we can compute  the triple intersection numbers amongst divisors:
\be\label{DDD gen}
D_i \cdot D_j \cdot D_k~, 
\ee
where at least one divisor is compact. In the following, we use this simple method to compute the M-theory prepotential \eqref{F from geom} in numerous examples.

Finally, one might wonder whether one can make sense of the triple intersection number \eqref{DDD gen} when all three divisors are non-compact. It is not well-defined by itself, but we shall briefly discuss a possible regularization in Appendix~\ref{app: intersect}.

\paragraph{Example.} As a simple example, consider the resolved $\C^3/\Z_3$ orbifold of Figure~\ref{fig:examples toric dig}. We have a curve $\CC\cong \CC_{10}$, and the two triangles have vertices $w_1, w_0, w_2$ and $w_1, w_0, w_3$. Then, we have \eqref{glsm inter construct} with $q_1=1$ and $q_0=-3$, so that:
\be
\begin{tabular}{l|cccc|c}
 & $D_1$ & $D_2$ & $D_3$ & $\bE_0$  &  \\
 \hline
    $\CC$  &$1$ &$1$ &$1$ &$-3$ &  $\xi_{\CC_{10}}$\\
\end{tabular}~. 
\ee
Here, we reordered the divisors in the standard order, and used the notation $D_0 = \bE_0$. We can also check that $\CC_{20}\cong \CC_{30}\cong \CC$. This is the standard GLSM description of a local $\mathbb{P}^2$, with $\CC \cong H$ the hyperplane class. We also have the linear relations $D_1\cong D_2 \cong D_3$ amongst toric divisors. The triple intersection numbers are:
\be
D_1^2 \bE_0= \CC D_1=1 ~, \qquad  D_1 \bE_0^2= \CC \bE_0=-3~, \qquad  
\bE_0^3 = -3 D_1 \bE_0^2 = 9~.
\ee
We then have $S= \nu \bE_0$ and the M-theory prepotential $\CF=-{1\ov 6}S^3= -{3\ov 2} \nu^3$.

\subsection{Toric threefold $\widehat\MG$ as a $U(1)_M$ fibration}\label{subsec: U1Mfib}
Now, let us pick a $U(1)_M$ inside the $U(1)^3$ toric action. We would like to view the full $\wh\MG$ as a $U(1)_M$ fibration over a five-dimensional space $\CM_5$:
\be\label{M5 hX fibration}
U(1)_M  \longrightarrow \wh\MG \longrightarrow \CM_5~.
\ee
Viewing $U(1)_M$ as the M-theory circle, we then have a type-IIA description. Not every $U(1)_M$ gives us a well-understood IIA configuration, however. In the following, we discuss this $U(1)_M$ fibration structure and the conditions for an ``allowed'' IIA reduction. This approach was first introduced in \cite{Aganagic:2009zk} and developed in \cite{Benini:2009qs, Jafferis:2009th, Benini:2011cma, Closset:2012ep, Closset:2012eq} in the context of M-theory on CY fourfold singularities.

\paragraph{GLSM reduction.} Let us first discuss the circle fibration structure \eqref{M5 hX fibration} from the point of the view of the GLSM \cite{Aganagic:2009zk}.  We consider the following non-standard parameterization of the GLSM \eqref{CY3 glsm}, as:
\be\label{CY2 from M}
\begin{tabular}{l|c|c}
 & $z_i$ & FI \\
 \hline
    $U(1)_{\bf a}$  &$Q^{\bf a}_i$ &  $\xi_{\bf a}$\\
    \hline
    $U(1)_M$& $Q^M_i$ &  $r_0$
\end{tabular}~, \qquad\qquad \sum_{i=1}^n Q^M_i =0~.
\ee
Here, we introduced an additional complex parameter $r_0+ i \theta_0$, together with a new gauge symmetry: 
\be\label{Mtheory mm}
\sum_{i=1}^n Q_i^M |z_i|^2= r_0~,\qquad
\theta_0 \sim \theta_0 + \alpha_M~, \qquad z_i \sim e^{i \alpha_M Q^M_i} z_i~.
\ee
This construction describes the same three-fold $\wh \MG$, since we can solve for $r_0$ using \eqref{Mtheory mm} and gauge fix $\theta_0$ to zero. On the other hand, we may consider the further projection of the CY$_3$ to a CY {\it two-fold} by ``forgetting'' $\theta_0$ and fixing $r_0$. In this picture, we interpret $\theta_0$ as the M-theory circle  coordinate. At fixed $r_0$, the GLSM \eqref{CY2 from M} is a redundant description of a toric CY$_2$ variety, denoted by $\wh {\bf Y}$. 
The parameter $r_0 \in \R$ is itself interpreted as the $x^9$ coordinate in type IIA string theory, over which $\wh {\bf Y}$ is fibered:
\be\label{M5 fibered}
\wh {\bf Y}(r_0) \longrightarrow \CM_5 \longrightarrow \R \cong \{ r_0\}~.
\ee
The only two-dimensional toric CY singularity is the $A$-type orbifold:
\be
A_{M-1} \cong \C^2/\Z_M~,
\ee
which can be resolved to the ALE space $\wh {\bf Y}$. The corresponding one-dimensional toric diagram is simply a line of $M+1$ points. Let us denote by $t_k$, $k=0, \cdots, M$, the corresponding GLSM fields. We have:
\be\label{CY2 Y}
\begin{tabular}{l|c|c}
 & $t_k$ & FI \\
 \hline
    $U(1)_s$  & $ - 2 \delta_{ks} + \delta_{k, s+1}+ \delta_{k, s-1}$ &  $\chi_s(r_0)$   
   \end{tabular}~, \qquad\quad
 s=1, \cdots, M-1~.
\ee
At any fixed $r_0$, we can derive the CY$_2$ GLSM \eqref{CY2 Y} from \eqref{CY2 from M}, by eliminating the redundant $z_i$ variables using the D-term constraints \eqref{Mtheory mm}. This defines a projection map:
\be 
z_i \mapsto t_k= z_k(r_0)~,
\ee 
where $t_k$ denotes the independent $z_i$ variables. 
The K\"ahler parameters $\chi_s(r_0)$ of $\wh {\bf Y}$ are the volumes of the $\mathbb{P}^1$'s in the ALE resolution:
\be\label{chi s def}
\chi_s(r_0) = \int_{\mathbb{P}^1_s} J_{\wh{\bf Y}}~.
\ee 
Note that the toric divisors $D_k^{\wh{\bf Y}} \cong \{ t_k=0\}$ of $\wh{\bf Y}$ for $k=1, \cdot, M-1$ are precisely the  exceptional curves:
\be
D_s^{\wh{\bf Y}} \cong \mathbb{P}^1_s~.
\ee
They intersect amongst themselves according to the $A_{M-1}$ Dynkin diagram, as is clear from \eqref{CY2 Y}. The toric divisors $D_0^{\wh{\bf Y}}$ and $D_M^{\wh{\bf Y}}$ are non-compact, on the other hand.

The M-theory circle is non-trivially fibered over $\CM_5$. By the M-theory/IIA duality, this implies the presence of non-trivial RR two-form flux $F_2^{\rm RR}= dC_1$ in type IIA. Moreover, whenever the M-theory circle degenerates in real-codimension four, there is a magnetic source for $C_1$ in type IIA, namely a number of D6-branes.

\begin{figure}[t]
\begin{center}
\subfigure[\small Allowed.]{
\includegraphics[height=4.5cm]{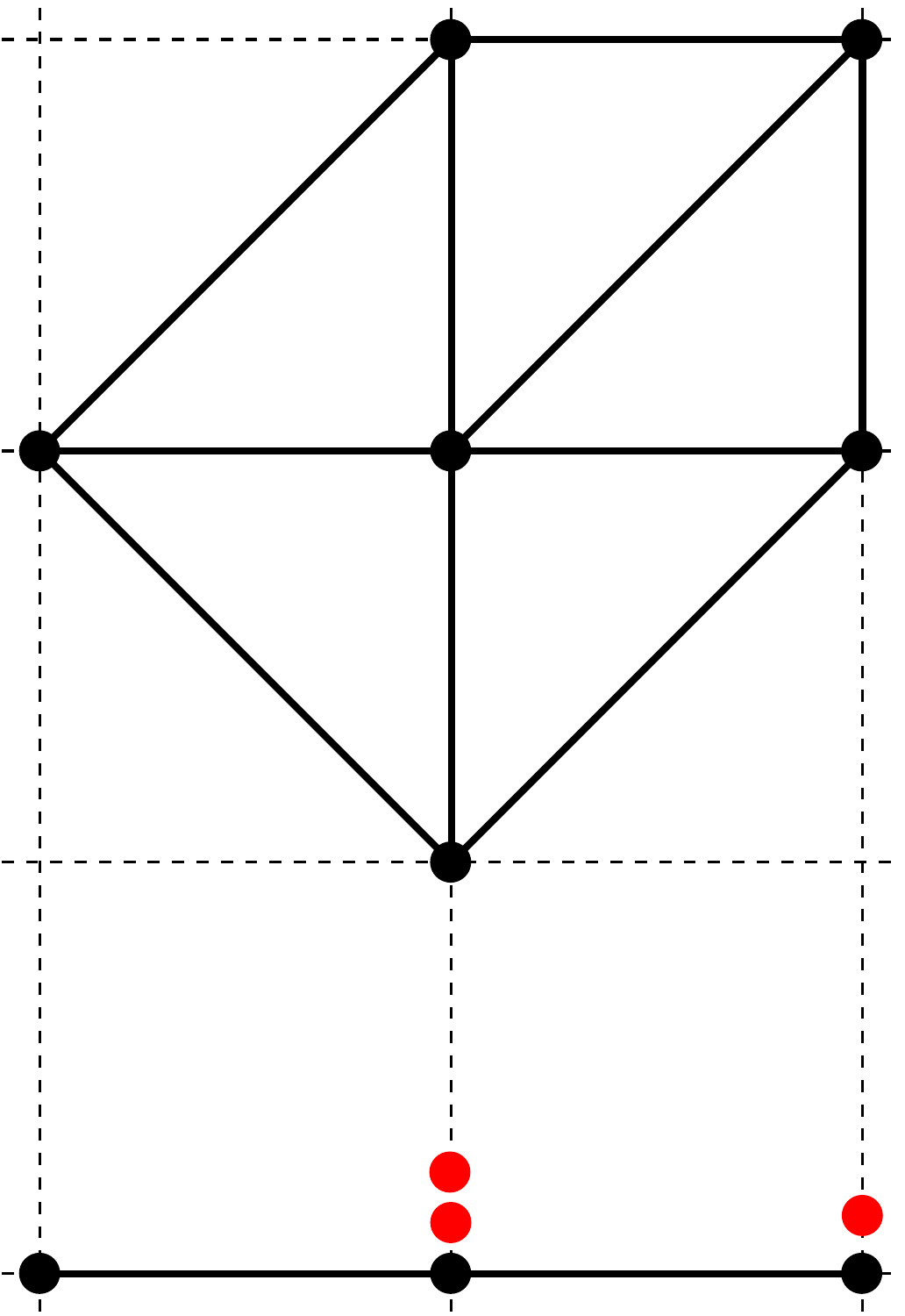}\label{fig:red exp1}}\qquad\qquad\qquad
\subfigure[\small Disallowed.]{
\includegraphics[height=4.5cm]{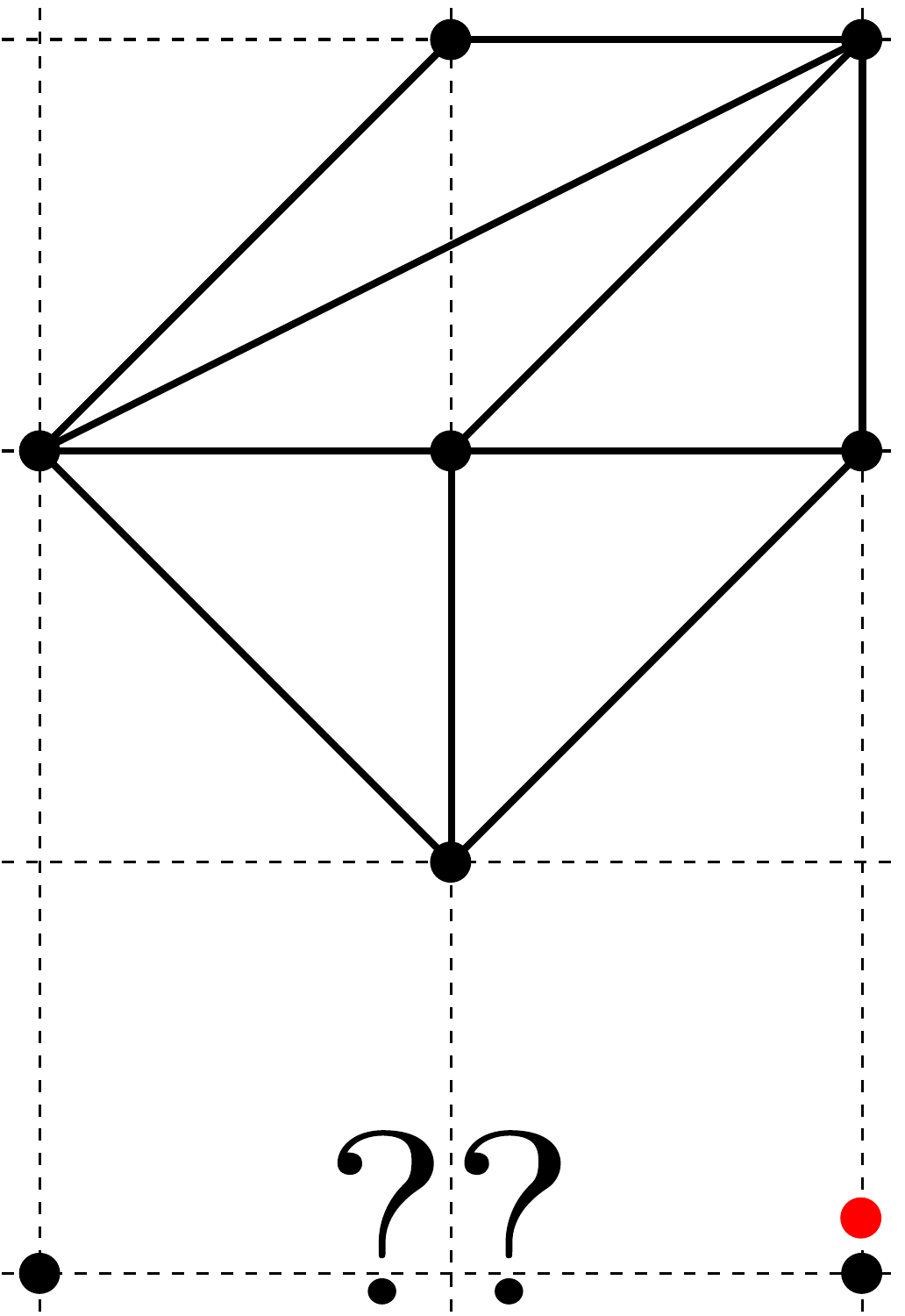}\label{fig:red exp2}}
\caption{{\bf Left:} An example of an allowed vertical reduction of a toric diagram $\Gamma$, leading to a resolved $A_1$ singularity $\wh {\bf Y}$, pictured by the 1d toric diagram at the bottom. The vertical lines in the middle of $\Gamma$ are mapped to a single exceptional curve $D^{\wh {\bf Y}}_1\cong \mathbb{P}^1 \subset \wh {\bf Y}$ wrapped by two D6-branes. The vertical line on the right of $\Gamma$ is mapped to a single D6-brane along the non-compact divisor $D^{\wh {\bf Y}}_2\cong \C \subset \wh {\bf Y}$. {\bf Right:} An example of a triangulation without an allowed vertical reduction, because there is an edge with $\Delta v= (2,1,0)$.   \label{fig:examples vert red}}
 \end{center}
 \end{figure}

\paragraph{Toric diagram reduction.} We can obtain the one-dimensional toric diagram $\Gamma_{\wh{\bf Y}}$ of $\wh{\bf Y}$ from the two-dimensional toric diagram $\Gamma$ of $\wh\MG$ by a simple projection. 
The $U(1)_M$ charges $Q_i^M$ introduced in \eqref{CY2 from M} do not satisfy $\sum_i v_i Q_i^M=0$. Instead, let us define:
\be
v_M \equiv {\rm primitive} \sum_{i=1}^n v_i Q_i^M~, 
\ee
the primitive vector in $\Z^3$ parallel to $\sum_{i=1}^n v_i Q_i^M$. Then, the toric vectors $\t v_k\in \Z^2$ of $\wh{\bf Y}$ are obtained by projecting the toric vectors $v_i$ to the plane orthogonal to $v_M$. (In general, there will be several $v_i$'s mapping to a single $\t v_k$. This corresponds to the map $z_i \mapsto t_i$ in the GLSM.) Due to the CY condition, we must have $v_M= (w_M, 0)$. It is convenient to choose:
\be\label{}
v_M= (0,1,0)~.
\ee
Reducing along this particular $U(1)_M \subset U(1)^3$ corresponds to a  ``vertical reduction'' of the toric diagram; an example of this is shown in Figure~\ref{fig:red exp1}. Of course, we may (and shall) also consider the ``horizontal reduction'' of $\Gamma$, or any other projection related to $v_M$ by an $SL(2, \Z)$ transformation of the toric diagram.

Given the CY$_3$ toric variety $\wh\MG$, we can understand how the exceptional divisors $\bE_j$ and the 2-cycles $\CC_{ij}$ (compact or non-compact) are projected onto $\CM_5$ by the vertical reduction along $v_M$. Let us consider the 2-cycle $\CC_{ij}$ corresponding to the edge $E_{ij}$ (internal or external) in $\Gamma$. We define:
\be
\Delta v\equiv v_j- v_i= (m, n, 0)~.
\ee
Without loss of generality, we choose $n>0$. Along $\CC_{ij}$, the $T^2_{(ij)}$ given by $U(1)_i \times U(1)_j \subset T^3$ degenerates, while the orthogonal $U(1)_\varphi \subset T^3$ is the local angular coordinate on $\CC_{ij}$.~\footnote{For an internal edge in $\Gamma$, this is a $U(1)_\varphi$ fibered over an interval, giving rise to a genus-zero curve. For an external edge of $\Gamma$, this corresponds to a rotation along a non-compact 2-cycle $\CC\cong \C$.} Note that $T^2_{(ij)}$ corresponds to the span of $\Delta v$ and $(0,0,1)$.
%%%%%%%%%%%%%%%
 %%%%%%%%%%
There are three cases:
\begin{itemize}
\item[(i)] {\bf Vertical edge:} If $\Delta v= (0, n, 0)$, we have a vertical edge of length $n$ on the toric diagram, which is really a set of $n$ edges of length one, each corresponding to a distinct 2-cycle $\CC_{ij}^{(q)}$, $q=1, \cdots, n$ with angular direction $U(1)_\varphi$ corresponding to $v_\varphi= (1,0,0)$. Since $\Delta v$ is parallel to $v_M$, the M-theory circle degenerates in real-codimension four along each $\CC_{ij}^{(q)}$, corresponding to a D6-brane wrapping a curve in type IIA \cite{Sen:1997kz}. Thus, each 2-cycle $\CC_{ij}^{(q)}$ in $\wh \MG$ projects to a D6-brane wrapping a single 2-cycle $D^{\wh{\bf Y}}$ in $\wh{\bf Y}$, but generally at different values of $r_0$. The D6-branes can be compact (wrapping a $\mathbb{P}^1$) or non-compact (along $D^{\wh{\bf Y}}\cong \C$), depending on whether the vertical edges are internal or external in $\Gamma$. Moreover, an M2-brane wrapped over  $\CC_{ij}^{(q)}$ maps to a  D2-brane wrapped over $D^{\wh{\bf Y}}$.

\item[(ii)] {\bf Allowed oblique edge:} If $\Delta v= (\pm 1,n,0)$, including the case of an horizontal edge ($n=0$), we have a two-cycle $\CC_{ij}$ with angular direction $v_\varphi = (n, \mp 1,0)$, which is projected out entirely by the vertical reduction. The curve $\CC_{ij}$ maps to an interval along $\R\cong \{r_0\}$ and to a point inside $\wh{\bf Y}$ at the intersection of the two adjacent divisors $D^{\wh{\bf Y}}_k$ and $D^{\wh{\bf Y}}_{k+1}$ to which $v_i$ and $v_j$ map. An M2-brane wrapped over $\CC_{ij}$ maps to a fundamental string stretched along the interval.

\item[(iii)] {\bf Disallowed oblique edge:} The last case is $\Delta v= (m,n,0)$ with $|m|>1$. Without loss of generality, we assume that $m$ and $n$ are mutually prime, so that we have a single edge and a single 2-cycle $\CC_{ij}$. The angular direction on the 2-cycle is $v_\varphi= (n, -m,0)$. In this case, a subgroup $\Z_m$ of $U(1)_M$ degenerates along the curve, and we do not have any simple interpretation of this degeneration in  type IIA string theory.
\end{itemize}

\noindent In the following, we will say that a given triangulated toric diagram $\Gamma$ has an ``allowed vertical reduction'' when all its edges (internal or external) are of type (i) or (ii). An example of a toric diagram without an allowed vertical reduction is given in Figure~\ref{fig:red exp2}. We will further comment on the physical interpretation of the ``disallowed'' edges in subsection~\ref{subsec:pqwebs} below.

\subsection{Reading off the 5d $\CN=1$ gauge theory from type IIA}\label{subsec: gaugefromIIA}
Given a toric resolved CY singularity $\wh\MG$ with its triangulated toric diagram $\Gamma$, we associate a low-energy gauge-theory phase to a given allowed IIA reduction. By an $SL(2, \Z)$ transformation of the toric diagram, we can always view it as a ``vertical reduction,'' as discussed above.

The type IIA configuration consists of a resolved  $A_{M-1}$ singularity $\wh{\bf Y}$, where $M$ is the horizontal length of $\Gamma$. (For instance, the toric diagram of Fig.~\ref{fig:red exp1} has $M=2$.) 
The ALE space $\wh{\bf Y}$ is fibered over the line $\{r_0\}$. As we will see in many examples, the resolution parameters $\chi_s(r_0)$ of the $M-1$ exceptional curves, \eqref{chi s def}, are piece-wise linear, continuous functions of $r_0$ \cite{Benini:2009qs, Benini:2011cma}. The jumps in the first derivative of $\chi_s(r_0)$ occurs at the locations of D6-brane sources. In fact, supersymmetry imposes the relation:
\be\label{chi to F}
\chi_s'(r_0) = {1\ov 2 \pi} \int_{\mathbb{P}^1_s} F_2^{\rm RR}
\ee
between the first derivative of $\chi_s(r_0)$ and the RR 2-form flux through the exceptional curve. 

For every vertical edge in $\Gamma$---that is, with $\Delta v= (0,1,0)$---, there is a D6-brane wrapping the curve $D^{\wh{\bf Y}}_k$ to which the vertical edges maps, at some particular value of $r_0$. Internal vertical edges give rise to D6-branes wrapped over the exceptional curves $\mathbb{P}^1_s$, while external vertical edges give rise to non-compact D6-branes. We refer to these two types of D6-branes as ``gauge'' and ``flavor'' D6-branes, respectively. 

As we cross a D6-brane wrapped over $D^{\wh{\bf Y}}_k$ at some particular $r_0= \xi_{{\rm D6};\, k}$, the RR fluxes through the exceptional cycles jump. Due to \eqref{chi to F}, the slope of $\chi_s(r_0)$ jumps accordingly, with: 
\be
\chi_s'(\xi_{{\rm D6};\, k}+\epsilon) -\chi_s'(\xi_{{\rm D6};\, k}-\epsilon) =- \big(\mathbb{P}^1_s \cdot D^{\wh{\bf Y}}_k\big)~,
\ee
for some small $\epsilon>0$.
For instance, for a gauge D6-brane wrapping $\mathbb{P}^1_s$ at $r_0=\xi_{{\rm D6};\, s}$, the slope of $\chi_s(r_0)$ jumps by $+2$ when crossing the brane, since $\mathbb{P}^1_s$ has self-intersection $-2$ inside $\wh {\bf Y}$.

\paragraph{5d $\CN=1$ quiver from D6-branes.} Given the above configuration of D6-branes wrapped along 2-cycles inside $\CM_5$, we can read off the 5d $\CN=1$ gauge theory along the transverse $\R^{1,4}$. For each vertical internal line with $\Delta v=(0,n,0)$ in the toric diagram, we (naively) have a gauge group $U(n)$. By the standard rules for branes at ADE singularities \cite{Douglas:1996sw}, we then obtain a $A$-type gauge-theory quiver of the form:
\be\label{Un quiver}
[U(n_0)]\; \rule[2pt]{0.8cm}{.6pt}\; U(n_1)_{k_1} \; \rule[2pt]{0.8cm}{.6pt}\; \quad\cdots\quad \; \rule[2pt]{0.8cm}{.6pt}\; U(n_{M-1})_{k_{M-1}} \; \rule[2pt]{0.8cm}{.6pt}\; [U(n_M)]
\ee
Here, each line represents an hypermultiplet in a bifundamental representation, and the bracketed groups on either ends are flavor groups, corresponding to the flavor D6-branes. Each gauge group may also have a non-trivial Chern-Simons term at level $k_s$, as we explain below.

The $U(n_s)$ gauge group is realized when the $n_s$ D6-branes wrapped over  $\mathbb{P}^1_s$ are brought on top of each other along the $r_0$ direction; this corresponds to a singular limit of $\wh\MG$. In general, $\wh\MG$ corresponds to a particular point on the Coulomb branch of the quiver \eqref{Un quiver}, where the distances between gauge D6-branes are the Coulomb branch VEVs $\langle \varphi \rangle$.
Moreover, the low-energy gauge group is actually:
\be\label{SUn quiver}
[U(n_0)]\; \rule[2pt]{0.8cm}{.6pt}\; SU(n_1)_{k_1} \; \rule[2pt]{0.8cm}{.6pt}\; \quad\cdots\quad \; \rule[2pt]{0.8cm}{.6pt}\; SU(n_{M-1})_{k_{M-1}} \; \rule[2pt]{0.8cm}{.6pt}\; [U(n_M)]
\ee
Namely, each overall $U(1) \subset U(n_s)$ is massive and decoupled. In M-theory, the gauge fields in the Cartan subalgebra of $SU(n)$ come from the periods of the three-form $C_3$ on the $n-1$ ``vertical'' curves.
The existence or not of the overall $U(1)$, on the other hand, depends on the precise boundary conditions for the ``KK monopole.''  In the present case, it is absent.~\footnote{This comes from the difference between the multi-Taub-NUT and the ALE metric. A multi-TN of charge $n$ has $\R^3 \times S^1$ asymptotics; reducing M-theory along the $S^1$ gives us a stack of $n$ flat D6-branes in type IIA. In that case, the overall $U(1)$ comes from the reduction of $C_3$ along a normalizable anti-self-dual two-form of TN$_n$. The ALE metric, on the other hand, describes the center region of TN$_n$, with $\C^n/\Z_n$ asymptotics, and does not support the $U(1)$ mode \protect\cite{Witten:2009xu}.}  The corresponding mechanism in type IIA in not entirely clear to us, however.

%%%%%%%%%%%
\paragraph{``Effective'' CS levels.} The presence of the RR flux in $\CM_5$ induces Chern-Simons interactions on the gauge D6-branes due to the Wess-Zumino term \cite{Aganagic:2009zk}:
\be
\int_{\R^{1,4} \times \mathbb{P}^1_s}  C_1 \wedge F\wedge F\wedge F~,
\ee
By integration by part, this induces a 5d Chern-Simons level along $\R^{1,4}$ for a probe D6-brane wrapped on $\mathbb{P}^1_s$:
\be
k_s(r_0)= -{1\ov 2\pi} \int_{\mathbb{P}^1_s} F^{\rm RR}_2(r_0)~.
\ee
Due to \eqref{chi to F}, the CS levels can thus be read off from the slopes of the IIA profiles. 
For each $U(n_s)$ factor in \eqref{Un quiver}, the effective CS level $k_s$ can be computed as follows \cite{Benini:2011cma}. Let us denote by $\chi'_{s, \, \pm}$ the slope to the right and left of the IIA profile $\chi_{s}$, respectively. That is, for each exceptional curve $\P^1_s$, we define:
\be
\chi'_{s,\, \pm} = \lim_{r_0 \rightarrow \pm \infty }\chi_s'(r_0)~.
\ee
Then, the effective Chern-Simons level  $k_s$ is given by minus the average of the slopes:
\be\label{k eff from IIA}
k_{s} = -\half \left(\chi'_{s,\, -} + \chi'_{s,\, +}  \right)~. 
\ee
In fact, this ``CS level'' (in the common parlance) can be half-integer, and corresponds to the CS contact term $\kappa$, including the half-integer contributions from matter fields---see equation~\eqref{k eff app} in Appendix. We will generally denote $k_s$ in \eqref{k eff from IIA} by $k_{s, {\rm eff}}$, as it is an ``effective'' CS level. 

\paragraph{Particle states from IIA.}  The half-BPS particles on the Coulomb branch are easily read off from the IIA configuration. For $\wh \MG$ in a given K\"ahler chamber, we have the gauge and flavor D6-branes at points along the $x^9=r_0$ direction. Let us denote these D6-branes by D6$_{k, (a)}$, where $k=0, \cdots, M$ and $a=1, \cdots, n_k$ for each $k$, and let $r_0=\xi_{k, (a)}$ be their $r_0$ positions. The perturbative particles simply arise from open string stretched between two D6-branes:

\begin{itemize}
\item The W-bosons of the $SU(n_s)$ gauge group are realized as open strings connecting the D6-branes wrapped over $\mathbb{P}^1_s$. Their masses are given by the distances between any two such gauge D6-branes along $r_0$:
\be
M(W_{s; i,j})= |\xi_{s, (a_i)}-\xi_{s, (a_j)}|~.
\ee
\item The bifundamental hypermultiplets are given by the open strings stretched between two gauge branes wrapping the intersecting curves $\mathbb{P}^1_s$ and $\mathbb{P}^1_{s+1}$, with masses:
\be
M(\CH_{s, s+1; i,j})= |\xi_{s, (a_i)}-\xi_{s+1, (a_j)}|~.
\ee
\item The fundamental hypermultiplets are the open string stretched between a flavor D6-brane along $D^{\wh{\bf Y}}_0$ and a gauge branes on $\mathbb{P}^1_1$, or between a gauge brane on $\mathbb{P}^1_{M-1}$ and the flavor brane along $D^{\wh{\bf Y}}_M$. Their mass is given similarly by the separation distance.
\end{itemize}

\noindent
In addition, we have the instanton particles, which arise as D2-branes wrapped over the exceptional curves $\mathbb{P}^1_s$ on top of a D6-brane at $r_0= \xi_{s,(a)}$. Their mass is given by the size of the curve they wrap:
\be
M(\CI_{s, (a)}) = \chi_s(\xi_{s,(a)})~.
\ee
This should be compared to \eqref{I mass heuristic} in the field-theory description. 
As mentioned there,  for every $s$, the particle of lowest mass can be viewed as ``the'' instanton particle $\CI_s$, while the other ones are bound state of $\CI_s$ with perturbative particles. The ``mixed'' instanton particle, corresponding to a D2-brane wrapping a curve  $\mathbb{P}^1_s$ at a location   $r_0= \xi_{k,(a)}$ with $k\neq s$, are also interpreted as a bound state of $\CI_s$ with other perturbative particles in the quiver.

\paragraph{String states from IIA.} Finally, we should discuss the monopole strings, which correspond to M5-branes wrapped over the exceptional compact divisors $D_j = \bE_j \subset \wh \MG$. Let $v_j$ denote the corresponding vector in the toric fan. If there is an allowed vertical reduction, $v_j$ is  part of a vertical line $v_{i}, v_j, v_{k}$ in the toric diagram, with the two curves $\CC_{ij} = D_i \cdot D_j$ and $\CC_{jk} = D_j \cdot D_j$ projecting down to two D6-branes wrapped over the same $\mathbb{P}^1_s$ (for the appropriate $s$) at  some locations $r_0= \xi_{s, (a)}$ and $\xi_{s, (a+1)}$, respectively. The compact four-cycle $D_j$ then maps to a 2-cycle $\mathbb{P}^1_s$ fibered over the interval $[\xi_{s, (a)}, \xi_{s, (a+1)}]$.

The monopole strings for each $SU(n_s)$ correspond to D4-branes wrapped over $\mathbb{P}^1_s$ and stretched between two subsequent gauge D6-branes. The corresponding string tensions can be computed as the integral of the IIA profile over the interval:
\be
T_{s, (a)} = \int_{\xi_{s, (a)}}^{\xi_{s, (a+1)}} dr_0\, \chi(r_0)~.
\ee

As we will see in a number of examples, this type IIA perspective on the 5d $\CN=1$ gauge-theory phases of 5d SCFTs allows us to read off the precise map between field-theory parameters and the K\"ahler parameters of $\wh\MG$ systematically.

\subsection{Global symmetries}
To conclude this section, let us briefly discuss global symmetries of the 5d $\CN=1$ theory from the M-theory/IIA perspective. 

\paragraph{Parity.} First of all, the 5d parity operation can be realized as a geometric operation on $\MG$. At the level of the toric diagram, it corresponds to an invertion of the toric vectors:
\be\label{C0 on w Parity}
w_i \mapsto w_i \cdot C_0  =- w_i~,\qquad C_0= \mat{-1 & 0\\ 0 & -1}~,
\ee
where $C_0\equiv S^2$ is the non-trivial central element of $SL(2, \Z)$. 
In the IIA geometry \eqref{M5 fibered}, parity can be realized as a reflexion of the $x^9= r_0$ coordinate. In particular, we see from \eqref{k eff from IIA} that the operation $\chi(r_0) \rightarrow \chi(-r_0)$ flips the signs of the effective CS levels. 

\paragraph{Global symmetry.} The rank of the flavor symmetry group $\GG_F$ can be read off from the toric geometry as the number of external points in the toric diagram minus 3:
\be
{\rm rank}(\GG_F)=f= n_E-3~,
\ee
namely,  the number of linearly-independent non-compact toric divisors.
 In the IIA setup, the global non-abelian flavor group $\GG_\CH \subset \GG_F$ acting on the hypermultiplets arises from coincident non-compact D6-branes. At the SCFT point, $\GG_F$ is sometimes enhanced to a larger global symmetry group \cite{Seiberg:1996bd}, although it is not obvious to see this directly from the singular geometry $\MG$.

Finally, the $SU(2)_R$ symmetry of the 5d $\CN=1$ gauge theory (which is preserved on its Coulomb branch) is realized geometrically in type IIA as the standard $SU(2)_R$ action on the hyper-K\"ahler ALE geometry $\wh {\bf Y}$ (see {\it e.g.} \cite{Johnson:1996py}).

%%%%%%%%%
 \begin{figure}[t]
\begin{center}
\includegraphics[height=3.6cm]{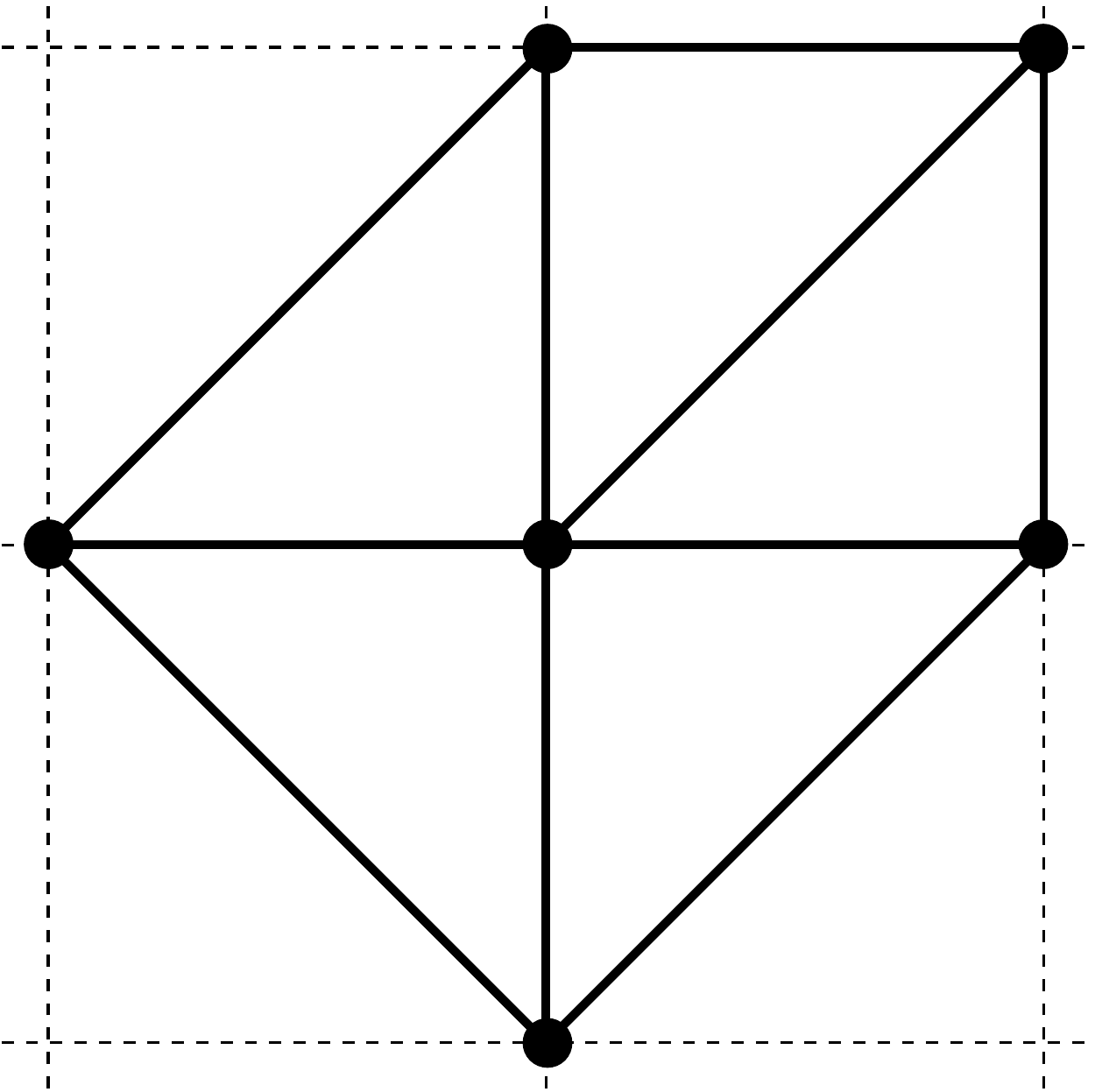}\qquad\qquad\qquad
\includegraphics[height=4.5cm]{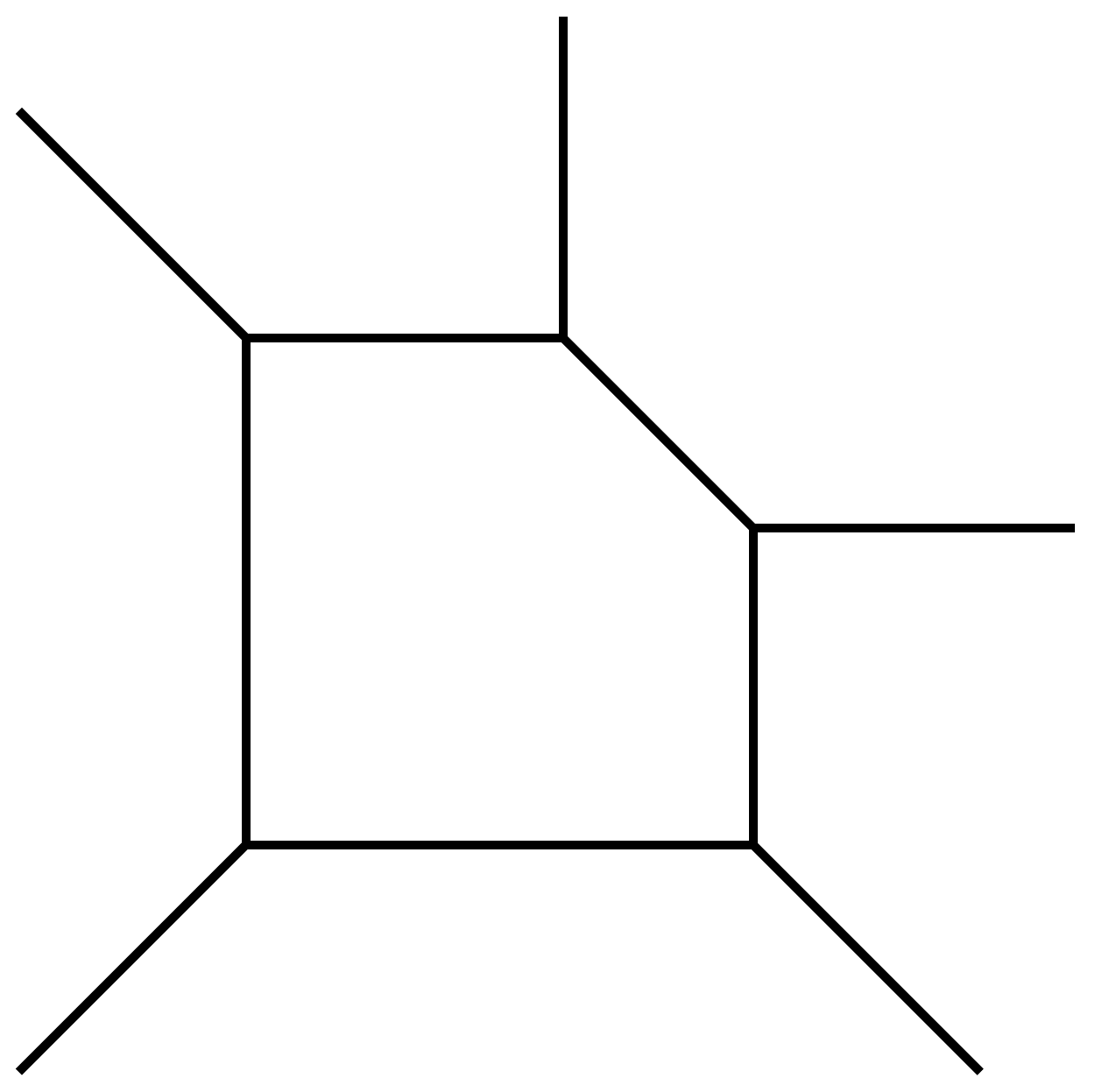}
\caption{A triangulated toric diagram and its dual $(p,q)$-web.\label{fig:pqweb}}
 \end{center}
 \end{figure} 
 %%%%%%%%%%%
\subsection{Comparing to $(p,q)$-webs}\label{subsec:pqwebs} 
To conclude this discussion of the IIA perspective on toric CY three-fold singularities in M-theory, it is interesting to compare it to the $(p,q)$-web description in type IIB \cite{Aharony:1997bh}. Recall that a five-brane $(p,q)$-web consists of a network of $(p,q)$-fivebranes, parallel along the $x^{0,1,2,3,4}$ directions and forming a web on the $(x^5, x^6)$ plane. The $SO(3)$ rotation group acting on the transverse directions $x^{7,8,9}$ becomes the $SU(2)_R$ R-symmetry of the 5d $\CN=1$ theory.~\footnote{Assuming that all the 5-branes sit at $x^{7,8,9}=0$. Moving the branes away from each other in the transverse directions corresponds to probing the Higgs branch, thus breaking the R-symmetry \protect\cite{Aharony:1997bh}.} 

The $(p,q)$-web is the dual graph to the triangulated toric diagram for $\wh \MG$, as shown in a example in Figure~\ref{fig:pqweb}. Let us choose an $SL(2,\Z)$ duality frame in which the D5-brane---corresponding to $(p,q)=(1,0)$---is represented by an horizontal line in the $(p,q)$-web. This corresponds to the ``vertical reduction'' of the toric diagram to type IIA. Thus, we have a simple correspondence between the 2-cycles (compact and non-compact) in $\wh \MG$ and the $(p,q)$-branes in the web:
\be
 \quad E_{ij} \in \Gamma\;\; \text{with} \;\; \Delta v= (m, n,0)  \;  \quad \leftrightarrow \quad \text{fivebrane with}\;\; (p,q)=(n,-m)~.
\ee
Of course, a vertical edge in the toric diagram corresponds to a D5-brane in the type-IIB $(p,q)$-web, which is indeed T-dual to a D6-brane in type IIA. Similarly, an horizontal edge corresponds to an NS5-brane, which is consistent with the T-duality between a stack of $M$ NS5-branes and the $A_{M-1}$ singularity the in type-IIA description. 

The $(p,q)$-web picture also gives us a complementary perspective on the nature of the ``disallowed'' edges discussed at the end of subsection~\ref{subsec: U1Mfib}: an oblique edge with $\Delta v= (m,n, 0)$ and $\gcd(m,n)=1$ corresponds to an $(n, -m)$-fivebrane, which does not have a perturbative string-theory description if $|m|>1$. Upon vertical reduction of  $\wh \MG$ to type IIA, we then expect to find a non-perturbative bound state of D6-branes and geometry. It would be interesting to explore this point of view further.~\footnote{Similar ``non-perturbative'' degenerations of the M-theory circle were briefly discussed in \protect\cite{Jafferis:2009th}.}

%%%%%%%%%%%%%%%%%%%%%%%%%%%%%%%%%%%%%%%%%%%%%%%
%%%%%%%%%%%%%%%%%%%%%%%%%%%%%%%%%%%%%%%%%%%%%%%
\section{Rank-one examples: $SU(2)$ gauge theories}\label{sec:SU2 expls}
\begin{figure}[t]
\begin{center}
\subfigure[\small $E_0$.]{
\includegraphics[height=2.5cm]{E0_sing.pdf}\label{fig:E0 sing}}\quad
\subfigure[\small $E_1$.]{
\includegraphics[height=2.5cm]{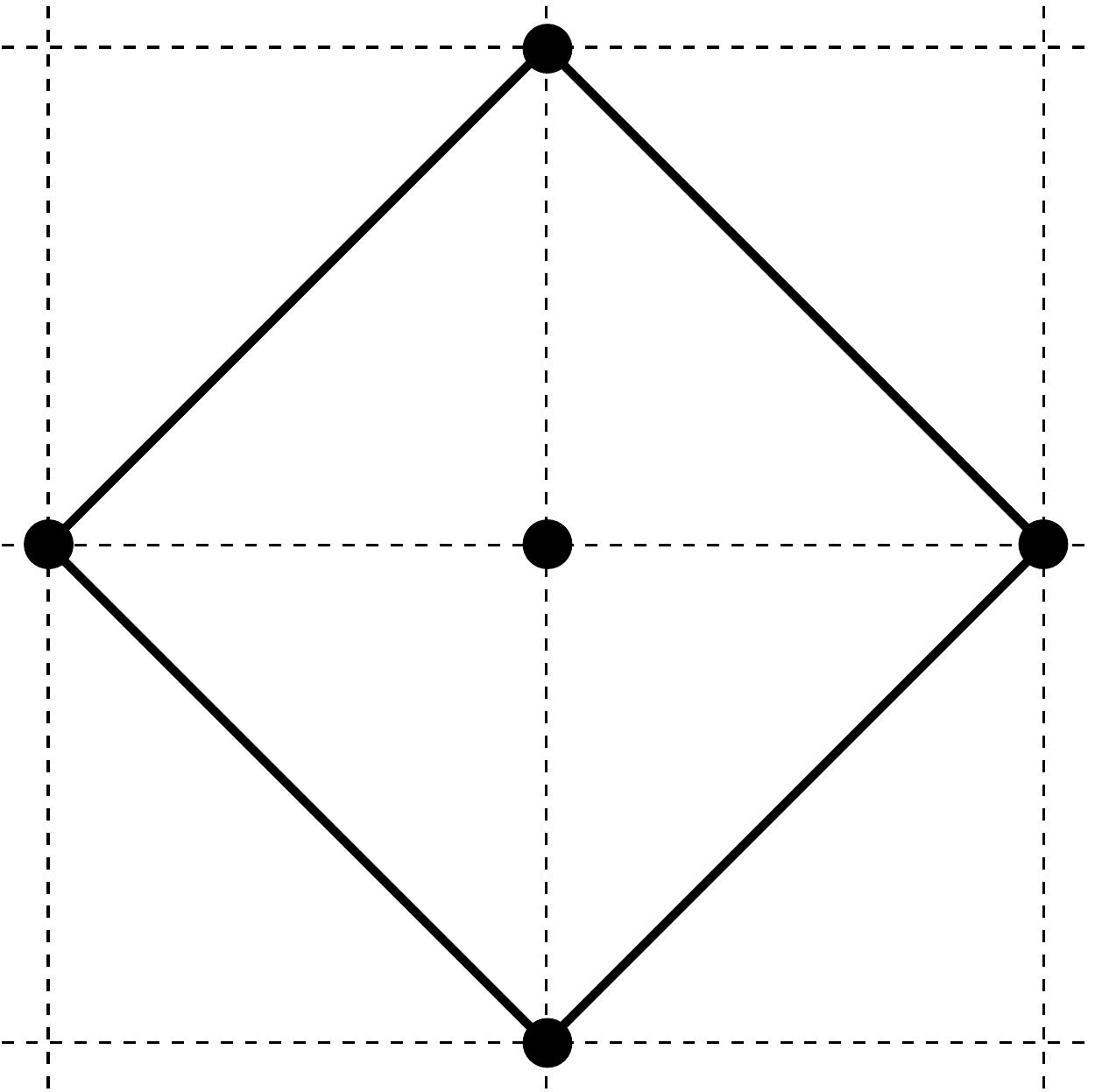}\label{fig:E1 sing}}\quad
\subfigure[\small $\t E_1$.]{
\includegraphics[height=2.5cm]{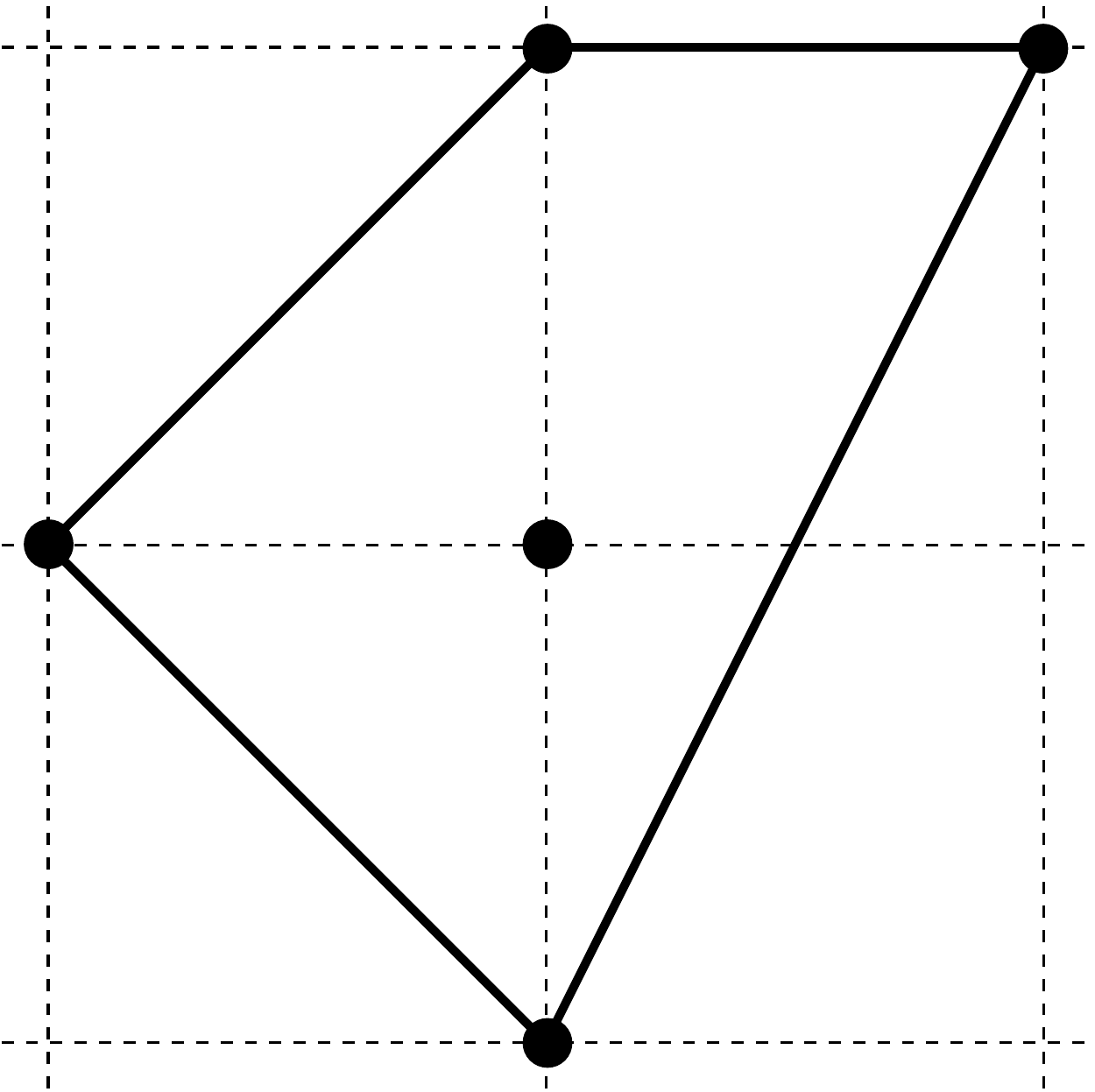}\label{fig:E1t sing}}\quad
\subfigure[\small $E_2$.]{
\includegraphics[height=2.5cm]{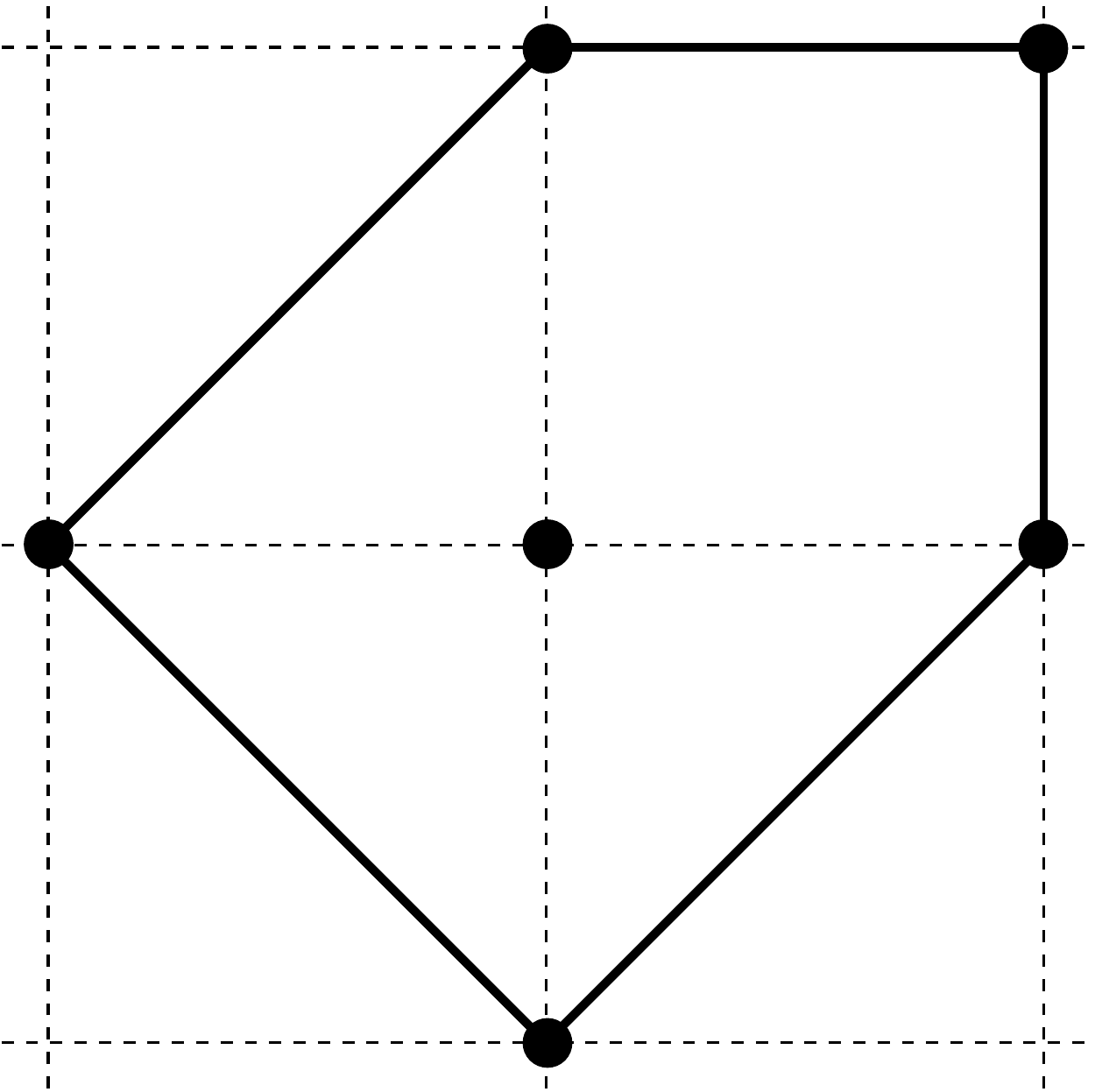}\label{fig:E2 sing}}\quad
\subfigure[\small $E_3$.]{
\includegraphics[height=2.5cm]{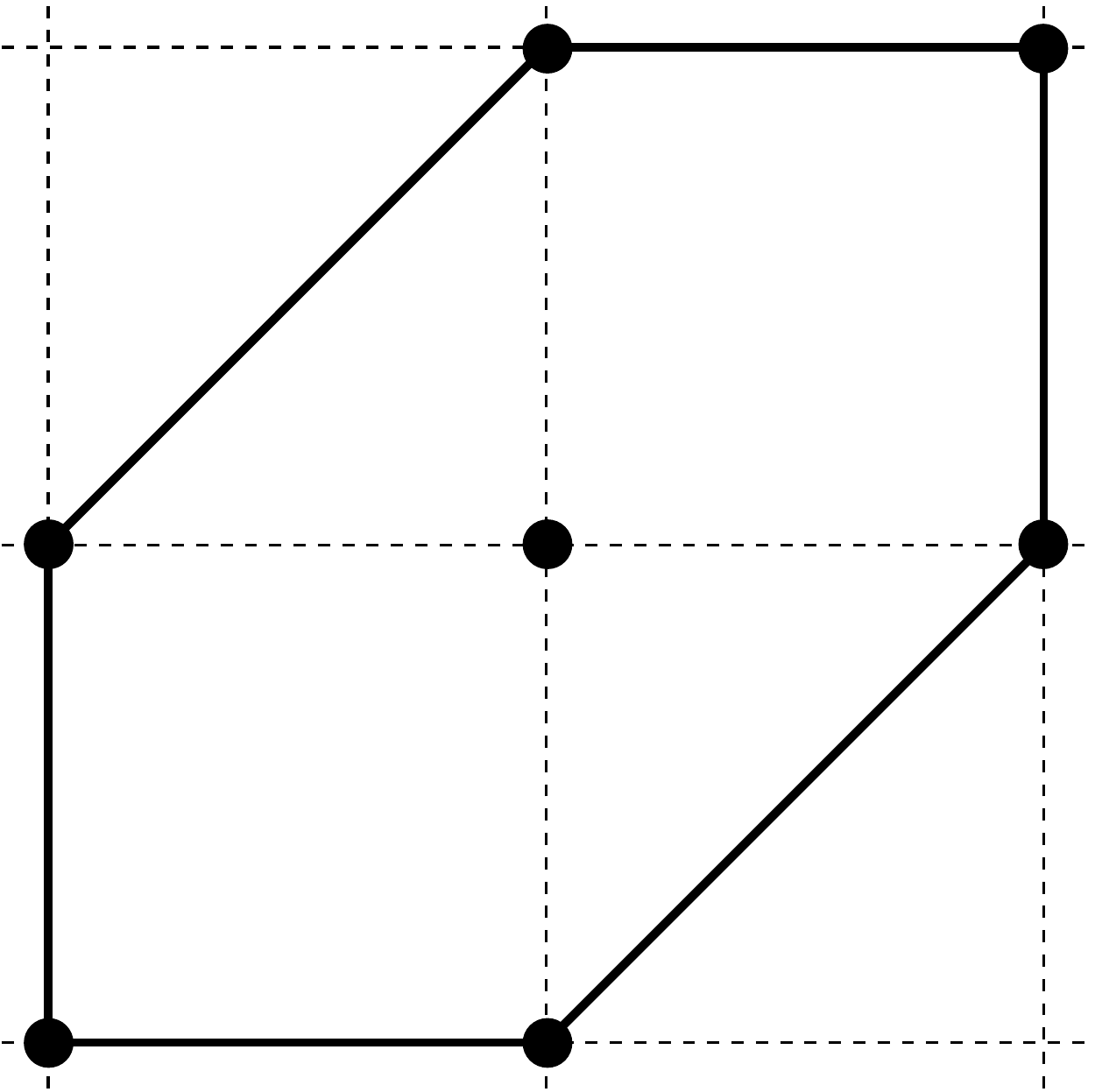}\label{fig:E3 sing}}\
\caption{Toric diagrams for the toric CY$_3$ singularities that engineer the $E_{N_f+1}$ 5d SCFTs. \label{fig:SU2 geoms}}
 \end{center}
 \end{figure}

%%%%%%%%%%%%%%%%%%%%%%%%%%%%%%%%%%%%%%%%%%%%%%%
In this section, in order to illustrate our method, we study the well-known $E_{N_f+1}$ series of 5d SCFTs \cite{Seiberg:1996bd}. These are five-dimensional superconformal theories with $E_{N_f+1}$ global symmetry, for $N_f < 8$. They can be engineered by considering M-theory on a CY$_3$ singularity ${\bf X}_{(N_f+1)} \equiv E_{N_f+1}$, obtained from the collapse of a del Pezzo surface, $dP_{N_f+1}$, inside a CY threefold. The Coulomb branch of this theory then corresponds to the local del Pezzo threefold:
\be\label{XNf}
\wh \MG_{(N_f+1)} ={\rm Tot}\big(\CK \rightarrow dP_{N_f+1} \big)~,
\ee
which is a resolution of the singular variety ${\bf X}_{(N_f+1)}$. This variety is toric if and only if $N_f \leq 2$, therefore we restrict ourselves to those cases.~\footnote{In section~\protect\ref{sec:nonTO}, we will explore some non-isolated toric singularities which give access to a subspace of the parameter space of the $E_{N_f+1}$ theory for $N_f=3,4, 5$.} The toric diagrams of the corresponding singularities are summarized in Figure~\ref{fig:SU2 geoms}. Note that $dP_0 \cong \mathbb{P}^2$, and that for $N_f=0$ we have two distinct singularities, corresponding to $\mathbb{F}_0 \cong \mathbb{P}^1 \times \mathbb{P}^1$ and $dP_1$, respectively; the corresponding SCFTs are denoted by $E_0$, $E_1$ and $\t E_1$ \cite{Morrison:1996xf}.  Note also that the singularities $E_1$ ($\mathbb{F}_0$)  and $E_3$ ($dP_3$) are ``parity invariant'' in the sense of \eqref{C0 on w Parity}, while the other three singularities in Fig.~\ref{fig:SU2 geoms} are not.

For $N_f \geq 0$, the $E_{N_f+1}$ SCFT admits a relevant deformation to an $SU(2)$ gauge theory with $N_f$ fundamental flavors. The gauge theory preserves only an $SO(2N_f)\times U(1)_T$ flavor symmetry, which is enhanced to $E_{N_f+1}$ at the UV fixed point \cite{Seiberg:1996bd, Morrison:1996xf}. Its prepotential reads:
\be\label{F SU2 Nf}
\CF_{SU(2), N_f} = h_0 \varphi^2 + {4\ov 3} \varphi^3 -{1\ov 6} \sum_{i=1}^{N_f} \sum_\pm \Theta(\pm \varphi+ m_i) (\pm\varphi+ m_i)^3~.
\ee
Here, we choose the $SU(2)$ Weyl chamber $\varphi\geq 0$. The single Coulomb branch parameter $\varphi$ corresponds to the single exceptional divisor in $\wh \MG_{(N_f+1)}$, which is $dP_{N_f+1}$ itself---in terms of the toric diagrams in Figure~\ref{fig:SU2 geoms}, this is the single internal point. The $N_f+1$ parameters $\mu=(h_0, m_i)$ correspond to the $N_f+1$ linearly-independent non-compact toric divisors---this is $f=n_E-3=N_f+1$ on the toric diagram, with $n_E= N_f+4$ the number of external points.

For later convenience, we also introduce the prepotential of the associated $U(2)_{k-\half N_f}$ theory with $N_f$ flavors:
\be\label{F U2 Nf}
\CF_{U(2), N_f} = \sum_{a=1}^2 \left( {h_0\ov 2} \phi_a^2 +{k\ov 6} \phi_a^3  -{1\ov 6} \sum_{i=1}^{N_f}  \Theta(\phi_a+ m_i) (\phi_a+ m_i)^3\right) + {1\ov 6} (\phi_1- \phi_2)^3~,
\ee
as discussed in section~\ref{subsec: instanton}. This reduces to \eqref{F SU2 Nf} for $\phi_1=- \phi_2= \varphi$.

By inspection of Figure~\ref{fig:SU2 geoms}, it is clear that the $E_1$, $\t E_1$, $E_2$ and $E_3$ singularities all admit some ``vertical reductions,'' depending on the partial resolution, as explained in section~\ref{sec: IIA perspective}. We will explore the corresponding gauge-theory phases in the following. On the other hand, the $E_0$ singularity in Fig.~\ref{fig:E0 sing} admits neither vertical nor horizontal (nor any other) reduction; its $(p,q)$-web contains a $(-1,2)$-fivebrane, and in that sense it is indeed a ``non-Lagrangian'' theory \cite{Morrison:1996xf, Aharony:1997ju}.

\subsection{$E_1$ SCFT and $SU(2)_0$ gauge theory}\label{subsec: E1 gauge theory}
\begin{figure}[t]
\begin{center}
\subfigure[\small Resolved $E_1$.]{
\includegraphics[height=4cm]{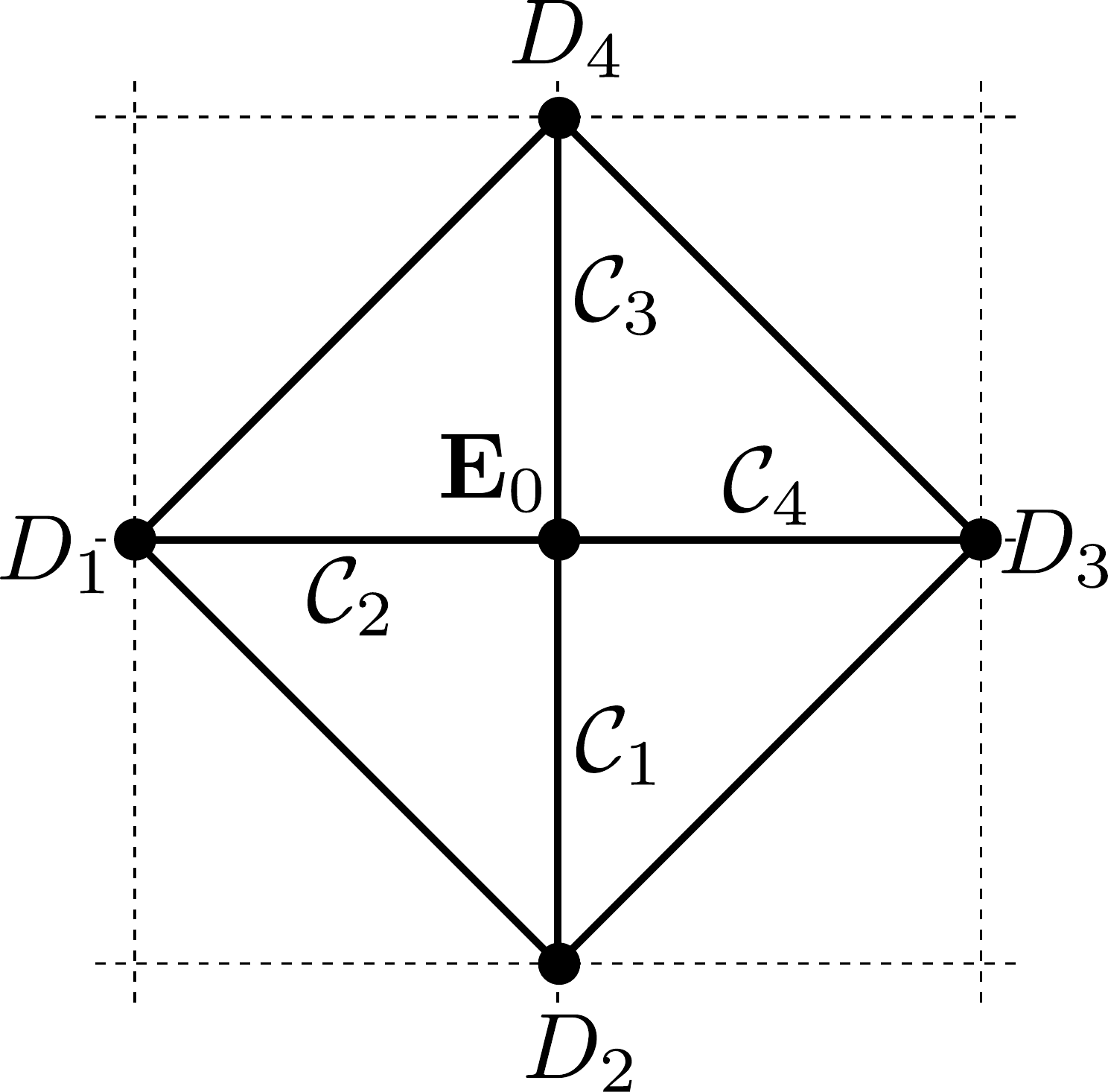}\label{fig:E1 res}}\qquad
\subfigure[\small Profile of $\chi(r_0)$ in IIA.]{
\includegraphics[height=4.5cm]{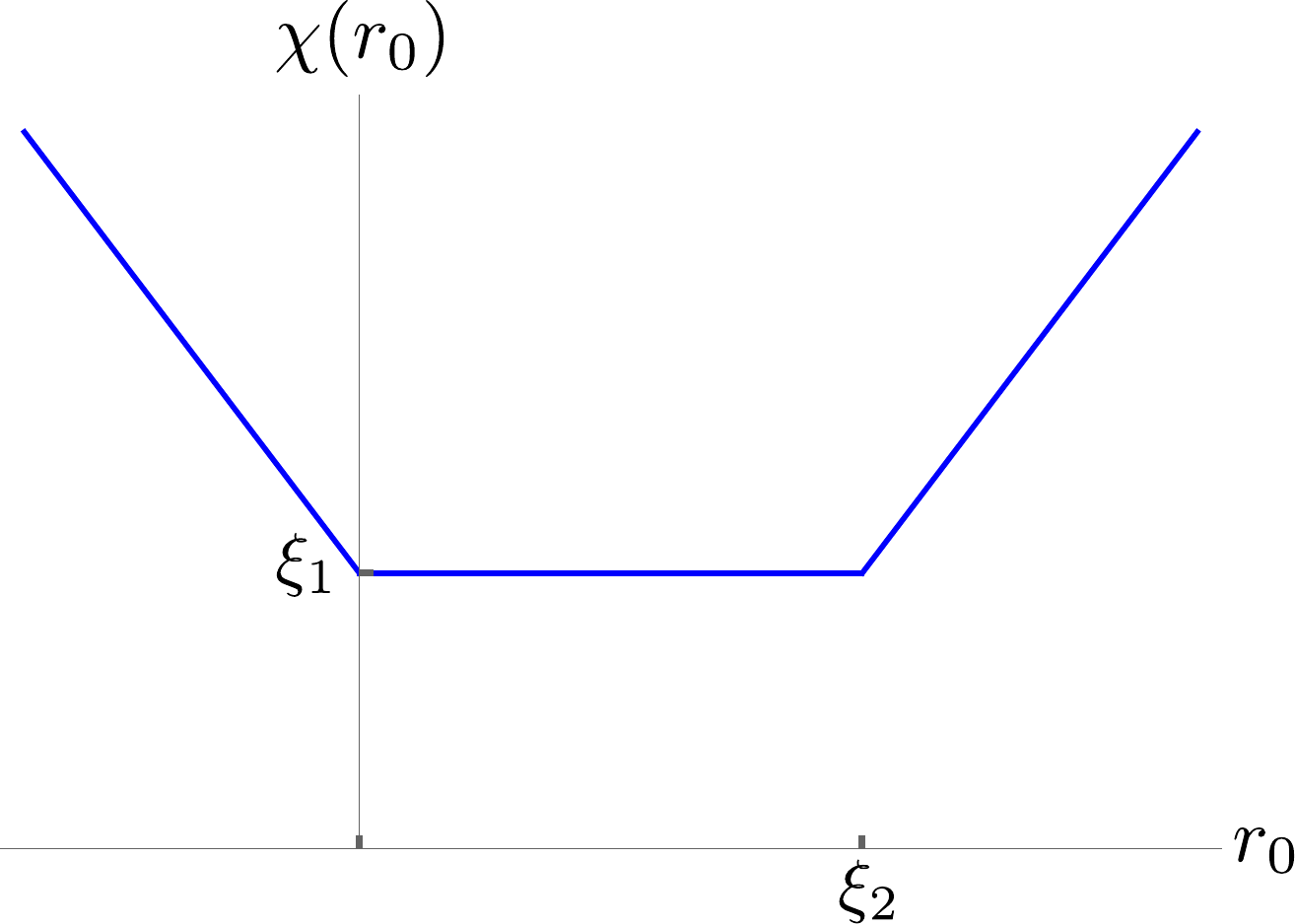}\label{fig:E1 IIA}}
\caption{The resolved $E_1$ singularity and its vertical reduction. We have $\xi_2= 2 \varphi$ in the gauge-theory language, and the $SU(2)$ gauge symmetry is restored at $\xi_2=0$. \label{fig:SU2_0}}
 \end{center}
 \end{figure}

Consider the toric singularity $E_1$, namely the complex cone over $\mathbb{F}_0\cong \mathbb{P}^1\times \mathbb{P}^1$. It has a unique resolution, corresponding to blowing up $\mathbb{F}_0$, as shown in Figure~\ref{fig:E1 res}. The corresponding GLSM description gives:
\be\label{CY2 F0}
\begin{tabular}{l|ccccc|c}
 & $D_1$ &$D_2$& $D_3$&$D_4$& ${\bf E}_0$ & ${\rm vol}(\CC)$\\
 \hline
$\CC_1$    &$1$ &$0$ & $1$& $0$&$-2$& $\xi_1$\\
 $\CC_2$   &$0$ &$1$ & $0$& $1$&$-2$& $\xi_2$\\
    \hline\hline
    $U(1)_M$    &$0$ &$1$ & $0$& $0$&$-1$ &$r_0$
\end{tabular}
\ee
We have four non-compact toric divisors $D_i$, and one compact toric divisor ${\bf E}_0 \cong \mathbb{F}_0$, with the following linear relations:
\be\label{D rels E1}
D_3 \cong D_1~, \qquad D_4 \cong D_2~, \qquad {\bf E}_0\cong -2 D_1 -2 D_2~.
\ee
The curves $\CC$ are given as intersections of pairs of divisors according to:
\be
\CC_1 = D_2\cdot {\bf E}_0~, \qquad
\CC_2 = D_1\cdot {\bf E}_0~, \qquad
\CC_3 = D_4\cdot {\bf E}_0~, \qquad
\CC_4 = D_3\cdot {\bf E}_0~, 
\ee
and we have the linear equivalences $\CC_3\cong \CC_1$ and $\CC_4 \cong \CC_2$. The volume of the curves are non-negative, $\xi_1 \geq 0$ and $\xi_2 \geq 0$. 
%%%%%%%%%
%%%%%%%%%

\paragraph{Geometric prepotential.} Let us first compute the geometric prepotential from M-theory. We may parametrize the K\"ahler cone by:
\be
S=  \mu D_1 + \nu \bE_0~.
\ee
The parameters $(\mu, \nu)$ are related to the FI parameters as:
\be\label{xi to munu su20}
\xi_1 = \mu- 2\nu \geq 0~,\qquad\quad  \xi_2 = -2 \nu \geq 0~.
\ee
One can easily compute the relevant triple-intersection numbers:
\be\label{DDD E1}
D_1^3=0~, \qquad D_1^2 \bE_0= 0~, \qquad D_1 \bE_0^2 = -2~, \qquad \bE_0^3=8~. 
\ee
Here, the result for $D_1^3$ for the triple-intersection of the non-compact divisor $D_1$ depends on a choice of regulator, as we discuss in Appendix~\ref{app: intersect}, and we can choose $D_1^3=0$ for future convenience---see Appendix~\ref{app: internumb E1}. The other intersection numbers are non-ambiguously defined, for any smooth resolution strictly inside the K\"ahler cone (that is, for $\xi_i >0$). This directly gives:
\be\label{F geom su20}
\CF(\nu, \mu) =-{1\ov 6} S^3 = \mu \nu^2 -{4\ov 3} \nu^3~.
\ee
This is a purely geometric result, which gives a prepotential for some dynamical  $U(1)$  5d $\CN=1$ vector multiplet (including the real scalar field $\nu$) in M-theory on the local CY threefold $\wh\MG$. To make contact with the non-abelian gauge-theory description of the low-energy theory, we need to choose a type-IIA string theory reduction.

\paragraph{Type IIA reduction and gauge-theory parameters.}
Let us consider the ``vertical reduction'' of the toric diagram of Fig.~\ref{fig:E1 res}, with the $U(1)_M$ charges indicated in the last line of \eqref{CY2 F0}. The type IIA background is a resolved $A_1$ singularity (that is, the resolved $\C^2/\Z_2$ orbifold), fibered over the $x^9= r_0$ direction. The three vertical points in the toric diagram give rise to two D6-branes wrapping the exceptional $\mathbb{P}^1$ in the resolved $A_1$ singularity, thus realizing a pure $SU(2)$ gauge theory.  

The exceptional  $\mathbb{P}^1$ corresponds to the curve $\CC_1$ in the M-theory description, but its volume $\chi$ varies along $r_0$, in a piecewise-linear fashion, as we explained in section~\ref{subsec: U1Mfib}. By reducing the GLSM  \eqref{CY2 F0}, we easily find the GLSM of the $A_1$ singularity:
\be
\begin{tabular}{l|ccc|c}
 & $t_1$ &$t_2$& $t_0$ &  \\
 \hline
$\mathbb{P}^1$    &$1$ &$1$ & $-2$& $\chi(r_0)$\\
 \end{tabular}~,
\ee
with:
\be
\chi(r_0) = \begin{cases}  \xi_1 + 2r_0- 2\xi_2 \qquad &\text{if}\;\;  r_0 \geq \xi_2\\
\xi_1 &\text{if}\;\;  0\leq r_0 \leq \xi_2\\
\xi_1- 2 r_0 &\text{if}\;\;  r_0 \leq 0
 \end{cases}
\ee
This profile is shown in Figure~\ref{fig:E1 IIA}.
The kinks of $\chi(r_0)$, where the slope jumps by $+2$ from left to right, indicate the $r_0$ positions of the D6-branes wrapped over the exceptional $\mathbb{P}^1$. 

From this IIA picture, we directly read off the gauge theory description, and the map between geometry and field-theory parameters. First of all, when $\xi_2=0$, the two wrapped D6-branes realize a 5d $SU(2)$ gauge group at $r_0=0$. The $SU(2)$ inverse  gauge coupling is then given by the size of the $\mathbb{P}^1$ at $r_0=0$, thus $h_0=\xi_1$ when $\xi_2=0$. The effective CS level \eqref{k eff from IIA} vanishes,~\footnote{For $SU(2)$, the CS level $k_{SU(2)} \in \Z$ mod $2$ in interpreted as a $\Z_2$-valued $\theta$-angle.} and the theory is parity-invariant.

Separating the D6-branes in the $r_0$ direction corresponds to going onto the Coulomb branch. The open strings stretched between the two wrapped D6-branes give us the W-bosons and their superpartners, with mass equal to the separation $\xi_2$. Finally, the instantonic particle on the Coulomb branch corresponds to a D2-brane wrapped over the $\mathbb{P}^1$, at either $r_0=0$ or $r_0=\xi_2$, and its mass is therefore equal to $\xi_1$. In the gauge theory, we have the particle masses:
\be\label{M to xi E1}
 M(W_\alpha) = 2\varphi=\xi_2~, \qquad \qquad  M(\CI_1)= M(\CI_2)= h_0 +2 \varphi= \xi_1~,
\ee
which are mapped to the K{\"a}hler  parameters $\xi_1, \xi_2$ as indicated. While this is a well-known result from the $(p,q)$-web point of view \cite{Aharony:1997bh}, the IIA perspective allow us to perform this computation systematically in complicated examples.~\footnote{By contrast, the particle states in 5-brane $(p,q)$-webs are generally themselves complicated string-webs, which can be rather more subtle to understand \protect\cite{Aharony:1997bh}.} 
The masses of the $SU(2)$ instantons were computed from \eqref{F U2 Nf}, following the prescription of section~\ref{subsec: instanton}.  Using \eqref{xi to munu su20}, the relations \eqref{M to xi E1} are equivalent to:
\be
\mu= h_0~, \qquad\qquad
\nu= - \varphi~.
\ee
Plugging these relations into the geometric prepotential \eqref{F geom su20}, we indeed recover the $SU(2)$ prepotential:
 \be\label{FSU2}
 \CF= h_0 \varphi^2+ {4\ov 3 } \varphi^3~.
 \ee
As another consistency check of these identification, let us also compute the string tension. From the field theory, we have:
\be\label{T su20}
T= \d_\varphi \CF = 2\varphi (h_0 + 2\varphi)~.
\ee
From the IIA geometry, a string is a D4-brane wrapped over the $\mathbb{P}^1$ and stretched between the wrapped D6-branes. Its tension is therefore given by $T=\xi_1 \xi_2$, in agreement with \eqref{T su20}. In M-theory, this corresponds to an M5-brane wrapping the exceptional divisor $\bE_0$, whose volume is indeed $\xi_1 \xi_2$.

Notice that in this simple case, along the magnetic wall $T=0$, at least one BPS particle is getting massless, also for the deformed theory.

Finally, note that we could also have chosen the S-dual ``horizontal reduction'' of the $E_1$ singularity to type IIA. From the symmetry of the toric diagram, it is clear that this gives an isomorphic gauge theory description in terms of pure $SU(2)$, but with the roles of $\xi_1$ and $\xi_2$ interchanged.

\subsection{$\t E_1$ SCFT and $SU(2)_\pi$ gauge theory}\label{subsec: E1 sing}
\begin{figure}[t]
\begin{center}
\subfigure[\small Resolved $\t E_1$.]{
\includegraphics[height=4cm]{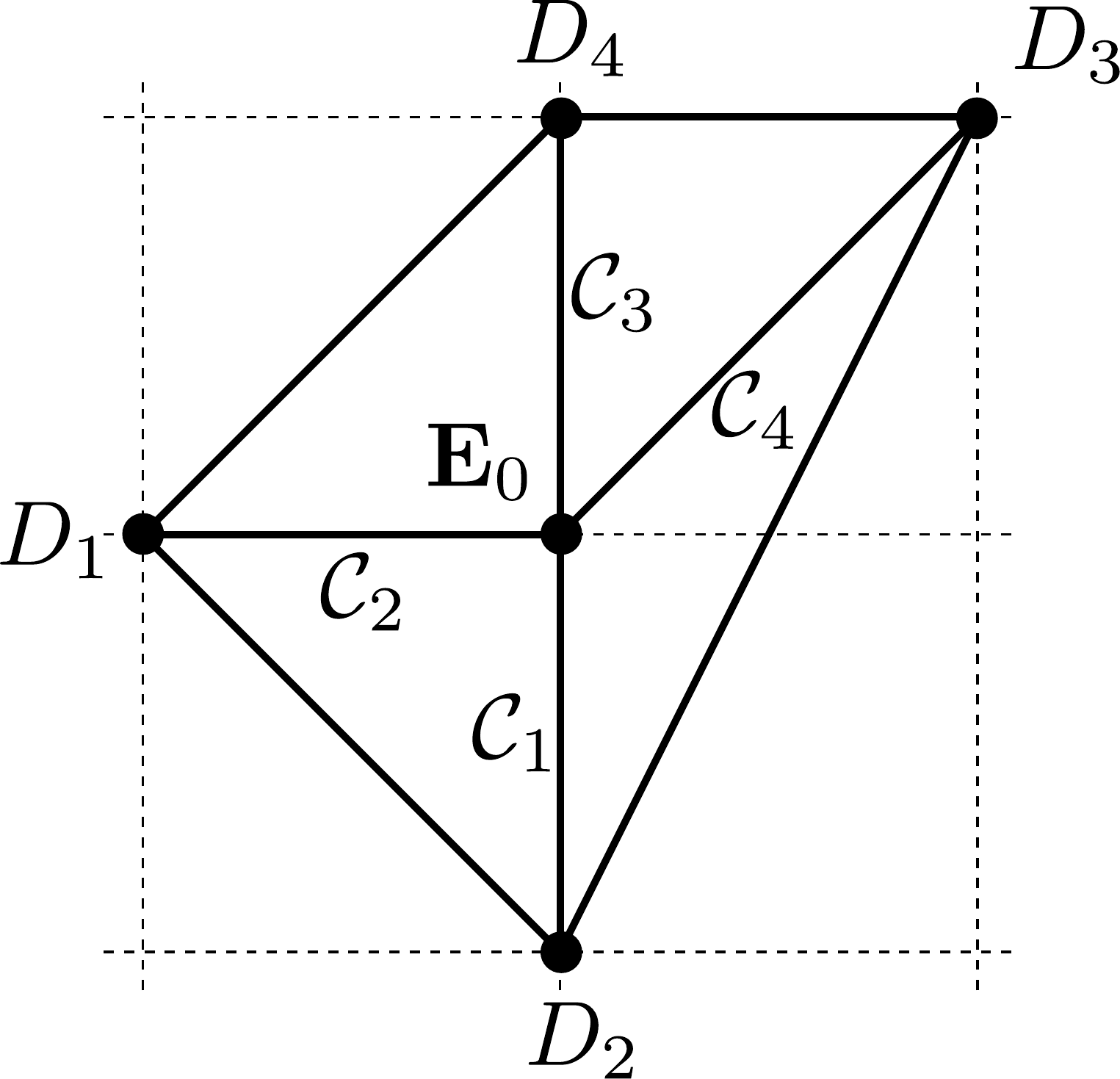}\label{fig:E1t res a}}\quad
\subfigure[\small Profile of $\chi(r_0)$ in IIA.]{
\includegraphics[height=4cm]{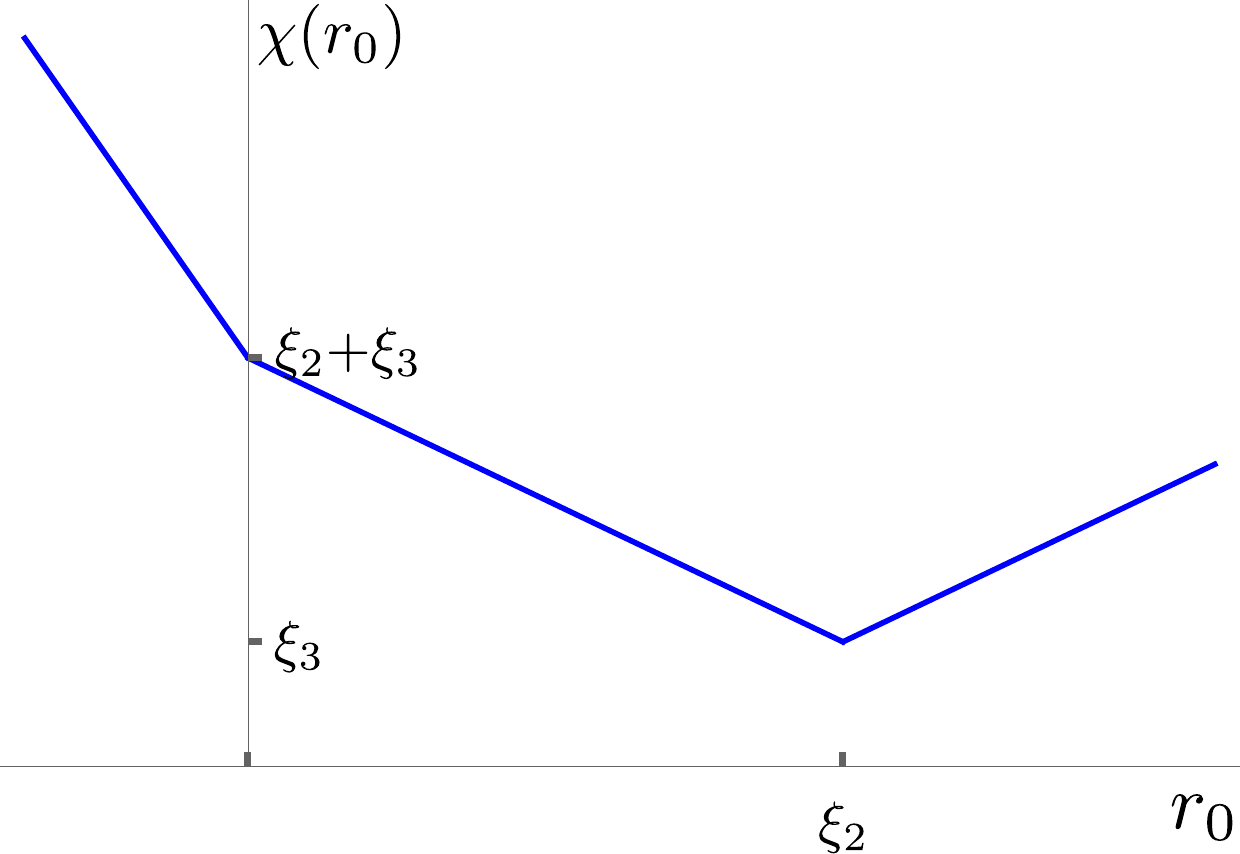}\label{fig:E1t IIA}}\quad
\subfigure[\small Flop on $\CC_3$.]{
\includegraphics[height=4cm]{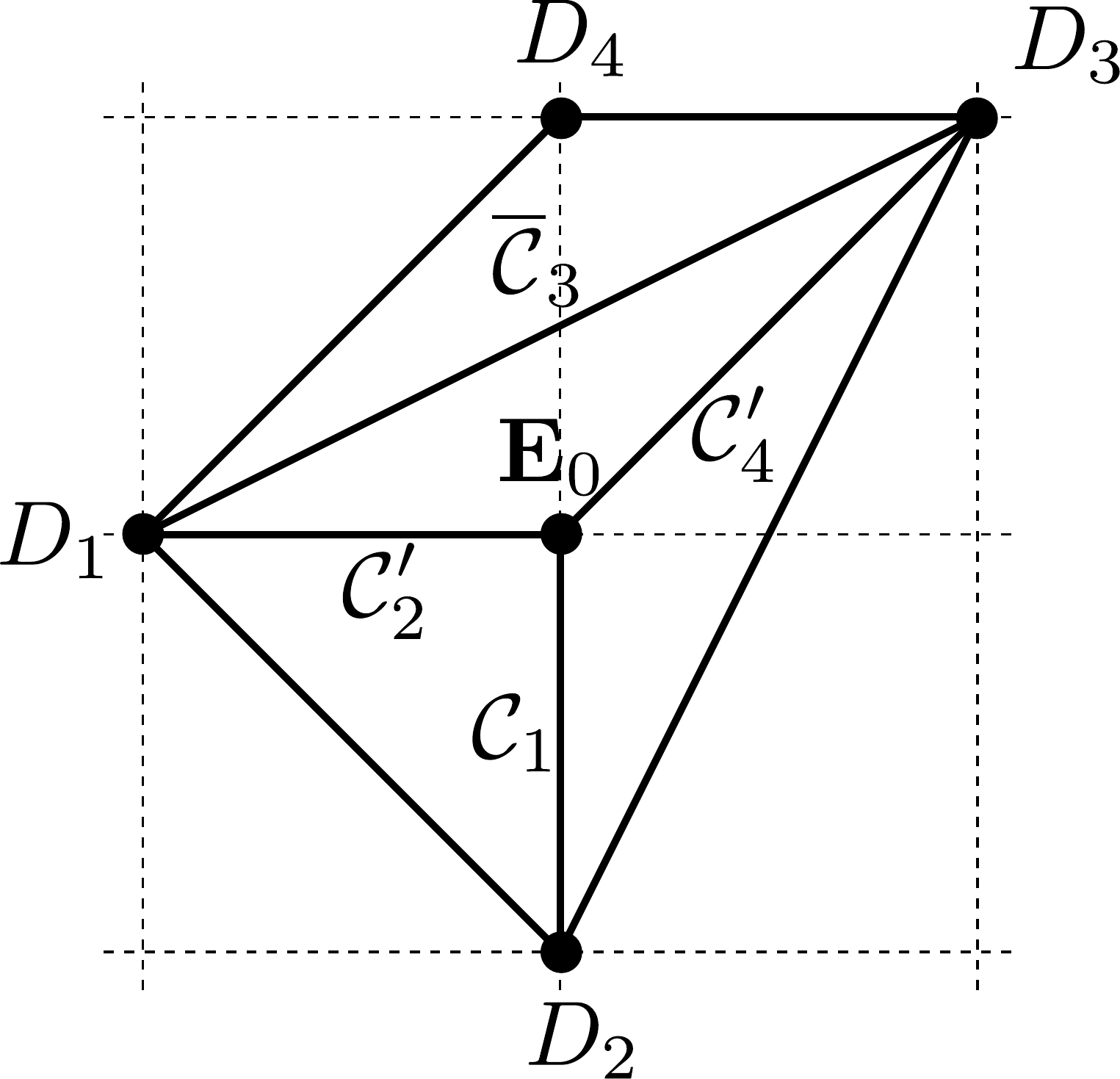}\label{fig:E1t res b}}
\caption{The resolved $\t E_1$ singularity and its vertical reduction. 
By flopping $\CC_3$ and sending ${\rm vol}(\b\CC_3)$ 
to infinity (that is, $\xi_3 \rightarrow -\infty$), one obtains the isolated $E_0$ theory. \label{fig: resdP1}}
 \end{center}
 \end{figure}

Our next example is the $\t E_1$ theory, corresponding to the local del Pezzo surface $dP_1$---that is, $\mathbb{P}^2$ blown up at one smooth point. The toric diagram of the threefold is shown in Figure~\ref{fig:E1t res a}, with  the blown-up $\P^1$ corresponding to $\CC_3$.
The GLSM can be chosen to be:
\be\label{CY2 dP1}
\begin{tabular}{l|ccccc|c}
 & $D_1$ &$D_2$& $D_3$&$D_4$& ${\bf E}_0$ & ${\rm vol}(\CC)$\\
 \hline
$\CC_2$    &$0$ &$1$ & $0$& $1$&$-2$& $\xi_2$\\
 $\CC_3$   &$1$ &$0$ & $1$& $-1$&$-1$& $\xi_3$\\
    \hline\hline
    $U(1)_M$    &$0$ &$1$ & $0$& $0$&$-1$ &$r_0$
\end{tabular}
\ee
The linear relations amongst toric divisors are:
\be
D_3 \cong D_1~, \qquad D_4\cong -D_1 + D_2~, \qquad \bE_0 \cong - D_1 -2 D_2~.
\ee
The linear relation amongst curves are:
\be\label{lin rel C su2pi}
\CC_4\cong \CC_2~, \qquad \CC_1 \cong \CC_2 + \CC_3~.
\ee
In particular, the curves $\{\CC_2, \CC_3$ can be chosen as generators of the Mori cone.

\paragraph{Geometric prepotential.} Let us consider: 
\be
S=  \mu D_1 + \nu \bE_0~.
\ee
with:
\be\label{xi to munu su2pi}
\xi_2 = - 2 \nu\geq 0~, \qquad\qquad \xi_3= \mu- \nu\geq 0~.
\ee
Here, the inequalities are the ones that define the K\"ahler chamber for the resolution of Figure~\ref{fig:E1t res a}.
For that resolution, the relevant intersection numbers are $D_1^3=0$, $D_1^2 \bE_0= 0$, $D_1 \bE_0^2 = -2$ and $\bE_0^3=8$, and therefore the prepotential is the same as in \eqref{F geom su20}, namely:
\be
\CF(\nu, \mu) =\mu \nu^2 -{4\ov 3} \nu^3~.
\ee
The $\t E_1$ geometry admits another resolution, shown in \ref{fig:E1t res b}, which admits no type-IIA reduction. More details on the intersection numbers are given in Appendix~\ref{app: t E1}.

%%%%%%%%%
%%%%%%%%%

\paragraph{Gauge theory description.} Consider the vertical reduction of Fig.~\ref{fig:E1t res a}. We again obtain a pure $SU(2)$ theory, from D6-branes wrapping the resolved $A_1$ singularity, but the IIA profile is different from the $E_1$ case, as shown in Fig.~\ref{fig:E1t IIA}. We now have:
\be
\chi(r_0) = \begin{cases}  r_0- \xi_2+ \xi_3 \qquad &\text{if}\;\; \xi_2\leq r_0\\
-r_0+ \xi_2+ \xi_3 &\text{if}\;\;  0\leq r_0 \leq \xi_2\\
- 3 r_0+ \xi_2 + \xi_3 &\text{if}\;\;  r_0 \leq 0
 \end{cases}
\ee
According to \eqref{k eff from IIA}, we naively have the Chern-Simons level:
\be
k_{SU(2)}= 1~,
\ee
by abuse of notation.
More precisely, we would obtain a $U(2)$ theory with level $k=1$. In the $SU(2)$ theory, this is interpreted as having a non-zero $\theta$-angle valued in $\pi_4(SU(2))= \Z_2$. Thus, the gauge-theory phases of the $E_1$ and $\t E_1$ theories correspond to $\theta= 0$ and $\theta= \pi$, respectively \cite{Intriligator:1997pq}.
We denote the $\theta=\pi$ theory by $SU(2)_\pi$. Note that the $\theta$-angle, like the CS level, breaks 5d parity. Interestingly, when considering the ``vertical reduction,'' parity in the gauge theory corresponds to a reflection $r_0\rightarrow -r_0$ direction. Indeed the main difference between the two IIA profiles for $E_1$ and $\t E_1$ in Figures~\ref{fig:E1 IIA} and \ref{fig:E1t IIA}, respectively, in the non-abelian limit $\xi_2=0$,  is that one is parity-symmetric and the other is not.

The masses of the W-boson and instanton particles are:
\be\label{M to xi E1t}
 M(W_\alpha) = 2\varphi=\xi_2~, \qquad
 M(\CI_1)= h_0 +3 \varphi= \xi_2 + \xi_3~, \qquad 
  M(\CI_2)= h_0 + \varphi= \xi_3~.
 \ee
Comparing to \eqref{xi to munu su2pi}, we see that $\mu= h_0$ and $\nu=-\varphi$, therefore the gauge theory prepotential is given by \eqref{FSU2}, and it is unaffected by the non-zero $\theta$-angle. On the other hand, the spectrum of instantonic particles $\CI_a$ is different.~\footnote{To obtain \protect\eqref{M to xi E1t}, we use the prescription \protect\eqref{I mass heuristic} for a $U(2)_{1}$ theory.}  That spectrum was first obtained in the $(p,q)$-web language in type IIB \cite{Aharony:1997bh}, and the type IIA perspective of course gives the same answer. Note also that we have:
\be
M(\CI_1) = M(\CI_2)+ M(W_\alpha)~.
\ee
In M-theory, the particles $\CI_1, \CI_2$ and $W_\alpha$ correspond to M2-branes wrapped over $\CC_1$, $\CC_3$ and $\CC_2$, respectively. Then the marginal bound state relation $\CI_1\cong \CI_2 + W_\alpha$ corresponds to the second linear relation in \eqref{lin rel C su2pi} amongst the curves.

Since the prepotential for $SU(2)_\pi$ is the same as for $SU(2)_0$, the string tension is also the same and given by \eqref{T su20}. The IIA prescription for the wrapped D4-brane tension gives:
\be
T= \int_{0}^{\xi_2} \chi(r_0) dr_0 = \xi_2 \Big(\half \xi_2+ \xi_3\Big)~,
\ee
which indeed reproduces \eqref{T su20} upon using \eqref{M to xi E1t}.

The second resolution of the $\t E_1$ singularity, shown in Fig.~\ref{fig:E1t res b}, can be obtained from the one in Fig.~\ref{fig:E1t res a} by flopping the curve $\CC_3$. This corresponds to sending the GLSM parameter $\xi_3$ through $\xi_3=0$ and taking it negative. In particular, one can consider the limit $\xi_3 \rightarrow - \infty$, which leads to the $E_0$ singularity. Since this would correspond to $h_0 <0$, we cannot describe this flow in the $SU(2)_\pi$ language. Its end point is the SCFT known as $E_0$, corresponding to a collapsing $\P^2$ \cite{Morrison:1996xf}.

\paragraph{Magnetic wall.} For this theory the magnetic wall, $T=0$ splits into two loci
\be
(I) \colon \xi_2 = 0 \quad\text{and}\quad (II)\colon \half \xi_2+ \xi_3 =0\,.
\ee
From the spectrum of BPS particle masses we see that the region $(I)$ indeed coincides with the hard-wall where the W-boson becomes massless. The region $(II)$, on the other hand, is not part of the K\"ahler chamber (a). We note that the BPS instanton $\CI_3$ can become massless at $\xi_3 = 0$, away from any magnetic wall, giving rise to a traversable instantonic wall. The theory in this case flows to a chamber that does not have a gauge theory interpretation. 

\subsection{$E_2$ SCFT and $N_f=1$ $SU(2)$ gauge theory}\label{subsec: E2 sing}
Consider the $E_2$ singularity. It has five distinct resolutions, shown in Figure \ref{fig:E2 res}. Only the first three of them admit a vertical reduction to type IIA. The resolutions (b), (c), (d) admit a horizontal reduction, giving us an S-dual gauge theory description. The resolution (e) admit no gauge theory description. We will focus here on the vertical reduction. 
%%%%%%%%%
 \begin{figure}[t]
\begin{center}
\subfigure[\small ]{
\includegraphics[height=2.5cm]{dP2_res_a.pdf}\label{fig:dP2 a}}\quad
\subfigure[\small ]{
\includegraphics[height=2.5cm]{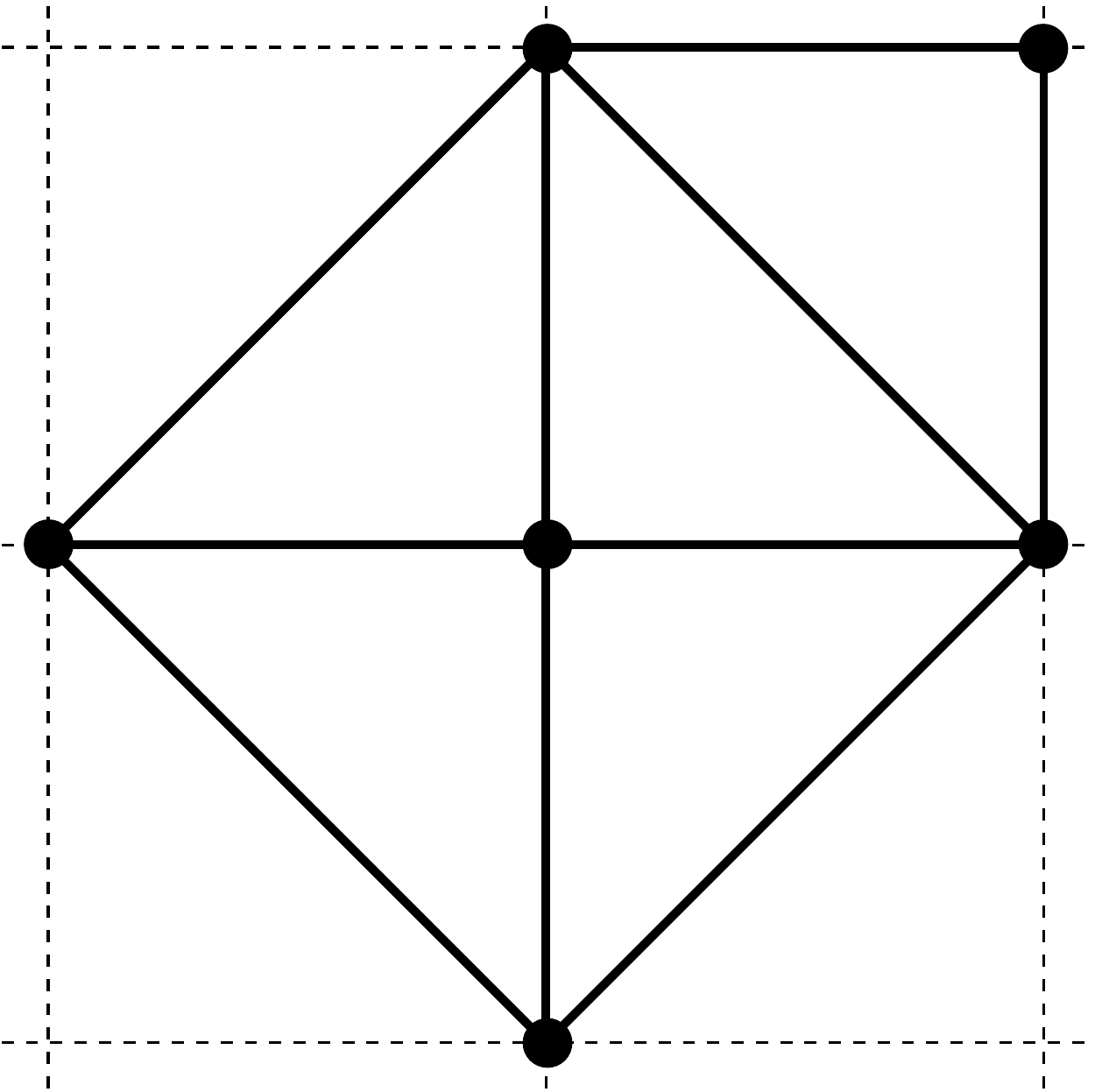}\label{fig:dP2 b}}\quad
\subfigure[\small ]{
\includegraphics[height=2.5cm]{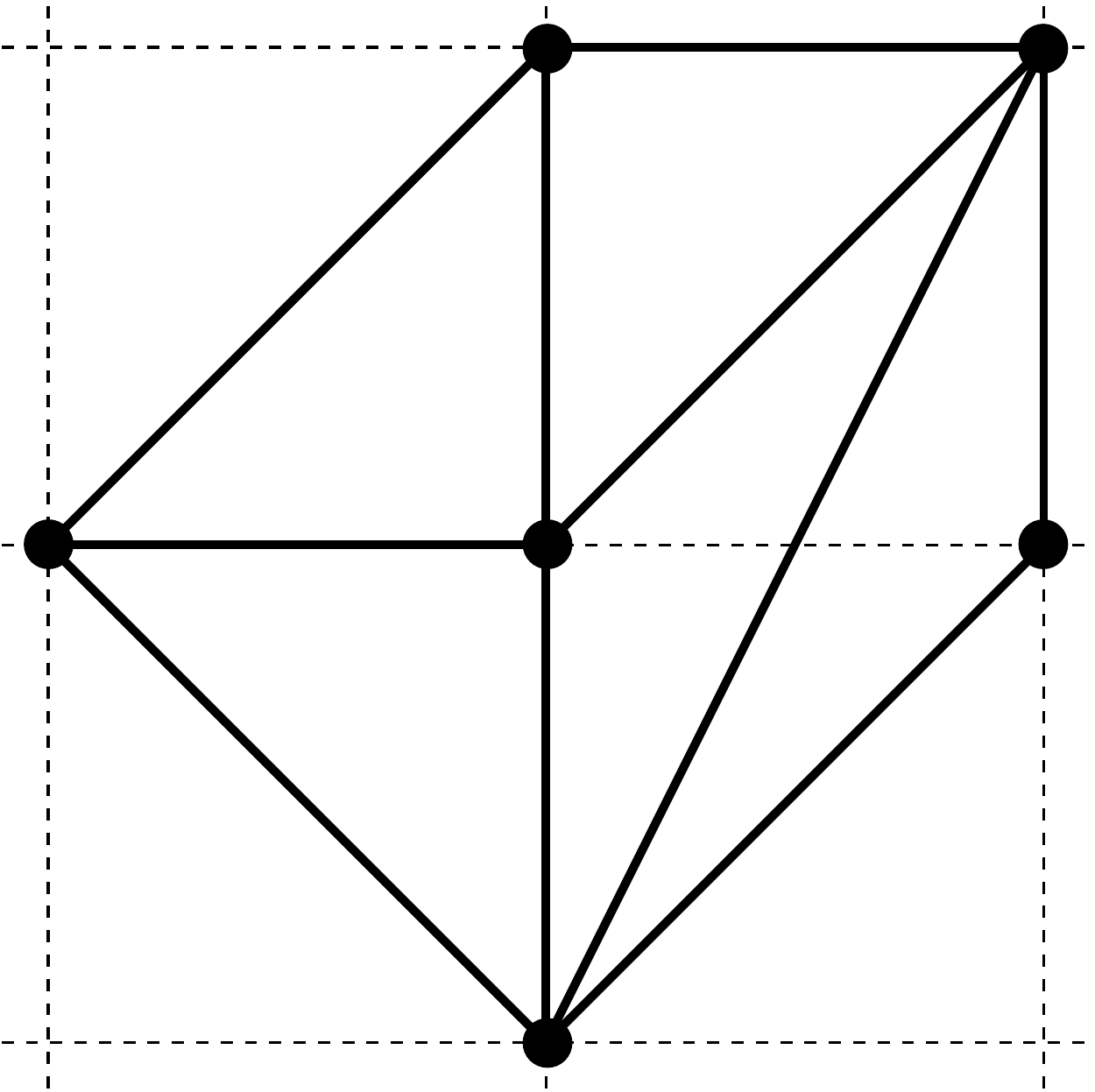}\label{fig:dP2 c}}\quad
\subfigure[\small ]{
\includegraphics[height=2.5cm]{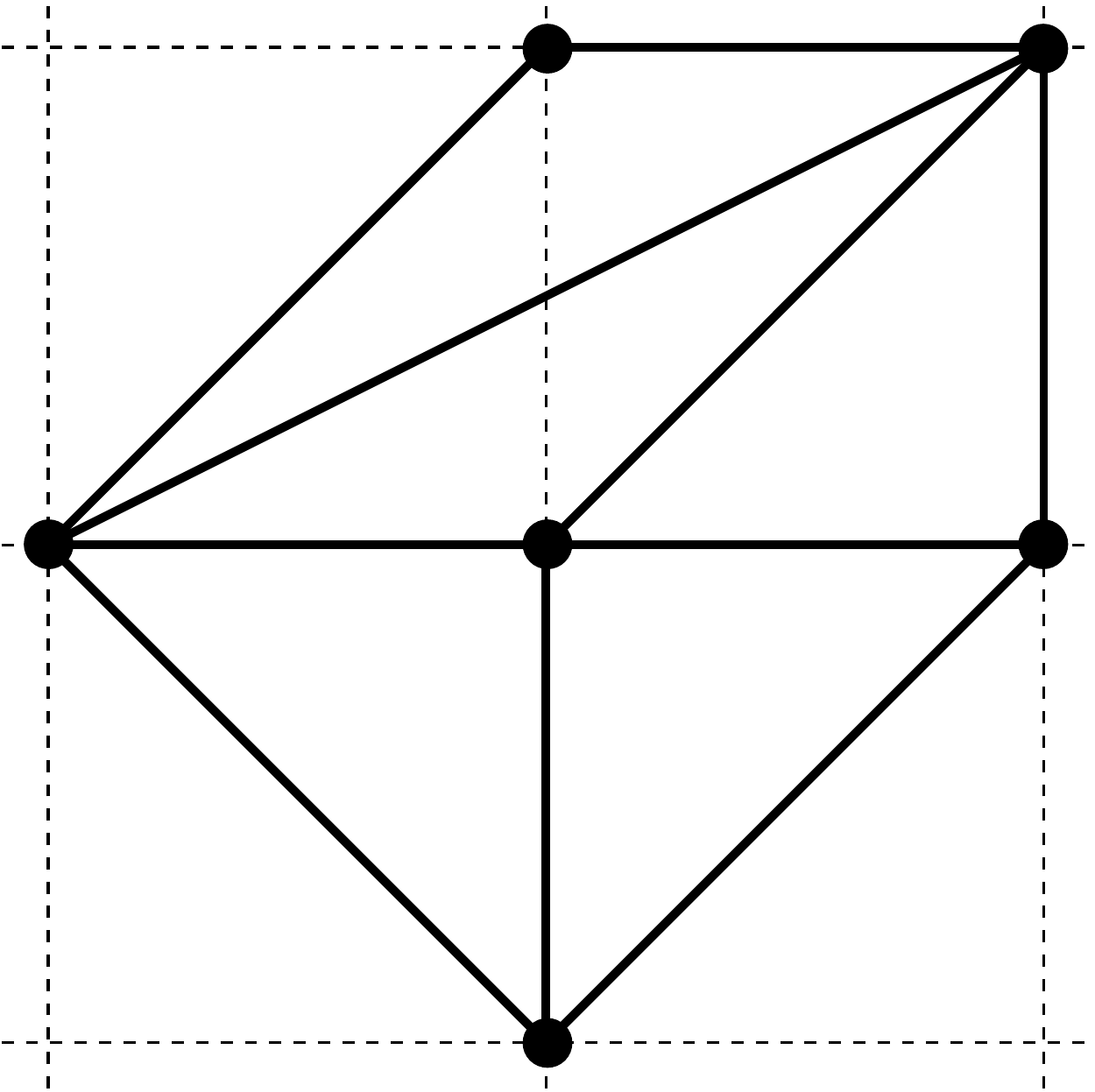}\label{fig:dP2 d}}\quad
\subfigure[\small ]{
\includegraphics[height=2.5cm]{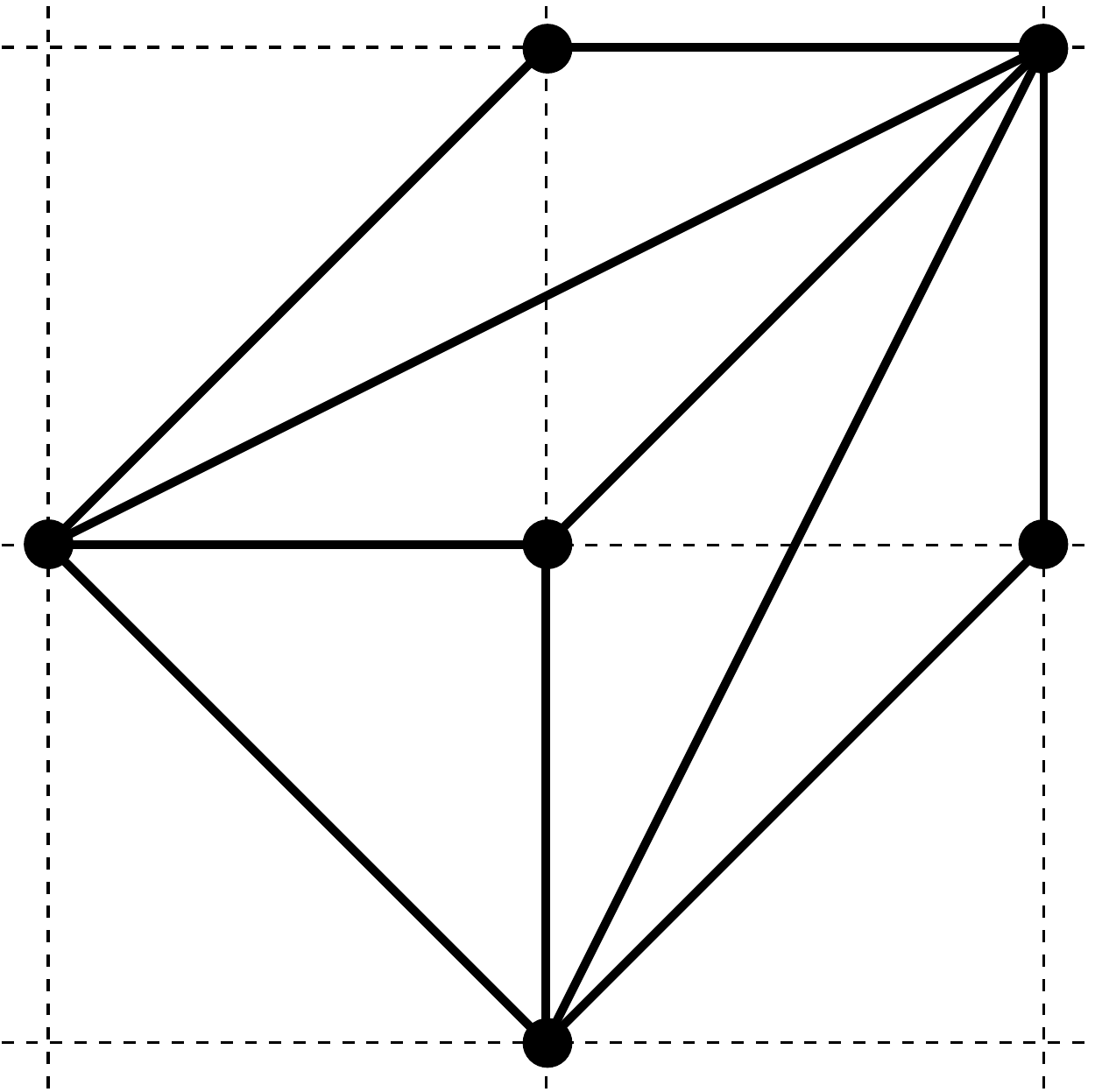}\label{fig:dP2 e}}
\caption{The five resolutions of the $E_2$ singularity.\label{fig:dP2 res}}
 \end{center}
 \end{figure} 
 %%%%%%%%%%%
 The GLSM describing the $E_2$ singularity can be chosen to be:
\be\label{intersect dP2a}
\begin{tabular}{l|cccccc|c}
 & $D_1$ &$D_2$& $D_3$&$D_4$& $D_5$ & $\bE_0$& \\
 \hline
$\CC_1$    &$1$&$0$ &$1$ & $0$& $0$&$-2$& $\xi_1$\\
$\CC_2$    &$0$&$1$ &$0$ & $1$& $0$&$-2$& $\xi_2$\\
$\CC_5$    &$0$&$0$ &$1$ & $1$& $-1$&$-1$& $\xi_5$\\
\hline\hline
$U(1)_M$&$0$&$1$ &$0$ & $0$& $0$&$-1$& $r_0$
 \end{tabular}
\ee
 Here, the rows are labelled by the curves $\CC_1$, $\CC_2$ and $\CC_5$ in resolution (a), as shown in Figure~\ref{fig:E2 res a}. They have positive volume if $\xi_1>0$, $\xi_2>0$ and $\xi_5>0$. By allowing more general values of the ``FI parameters'' $\xi_1$, $\xi_2$ and $\xi_5$, the same GLSM also describes all the resolutions shown in Figure~\ref{fig:dP2 res}.  Note also the linear relations amongst toric divisors:
\be
D_1 \cong D_3+ D_5~, \qquad D_2\cong D_4 + D_5~, \qquad \bE_0 \cong -2 D_1 -2 D_2 + D_5~,
\ee
  which hold independently of the partial resolution.
 
 The resolutions (a), (b) and (c) give rise to an $SU(2)$ gauge theory with a single fundamental flavor ($N_f=1$). Depending on the values of the parameters, we have three distinct chambers on the Coulomb branch. The gauge-theory prepotential \eqref{F SU2 Nf} with $N_f=1$ takes the values:
 \be\label{E2 F 3 chambers}
 \CF =\begin{cases} \left(h_0-{m\ov 2}\right) \varphi^2 + {7\ov 6} \varphi^3- \half m^2 \varphi- {1\ov 6} m^3  \quad & \text{if}\;\;\; \varphi +m>0,\;\;  -\varphi+m <0,\\
(h_0-m)\varphi^2 + {4\ov 3} \varphi^3- {1\ov 3}m^3 \quad & \text{if}\;\; \;\varphi +m>0,\;\;  -\varphi+m >0,  \\
h_0\varphi^2 + {4\ov 3} \varphi^3\quad & \text{if}\;\; \;\varphi +m<0,\; \; -\varphi+m <0. 
  \end{cases}
 \ee
 As we will show, these three Coulomb-branch chambers correspond to the first three resolutions in Fig.~\ref{fig:dP2 res}:
 \bea
 {\rm(a)}\qquad \leftrightarrow\qquad  \varphi +m>0~, \quad  -\varphi+m <0~, \cr
 {\rm(b)}\qquad \leftrightarrow\qquad  \varphi +m>0~, \quad  -\varphi+m >0~, \cr
  {\rm(c)}\qquad \leftrightarrow\qquad  \varphi +m<0~, \quad  -\varphi+m <0~. \cr
 \eea
Recall that we have $\varphi>0$, while the real mass $m$ can take both signs.

\paragraph{Geometric prepotential.} The M-theory prepotential $\CF= -{1\ov 6} S^3$ depends on the partial resolution we consider. Let us define:
\be
S= \mu_1 D_1 + \mu_5 D_5 +\nu \bE_0~.
\ee
The parameters $\mu, \nu$ are related to the FI parameters in \eqref{intersect dP2a} by:
\be\label{xi to mu nu E2}
\xi_1 = \mu_1 - 2\nu~, \qquad \xi_2 = -2 \nu~, \qquad \xi_5= - \mu_5 - \nu~.
\ee
By a direct computation of the intersection numbers (see Appendix~\ref{app: E2 sing} for more details), 
we find the following result:~\footnote{Here, as before, we defined the ``non-compact'' intersection numbers in a convenient fashion.}
\bea\label{F geom su2Nf1}
&\CF_{(a)}& =&\;  -{7\ov 6}\nu^3 +(\mu_1 + \half \mu_5)\nu^2+ \half \mu_5^2 \nu -{1\ov 6} \mu_5^3~, \cr
&\CF_{(b)}& =&\;  -{4\ov 3}\nu^3 +\mu_1\nu^2 -{1\ov 3} \mu_5^3~, \cr
&\CF_{(c)}& =&\;   -{4\ov 3}\nu^3 +(\mu_1+\mu_5)\nu^2~, 
\eea
for the first three resolutions. This matches perfectly with the field-theory result \eqref{E2 F 3 chambers}, given the map of parameters:
\be\label{mu to gauge E2}
\mu_1 = h_0 -m~, \qquad \mu_5= m~, \qquad \nu= - \varphi~,
\ee
To derive \eqref{mu to gauge E2}, we again turn to the IIA reduction.

%%%%%%%%%
 \begin{figure}[t]
\begin{center}
\subfigure[\small Chamber (a). The flavor D6-brane lies between the two gauge D6-branes.]{
\includegraphics[height=4cm]{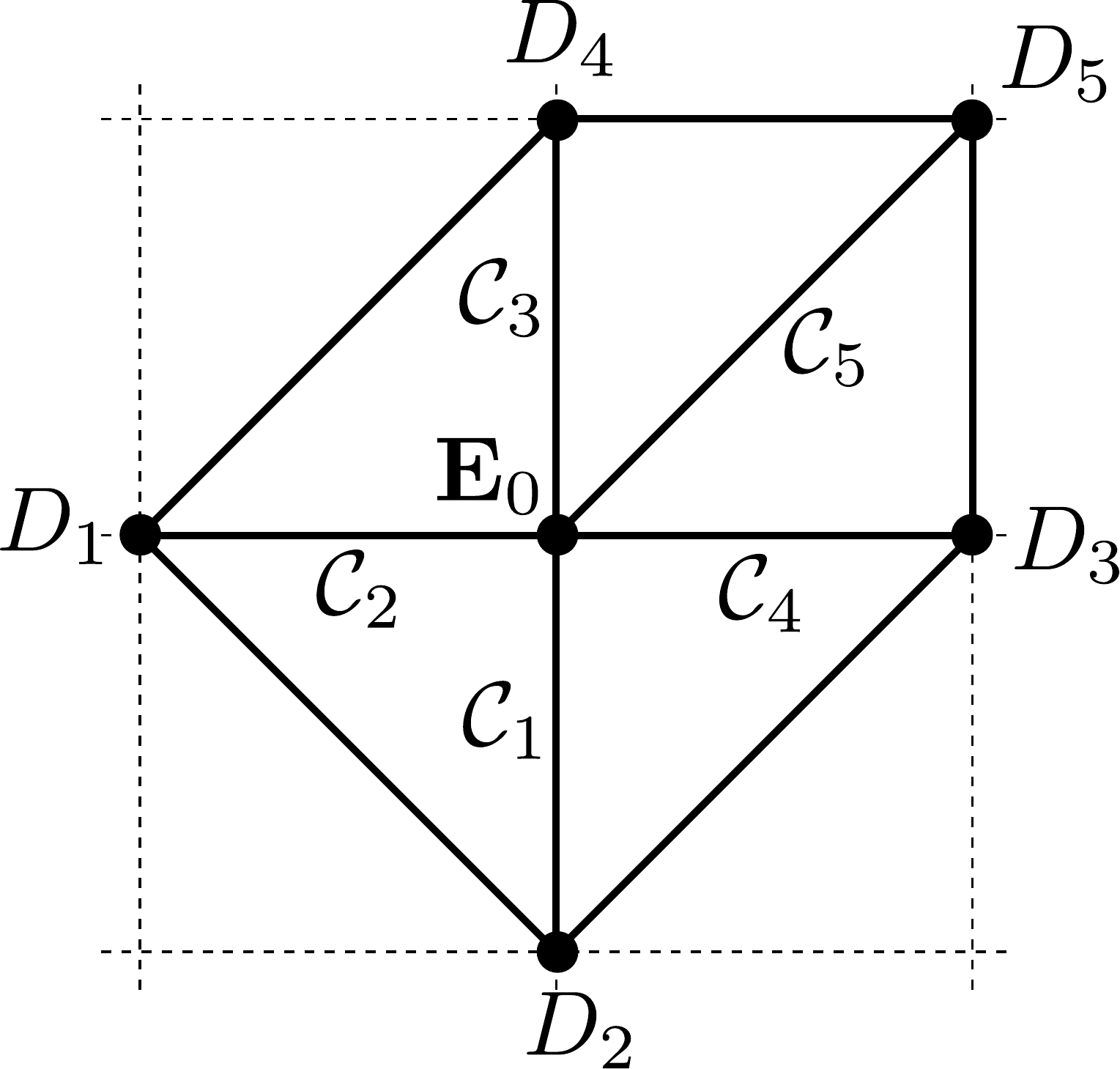}\label{fig:E2 res a}\qquad\qquad\qquad
\includegraphics[height=4cm]{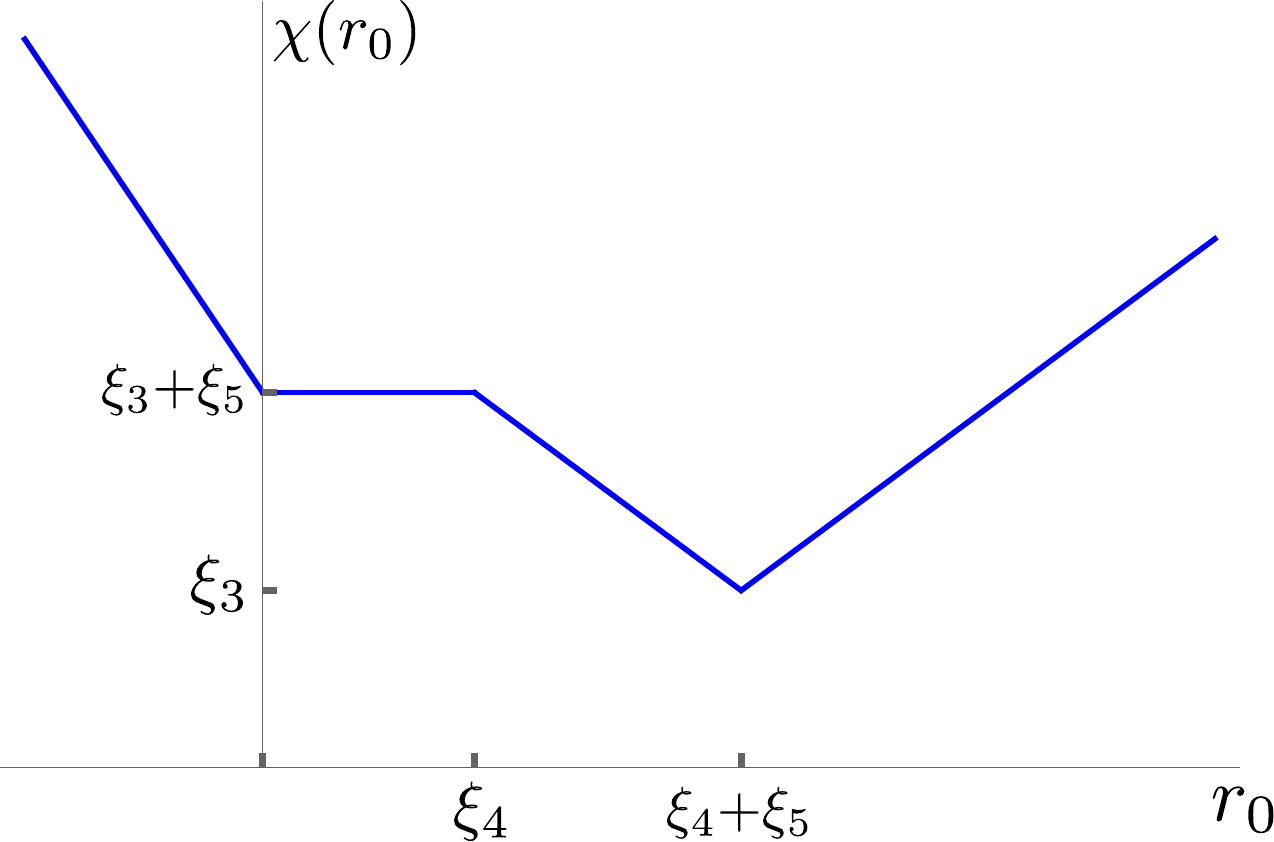}}\\
\subfigure[\small Chamber (b). The flavor D6-brane lies to the right of the gauge branes.]{
\includegraphics[height=4cm]{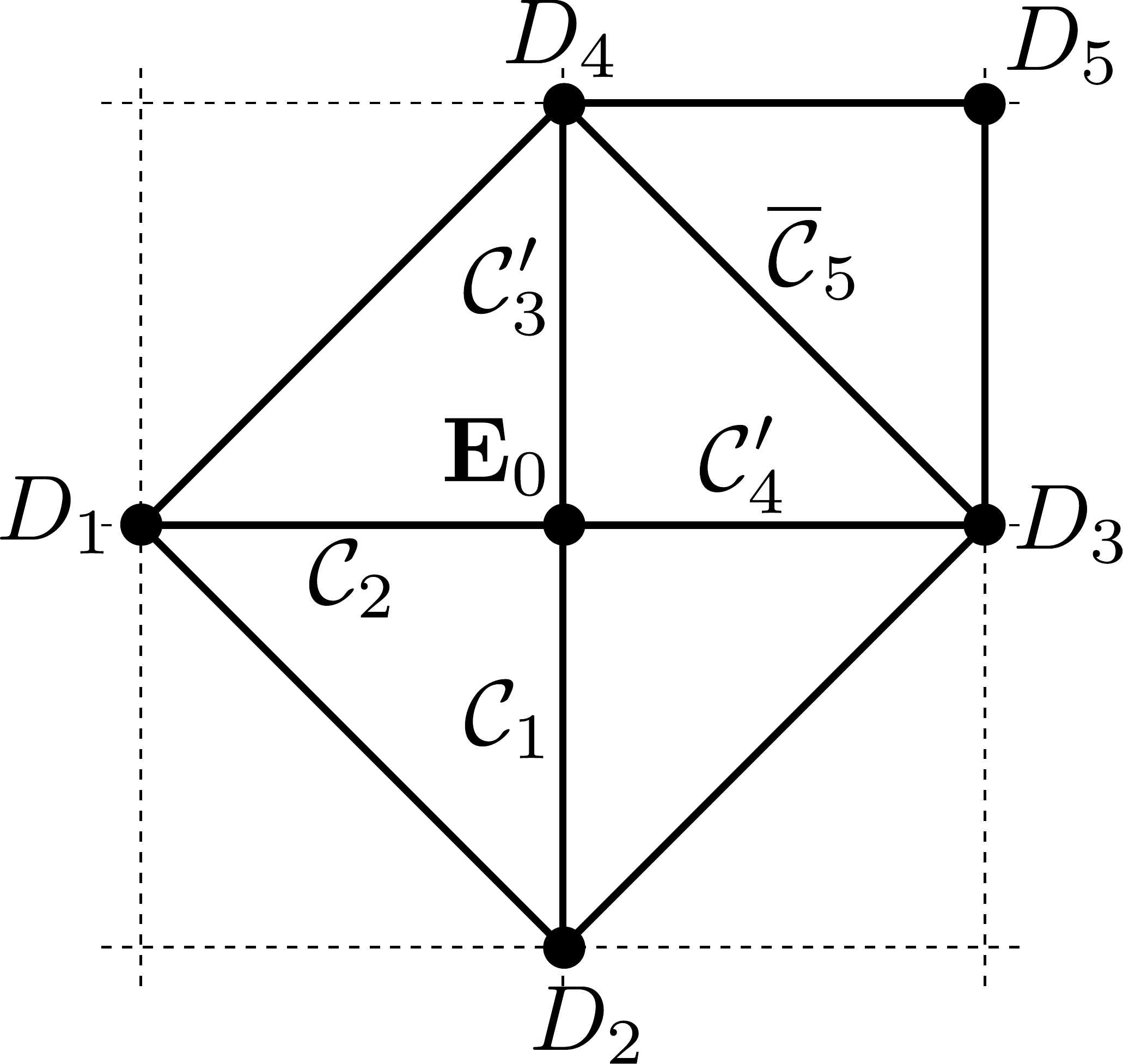}\label{fig:E2 res b}\qquad\qquad\qquad
\includegraphics[height=4cm]{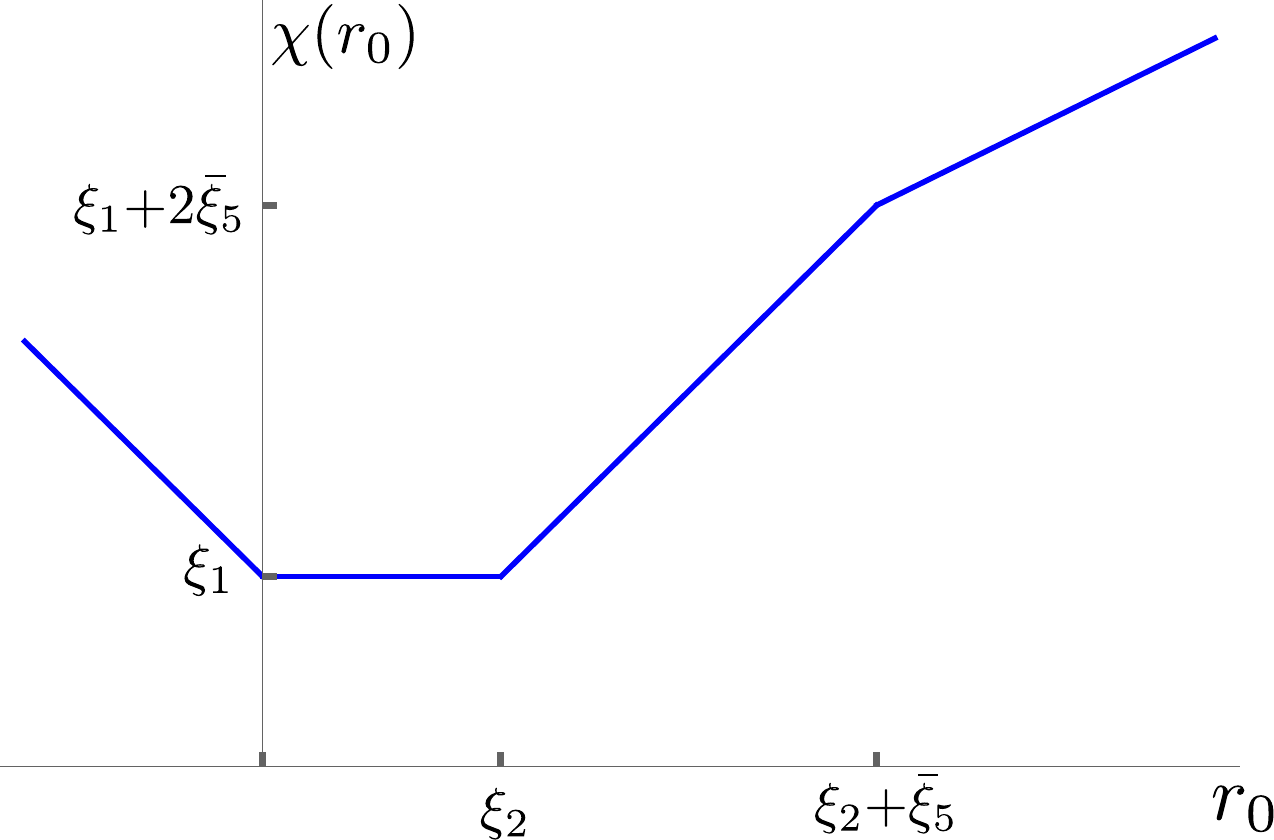}}\\
\subfigure[\small Chamber (c). The flavor D6-brane lies to the left of the gauge branes. ]{
\includegraphics[height=4cm]{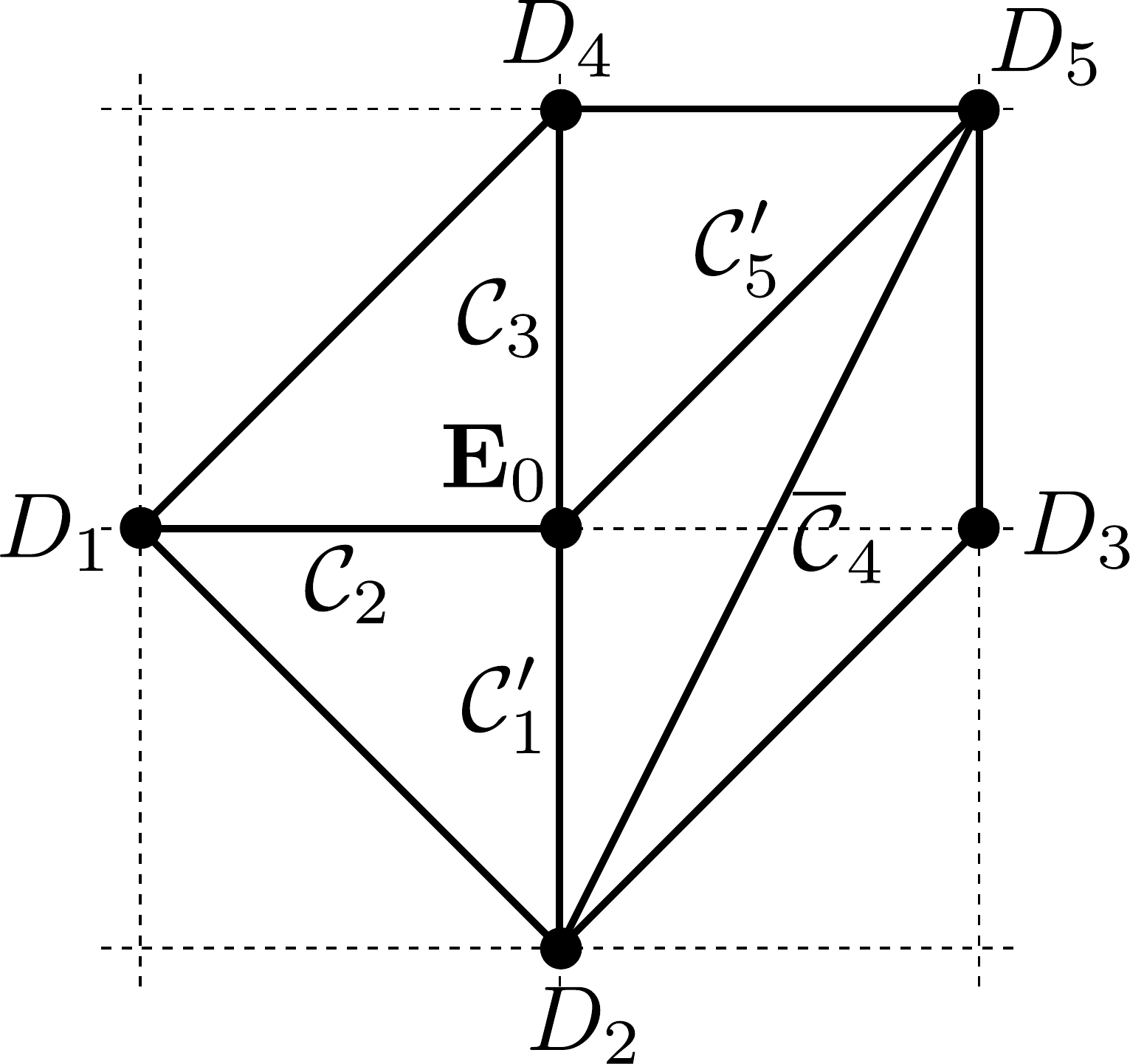}\label{fig:E2 res c}\qquad\qquad\qquad
\includegraphics[height=4cm]{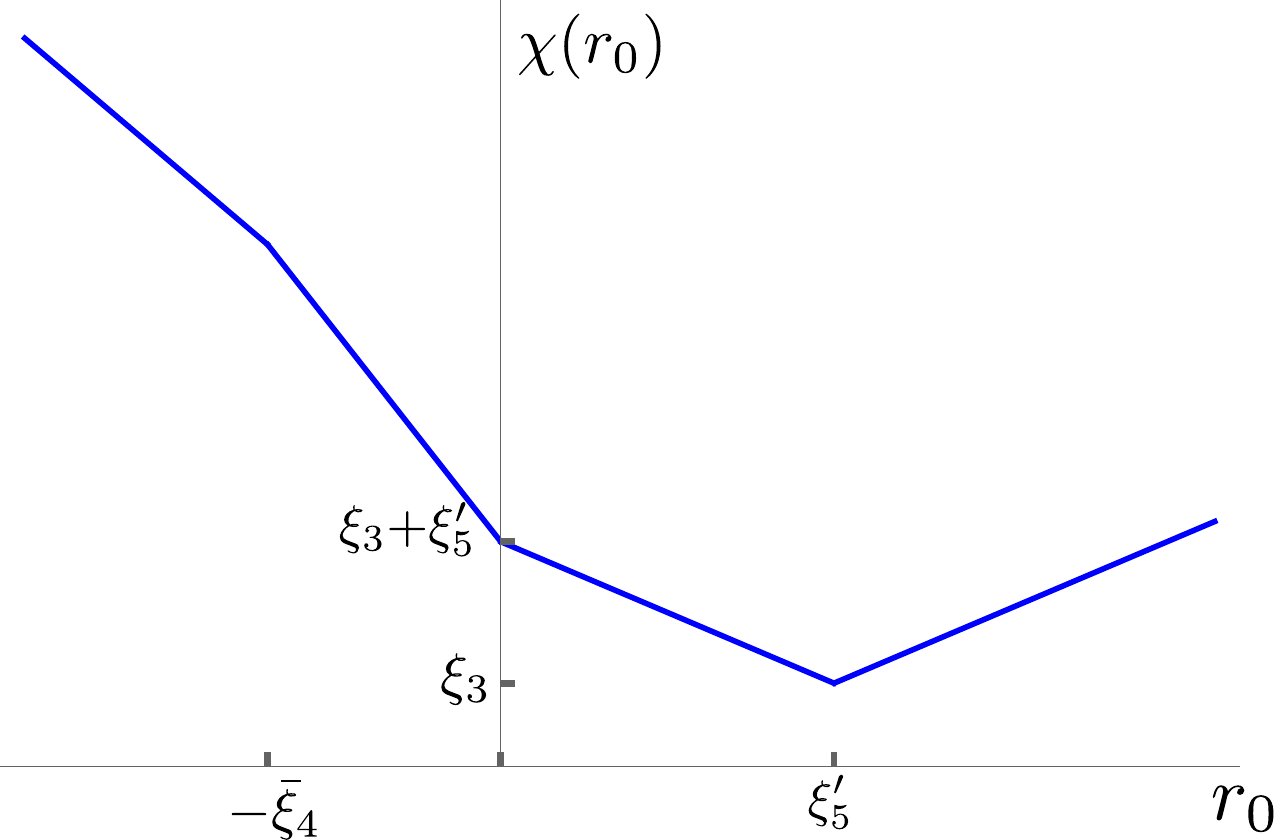}}
\caption{The three resolutions of the $E_2$ singularity with a ``vertical reduction,'' and the corresponding IIA profiles for $\chi(r_0)$. \label{fig:E2 res}}
 \end{center}
 \end{figure} 
 %%%%%%%%%%%

\paragraph{Comparing to the results obtained using the IMS prepotential.}
At this point, it may be useful to make some comments related to our new prescription for the prepotential.~\footnote{We thank the SciPost referee for this suggestion.} The IMS prepotential \cite{Intriligator:1997pq} in this theory reads:
\be\label{expl IMS F}
 \CF_{\rm IMS}=\t h_0 \varphi^2+{4\ov 3 } \varphi^3 + {1\ov 12} \left|\varphi+ m\right|^3+ {1\ov 12} \left|-\varphi+ m\right|^3 
= \begin{cases}
\t h_0 \varphi^2 + {7\ov 6} \varphi^3- \half m^2 \varphi~, \\
\left(\t h_0-{m\ov 2}\right)\varphi^2 + {4\ov 3} \varphi^3~,   \\
\left(\t h_0+{m\ov 2}\right)\varphi^2 + {4\ov 3} \varphi^3~,
  \end{cases}
  \ee
in the three chambers, up to the constant term. It is clear that, modulo the constant term, this agrees with \eqref{E2 F 3 chambers} upon redefining the gauge coupling parameter:
\be\label{rede h0t}
\t h_0 = h_0 - {m\ov 2}~.
\ee
Thus, the prepotential \eqref{expl IMS F} can be matched to the M-theory prepotential \eqref{F geom su2Nf1}, but only by using a different map between geometry and field theory parameters, with $\t h_0= \mu_1+\half \mu_5$ instead of $h_0= \mu_1+ \mu_5$ identified as the gauge coupling. Precisely this alternative parameterization was found in \cite{Aharony:1997bh} when matching the string tension computed as $T=\d_\varphi \CF_{\rm IMS}$ with the area of the face on the $(p,q)$-web.~\footnote{It is instructive to compare section III.D of \protect\cite{Aharony:1997bh} to our formalism in more detail. Their parameters $(1/g_0^2, m, \phi)$ for $SU(2), N_f=1$ correspond to our $(\t h_0, m, 2\varphi)$, respectively. The brane configuration in their Fig.16(a) and (b) correspond to our resolution (a) and (b), respectively. The `simple' and `subtle' quark states in their discussion correspond to $\CH_+$ and $\CH_-$, respectively, in our discussion around \protect\eqref{rel E2 a Masses} below.  They then identify the `simplest' instanton mass as $\t h_0 + 2 \varphi- {m\ov 2}$, in our notation; given the shift \protect\eqref{rede h0t}, this agrees with $M(\CI_2)$ which we find in \protect\eqref{rel E2 a Masses ii} below; the lightest instanton, $\CI_1$, would correspond to a D-string along the top D5-brane on the $(p,q)$-web.}
Similar comments hold for all the other examples below. 

\paragraph{Gauge-theory chamber (a).}  Consider the vertical reduction of resolution (a), which is shown in Fig.~\ref{fig:E2 res a}. In this geometry, we have the relations:
\be
\CC_1 \cong \CC_3 + \CC_5~, \qquad \CC_2 \cong \CC_4 + \CC_5~,
\ee
amongst the curves, and therefore $\{\CC_3, \CC_4, \CC_5\}$ generates the Mori cone. We denote by $\xi_i$ the volume of $\CC_i$, and therefore we have $\xi_3= \xi_1- \xi_5$ and $\xi_4= \xi_2- \xi_5$. 
The IIA reduction gives again a resolved $A_1$ singularity fibered over $r_0$. One finds:
\be\label{chi a dP2}
\text{(a)}\; : \qquad \chi(r_0) = \begin{cases}  
r_0+ \xi_3- \xi_4- \xi_5\qquad &\text{if}\quad  \xi_4 + \xi_5 \geq r_0 \cr 
-r_0+ \xi_3+\xi_4+\xi_5\qquad &\text{if}\quad \xi_4 \leq r_0\leq  \xi_4 + \xi_5 \cr 
 \xi_3+\xi_5\qquad &\text{if}\quad 0 \leq r_0\leq  \xi_4  \cr 
-2r_0  +\xi_3+\xi_5\qquad &\text{if}\quad  r_0\leq 0 \cr 
\end{cases}
\ee
as shown in Fig.~\ref{fig:E2 res a}. We see that there are now two wrapped D6-branes at $r_0=0$ and $r_0=\xi_4+\xi_5= \xi_2$, corresponding to the gauge group, and one non-compact ``flavor'' D6-brane at $r_0= \xi_4$. (At the location of the flavor brane, the slope of $\chi(r_0)$ decreases by $-1$ from left to right.) The ``effective CS level'' \eqref{k eff from IIA} is now given by:
\be
k_{SU(2),\, {\rm eff}} =\half~.
\ee
In our conventions, this is interpreted as a bare CS level $k=1$ plus the contribution $-\half$ from the hypermultiplet.

The open string between the gauge D6-branes provide the W-boson, and the open strings stretching from the gauge branes to the flavor brane provide the hypermultiplet modes. Therefore, we find:
\be\label{rel E2 a Masses}
M(\CH_+) = \varphi + m = \xi_4~, \qquad
M(\CH_-) = \varphi - m = \xi_5~, \qquad
M(W_\alpha) =2\varphi = \xi_4+ \xi_5~.
\ee
In particular, we see that, in this particular resolution, the hypermultiplet comes from M2-branes wrapped over $\CC_4$ and $\CC_5$ in M-theory. (This is to be contrasted with the rather more complicated description of the hypermultiplet states in the $(p,q)$-web \cite{Aharony:1997bh}.)
The masses of the instantonic particles are given by the volumes of the D2-branes wrapped at the $r_0$ locations of the gauge D6-branes. We find:
\be\label{rel E2 a Masses ii}
M(\CI_1) = h_0 + \varphi = \xi_3~, \qquad 
M(\CI_2) = h_0 +2 \varphi -m = \xi_3+ \xi_5~.
\ee
In particular, $\CI_1$ comes from an M2-brane wrapped over $\CC_3$. Together with \eqref{xi to mu nu E2}, the relations \eqref{rel E2 a Masses}-\eqref{rel E2 a Masses ii} imply the map \eqref{mu to gauge E2} between $(\mu_1, \mu_5, \nu)$ and the gauge-theory variables, as anticipated. The gauge-theory formula for the instanton masses in \eqref{rel E2 a Masses ii} is derived from \eqref{F U2 Nf} with the $U(2)$ CS level $k=1$.

As a consistency check, let us consider the string tension. The gauge-theory prepotential  implies:
\be\label{T a E2}
T^{(a)} = \d_\varphi \CF= {7\ov 2}\varphi^2 + (2h_0-m)\varphi - \half m^2~.
\ee
From the IIA D4-brane description, we find:
\be
T^{(a)} = \int_0^{\xi_4+ \xi_5} \chi(r_0) dr_0 =(\xi_3+ \xi_5)\xi_4  + \Big(\xi_3  + \half \xi_5\Big) \xi_5~.
\ee
Plugging in \eqref{rel E2 a Masses}-\eqref{rel E2 a Masses ii}, this precisely reproduces the gauge-theory result \eqref{T a E2}.

\paragraph{Magnetic wall.} In this case, we see that the tension does not factorize into a product of masses of BPS particles. We obtain:
\be\label{teeens}
T^{(a)} = (\xi_3 + \xi_5)(\xi_4 + \xi_5) - \tfrac12 \xi_5^2 = M(\CI_1) M(W_\alpha) - \tfrac12 (M(\CH_-))^2
\ee
A magnetic wall would correspond to solving $T=0$. Solving this equations for $\xi_5$ gives
$\xi_5 = - \xi_3 - \xi_4 \pm \sqrt{\xi_3^2+\xi_4^2}$, which is always negative in chamber (a). Hence, there is no magnetic wall in this chamber (except at the origin, corresponding to the SCFT).  

\paragraph{Gauge-theory chamber (b).} It is instructive to consider the other resolutions with a vertical reduction, as a consistency check. The toric diagram of resolution (b) and its IIA profile are shown in Figure~\ref{fig:E2 res b}. 
We have:
\be\label{chi b dP2}
\text{(b)}\; : \qquad \chi(r_0) = \begin{cases}  
r_0+ \xi_1- \xi_2+ \b\xi_5\qquad &\text{if}\quad  \xi_2 + \b\xi_5 \geq r_0 \cr 
2r_0+ \xi_1-2\xi_2 \qquad &\text{if}\quad \xi_2 \leq r_0\leq  \xi_2 + \b\xi_5 \cr 
 \xi_1 \qquad &\text{if}\quad 0 \leq r_0\leq  \xi_2  \cr 
-2r_0 + \xi_1 \qquad &\text{if}\quad  r_0\leq 0 \cr 
\end{cases}
\ee
Here, the parameters $\xi_1$ and $\xi_2$ are the same as above, and $\b\xi_5= - \xi_5$ (while $\xi_3'= \xi_3+\xi_5$ and $\xi_4'= \xi_4+\xi_5$). Indeed, resolution~(b) is obtained from resolution~(a) by flopping the curve $\CC_5$ to $\b\CC_5= - \CC_5$. In the gauge theory, this corresponds to changing the sign of the hypermultiplet mode $\CH_-$. Note that the flavor brane is to the right of the gauge branes along the $r_0$ direction.
We now have:
\be\label{rel E2 b Masses}
M(\CH_+) = \varphi + m = \xi_2+ \b\xi_5~, \quad
M(\CH_-) = -\varphi + m =\b\xi_5~, \quad
M(W_\alpha) =2\varphi = \xi_2~,
\ee
and:
\be\label{rel E2 b Masses ii}
M(\CI_1) =M(\CI_2) = h_0 -m +2 \varphi = \xi_1~.
\ee
The instanton mass is again obtained from a $U(2)_{1}$ description. We also have the string tension:
\be
T^{(b)} = \xi_1 \xi_2 = 2\varphi (h_0 - m +2 \varphi) = \d_\varphi\CF~.
\ee

Note that we can take the limit $m \rightarrow + \infty$ within this chamber. In the field theory, this corresponds to integrating out the hypermultiplet with positive mass, which gives the $SU(2)_0$ gauge theory.~\footnote{Since we start from a theory with $k_{\rm eff}= \half$, integrating out the hypermultiplet with positive mass gives rise to a theory with $k_{\rm eff}=0$.} This is clearly also what happens in the geometry: the $E_2$ geometry reduces to the $E_1$ geometry as size of $\b\CC_5$ is sent to infinity. (This is clear from the toric diagram of Fig.~\ref{fig:E2 res b}, since that limit effectively decouples the divisor $D_5$ from the rest of the geometry as we ``decompactify'' $\b\CC_5$.)

%\paragraph{Magnetic wall.} In this case the string tension clearly factors
%\be
%T^{(b)} = M(W_\alpha) (M(\CI_1) + M(\CI_2))/2
%\ee
%The magnetic wall at $T=0$ splits into two irreducible components. One coincides with the hard wall $M(W_\alpha) =0$, the other is where the two instantons become massless. This is consistent with the remark above that the $E_2$ geometry reduces to the $E_1$ case.

\paragraph{Gauge-theory chamber (c).}  The third resolution and its associated IIA profile are shown in Figure~\ref{fig:E2 res c}. The relations amongst curves are:
\be
\CC_1'= \CC_3+ \CC_2~, \qquad \CC_5'\cong \CC_2~.
\ee
We now have: 
\be\label{chi c dP2}
\text{(c)}\; : \qquad \chi(r_0) = \begin{cases}  
r_0+ \xi_3- \xi_5'+ \b\xi_5\qquad &\text{if}\quad   \xi_5' \geq r_0 \cr 
-r_0+ \xi_3+\xi_5' \qquad &\text{if}\quad 0 \leq r_0\leq   \xi_5' \cr 
 -3 r_0+\xi_3+ \xi_5' \qquad &\text{if}\quad -\b\xi_4 \leq r_0\leq 0  \cr 
-2r_0 + \xi_3+ \b\xi_4+ \xi_5' \qquad &\text{if}\quad  r_0\leq -\b\xi_4 \cr 
\end{cases}
\ee
Here we introduced the new parameters:
\be
\b\xi_4= - \xi_4~, \qquad \qquad \xi_5' = \xi_2~.
\ee
corresponding to the positive sizes of the curves $\b\CC_4$ and $\CC_5'$, respectively. This resolution can be obtained from resolution (a) by flopping the curve $\CC_4$.  Now the flavor brane is located to the left of the gauge branes.
We find:
\be\label{rel E2 c Masses}
M(\CH_+) =- \varphi - m = \b\xi_4~, \quad
M(\CH_-) = \varphi - m =\b\xi_4+ \xi_5'~, \quad
M(W_\alpha) =2\varphi = \xi_5'~,
\ee
and:
\be\label{rel E2 c Masses ii}
M(\CI_1) = h_0 +3 \varphi = \xi_3+ \xi_5'~, \qquad
M(\CI_2) = h_0 + \varphi = \xi_3~.
\ee
The string tension reads:
\be
T^{(c)} = \int_0^{\xi_5'} \chi(r_0) dr_0= \xi_5'\Big(\xi_3+ \half \xi_5'\Big)= 2\varphi (h_0 + 2\varphi)= \d_\varphi \CF~.
\ee

In this chamber, we can take the limit $m \rightarrow  -\infty$, corresponding to sending the size of $\b\CC_4$ to infinity. This leads to the parity-violating $SU(2)_\pi$ gauge theory discussed above (with ``$k_{\rm eff}=1$''), and correspondingly we recover the $\t E_1$ geometry in that limit. Note that the fact that we obtain distinct geometries depending on the sign of $m$ is a manifestation of the parity anomaly carried by this single hypermultiplet.

\medskip

In summary, we have verified that the three complete resolutions of the $E_2$ singularity with an allowed vertical reduction can be precisely matched to the gauge theory description, with all the mass parameters turned on. The perturbative RG flows also have a simple geometric interpretation as partial resolutions of the $E_2$ singularity which results in either the $E_1$ or the $\t E_1$ singularity.

\subsection{$E_3$ SCFT and $N_f=2$ $SU(2)$ gauge theory}\label{subsec:E3 SCFT}
\begin{figure}[ht]
%\vspace{-10pt}
\begin{center}
\subfigure[\small]{
\includegraphics[width=2.3cm]{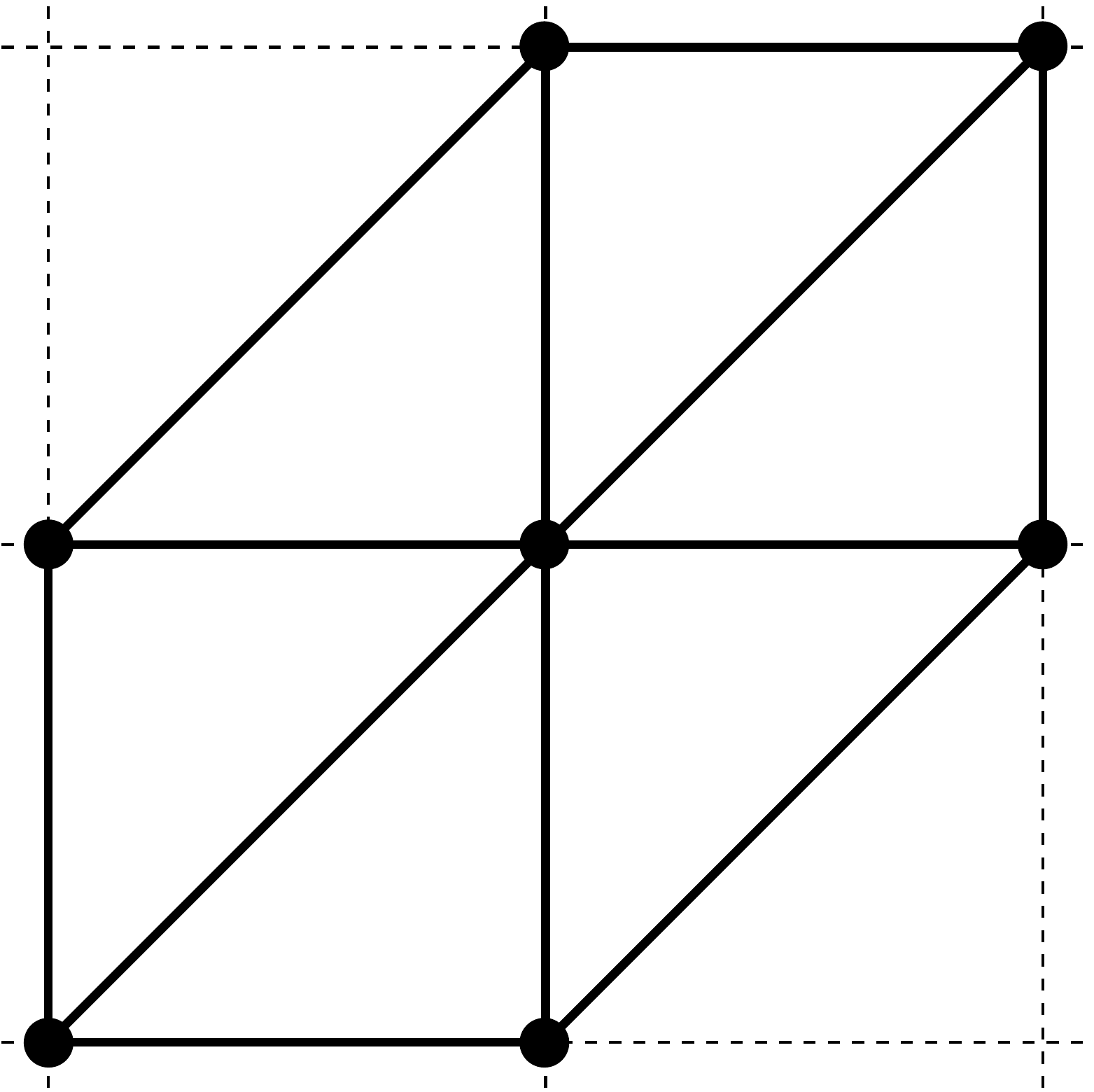}\label{fig:dp3-a}}\,
\subfigure[\small]{
\includegraphics[width=2.3cm]{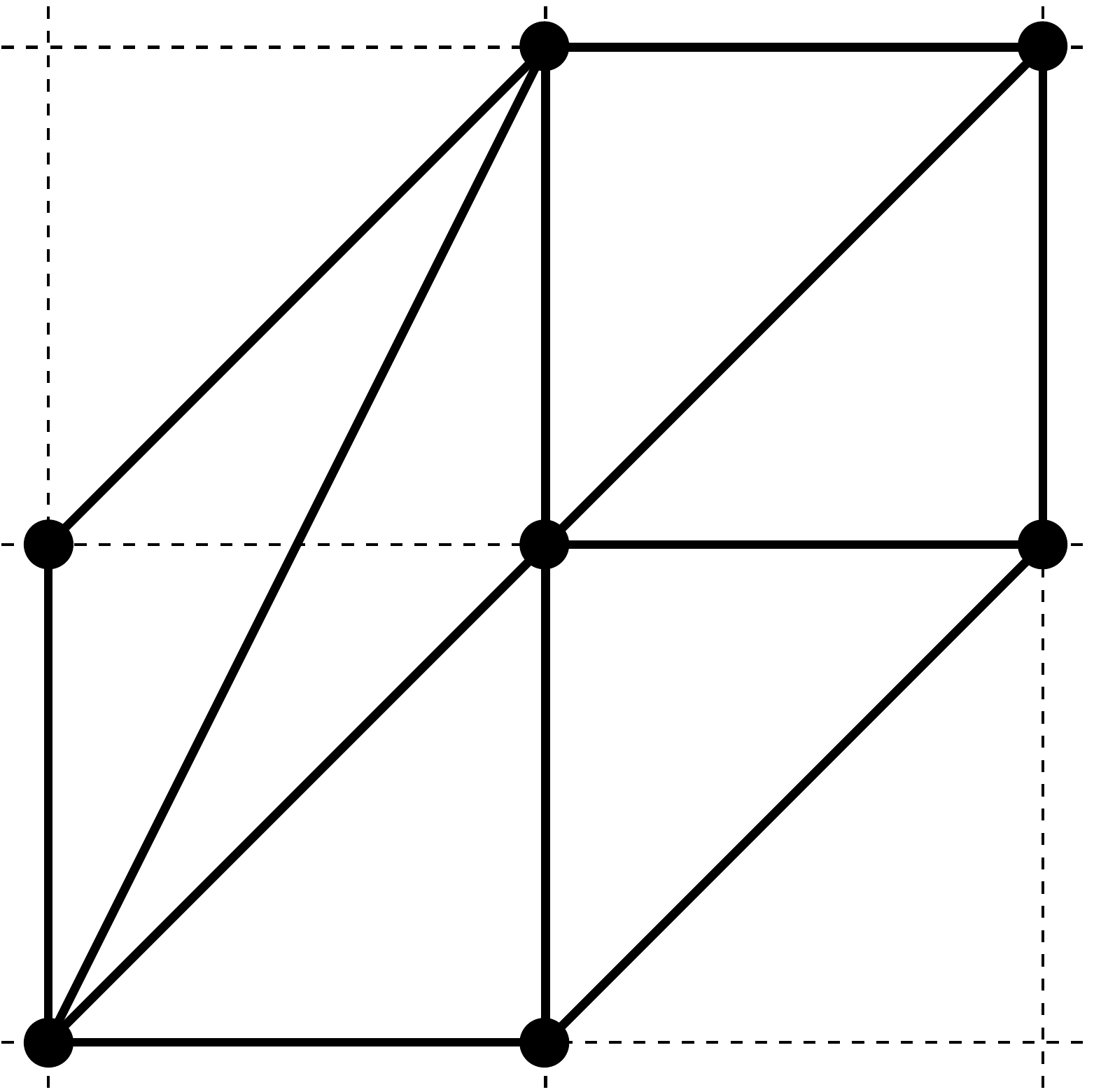}\label{fig:dp3-b}}\,
\subfigure[\small]{
\includegraphics[width=2.3cm]{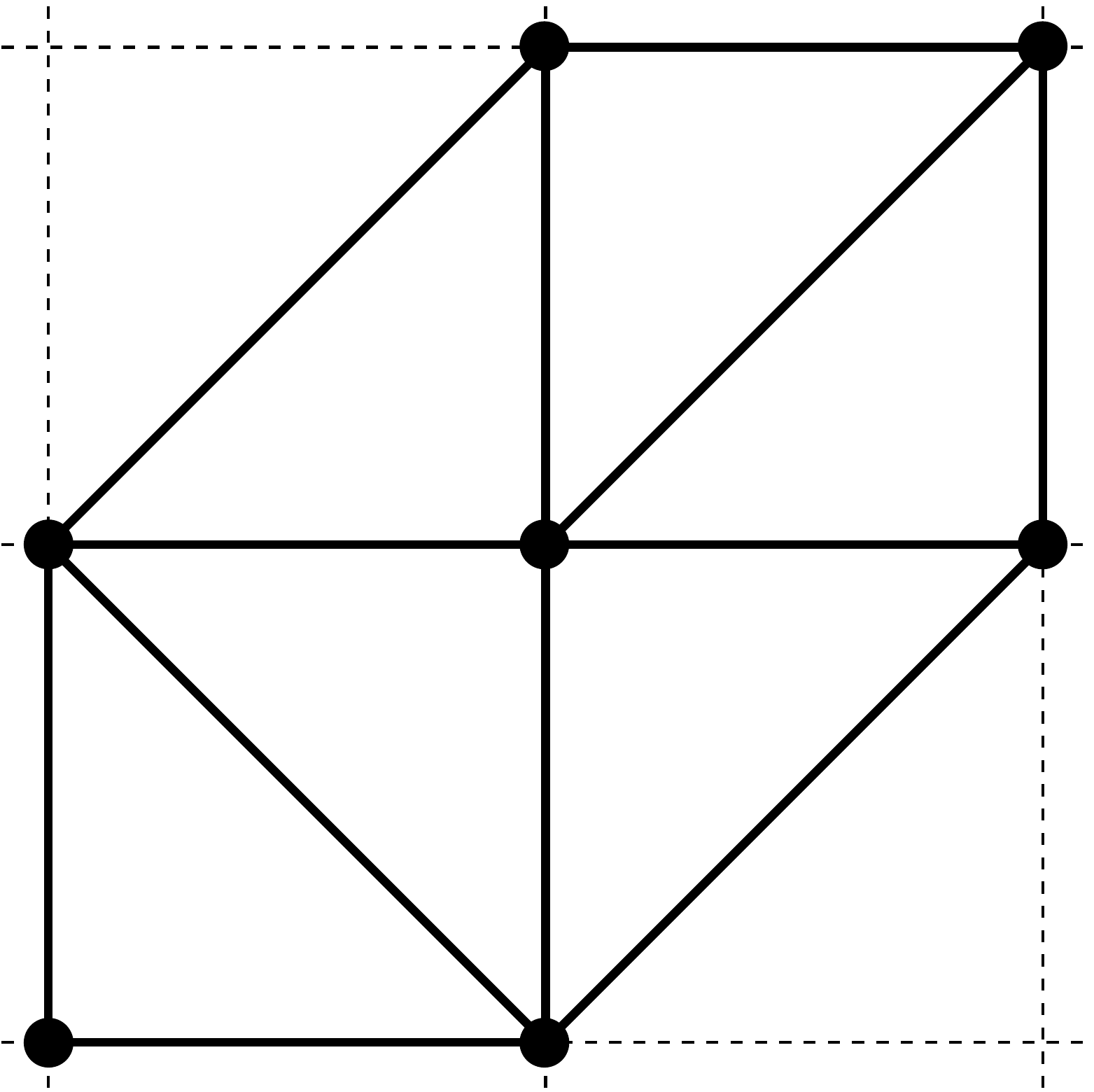}\label{fig:dp3-c}}\,
\subfigure[\small]{
\includegraphics[width=2.3cm]{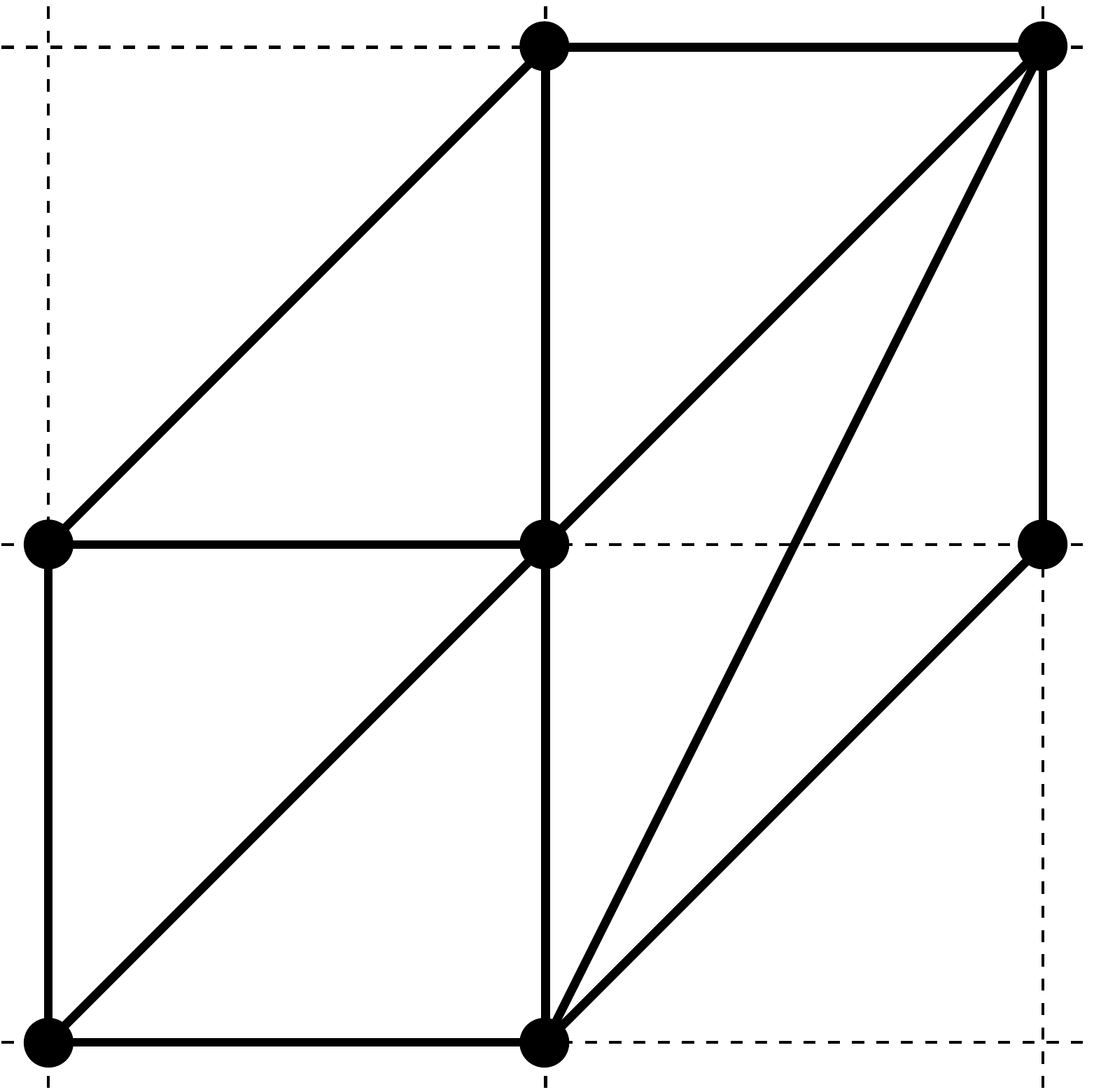}\label{fig:dp3-d}}\,%new d
\subfigure[\small]{
\includegraphics[width=2.3cm]{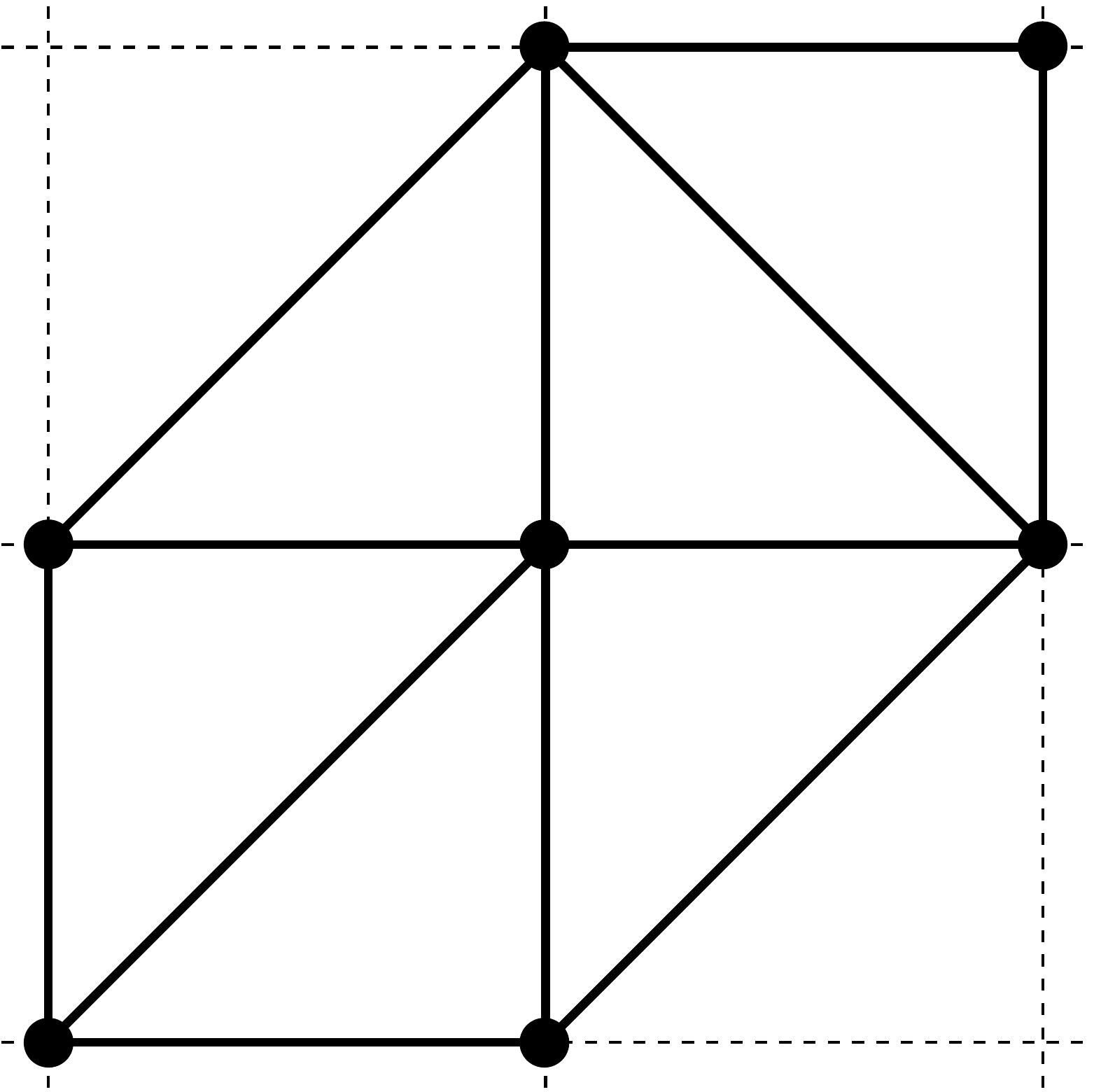}\label{fig:dp3-e}}\,%new e
\subfigure[\small]{
\includegraphics[width=2.3cm]{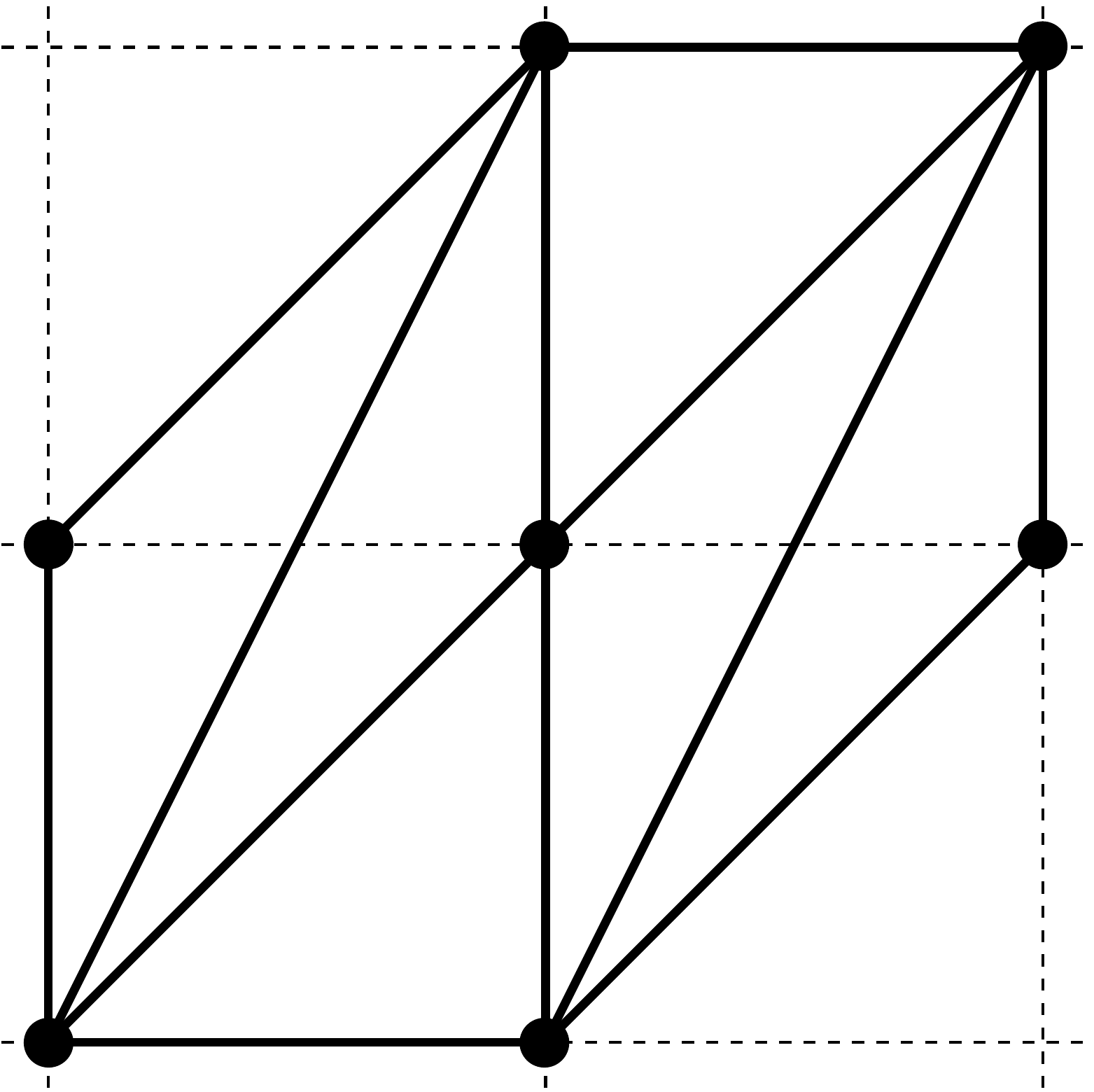}\label{fig:dp3-f}}\\%new f
%%%%%
\subfigure[\small]{
\includegraphics[width=2.3cm]{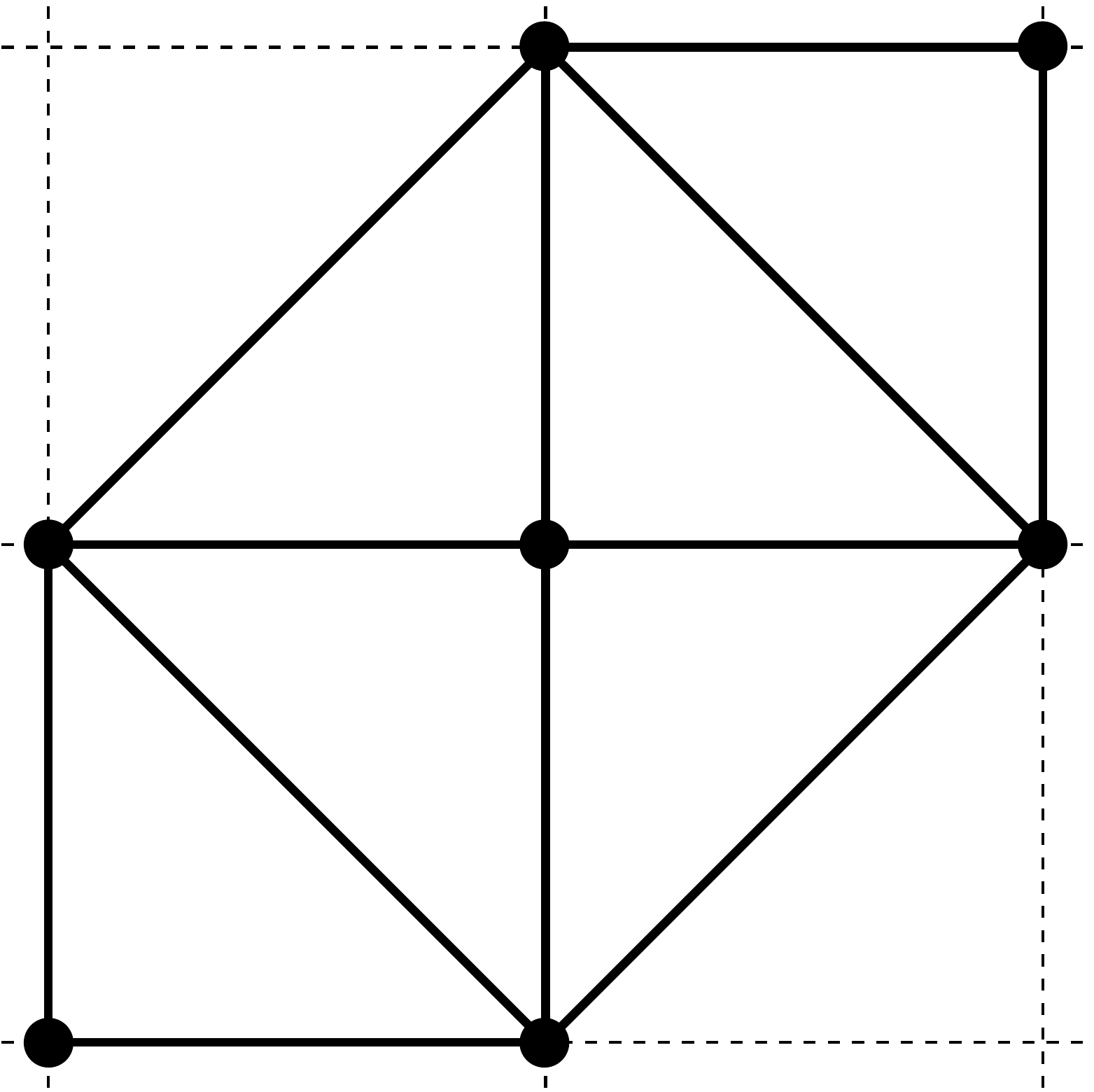}\label{fig:dp3-g}}\,%new g
\subfigure[\small]{
\includegraphics[width=2.3cm]{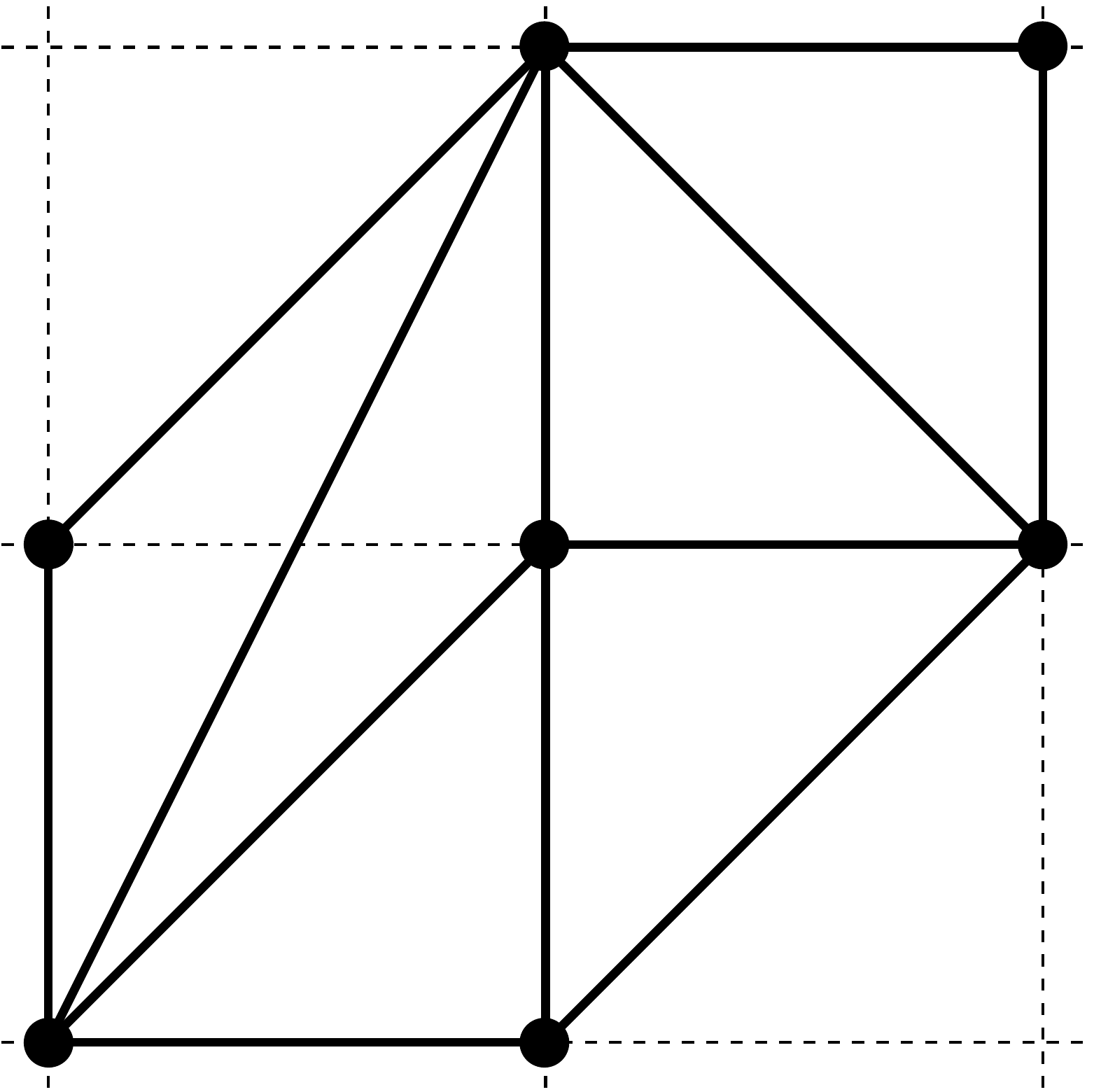}\label{fig:dp3-h}}\,%new h
\subfigure[\small]{
\includegraphics[width=2.3cm]{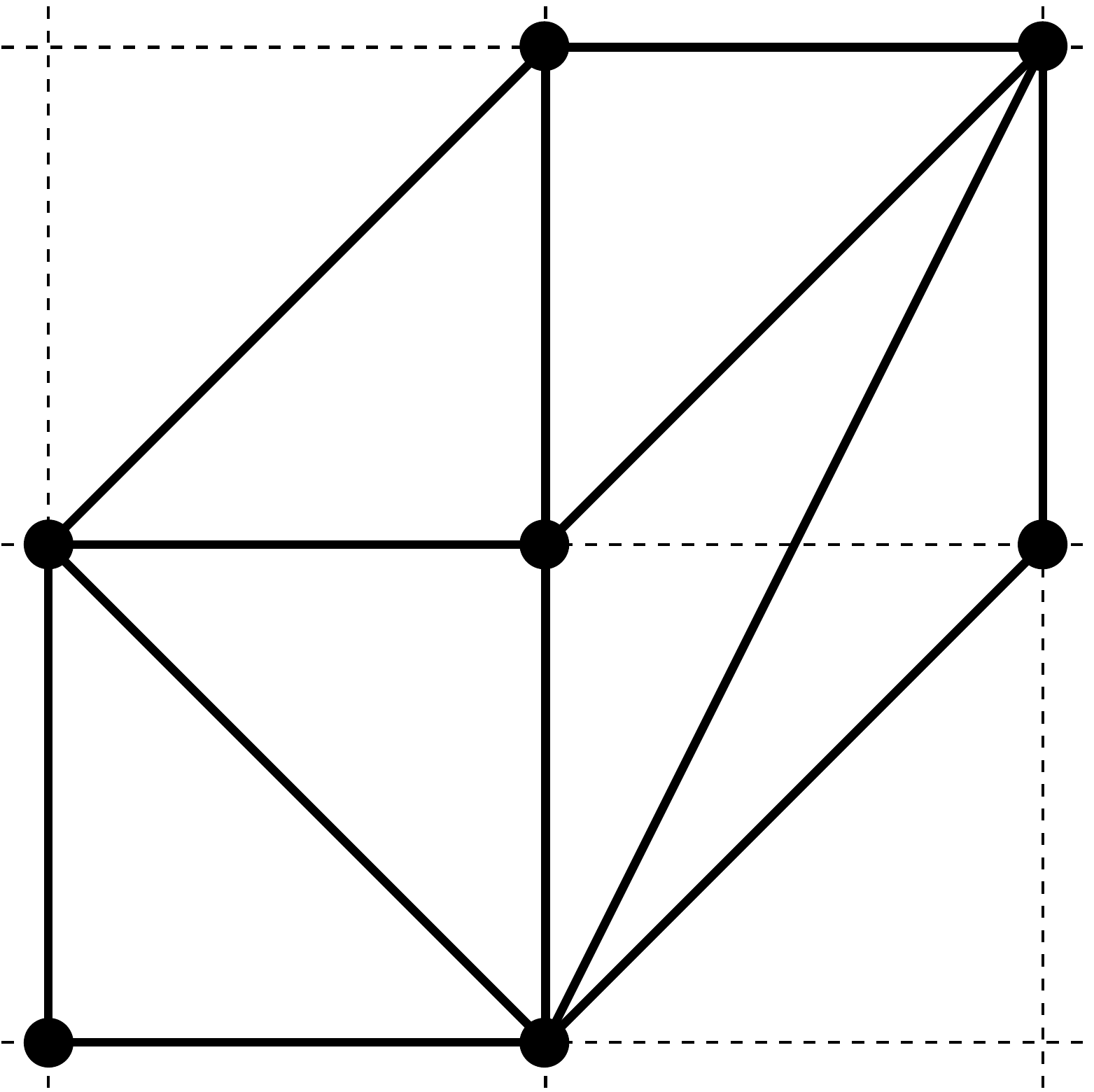}\label{fig:dp3-i}}\,%new i
\subfigure[\small]{
\includegraphics[width=2.3cm]{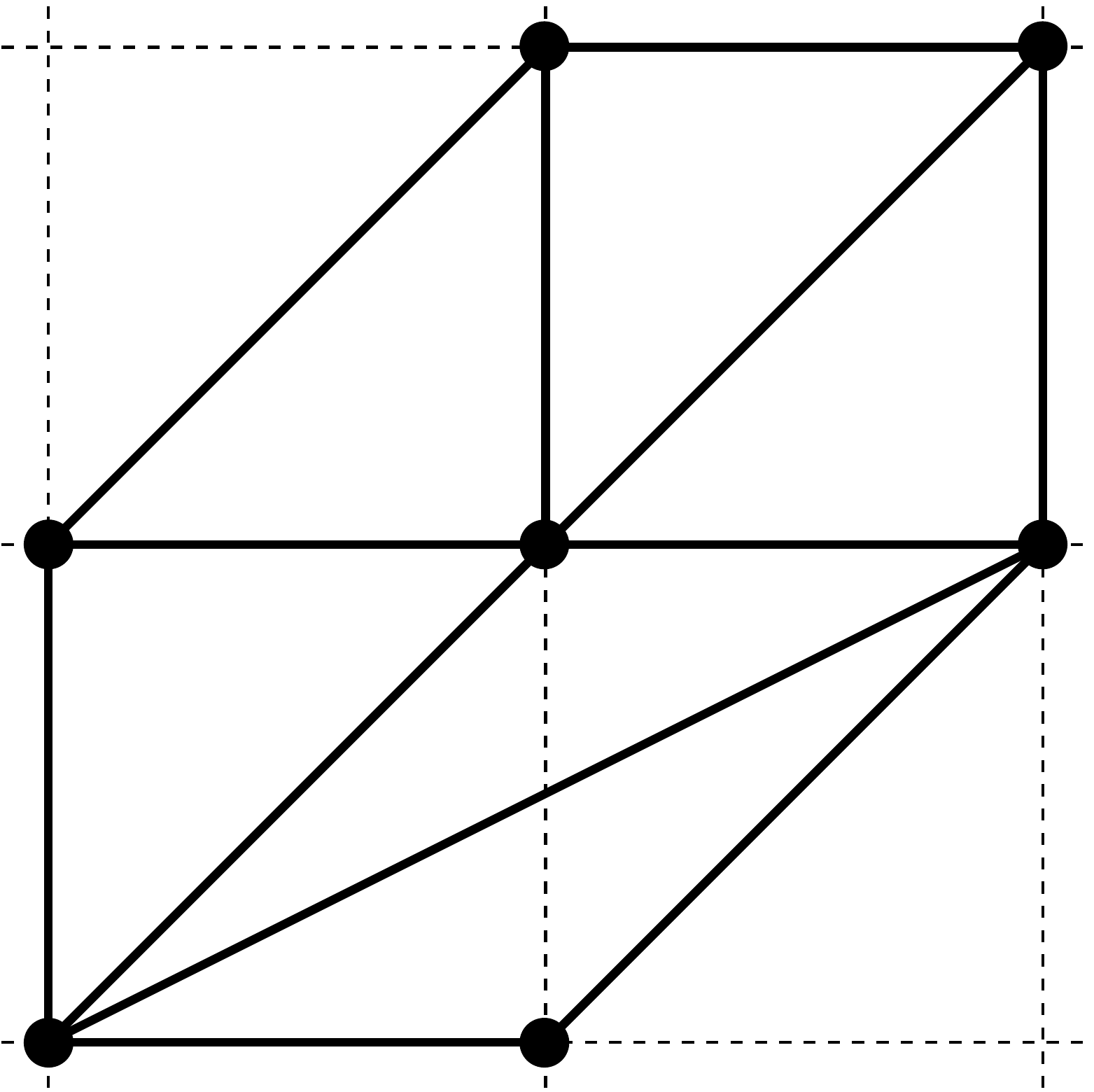}\label{fig:dp3-j}}\,%new j
\subfigure[\small]{
\includegraphics[width=2.3cm]{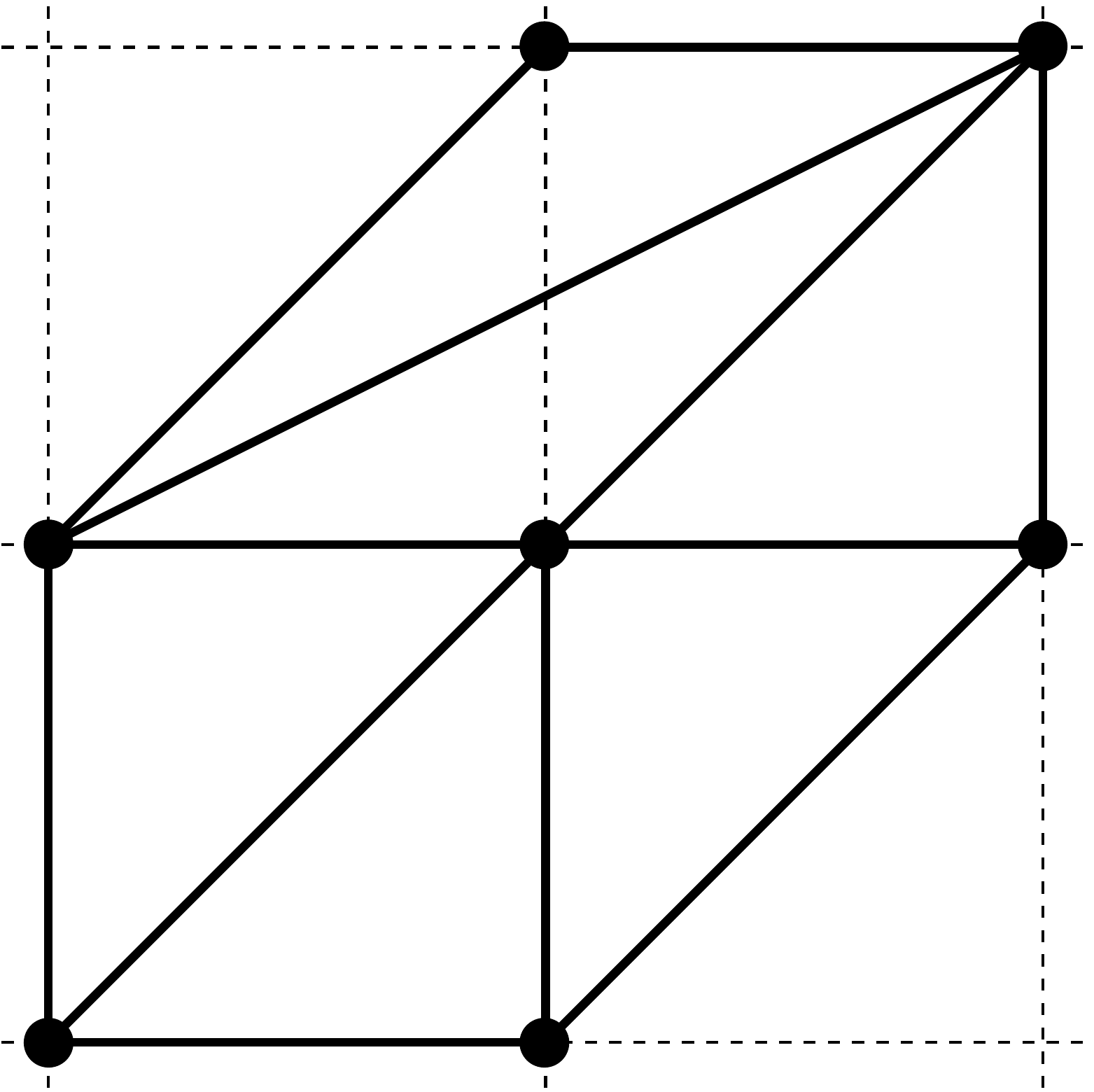}\label{fig:dp3-k}}\,%new k
\subfigure[\small]{
\includegraphics[width=2.3cm]{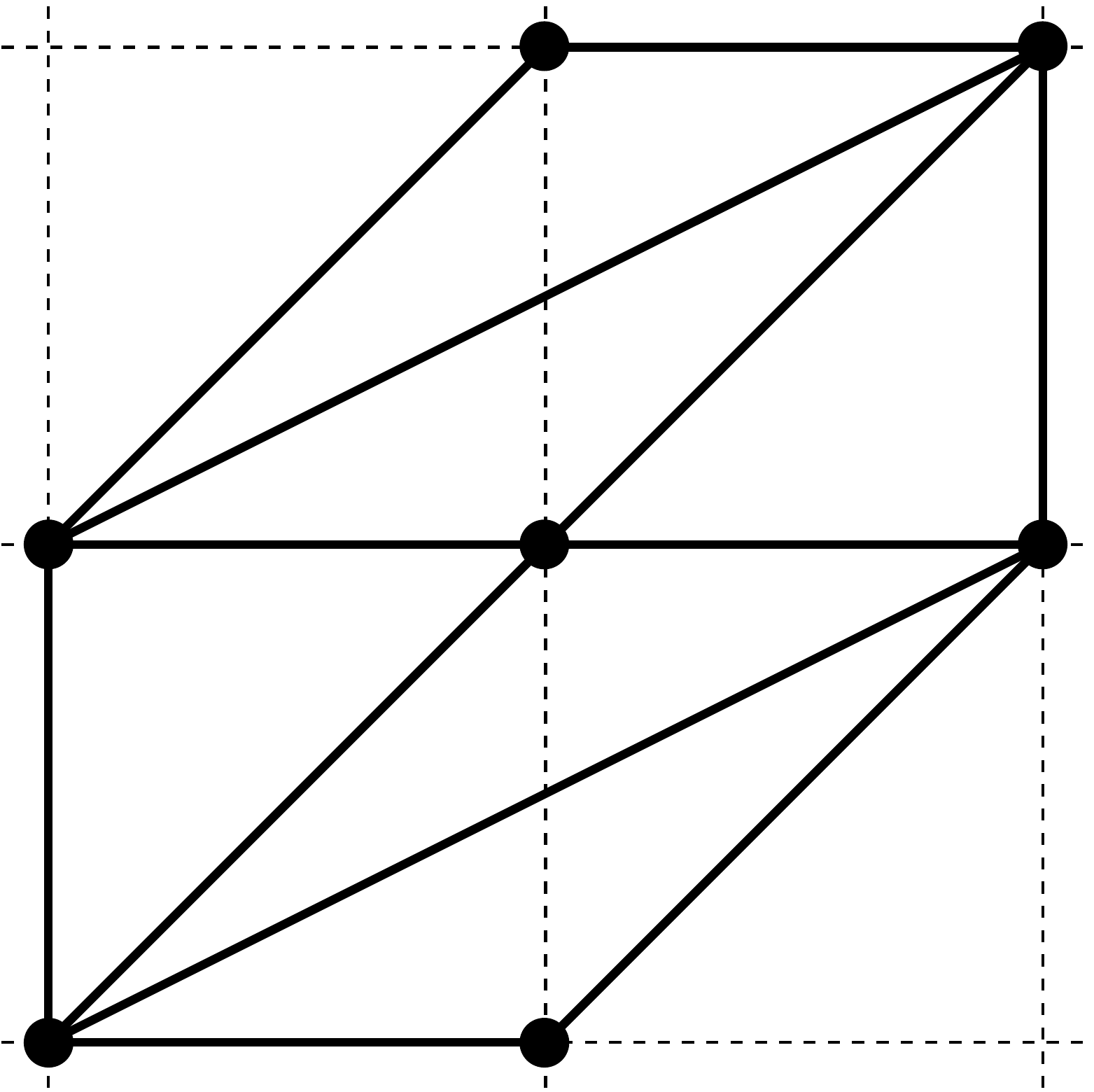}\label{fig:dp3-l}}\\%new l
\subfigure[\small]{
\includegraphics[width=2.3cm]{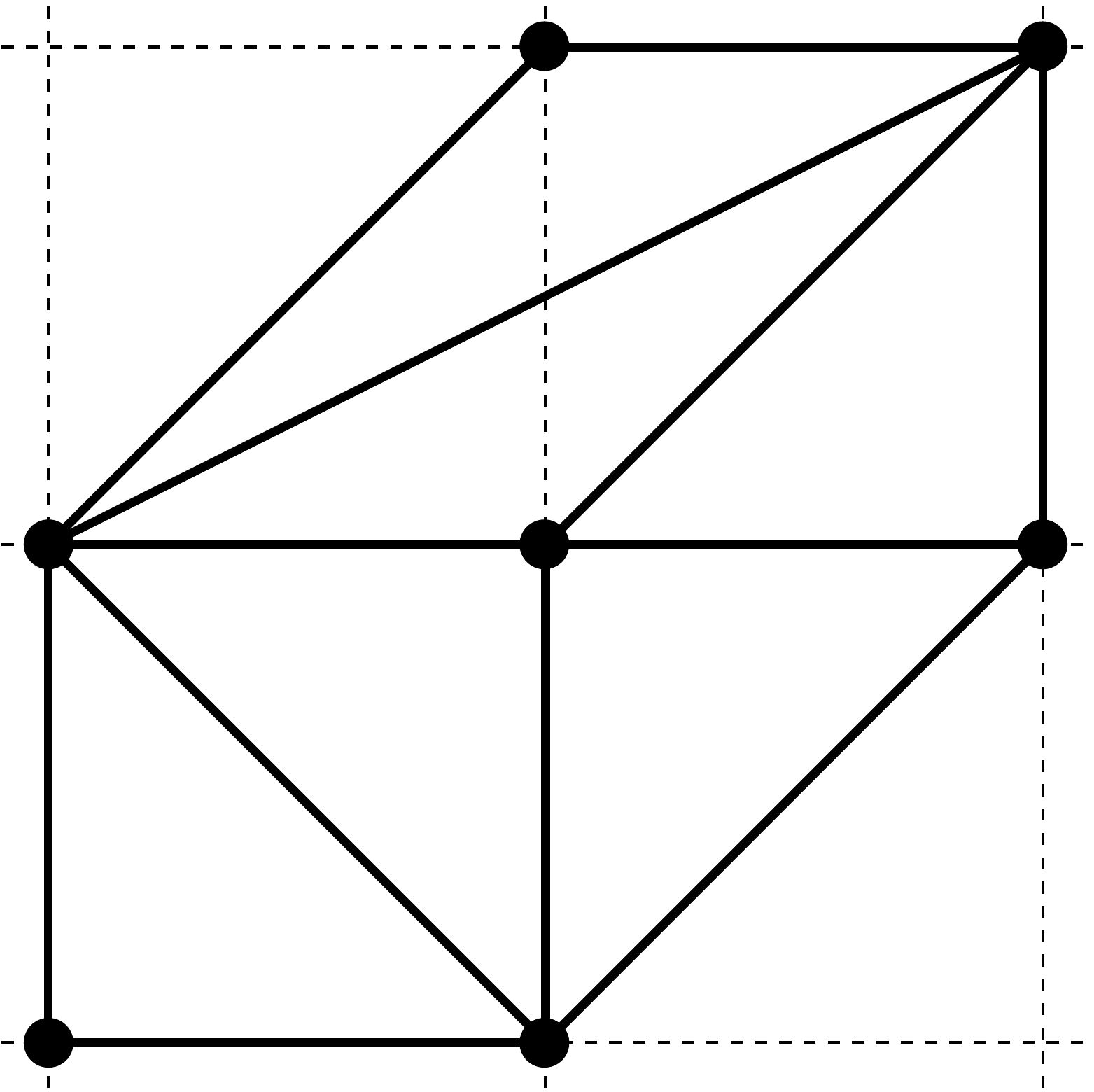}\label{fig:dp3-m}}\,%new m
\subfigure[\small]{
\includegraphics[width=2.3cm]{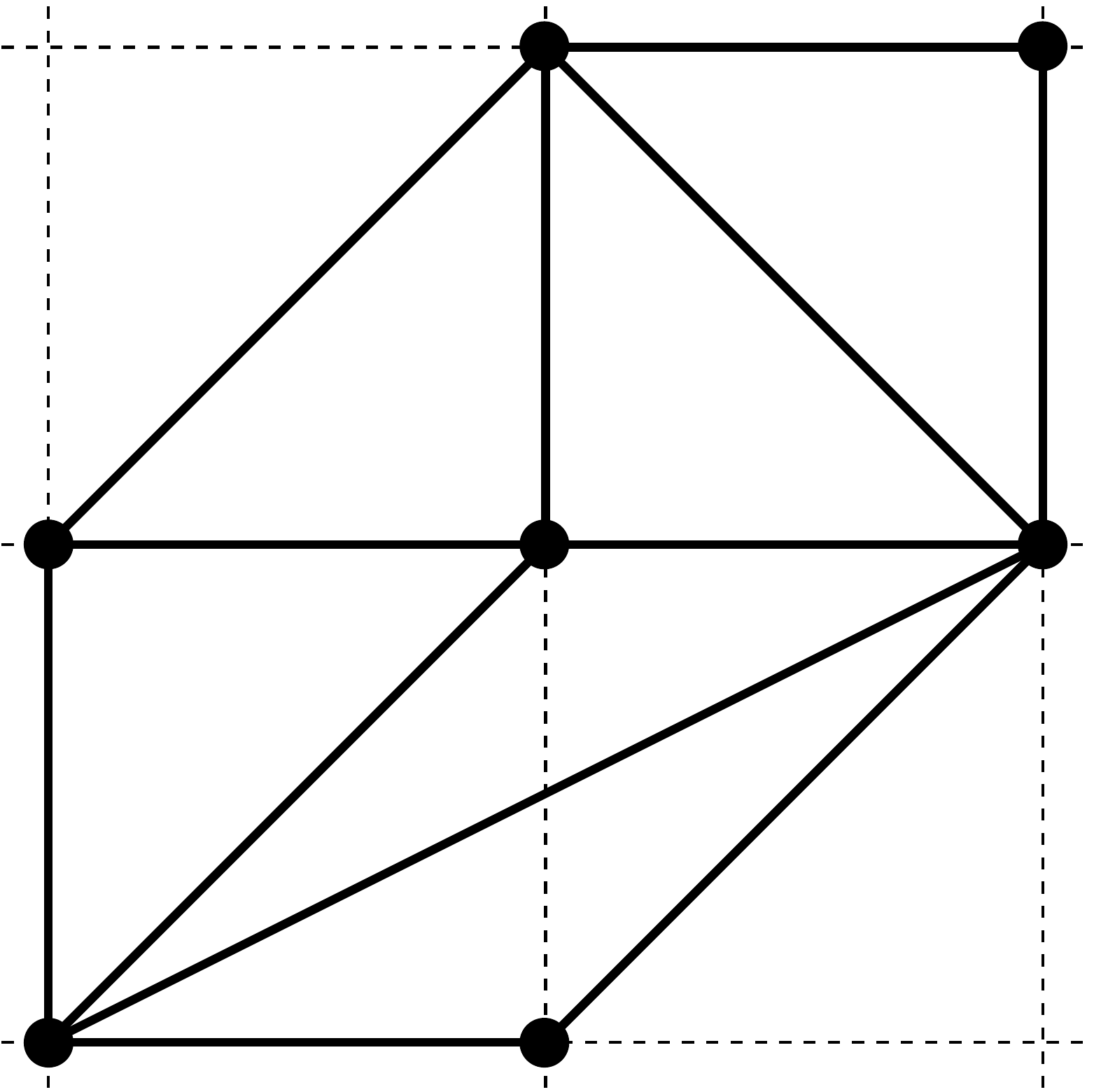}\label{fig:dp3-n}}\,%new n
\subfigure[\small]{
\includegraphics[width=2.3cm]{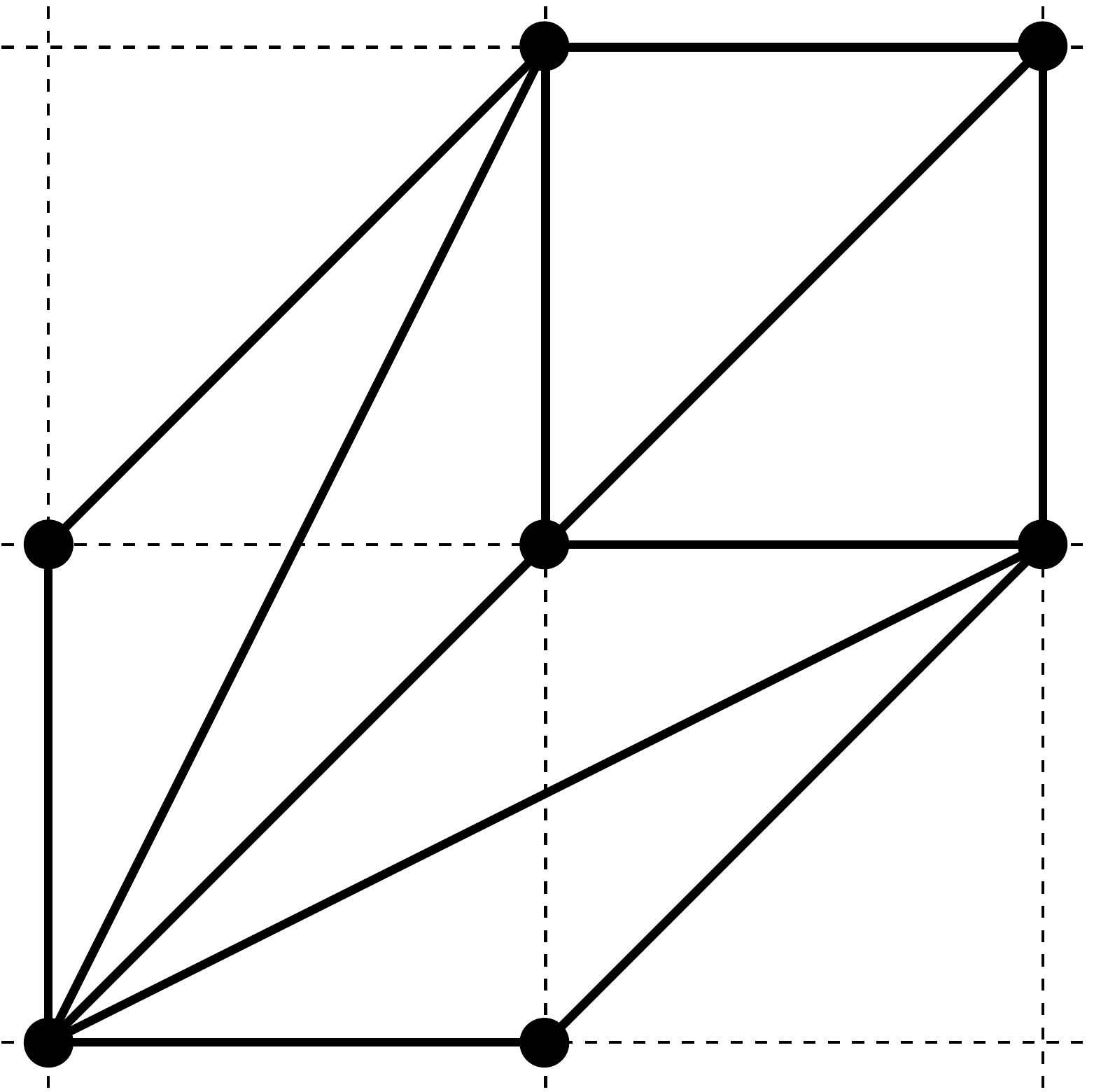}\label{fig:dp3-o}}\,%new o
\subfigure[\small]{
\includegraphics[width=2.3cm]{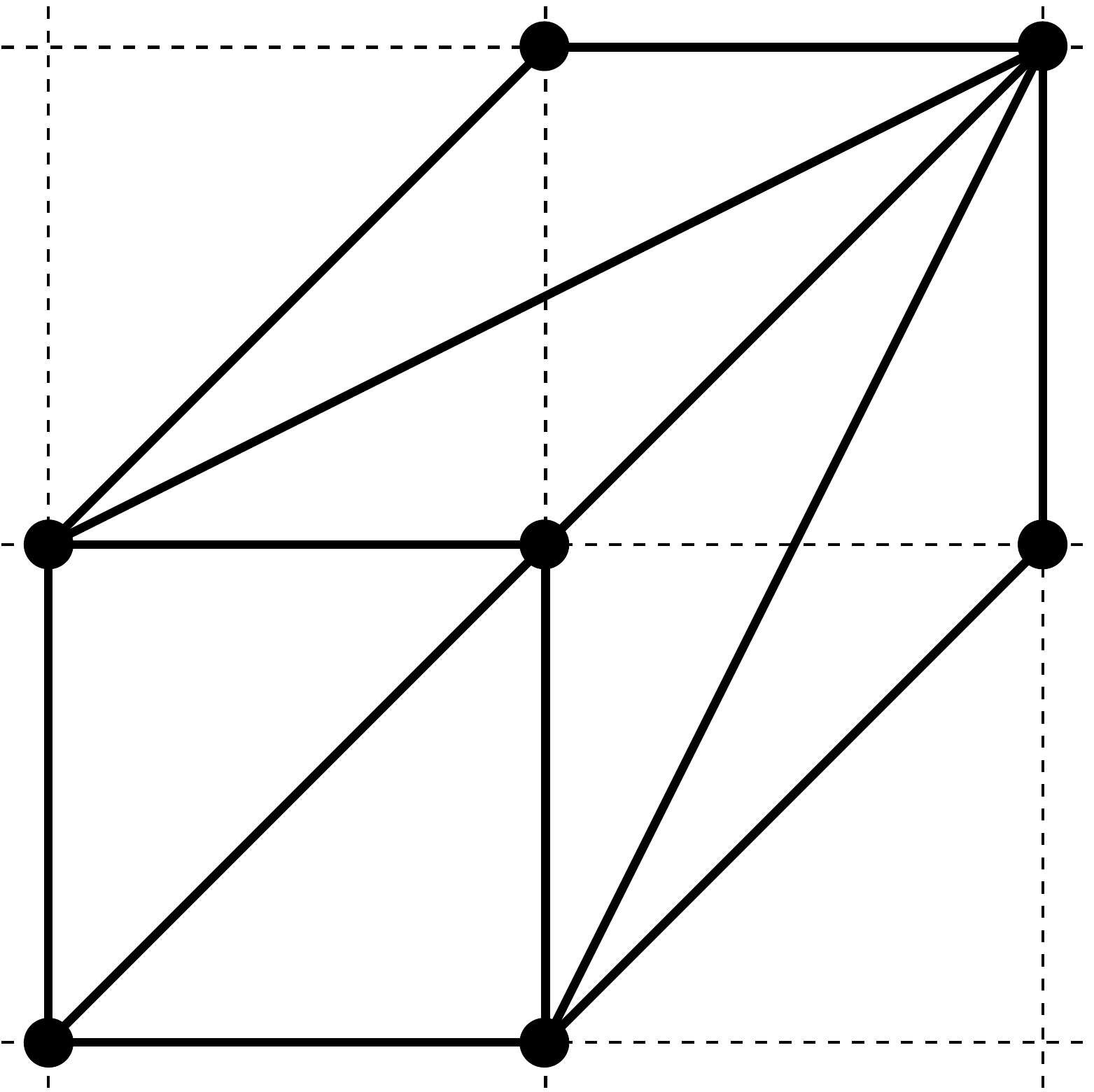}\label{fig:dp3-p}}\,%new p
\subfigure[\small]{
\includegraphics[width=2.3cm]{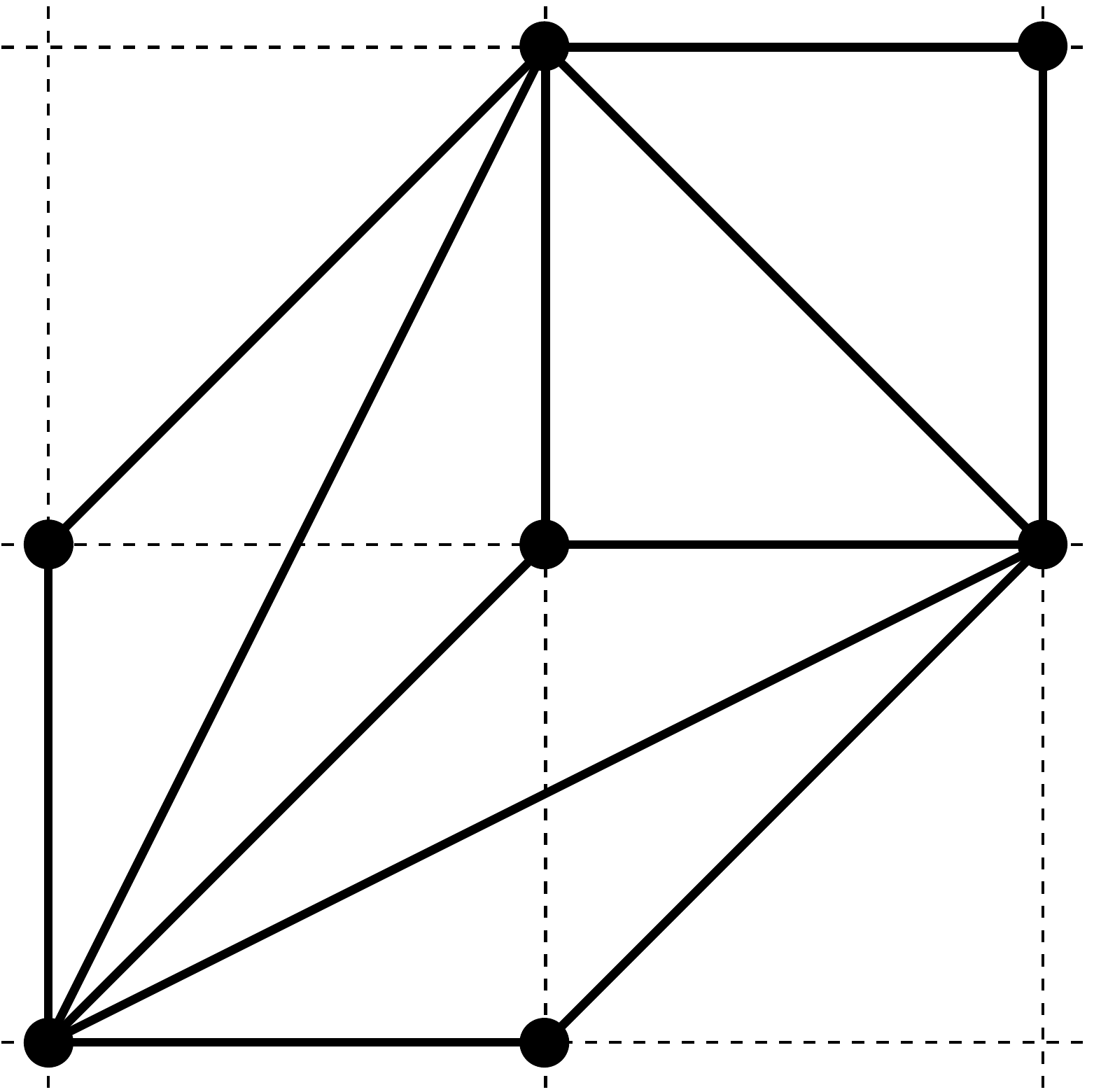}\label{fig:dp3-q}}\,%new q
\subfigure[\small]{
\includegraphics[width=2.3cm]{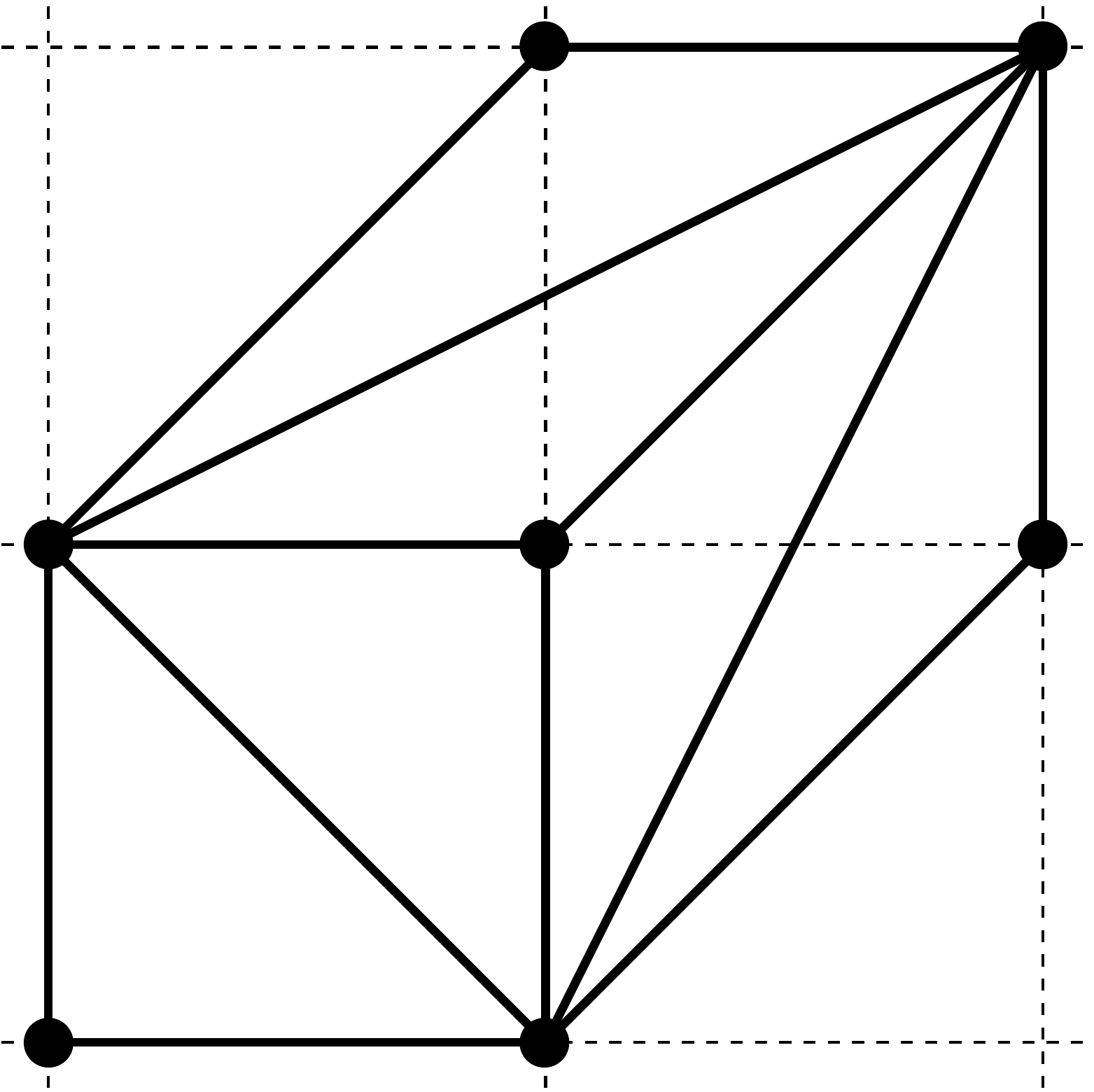}\label{fig:dp3-r}}%new r
\caption{The 18 resolutions of the $E_3$ singularity. The 9 resolutions (a)-(i) admit a vertical reduction. There are also 9 resolutions that admit an horizontal reduction (namely: (a), (c), (e), (g), (j)-(n)), and 9 resolutions that admit an ``oblique'' reduction (namely: (a), (b), (d), (f), (g)-(l), (o), (p)). The last three resolutions have no ``allowed'' IIA reduction. \label{fig:dP3 res}}
\end{center}
\end{figure}

\begin{figure}[h]
\begin{center}
\subfigure[\small  The two flavor D6-branes lie between the two gauge D6-branes.]{
\includegraphics[height=4.7cm]{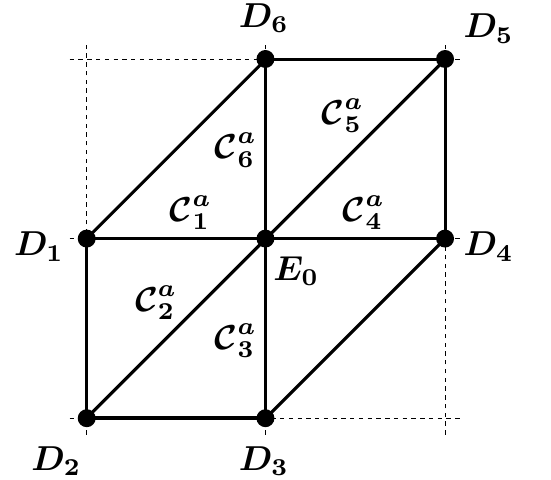}\label{fig:E3 res a}\qquad 
\includegraphics[height=5.2cm]{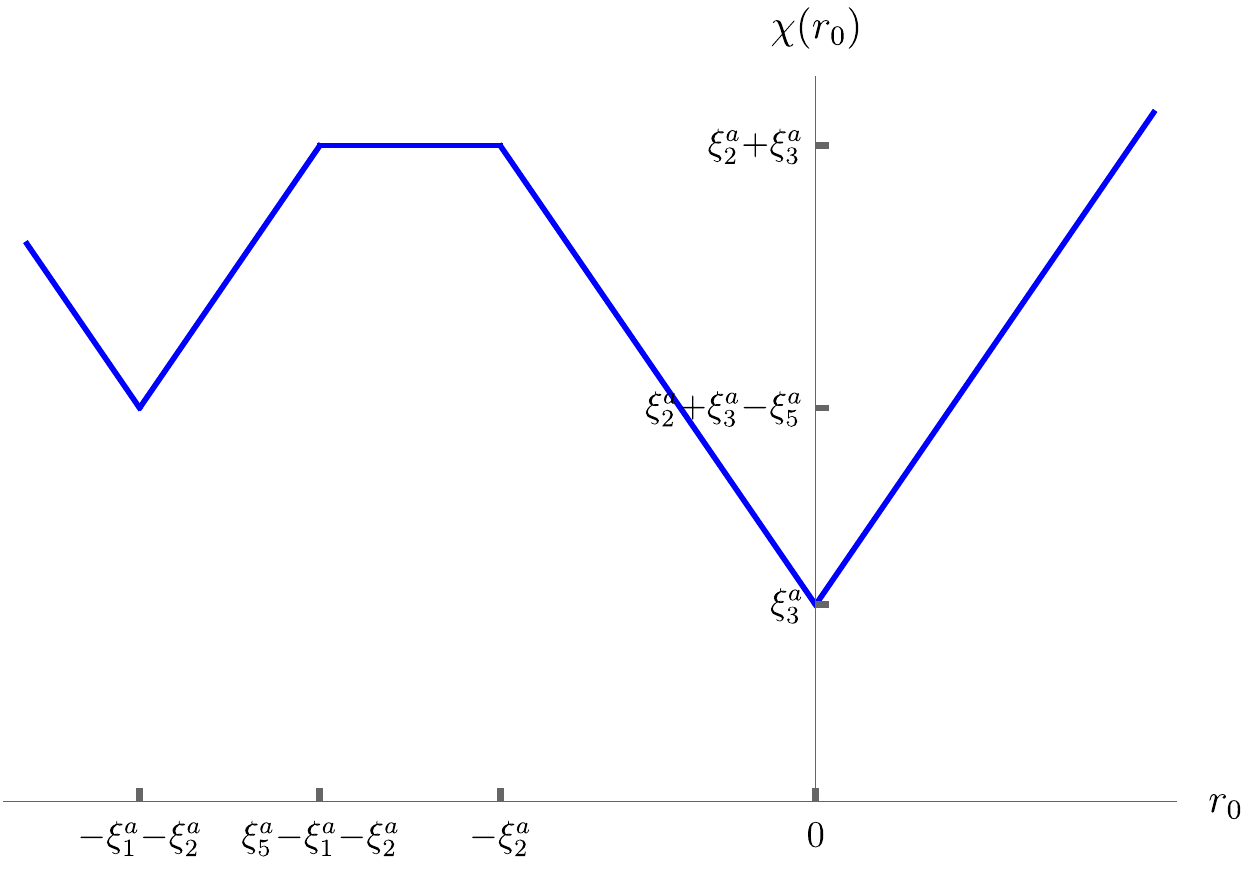}}\\
\subfigure[\small One flavor brane between the two gauge branes, and the other to the left.]{
\includegraphics[height=4.7cm]{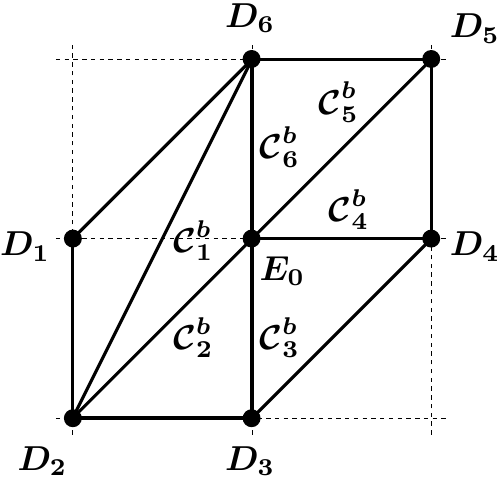}\label{fig:E3 res b}\qquad 
\includegraphics[height=5.2cm]{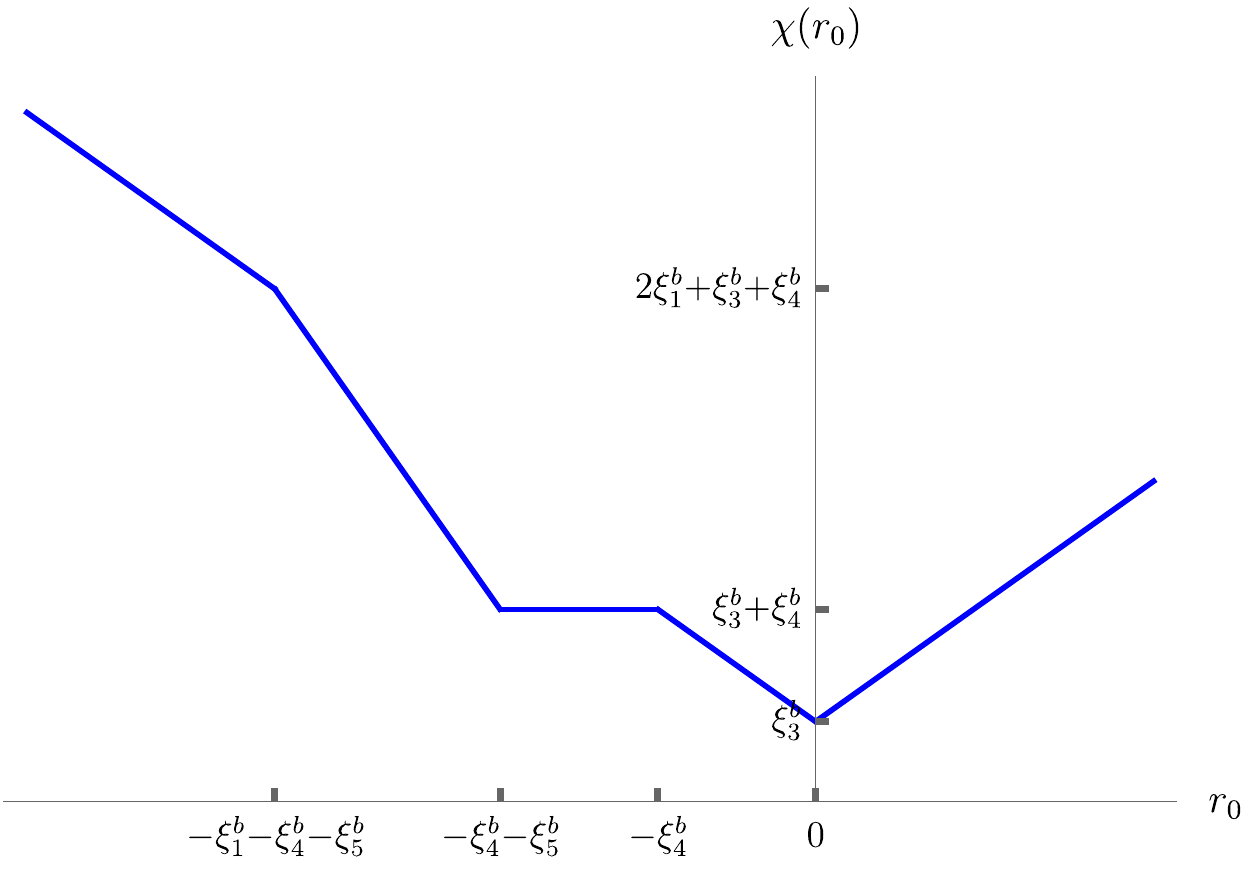}}\\
\subfigure[\small   One flavor brane between the two gauge branes, and the other to the right. ]{
\includegraphics[height=4.7cm]{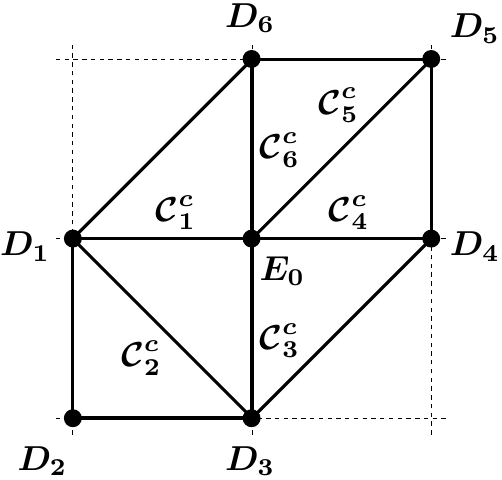}\label{fig:E3 res c}\qquad 
\includegraphics[height=5.2cm]{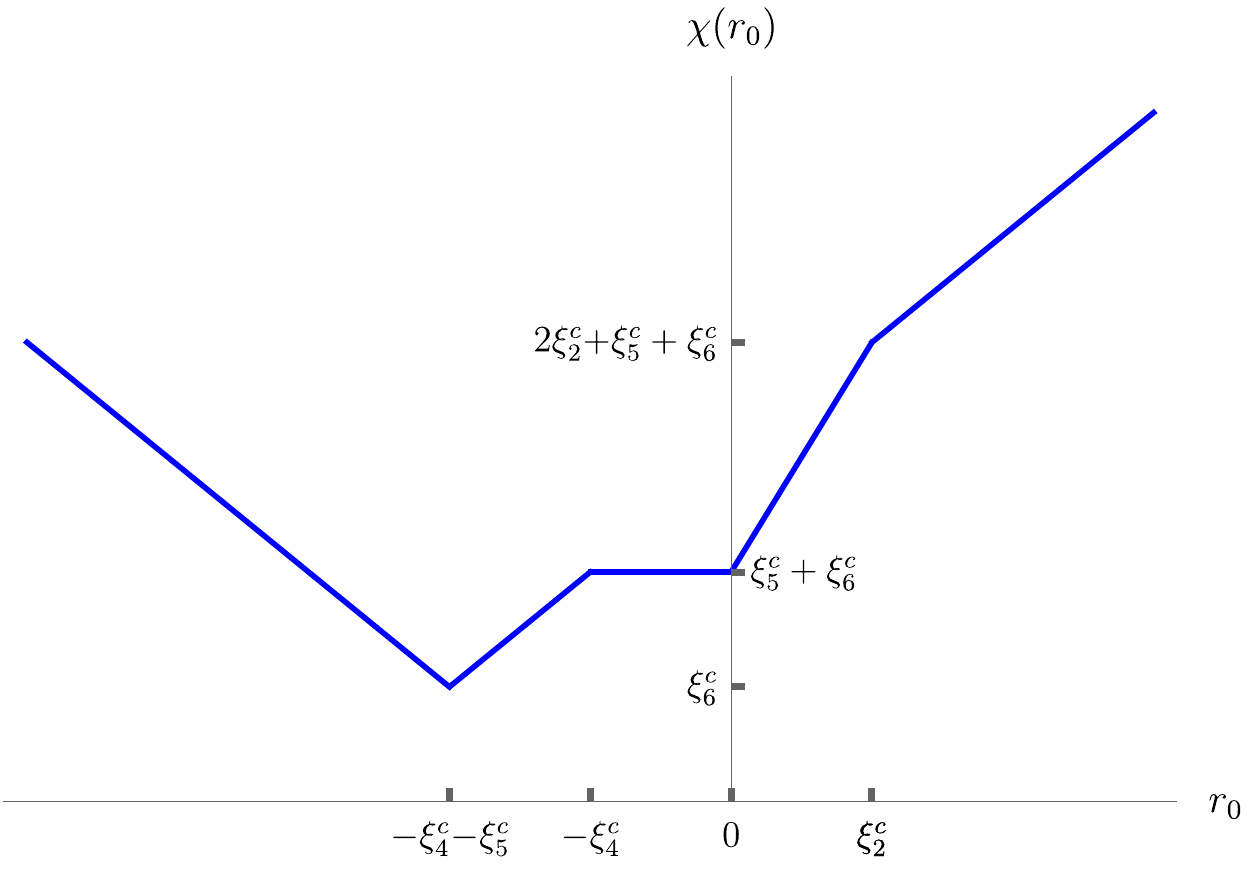}}
\caption{The first three resolutions, (a), (b) and (c), of the $E_3$ singularity and their IIA profiles $\chi(r_0)$ upon vertical reduction. \label{fig:E3 res} }
 \end{center}
 \end{figure}

Consider the $E_3$ singularity. It has 18 distinct resolutions, 9 of which admit a vertical reduction to type IIA, as shown in Figure \ref{fig:dP3 res}. Its GLSM can be chosen to be:
\be\label{intersect dP3a}
\begin{tabular}{l|ccccccc|c}
 & $D_1$ &$D_2$& $D_3$&$D_4$& $D_5$ & $D_6$ & $\bE_0$& \\
 \hline
$\CC_1$   & $-1$ & $1$ & $0$ & $0$ & $0$ & $1$   & $-1$ & $\xi_1$\\
$\CC_2$   & $1$  & $-1$ & $1$ & $0$ & $0$ & $0$  & $-1$ & $\xi_2$\\
$\CC_3$   & $0$  & $1$ & $-1$ & $1$ & $0$ & $0$  & $-1$ & $\xi_3$\\
$\CC_5$   & $0$  & $0$ & $0$ & $1$ & $-1$ & $1$ & $-1$ & $\xi_5$\\
\hline\hline
$U(1)_M$ & $0$ &$0$ &$-1$ & $0$ & $0$ & $0$ & $1$ & $r_0$
 \end{tabular}
\ee
The rows are labeled by the curves $\CC_1$, $\CC_2$, $\CC_3$ and $\CC_5$ in resolution (a), as shown in Figure \ref{fig:E3 res a}. They have positive volume if $\xi_1 > 0$, $\xi_2 > 0$, $\xi_3 > 0$ and $\xi_5 > 0$. By allowing more general values of the FI parameters, the same GLSM also describes all the resolutions shown in Figure \ref{fig:dP3 res}. Note also the linear relations among toric divisors:
\be \label{eq:dp3 lin eq}
D_4\cong D_1 + D_2 - D_5~, \;\;
 D_6 \cong D_2 + D_3 - D_5~, \;\;
 \bm{E}_0 \cong -2D_1-3D_2-2D_3+D_5~,
\ee
which hold independently of the partial resolution.

%%%%%%%%
%%%%%%%%
The 9 resolutions (a)-(i) give rise to an $SU(2)$ gauge theory with two fundamental flavors ($N_f = 2$). Depending on the values of the parameters, we have nine distinct chambers on the Coulomb branch. The gauge theory prepotential  \eqref{F SU2 Nf} with $N_f = 2$ takes the form:
\bea\label{Fabc FT dP3}
&\CF_{(a)} &=&\; \left(h_0-\frac{m_1+m_2}{2}\right)\varphi^2 + \varphi^3 - \frac{1}{2}(m_1^2+m_2^2)\varphi-\frac{1}{6}(m_1^3+m_2^3)~, \\
&\CF_{(b)} &=&\;   \left(h_0-\frac{m_1}{2}-m_2\right)\varphi^2 + \frac{7}{6}\varphi^3 - \frac{1}{2}m_1^2\varphi-\frac{1}{6}(m_1^3+m_2^3)~,  \\
&\CF_{(c)} &=&  \left(h_0-\frac{m_1}{2}\right)\varphi^2 + \frac{7}{6}\varphi^3 - \frac{1}{2}m_1^2\varphi-\frac{1}{6}m_1^3~,\; 
\eea
and:
\bea\label{Fabc FT dP3 bis}
&\CF_{(d)} &=&\;  \left(h_0-\frac{m_2}{2}\right)\varphi^2 + \frac{7}{6}\varphi^3 - \frac{1}{2}m_2^2\varphi - \frac{1}{6}m_2^3~, \\
&\CF_{(e)} &=&\;   \left(h_0-m_1-\frac{m_2}{2}\right)\varphi^2 + \frac{7}{6}\varphi^3 -\frac{1}{2}m_2^2\varphi-\frac{1}{6}(m_1^3+m_2^3)~, \\
&\CF_{(f)} &=&\;   \left(h_0-m_2\right)\varphi^2 + \frac{4}{3}\varphi^3 - \frac{1}{2}m_2^3~, \\
&\CF_{(g)} &=&\;   \left(h_0-m_1\right)\varphi^2 + \frac{4}{3}\varphi^3 -\frac{1}{3}m_1^3~,\\
&\CF_{(h)} &=&\;    \left(h_0-m_1-m_2\right)\varphi^2 + \frac{4}{3}\varphi^3 -\frac{1}{3}(m_1^3+m_2^3)~, \\
&\CF_{(i)} &=&\;    h_0\varphi^2 + \frac{4}{3}\varphi^3~.
\eea
Here, the subscripts correspond to the field theory chambers:{ \small
\bea\label{res to FT chambers dP3}
&{\rm (a)} \, \leftrightarrow\, \left\{\begin{array}{@{}c@{}}\varphi +m_1\geq 0,\;\;  -\varphi+m_1 <0\\ \varphi + m_2 \geq 0, \;\; -\varphi+m_2 < 0\end{array}\right. \qquad
&{\rm (b)} \,\leftrightarrow\, \left\{\begin{array}{@{}c@{}}\varphi +m_1\geq 0,\;\;  -\varphi+m_1 <0\\ \varphi + m_2 \geq 0, \;\; -\varphi+m_2 \geq 0\end{array}\right. \\
&{\rm (c)} \,\leftrightarrow\, \left\{\begin{array}{@{}c@{}}\varphi +m_1\geq 0,\;\;  -\varphi+m_1 <0 \\ \varphi + m_2 < 0, \;\; -\varphi+m_2 < 0\end{array}\right.  \qquad
&{\rm (d)} \,\leftrightarrow\,\left\{\begin{array}{@{}c@{}}\varphi +m_1<0,\;\;  -\varphi+m_1 <0\\ \varphi + m_2 \geq 0, \;\; -\varphi+m_2 < 0\end{array}\right. \\
&{\rm (e)} \,\leftrightarrow\,\left\{\begin{array}{@{}c@{}}\varphi +m_1\geq 0,\;\;  -\varphi+m_1 \geq 0\\ \varphi + m_2 \geq 0, \;\; -\varphi+m_2 < 0\end{array}\right.  \qquad
&{\rm (f)} \,\leftrightarrow\, \left\{\begin{array}{@{}c@{}}\varphi +m_1<0,\;\;  -\varphi+m_1 <0\\\varphi + m_2 \geq 0, \;\; -\varphi+m_2 \geq 0\end{array}\right. \\
&{\rm (g)} \,\leftrightarrow\, \left\{\begin{array}{@{}c@{}}\varphi +m_1\geq 0,\;\;  -\varphi+m_1 \geq 0\\ \varphi + m_2 < 0, \;\; -\varphi+m_2 < 0\end{array}\right.  \qquad
&{\rm (h)} \,\leftrightarrow\, \left\{\begin{array}{@{}c@{}}\varphi +m_1\geq 0,\;\;  -\varphi+m_1 \geq 0\\ \varphi + m_2 \geq 0, \;\; -\varphi+m_2 \geq 0\end{array}\right. \\
&{\rm (i)} \,\leftrightarrow\, \left\{\begin{array}{@{}c@{}}\varphi +m_1<0, \;\;  -\varphi+m_1 <0\\ \varphi + m_2 < 0, \;\; -\varphi+m_2 < 0\end{array}\right.
\eea}
Recall that we have $\varphi > 0$, while the real masses $m_1$ and $m_2$ can take both signs. Nonetheless, it is convenient to choose $m_1 >m_2$ (using the U(2) flavor symmetry), in which case the field theory chambers \eqref{res to FT chambers dP3} precisely correspond to the first 9 resolutions denoted by the same letters in Figure~\ref{fig:dP3 res}.

\paragraph{Geometric prepotential.} To compute the M-theory prepotential $\CF = -\frac{1}{6}S^3$, let us define:
\be
S  = \mu_1 D_1 + \mu_2 D_2 + \mu_3 D_3 + \nu \bE_0~.
\ee
The parameters $\mu, \nu$ are related to the FI terms of \eqref{intersect dP3a} by:
\be\label{eq:1p13}
 \mu_1 = \xi_2 + \xi_3-2\xi_5~, \;\; \mu_2 = \xi_1+\xi_2+\xi_3-3\xi_5~, \;\;  \mu_3 = \xi_1+\xi_2-2\xi_5~, \;\; \nu = -\xi_5~.
\ee
By direct computation of the intersection numbers (see Appendix~\ref{app: E3 sing}), we find:~\footnote{Here, for simplicity, we did not keep track of the ``non-compact'' intersection numbers, so those expressions are valid up to $\nu$-independent terms. Correspondingly, we only match to field theory up to a $\varphi$-independent term. The same comment will apply to the other geometries that we study in the rest of the paper.}
\bea\label{Fabc dP3 Mtheory}
\CF_{(a)} &= -\nu ^3+\left(\frac{\mu _1}{2}+\frac{\mu _2}{2}+\frac{\mu _3}{2}\right) \nu ^2+\left(\frac{\mu _1^2}{2}-\mu _2 \mu _1+\frac{\mu _2^2}{2}+\frac{\mu _3^2}{2}-\mu _2 \mu _3\right) \nu  	\\
\CF_{(b)} &= -\frac{7 \nu ^3}{6} + \left(\mu _2+\frac{\mu _3}{2}\right) \nu ^2+\left(\frac{\mu _3^2}{2}-\mu _2 \mu _3\right) \nu 	\\
\CF_{(c)} &= -\frac{7 \nu ^3}{6} + \left(\mu _1+\mu _3\right) \nu ^2-\mu _1 \mu _3 \nu 
\eea
for the first two resolutions of the $E_3$ singularity, up to a $\nu$-independent constant term.
As we will show, the geometric parameters are matched to the field-theory parameters according to:
\be\label{mu to FT dP3}
 \mu_1 = h_0 + m_1~, \quad \mu_2 = h_0 + 2m_1 - m_2~, \quad \mu_3 = 2m_1~, \quad \nu = -\varphi + m_1~.
\ee
Indeed, plugging \eqref{mu to FT dP3} into \eqref{Fabc dP3 Mtheory} reproduces the field-theory result \eqref{Fabc FT dP3}, up to the constant term. The same geometric computation can be carried out for the six other resolutions (d)-(i), and one finds perfect agreement with \eqref{Fabc FT dP3 bis}.

\paragraph{Gauge-theory chamber (a).} Consider the vertical reduction of resolution (a), which is shown in Figure \ref{fig:E3 res a}. In this geometry we have the relations
\begin{align}
\CC_4^a &\cong \CC_1^a + \CC_2^a - \CC_5^a~, \qquad \CC_6^a \cong \CC_2^a + \CC_3^a - \CC_5^a~,	
\end{align}
amongst the curves, and therefore $\{\CC_1, \CC_2, \CC_3, \CC_5\}$ generates the Mori cone. We denote by $\xi_i \equiv \xi_i^a$, the volume of $\mathcal{C}_i\equiv \CC_i^a$, so that we have $\xi_4 = \xi_1+\xi_2-\xi_5>0$ and $\xi_6 = \xi_2+\xi_3-\xi_5>0$. The IIA reduction gives again a resolved $A_1$ singularity fibered over $r_0$. One finds:
\be\label{chi a dP3}
\text{(a)}\; : \; \chi(r_0) = \begin{cases}  
  r_0 + \xi_3^a \qquad &\text{if}\quad  r_0\geq 0 \cr 
   - r_0 + \xi_3^a \qquad &\text{if}\quad -\xi_2^a \leq r_0 \leq 0 \cr 
 \xi_2^a + \xi_3^a \qquad &\text{if}\quad \xi_5^a-\xi_1^a-\xi_2^a \leq r_0 \leq -\xi_2^a  \cr 
r_0+ \xi_1^a + 2\xi_2^a+\xi_3^a-\xi_5^a \; &\text{if}\quad -\xi_1^a-\xi_2^a \leq r_0 \leq \xi_5^a-\xi_1^a-\xi_2^a \cr     
-r_0 -\xi_1^a + \xi_3^a - \xi_5^a &\text{if}\quad  r_0 \leq -\xi_1^a-\xi_2^a \cr 
\end{cases}
\ee 
as shown in Figure~\ref{fig:E3 res a}.  We directly see that there are two gauge D6-branes at $r_0 = 0$ and $r_0={-}\xi_1^a{-}\xi_2^a$, respectively,  and two non-compact ``flavor'' D6-branes at $r_0 =\xi_5^a{-}\xi_1^a{-}\xi_2^a$ and $r_0={-}\xi_2^a$. The effective CS level \eqref{k eff from IIA} vanishes in this theory, $k_{\rm eff}=0$. 
In our conventions, this is interpreted as a bare CS level $k = 1$ (for the ``naive'' $U(2)$ gauge group) plus the contribution $-\frac{1}{2}-\frac{1}{2}=-1$ from the two hypermultiplets.

%%%%%%%
 %%%%%%%%
The open string between the gauge D6-branes provide the W-boson, and the open strings stretching from the gauge branes to the two flavor branes provide the hypermultiplet modes. Therefore, we find:
\be
M(W_{\alpha}) = 2\varphi = \xi_1^a + \xi_2^a~,
\ee
and:
\bea\label{eq:hypermultiplet_mass_dp3_a}
& M(\mathcal{H}_{1,1}) = \varphi-m_1=\xi_5^a~,\qquad 
&& M(\mathcal{H}_{1,2})  = \varphi - m_2 = \xi_1^a\\
&M(\mathcal{H}_{2,1}) = \varphi+m_1=\xi_1^a+\xi_2^a-\xi_5^a~,\qquad 
&& M(\mathcal{H}_{2,2}) = \varphi+m_2 = \xi_2^a 
\eea
The masses of the instanton particles are given by the volumes of the wrapped D2-branes at the $r_0$ locations of the gauge D6-branes:
\be\label{eq:instanton_mass_dp3_a}
	M(\mathcal{I}_1) = h_0+\varphi-m_1-m_2=\xi_3^a, \qquad\quad
	 M(\mathcal{I}_2) = h_0+\varphi =\xi_2^a+\xi_3^a-\xi_5^a~.	
\ee
Here, as before, the instanton masses can be calculated using the prepotential for the $U(2)$ $N_f=2$ theory (at bare CS level $k=1$).  Note that the instanton particle $\mathcal{I}_1$ comes from an M2-brane wrapped over $\CC_3^a$, whereas $\mathcal{I}_2$ comes from an M2-brane wrapped over $\CC_6^a$. 
Plugging the above identifications into \eqref{eq:1p13}, we derive the geometry-to-field-theory map \eqref{mu to FT dP3}.  One can also check that the IIA computation of the string tension:
\be
 T = \int_{-\xi_1-\xi_2}^{0}\chi(r_0)dr_0 = \xi_1 \xi_2 + \xi_2 \xi_3 + \xi_3 \xi_1 +\half(\xi_2^2-\xi_5^2)~,	
\ee
agrees with the field-theory result, $T=\d_{\varphi} \CF_{(a)}$ in this chamber. As in previous examples, we see clearly that an instanton can become massless while the string tension is kept finite.

%%%%%%%%
\paragraph{Gauge-theory chamber (b).}
The toric diagram of resolution (b) and its IIA profile are shown in Figure \ref{fig:E3 res b}. 
In this geometry, we have the relations:
\begin{align}
\CC_2^b &\cong \CC_4^b + \CC_5^b~, \qquad \CC_6^b \cong \CC_3^b + \CC_4^b~,	
\end{align}
amongst the curves; we therefore pick $\{\CC_1^b, \CC_3^b, \CC_4^b, \CC_5^b\}$ as a basis. The K\"ahler volume of $\CC_i^a$,  denoted by $\xi_i^b$, are related to parameters from resolution (a) by:
\be
\xi_1^b = - \xi_1^a= - \xi_1~, \quad \xi_3^b = \xi_3^a= \xi_3~, \quad \xi_4^b= \xi_4^a=\xi_1+ \xi_2- \xi_5~, \quad \xi_5^b = \xi_5^a= \xi_5~.
\ee
Indeed, the resolution (b) is obtained from resolution (a) by flopping the curve $\mathcal{C}_1^a$ to $\mathcal{C}_1^b \cong -\mathcal{C}_1^a$.
We find the IIA profile:
\be\label{chi b dP3}
\text{(b)}\; : \; \chi(r_0) = \begin{cases}  
  r_0 + \xi_3^b \qquad &\text{if}\quad  r_0\geq 0 \cr 
 - r_0 + \xi_3^b \qquad &\text{if}\quad -\xi_4^b \leq r_0 \leq 0 \cr 
  \xi_3^b+\xi_4^b \qquad &\text{if}\quad -\xi_4^b-\xi_5^b \leq r_0 \leq -\xi_4^b  \cr 
-2r_0 + \xi_3^b - \xi_4^b - 2\xi_5^b \; &\text{if}\quad -\xi_1^b-\xi_4^b-\xi_5^b \leq r_0 \leq -\xi_4^b-\xi_5^b \cr 
-r_0 + \xi_1^b + \xi_3^b - \xi_5^b &\text{if}\quad  r_0 \leq -\xi_1^b-\xi_4^b-\xi_5^b \cr 
\end{cases}
\ee 
 In this case, the flavor D6-branes are at $r_0 = {-}\xi_1^b{-}\xi_4^b{-}\xi_5^b$ and $r_0 = {-}\xi_4^b$, whereas the gauge D6-branes are at $r_0 = {-}\xi_4^b{-}\xi_5^b$ and $r_0 = 0$. Thus, one flavor brane lies between the gauge D6-branes and the other flavor brane lies to the left of the gauge D6-branes along the $r_0$ direction. The W-boson mass is given by $M(W) = 2\varphi = \xi_4^b + \xi_5^b$ and the hypermultiplets masses are:
 \bea\label{eq:hypermultiplet_mass_dp3_b}
& M(\mathcal{H}_{1,1}) = \varphi-m_1=\xi_5^b~,\qquad 
&& M(\mathcal{H}_{1,2})  = -\varphi + m_2 = \xi_1^b\\
&M(\mathcal{H}_{2,1}) = \varphi+m_1=\xi_4^b~,\qquad 
&& M(\mathcal{H}_{2,2}) = \varphi+m_2 =  \xi_1^b + \xi_4^b + \xi_5^b~.
\eea
The instanton masses are:
\be\label{eq:instanton_mass_dp3_b}
	M(\mathcal{I}_1) = h_0 + \varphi - m_1 - m_2 = \xi_3^b~, \qquad\quad
	 M(\mathcal{I}_2) =  h_0 + 2\varphi - m_2  = \xi_3^b + \xi_4^b~.	
\ee
These results of course agree with \eqref{mu to FT dP3}. The string tension computed from IIA also agrees with $T= \d_{\varphi} \CF_{(b)}$.

\paragraph{Gauge-theory chamber (c).} The third resolution and its associated IIA profile are shown in Figure \ref{fig:E3 res c}. It can be obtained from resolution (a) by flopping the curve $\CC_2^a$.
The relations among curves are:
\begin{align}
	\mathcal{C}_1^c &\cong \mathcal{C}_4^c + \mathcal{C}_5^c~,\qquad \quad \mathcal{C}_3^c \cong \mathcal{C}_5^c + \mathcal{C}_6^c~.
\end{align}
We now have:
\be\label{chi c dP3}
\text{(c)}\; : \qquad \chi(r_0) = \begin{cases}  
  r_0 + \xi_2^c + \xi_5^c + \xi_6^c \qquad &\text{if}\quad  r_0\geq \xi_2^c \cr 
 2 r_0 + \xi_5^c + \xi_6^c \qquad &\text{if}\quad 0 \leq r_0 \leq \xi_2^c \cr 
 \xi_5^c+\xi_6^c \qquad &\text{if}\quad -\xi_4^c \leq r_0 \leq 0  \cr 
r_0 + \xi_4^c + \xi_5^c + \xi_6^c \qquad &\text{if}\quad -\xi_4^c-\xi_5^c \leq r_0 \leq -\xi_4^c \cr     
-r_0 - \xi_4^c - \xi_5^c + \xi_6^c &\text{if}\quad  r_0 \leq -\xi_4^c-\xi_5^c \cr 
\end{cases}
\ee 
Here, we have:
\be
\xi_1^c = \xi_1 + \xi_2~, \qquad \xi_2^c = -\xi_2~, \qquad \xi_3^c = \xi_2+ \xi_3~,\qquad \xi_{4,5,6}^c= \xi_{4,5,6}~.
\ee
We then find $M(W_\alpha)= 2\varphi = \xi_4^c + \xi_5^c$ for the W-boson, as well as: 
 \bea\label{eq:hypermultiplet_mass_dp3_c}
& M(\mathcal{H}_{1,1}) =  \varphi - m_1 = \xi_5^c~,\qquad 
&& M(\mathcal{H}_{1,2})  = \varphi - m_2 = \xi_2^c + \xi_4^c + \xi_5^c~, \\
&M(\mathcal{H}_{2,1}) =  \varphi + m_1 = \xi_4^c~,\qquad 
&& M(\mathcal{H}_{2,2}) =  -\varphi-m_2 = \xi_2^c~,
\eea
for the hypermultiplets, and:
\be\label{eq:instanton_mass_dp3_c}
	M(\mathcal{I}_1) =h_0 + 2\varphi - m_1 =  \xi_5^c + \xi_6^c~, \qquad\quad
	 M(\mathcal{I}_2) = h_0 + \varphi = \xi_6^c~,
\ee
for the instanton particles, again in perfect agreement with \eqref{mu to FT dP3}.

Note that resolutions (b) and (c) are compatible with the limits $m_2 \rightarrow \infty$ and 
$m_2 \rightarrow -\infty$, respectively. In the field theory, this leads to an $SU(2)$ $N_f=1$ low energy theory. This correspond to removing the toric divisors $D_1$ (resp., $D_2$) by sending the size of $\CC_1^b$ (resp., $\CC_2^c$) to infinity, giving us the resolved $E_2$ singularity (the $dP_2$ geometry).

\paragraph{Other resolutions and decoupling limits.} The vertical reduction of the other 6 resolutions (d) to (i) can be performed in exactly the same way.  The various geometric decoupling limits also agree with the field-theory RG flows. For instance, resolution (f)  is compatible with the limit $m_1 \rightarrow -\infty$, $m_2 \rightarrow \infty$, which leads to a pure $SU(2)_0$ theory. (That is, $k_{\rm eff}=0$ in the IR because the two flavors have real masses of opposite signs.) Indeed, we get the $E_1$ ($\mathbb{F}_0$) geometry upon decouling the divisors $D_1$ and $D_4$. A similar comment holds for resolution (g). On the other hand, resolutions (h) and (i) are compatible with the decoupling limits $m_{1,2} \rightarrow \infty$ and $m_{1,2} \rightarrow -\infty$, respectively, leading to the $SU(2)_\pi$ theory (that is, $k_{\rm eff}=\pm 1$) from the $\t E_1$ ($dP_1$) geometry.

%%%%%%%%%%%%%%%%%%%
\section{Higher-rank example: $SU(N)_k$}\label{sec: SUNk}
To illustrate our methods in some simple higher-rank example, let us consider a family of toric geometries that admit a resolution leading to an $SU(N)_k$ gauge theory \cite{Intriligator:1997pq, Aharony:1997bh}.

\subsection{The $SU(N)_k$ gauge theory and its Coulomb branch}
Consider a 5d $\CN=1$ gauge theory with gauge group $SU(N)$ and CS level $k\in \Z$. Let $\varphi_a$ ($a=1, \cdots, N-1$) denote the Coulomb branch scalars, and let $\phi_a$ denote the associated $U(N)$ parameters, as in \eqref{U to SU}. We denote by $A_{ab}=A_{ba}$ the Cartan matrix of $SU(N)$, namely:
\be\label{def Aab SUN}
A_{ab} =2 \delta_{a,b} - \delta_{a, b+1}- \delta_{a+1, b}~.
\ee
The $N-1$ simple roots are then given by:
\be
\alpha_{(b)}^a = A_{ab}
\ee
The prepotential on the Coulomb branch of the $SU(N)_k$ theory is easily computed. We have:
\be\label{F SUNk i}
\CF(\varphi, h_0)=\CF_{\rm kin}+ \CF_{\rm CS}+ \CF_{\oneloop}~,
\ee
with:
\bea\label{F SUNk ii}
&  \CF_{\rm kin}&=&\; \;\half h_0 \sum_{a=1}^N \phi_a^2 &=&\;\; \half h_0\, A^{ab} \varphi_a \varphi_b~,\\
&  \CF_{\rm CS}&=&\; \; {k\ov 6} \sum_{a=1}^N \phi_a^3 &=&\;\; {k\ov 2} \sum_{a=1}^{N-1} \big(\varphi_{a-1}^2 \varphi_a- \varphi_a \varphi_{a+1}^2\big)~,\\
&  \CF_{\oneloop}&=&\;\; {1\ov 6} \sum_{a<b} (\phi_a- \phi_b)^3 &=&\;\; {4\ov 3}\sum_{a=1}^{N-1} \varphi_a^3 + \half \sum_{a=1}^{N-1} (N-2 a)\big(\varphi_{a-1}^2 \varphi_a- \varphi_a \varphi_{a+1}^2\big)~.
\eea
The W-bosons associated to the simple roots have masses:
\be
M(W_{\alpha_{(b)}})= A^{ab} \varphi_a~,
\ee
and become massless at the boundaries of the fundamental Weyl chamber.

\subsection{$SU(N)_k$ gauge theory from a toric singularity}
\begin{figure}[t]
\begin{center}
\subfigure[\small $N=3$, $h=2$.]{
\includegraphics[width=3cm]{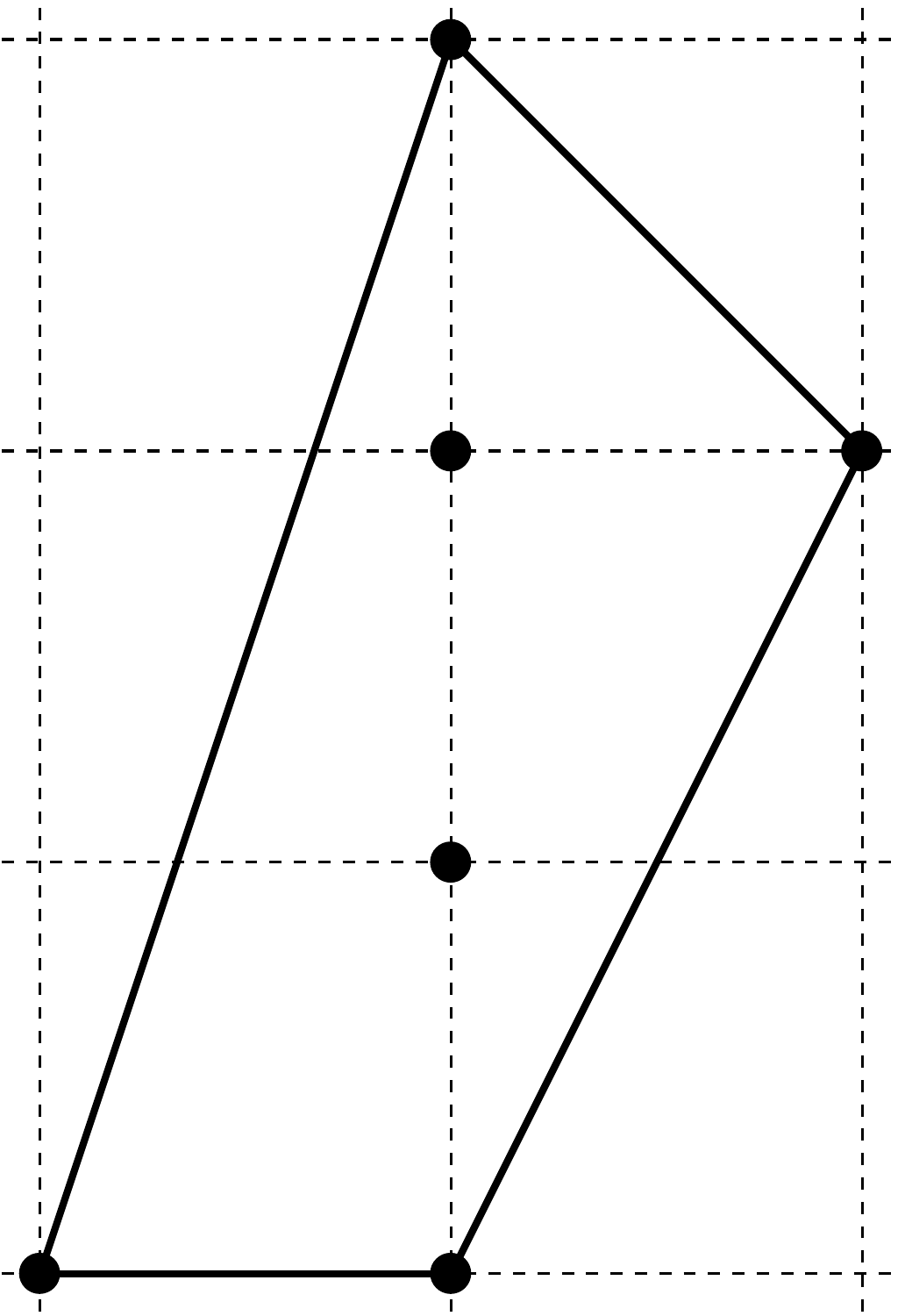}\label{fig:SUN sing 1}}\qquad
\subfigure[\small $N=4$, $h=4$.]{
\includegraphics[width=3cm]{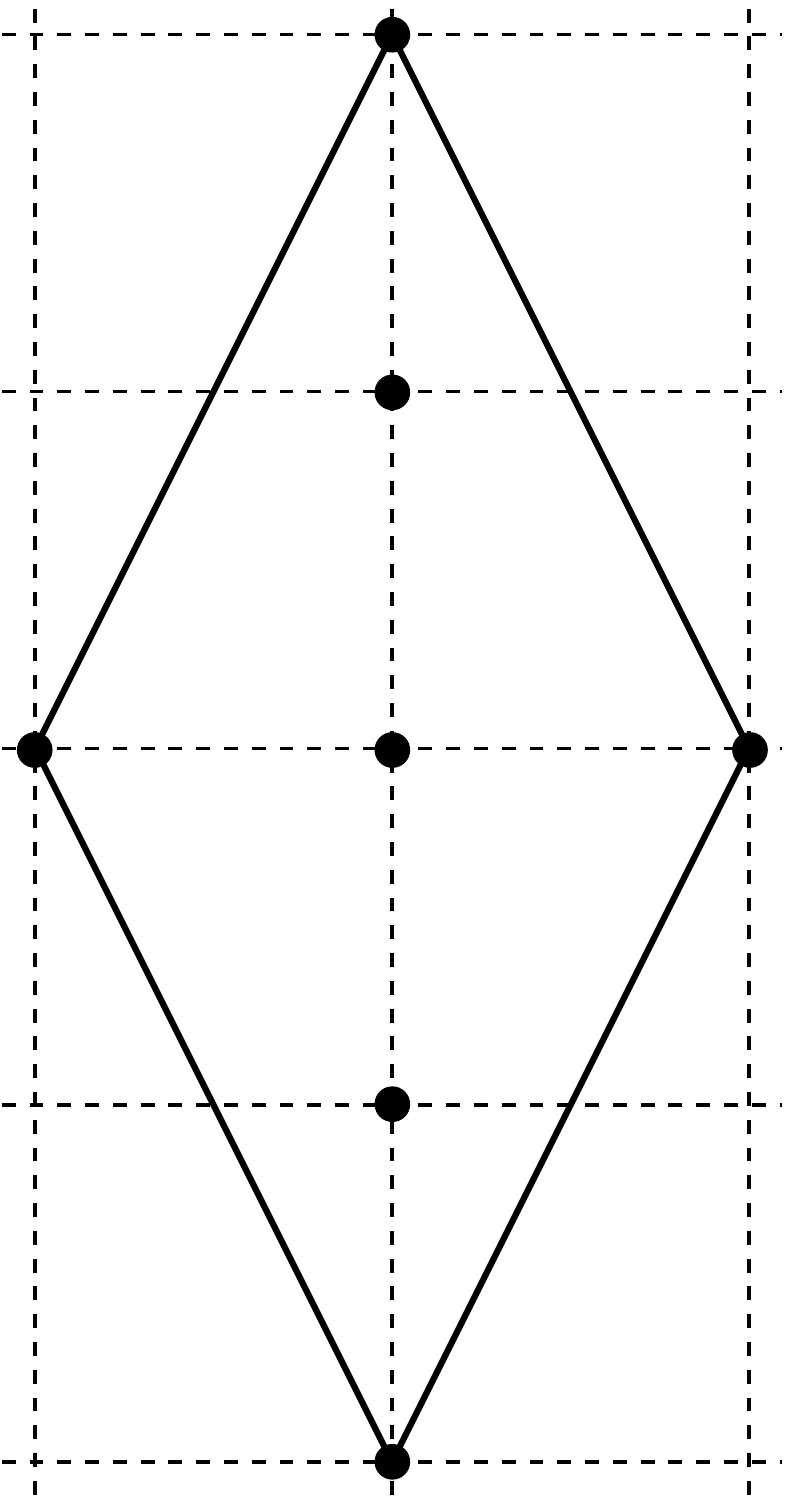}\label{fig:SUN sing 2}}\qquad
\subfigure[\small $N=4$, $h=5$.]{
\includegraphics[width=3cm]{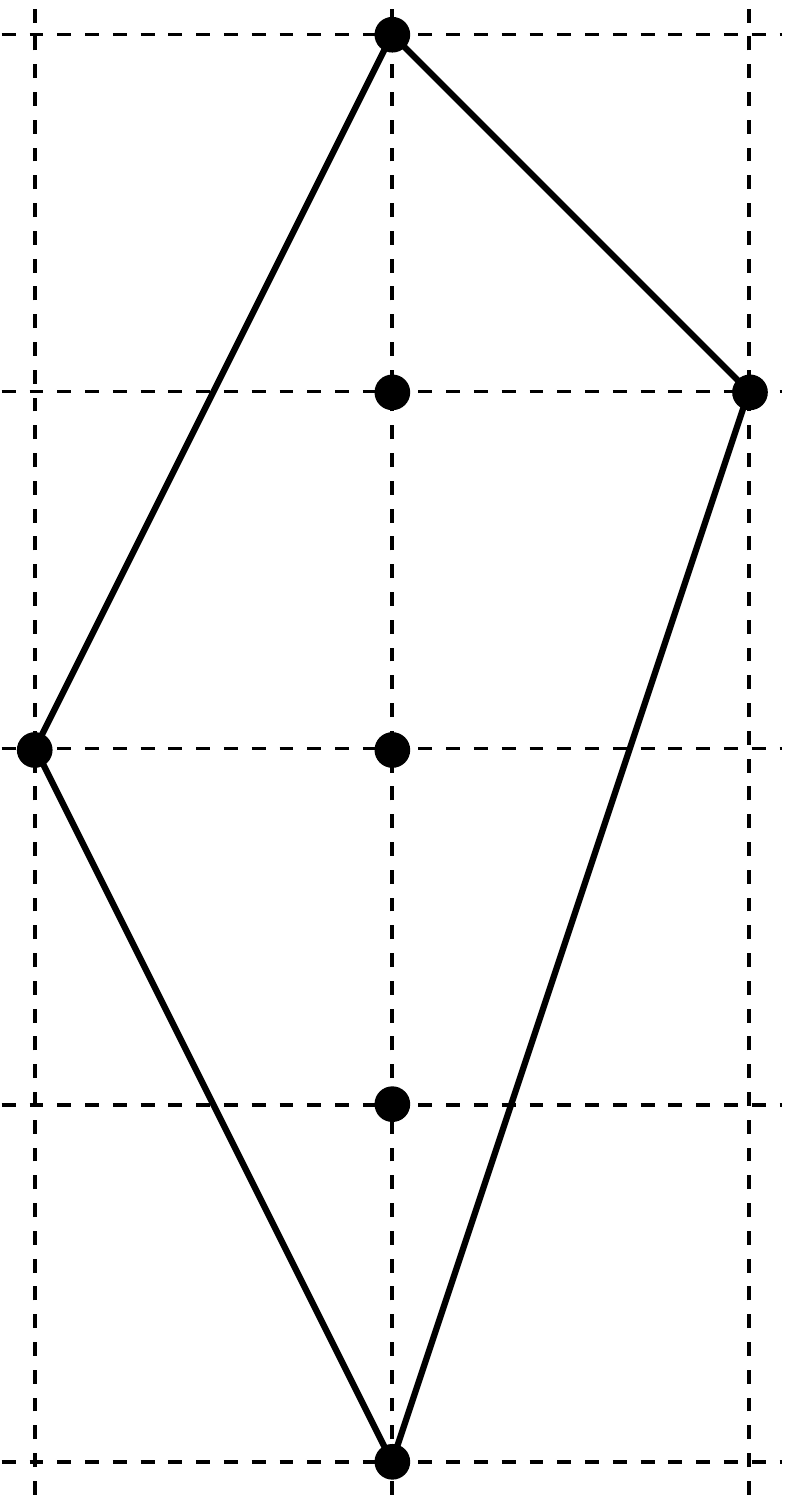}\label{fig:SUN sing 3}}\qquad
\subfigure[\small $N=4$, $h=7$.]{
\includegraphics[width=3cm]{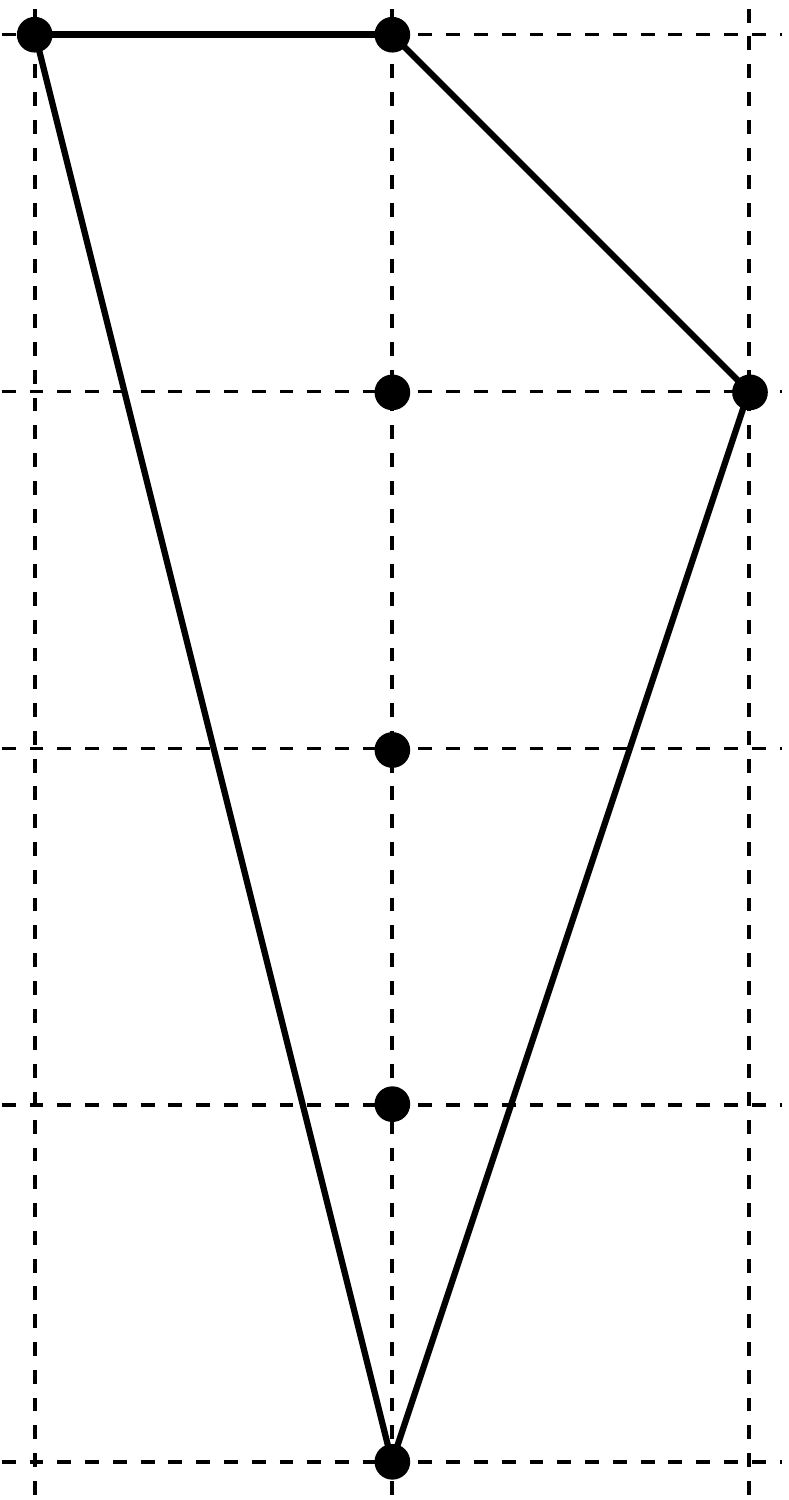}\label{fig:SUN sing 4}}
\caption{Examples of ``$SU(N)_k$ toric singularities.'' Upon vertical reduction, the corresponding gauge theories are $SU(3)_{-1}$, $SU(4)_0$, $SU(4)_1$ and $SU(4)_3$, respectively. \label{fig:SUN sings}}
 \end{center}
 \end{figure}

Consider a convex toric diagram with external points at:
\be\label{toric diag SUNk}
w_0 = (0, 0)~, \quad w_N= (0,N)~, \quad w_x= (-1, h_x)~, \quad w_y=(1, h_y)~,
\ee
with $h_x, h_y \in \Z$ and $N>2$. (The case $N=2$ has been discussed above.) We impose the condition:
\be\label{convex condi SUN toric}
0<h< 2N~, \qquad\qquad h\equiv h_x + h_y~,
\ee
to ensure that the toric diagram is strictly convex.
Some examples are shown in Figure~\ref{fig:SUN sings}. Note that the toric diagram preserves parity, in the sense of \eqref{C0 on w Parity}, if and only if $h=N$. This suggests that the $SU(N)$ CS level should be identified with $k= h-N$, up to the overall sign which is a matter of convention. We will confirm this expectation using the IIA setup.
%%%%%%%%%
 %%%%%%%%%%%

\begin{figure}[t]
\begin{center}
\subfigure[\small Toric diagram and toric divisors.]{
\includegraphics[width=4.5cm]{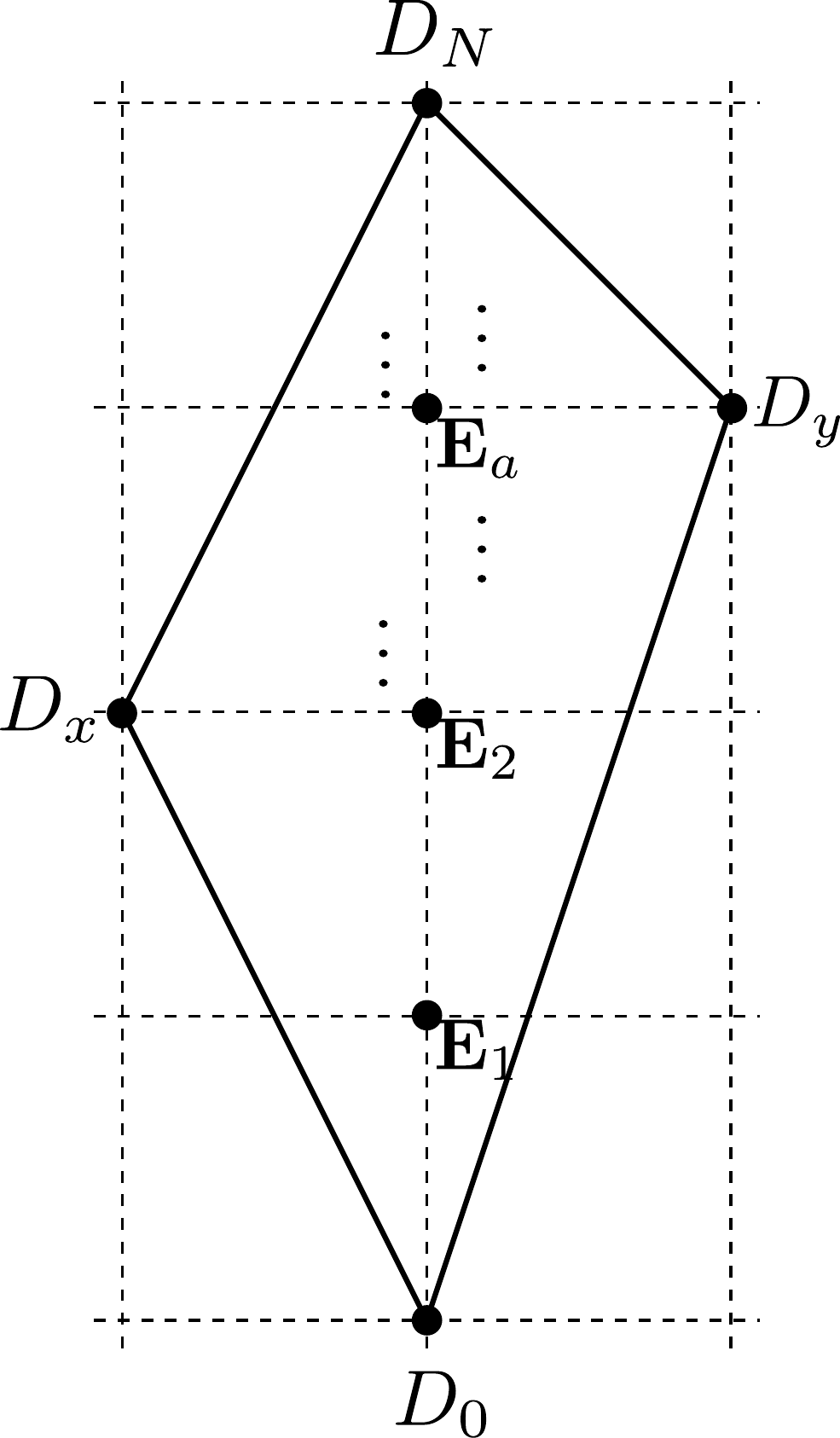}\label{fig:SUN sing full}}\qquad
\subfigure[\small Unique resolution with allowed vertical reduction.]{
\includegraphics[width=4.5cm]{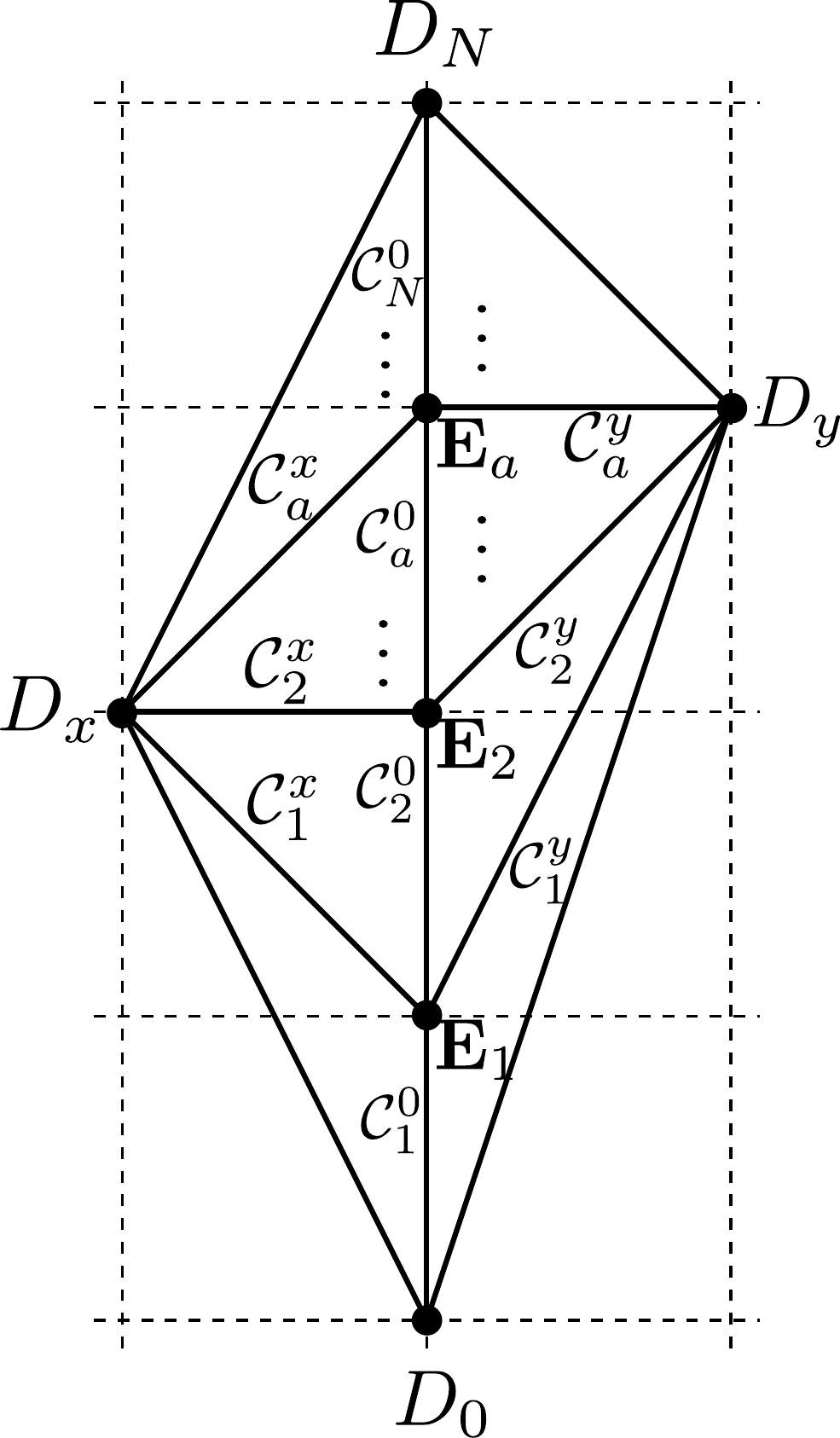}\label{fig:SUN res full}}
\caption{A rank-$(N{-}1)$ toric singularity and its unique resolution with an ``allowed'' vertical reduction, leading to an $SU(N)_k$ gauge theory on D6-branes in type IIA. \label{fig:SUNk res}}
 \end{center}
 \end{figure}

While this toric singularity may admit many distinct resolutions, there is a unique triangulation of the toric diagram with an allowed vertical reductions, as shown in Figure~\ref{fig:SUNk res}. Let us denote by $D_0$, $D_N$, $D_x$ and $D_y$ the toric divisors associated to the four external points \eqref{toric diag SUNk}. We also have the exceptional toric divisors $\bE_a$ ($a=1, \cdots, N-1$), corresponding to the internal points $w_a=(0, a)$ in the toric diagram. We have the linear equivalences:
\bea
&D_x \cong D_y~, \qquad &&D_0 \cong   (N-1) D_N + (h-2)D_x+\sum_{a=1}^{N-1} (a-1) \bE_a~,\\
&  &&  D_N \cong -D_0 - 2 D_x - \sum_{a=1}^{N-1} \bE_a~,
\eea
amongst toric divisors.
The resolution shown in Fig.~\ref{fig:SUN res full} contains the exceptional curves:
\bea
&\CC_a^x\cong D_x \cdot \bE_a~, \qquad\quad
&\CC_a^y\cong D_y \cdot \bE_a~, \qquad & a=1, \cdots N-1~,\\
&\CC_a^0 \cong \bE_{a-1} \cdot \bE_a~, &\qquad & a= 1, \cdots N~,
\eea
with $\bE_0\equiv D_0$ and $\bE_N\equiv D_N$.
One finds the following linear equivalences amongst them:
\be
\CC_a^x \cong \CC_a^y~,\quad \qquad \qquad 
\CC_a^0 - \CC_{a+1}^0 \cong (h-2 a) \CC_a^x~,\qquad a=1, \cdots, N-1~.
\ee
Thus we may consider the $N$ independent curves $\CC_a^x$ and $\CC_1^0$, for instance. According to our general discussion in section~\ref{subsec: gaugefromIIA}, M2-branes wrapped on $\CC_a^x$ and on $\CC_a^0$ map to the W-bosons $W_{\alpha_{(a)}}$ (associated to the simple roots $\alpha_{(a)}$) and to the instanton particles $\CI_a$, respectively.

%%%%%%%%%
 %%%%%%%%%%%

\subsubsection{Type IIA reduction}
\begin{figure}[t]
\begin{center}
\includegraphics[height=5cm]{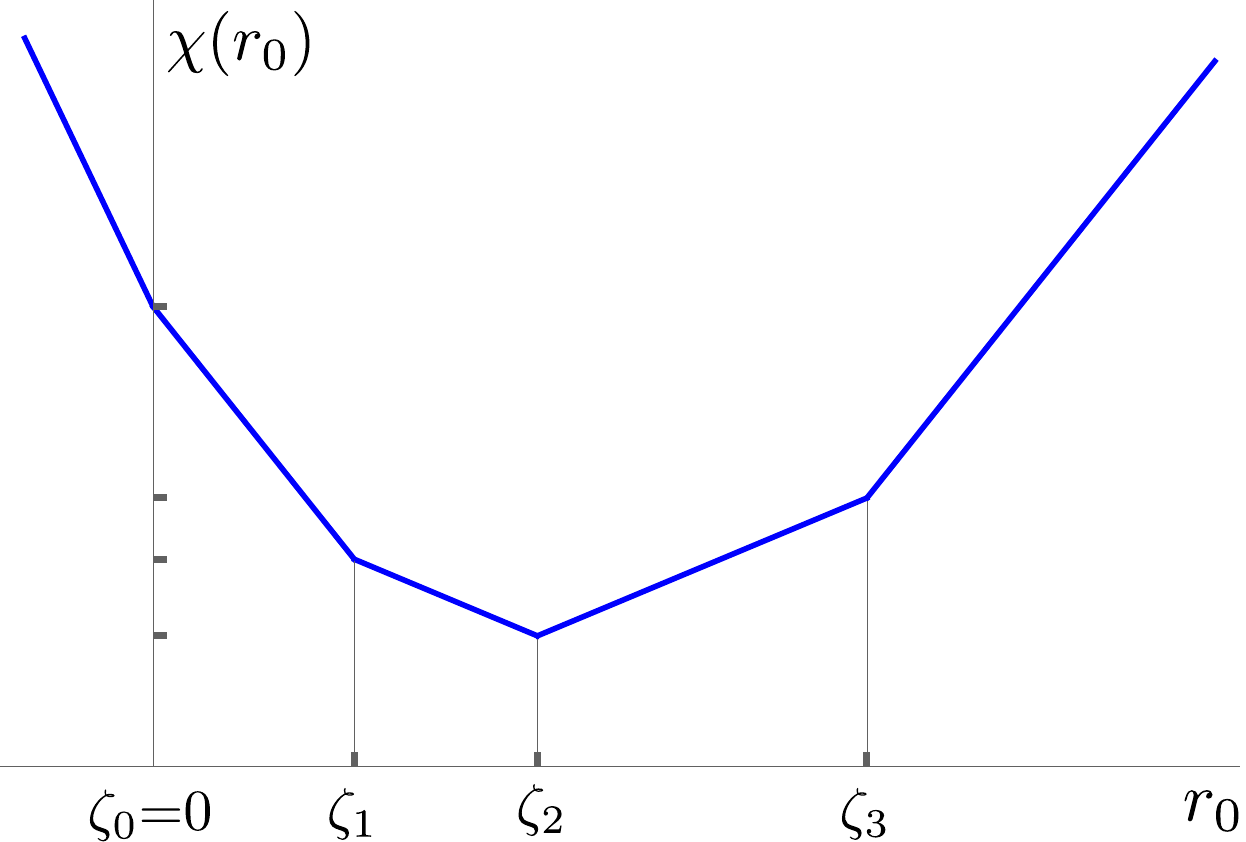}
\caption{Type IIA profile for $\chi(r_0)$ in the case of the $SU(4)_1$ singularity ($N=4, h=5$). Each area  under the curve between $\zeta_{a-1}$ and $\zeta_a$ corresponds to the tension $T_a$ of an elementary magnetic string.  \label{fig:SUNk profile}}
 \end{center}
 \end{figure}

Consider the vertical reduction of the triangulated toric diagram of Fig.~\ref{fig:SUN res full}. We have the GLSM description:
\be\label{GLSM SUNk}
\begin{tabular}{l|ccccc|c}
 & $D_0$ &$\bE_b$& $D_N$&$D_x$& $D_y$ & \\
 \hline
$\CC_1^0$    &$h-2$&$-h \delta_{b,1}$ &$0$ & $1$&$1$& $\eta$\\
$\CC_a^x$    &$\delta_{0,1}$&$-A_{ab}$ &$\delta_{a, N-1}$ & $0$& $0$& $\xi_a^x$\\
\hline\hline
$U(1)_M$&$1$&$-\delta_{b,1}$ &$0$ & $0$& $0$& $r_0$
 \end{tabular}
\ee
with $a, b = 1, \cdots, N-1$, and $A_{ab}$ as defined in \eqref{def Aab SUN}. It is convenient to introduce the parameters:
\be
\zeta_a= \sum_{b=1}^a \xi_b^x~, \qquad a=0, \cdots, N-1~,
\ee
with $\zeta_0=0$.
It is clear from the vertical projection of the toric diagram that the IIA background is a resolved $A_1 \cong \C^2/\Z_2$ singularity, fibered over $\R$, with $N$ D6-branes wrapped over the exceptional $\P^1$. By a direct computation, one finds the IIA profile:
\be\label{SUN IIA profile}
\chi(r_0) = \eta + (2 a- h) r_0 - 2 \sum_{b=0}^{a-1} \zeta_b~, \qquad {\rm if}\quad \zeta_{a-1}\leq r_0 \leq \zeta_a~, \qquad a=0, \cdots, N~,
\ee
with the understanding that $\zeta_{-1}\equiv -\infty$ and $\zeta_N\equiv \infty$. An example is shown in Figure~\ref{fig:SUNk profile}.
Here, each kink corresponds to a single wrapped D6-brane.
According to \eqref{k eff from IIA}, this gives rise to a CS level:
\be
k= h-N~,
\ee
as anticipated.
The convexity condition \eqref{convex condi SUN toric} on the toric diagram implies that:
\be
|k|< N~.
\ee
Thus, we recover the known fact that the $SU(N)_k$ theory with $|k|<N$ has a UV fixed point described by an isolated toric singularity \cite{Aharony:1997bh}. The limiting case $|k|=N$, on the other hand, corresponds to a non-isolated toric singularity (with a line of $A_1$ singularities that has been related to an enhanced $SU(2)$ global symmetry at strong coupling \cite{Bergman:2013aca}). Incidentally, we note that UV fixed points are expected for $SU(N)_k$ with $k>N$ as well \cite{Jefferson:2018irk, Cabrera:2018jxt}, but they cannot be realized in the strictly toric framework.

 %%%%%%%%%%%
 %%%%%%%%%%%
From the IIA profile \eqref{SUN IIA profile}, we can read off the map between the geometry and the field-theory parameters. The open strings stretched between two adjacent D6-branes give rise to the simple-root W-bosons---more generally, open strings stretched between any two distinct D6-branes span the full set of positive roots, in the obvious way. The simple-root W-bosons have masses:
\be
M(W_{\alpha_{(b)}})= A^{ab} \varphi_a = \zeta_{b}- \zeta_{b-1}= \xi_b^x~, \qquad b=1, \cdots, N-1~.
\ee
On the other hand, the instanton particles have masses:
\be
M(\CI_a) = h_0 +2 \varphi_a + (k+N-2a)(\varphi_a -\varphi_{a-1})~, \qquad a=1, \cdots, N~,
\ee
according to \eqref{I mass heuristic}, which should be identified with the volumes of the wrapped D2-branes at the kinks of \eqref{SUN IIA profile}: 
\be
M(\CI_a)=\eta + (2a-h)\zeta_{a-1} - 2 \sum_{b=0}^{a-1} \zeta_b~.
\ee
This implies the relations:
\be\label{geom to ft SUNk}
 \xi_a^x= A^{ab} \varphi_b~, \qquad \eta= h_0 + (N+k) \varphi_1~.
\ee
Note that $\eta$ is also the volume of the curve $\CC_1^0$ in M-theory. More generally, the instanton $\CI_a$ corresponds to an M2-brane wrapped over $\CC_a^0$. 
Moreover, given the geometry/gauge-theory map \eqref{geom to ft SUNk}, we can easily check that the type-IIA and gauge-theory computations of the string tensions also agree, with:
\be
T_a= \d_{\varphi_a} \CF = \int_{\zeta_{a-1}}^{\zeta_a}  \chi(r_0)~,
\ee
in terms of the prepotential \eqref{F SUNk i}-\eqref{F SUNk ii} and of the IIA profile \eqref{SUN IIA profile}, respectively.

\subsubsection{Prepotential from M-theory}
Finally, let us also compute the prepotential directly from M-theory. We have:
\be
S= \mu_x D_x + \sum_{a=1}^{N-1} \nu_a \bE_a~.
\ee
The parameters $\mu_x$ and $\nu_a$ are related to the GLSM FI parameters by:
\be
\eta= \mu_x - h \nu_1~, \qquad\qquad  \xi_a^x = - A^{ab} \nu_b~.
\ee
This follows from \eqref{GLSM SUNk}. Comparing to \eqref{geom to ft SUNk}, we directly see that:
\be
\nu_a = - \varphi_a~, \qquad\qquad \mu_x = h_0~.
\ee
Therefore, one finds:
\be\label{CF geom SUNk}
\CF= -{1\ov 6} S^3=   \half h_0^2 \varphi_a (D_x^2 \bE^a) - \half h_0 \varphi_a \varphi_b (D_x \bE^a \bE^b) + {1\ov 6} \varphi_a \varphi_b \varphi_c (\bE^a\bE^b\bE^c)
\ee
up to an ambiguous $\varphi$-independent term, which we can set to zero in conventions in which $D_x^3=0$. One can check that, for the resolution of Fig.~\ref{fig:SUN res full}, the only non-zero triple-intersection numbers are:
\be
D_x \bE_a \bE_b = - A_{ab}~, \quad \bE_a^3=8~, \quad \bE_{a-1}^2 \bE_a = h-2 a~, \quad 
\bE_{a-1} \bE_a^2 = 2 a-2 -h~.
\ee
Plugging this into \eqref{CF geom SUNk}, one reproduces the field theory prepotential \eqref{F SUNk i}-\eqref{F SUNk ii}.

%%%%%%%%%%%%%%%%%%%
\section{S-duality for 5d $\CN=1$ gauge theories}\label{sec:SDUL}
A given toric CY$_3$ singularity $\MG$ may admit several type-IIA reductions, giving rise to several distinct gauge-theory ``phases.'' This phenomenon is sometimes called ``UV duality,'' since different 5d $\CN=1$ gauge theories can have the same UV completion as a 5d SCFT \cite{Aharony:1997bh, Bao:2011rc, Bergman:2013aca, Taki:2014pba, Bergman:2014kza, Zafrir:2014ywa, Mitev:2014jza}. In our setup, we have an $SL(2, \Z)$-worth family of potential choices of the ``M-theory circle $U(1)_M$'' along which to reduce to type IIA, corresponding exactly to doing an $SL(2, \Z)$ transformation of the toric diagram before peforming the ``vertical reduction.''  In type IIB string theory, this corresponds to S-duality on the $(p,q)$-web.

In this section, we study examples of 5d ``UV dualities'' with all the possible mass parameters $\mu_j$ turned on. After briefly discussing the rank-one case, we will focus on one particularly nice rank-two example, the ``beetle'' singularity.

As anticipated in the introduction, when sending all the massive deformation parameters to zero, $\mu_i \to 0$, while keeping $\nu_i\neq 0$ finite, we obtain a partial resolution of the singularity which is ``universal'' across gauge-theory descriptions. Most IR phases disappear in that limit, and the gauge theory phases coalesce in a much simpler geometry. This universality indicates that we can identify the resulting smaller extended K\"ahler cone with the genuine Coulomb branch of the 5d SCFT, $\CM_\FT^C$. We stress that, in that limit, the gauge theory interpretation breaks down as the gauge coupling is sent to infinity. Nevertheless, perhaps surprisingly, the extrapolated gauge theory result agrees perfectly with the geometric analysis.

\subsection{``Self-duality'' of $SU(2)$ gauge theories}
Consider first the rank-one examples. It is clear that the toric diagrams for $E_{N_f+1}$ in Figure~\ref{fig:SU2 geoms} can have both an horizontal and a vertical reduction. We then have a self-similar S-duality:
\be\label{self S dual}
SU(2),  N_f \; \text{hypers}   \qquad \longleftrightarrow\qquad SU(2),  N_f \; \text{hypers} ~,
\ee
for $N_f=0,1,2$.

As the simplest example, consider the $E_1$ singularity, which is resolved to a local $\mathbb{F}_0\cong \mathbb{P}^1\times \mathbb{P}^1$. As discussed in section~\ref{subsec: E1 gauge theory}, we have a single K\"ahler chamber with two exceptional curves $\CC_1$, $\CC_2$ of non-negative volume, $\xi_1\geq 0$, $\xi_2\geq 0$. Upon vertical reduction, one finds the relations:
\be
\xi_1= h_0 + 2\varphi~, \qquad \xi_2 =  2\varphi~,
\ee
between the K\"ahler parameters and the field-theory parameters of the pure $SU(2)_0$ theory. If we perform an horizontal reduction instead, we have:
\be\label{xi12 SU2 S dual}
\xi_1= 2\varphi^D~, \qquad \xi_2 =  h_0^D+2\varphi^D~,
\ee
where $h_0^D$ and $\varphi^D$ are the parameters of a {\it dual} pure $SU(2)_0$. This S-duality exchanges the $W$-boson and the instanton particle. Interestingly, strictly speaking, the first $SU(2)$ description is only valid for $\xi_1 >\xi_2$, so that $h_0 >0$. To explore the rest of the K\"ahler chamber using a low-energy gauge-theory perspective, one should use the S-dual description with the identifications \eqref{xi12 SU2 S dual} instead. The loci $\xi_1= \xi_2$ is at the ``strong coupling'' limit $h_0=0$, thus corresponding to the Coulomb branch of the $E_1$ SCFT.  (It is not particularly special from the point of view of geometry, however.) The limit $\xi_2\rightarrow 0$ corresponds to a non-traversable K\"ahler wall, where the W-boson in the first $SU(2)$ description becomes massless \cite{Witten:1996qb}, while the limit $\xi_2\rightarrow 1$ corresponds to an instantonic wall, where the instanton particle in the first $SU(2)_0$ description becomes massless. In the S-dual description, the instantonic wall is simply the Weyl chamber wall of the dual $SU(2)$ gauge group, and it is therefore non-transversable.

Similar comments holds about the S-duality \eqref{self S dual}  for $N_f>0$. 
We should also note that the $E_2$ and $E_3$ strongly-coupled SCFT Coulomb branches, $\CM_\FT^C$, with $h_0=0$ and $m_i=0$ (equivalently, $\mu=0$ and $\nu= - \varphi< 0$), are only accessible in the ``star-shaped'' triangulations, Figures~\ref{fig:E2 res a} and \ref{fig:E3 res a}, respectively. The SCFT prepotential on $\CM_\FT^C$ reads:
\be
\CF= {8-N_f\ov 6}\varphi^3~.
\ee

\subsection{The beetle geometry and its gauge-theory phases}\label{sec: beetle}
\begin{figure}[t]
\begin{center}
\includegraphics[height=4cm]{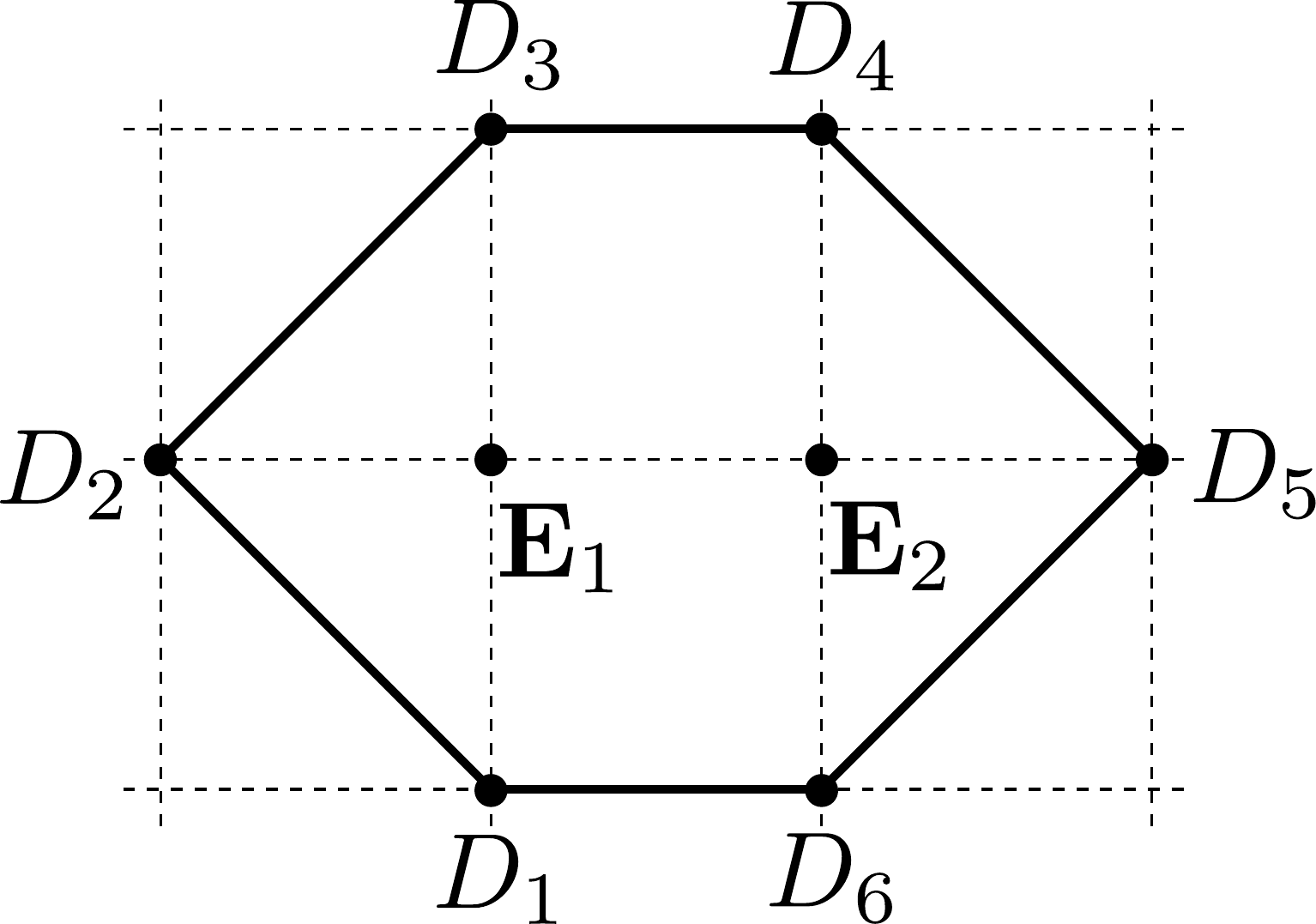}
\caption{The toric diagram of the beetle geometry \label{fig: beetle sing}}
 \end{center} 
 \end{figure}

\begin{figure}[t]
\begin{center}
%%%su2su2
\subfigure[\small ]{\includegraphics[width=2.3cm]{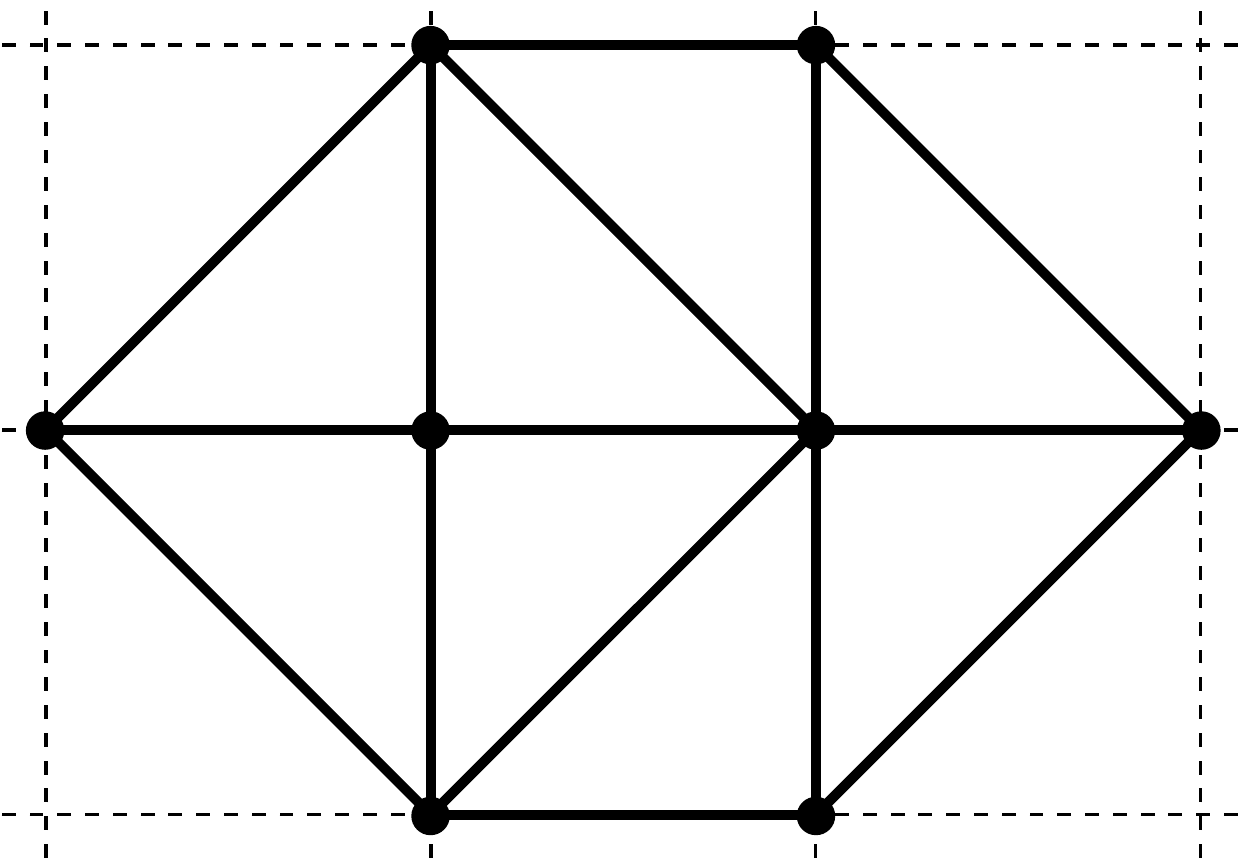}\label{fig:beetle res a}}\;
\subfigure[\small ]{\includegraphics[width=2.3cm]{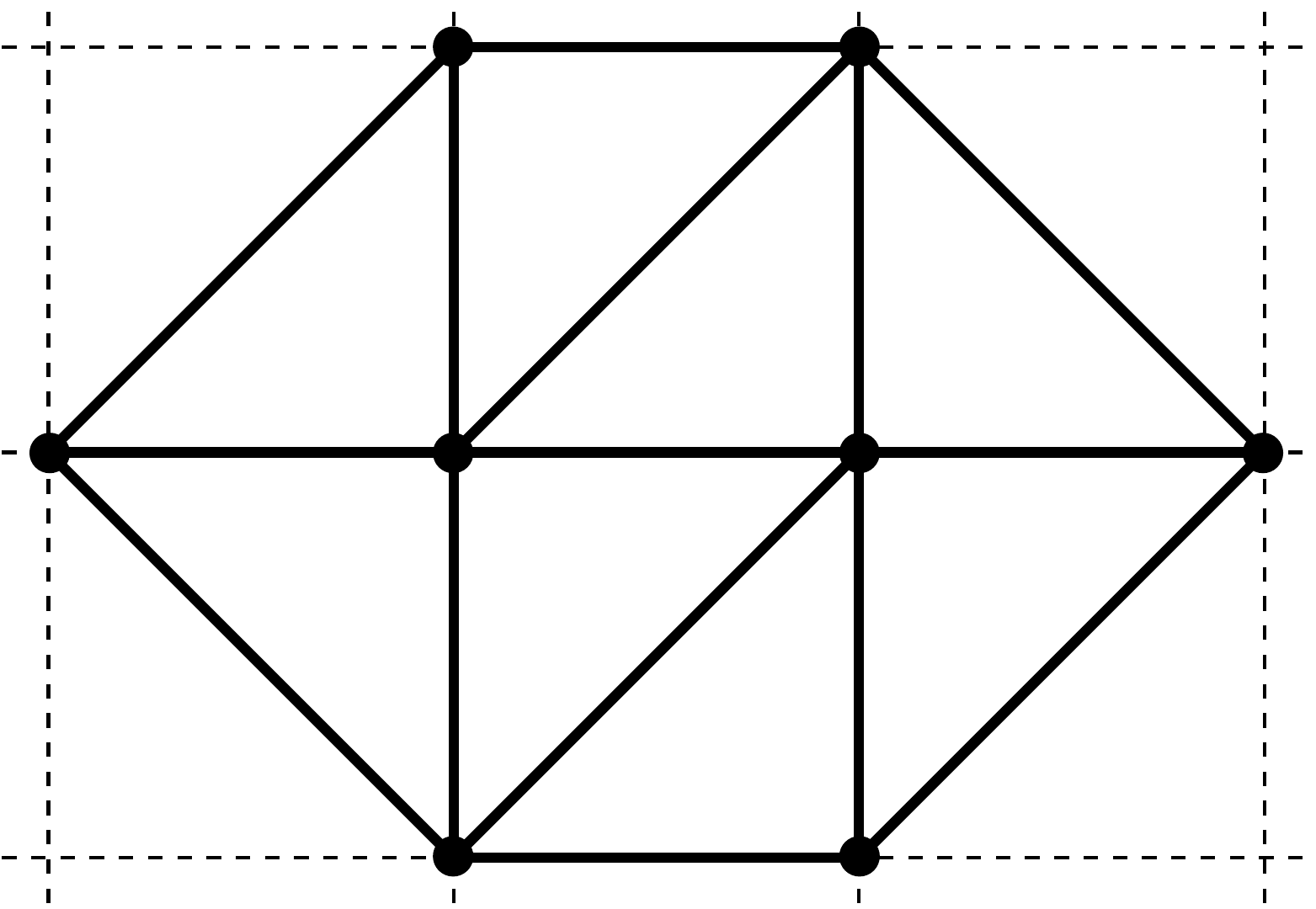}\label{fig:beetle res b}}\;
\subfigure[\small ]{\includegraphics[width=2.3cm]{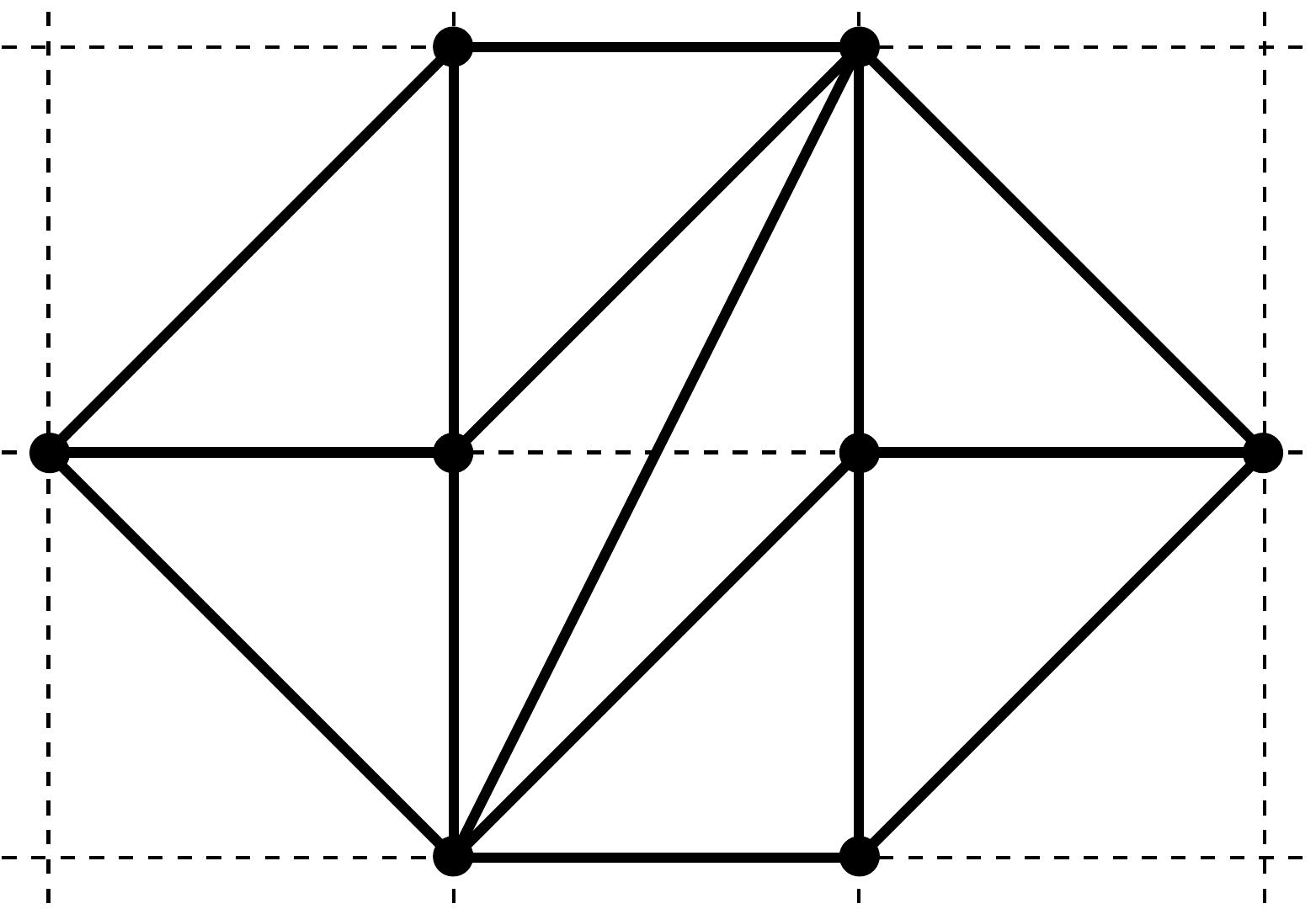}\label{fig:beetle res c}}\;
\subfigure[\small ]{\includegraphics[width=2.3cm]{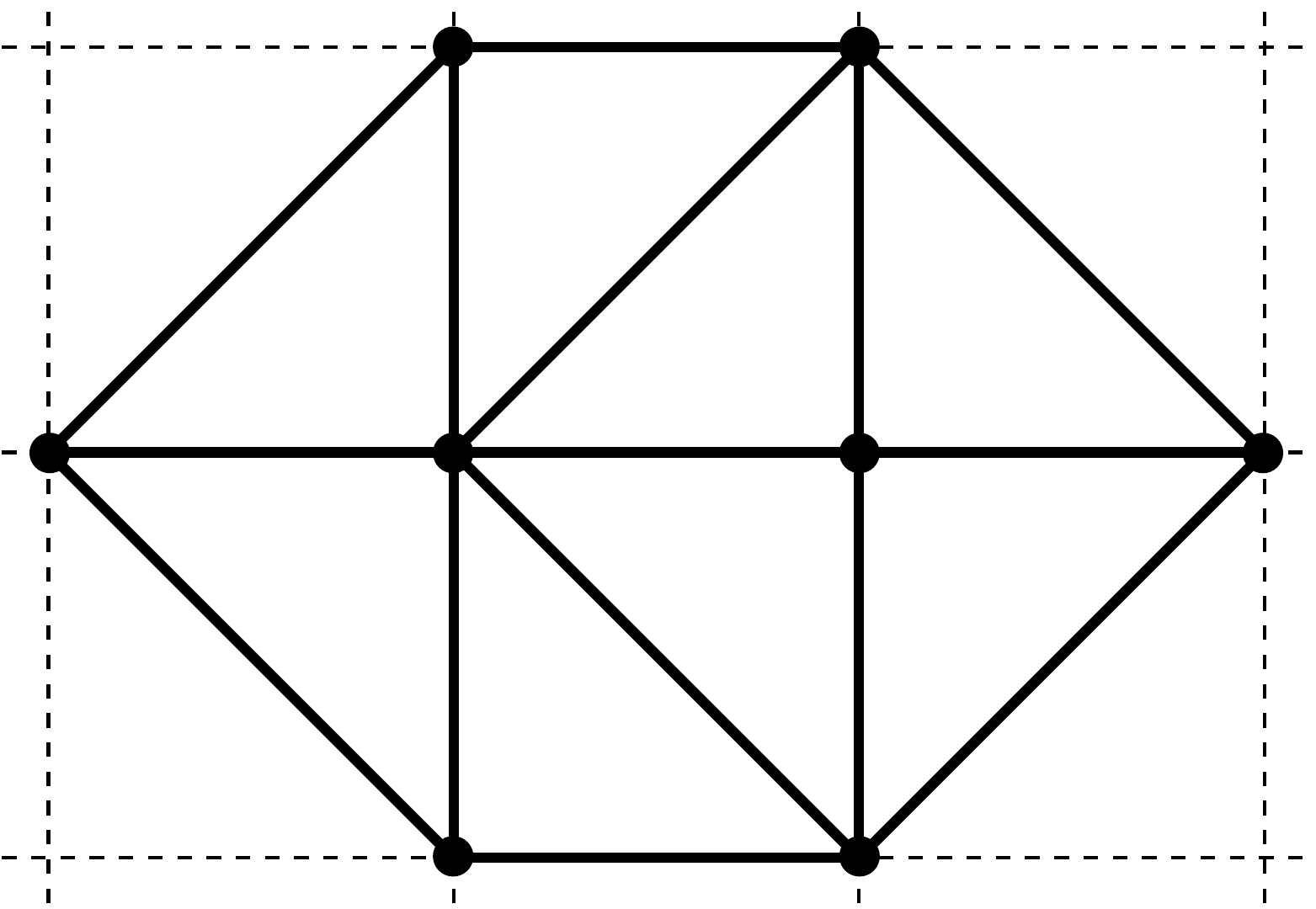}\label{fig:beetle res d}}\;
\subfigure[\small ]{\includegraphics[width=2.3cm]{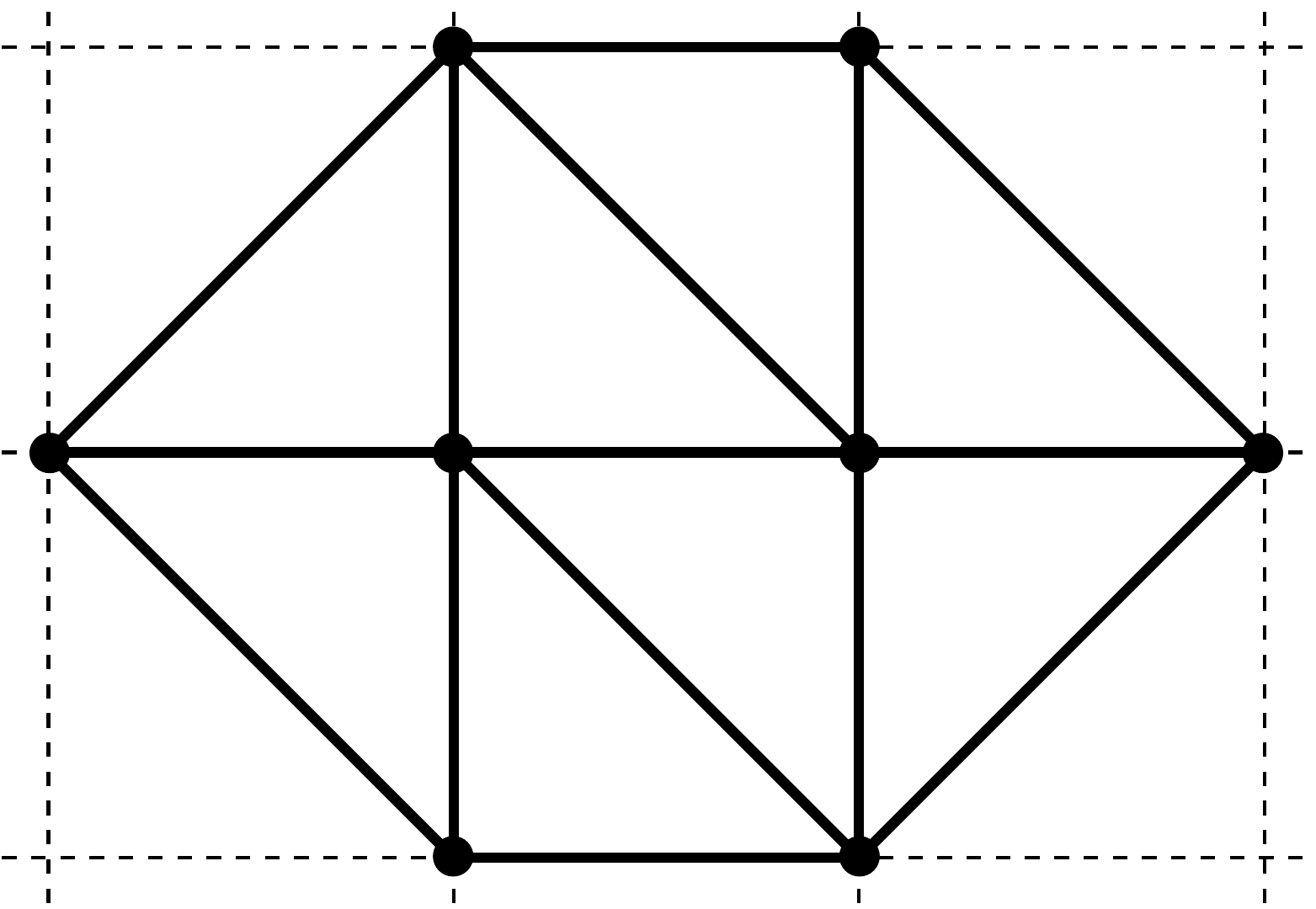}\label{fig:beetle res e}}\;
\subfigure[\small ]{\includegraphics[width=2.3cm]{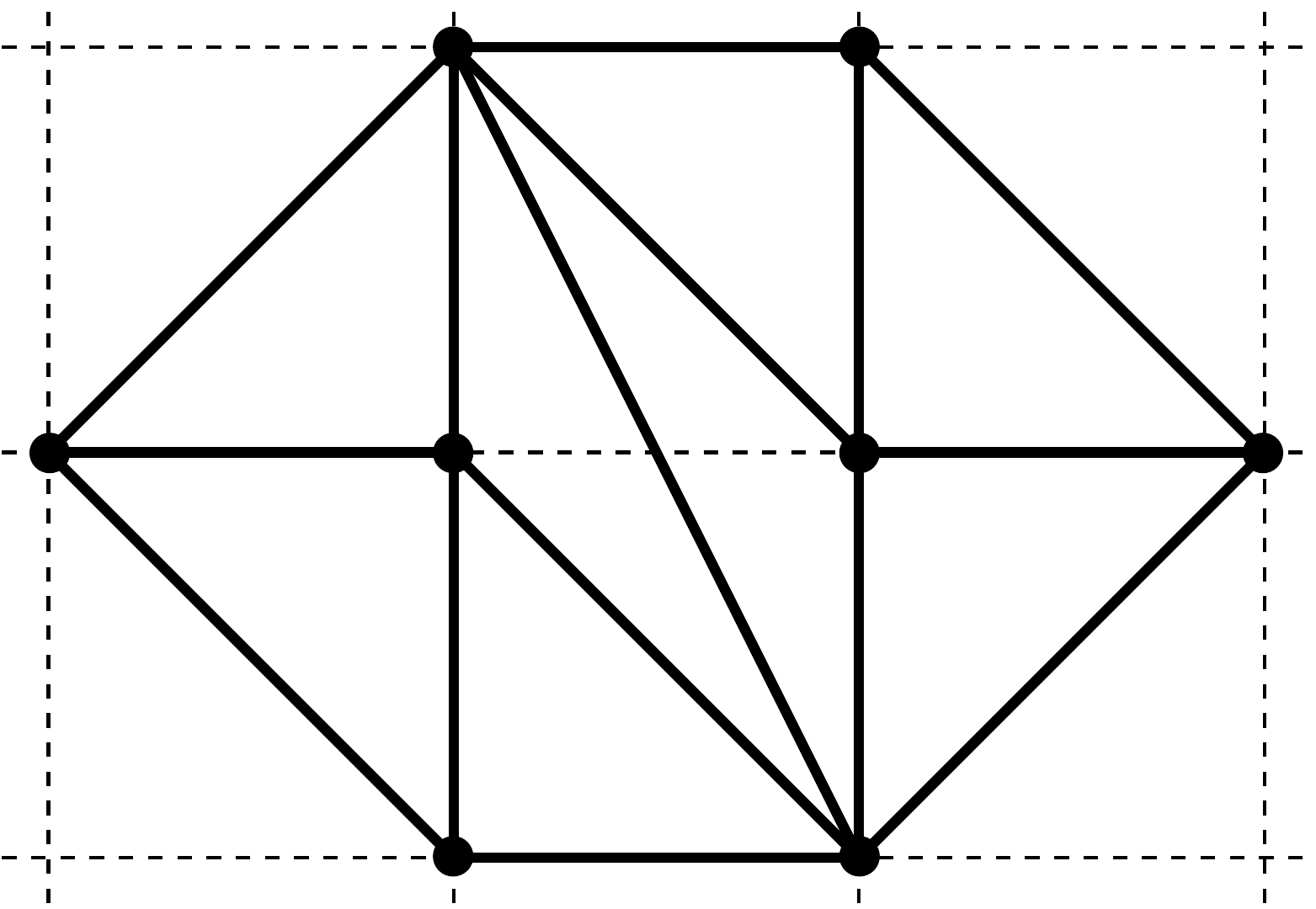}\label{fig:beetle res f}}
\\
%%%%%%su3
\subfigure[\small ]{\includegraphics[width=2.3cm]{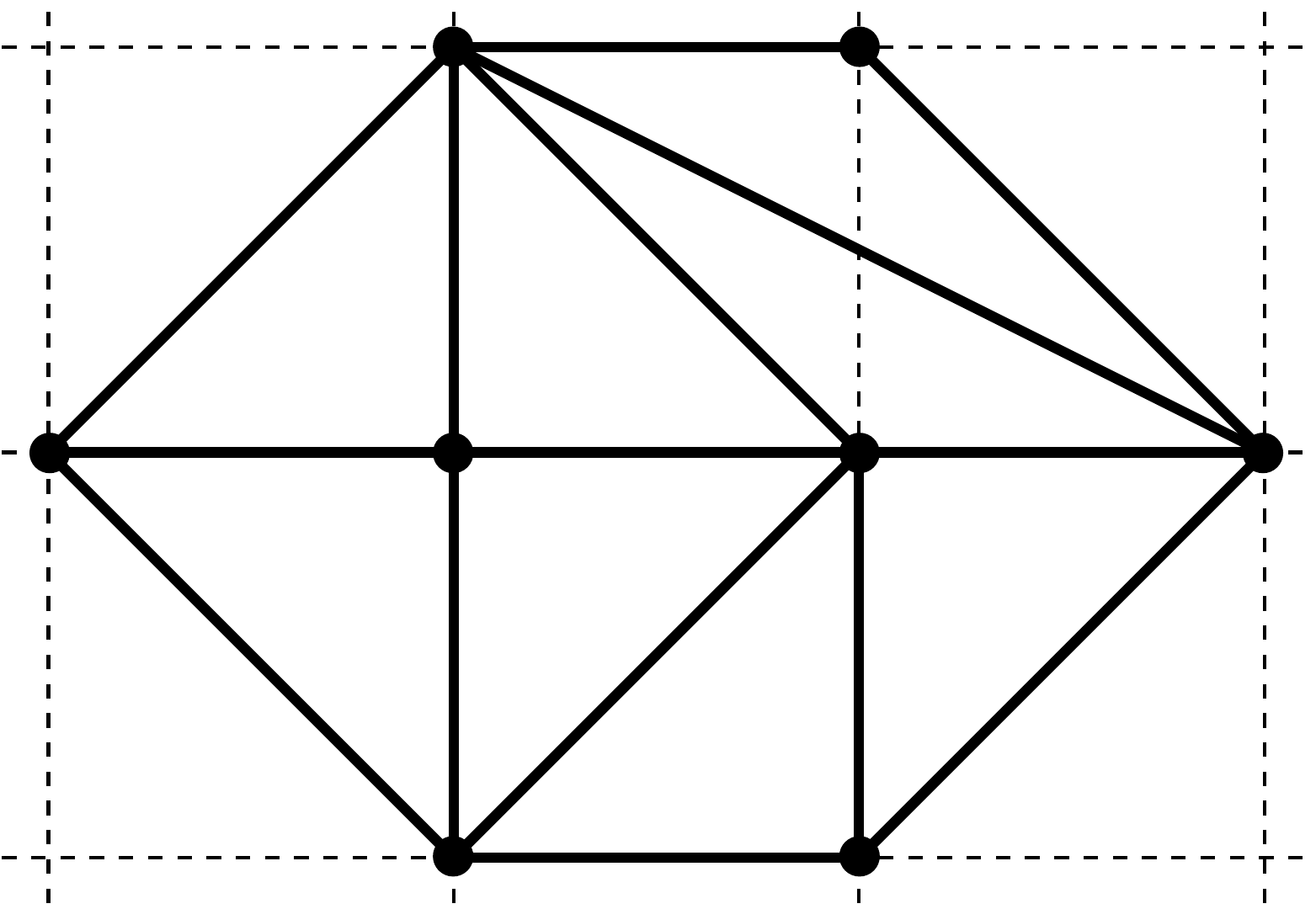}\label{fig:beetle res g}}\;
\subfigure[\small ]{\includegraphics[width=2.3cm]{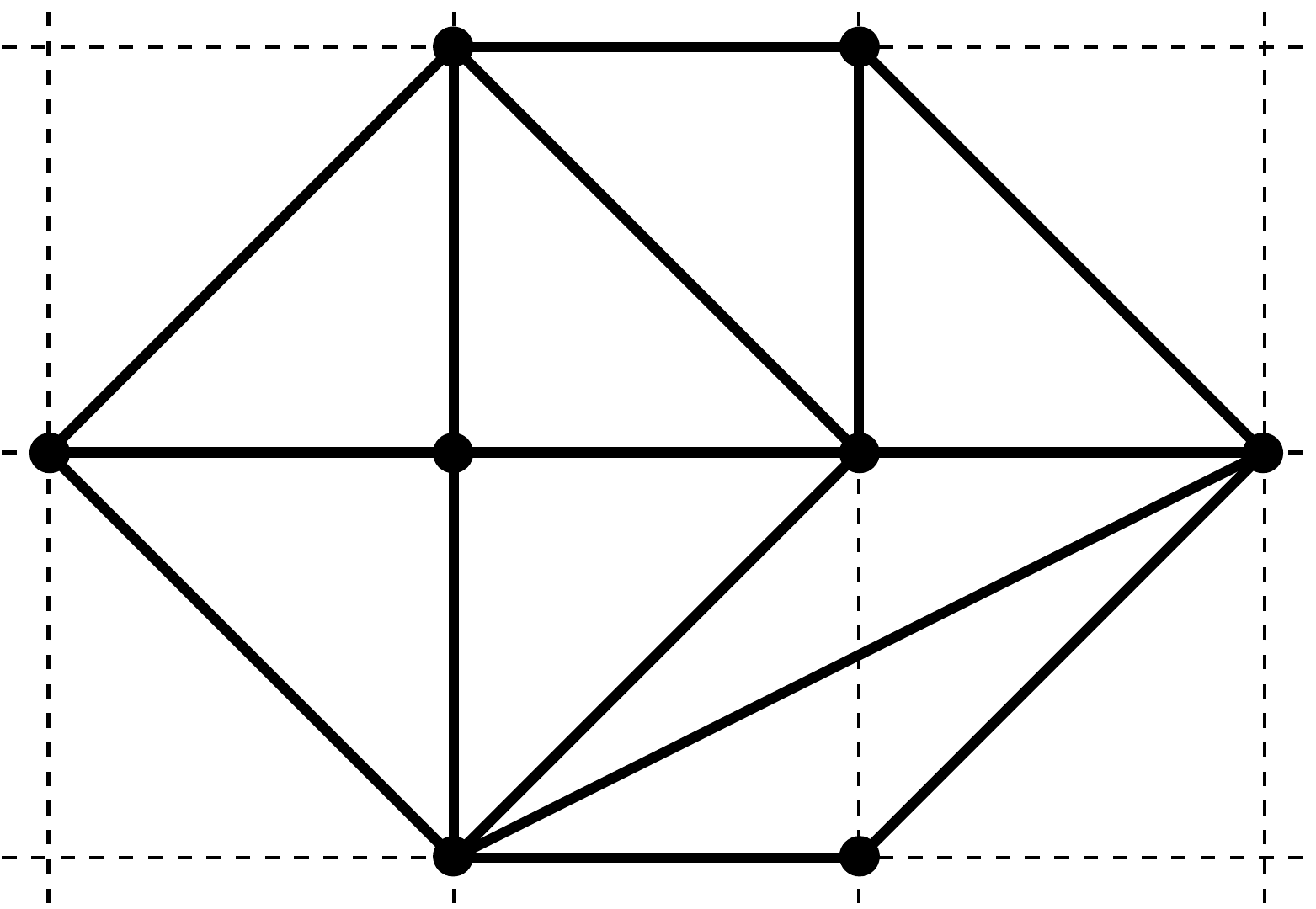}\label{fig:beetle res h}}\;
\subfigure[\small ]{\includegraphics[width=2.3cm]{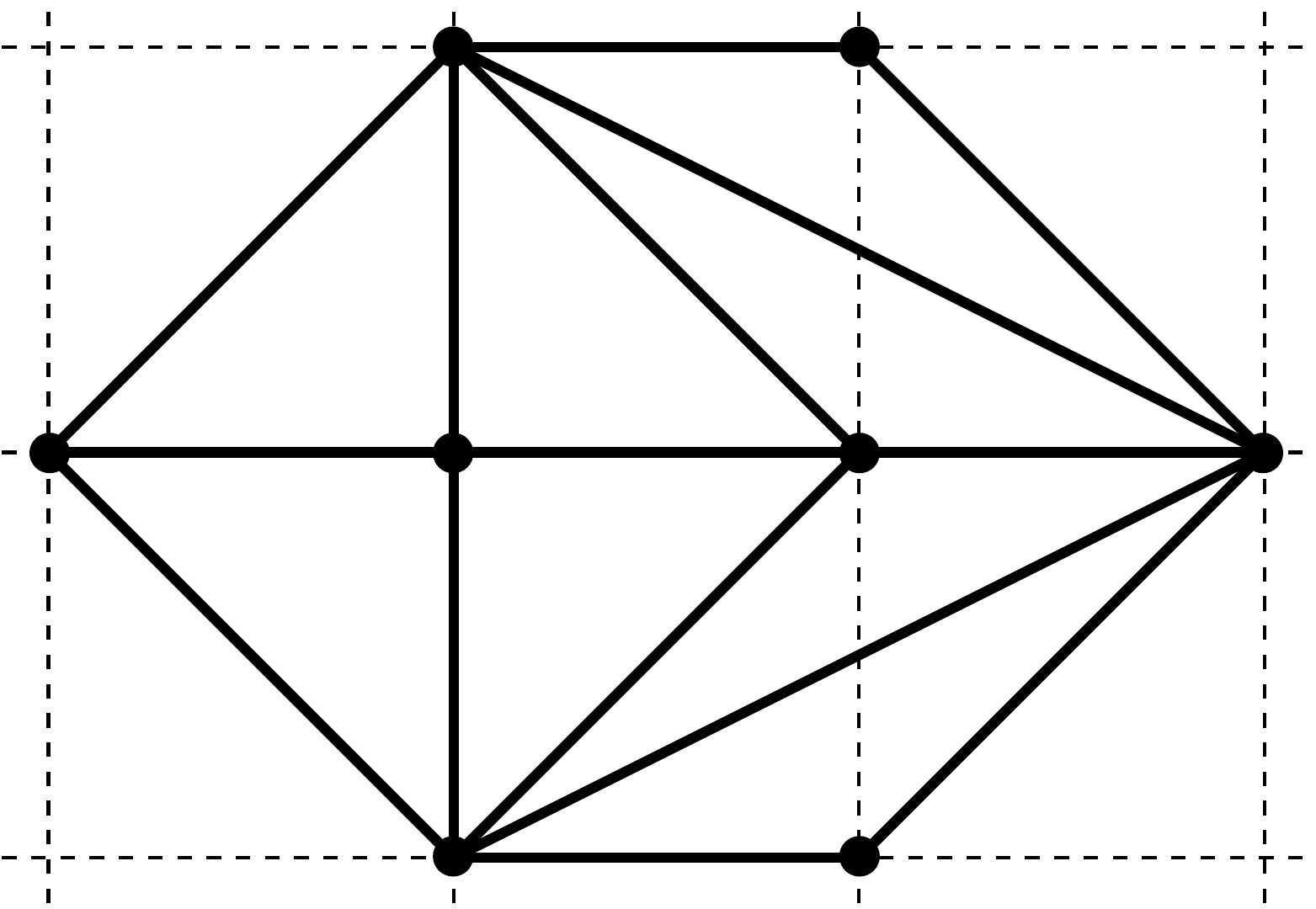}\label{fig:beetle res i}}\;
\subfigure[\small ]{\includegraphics[width=2.3cm]{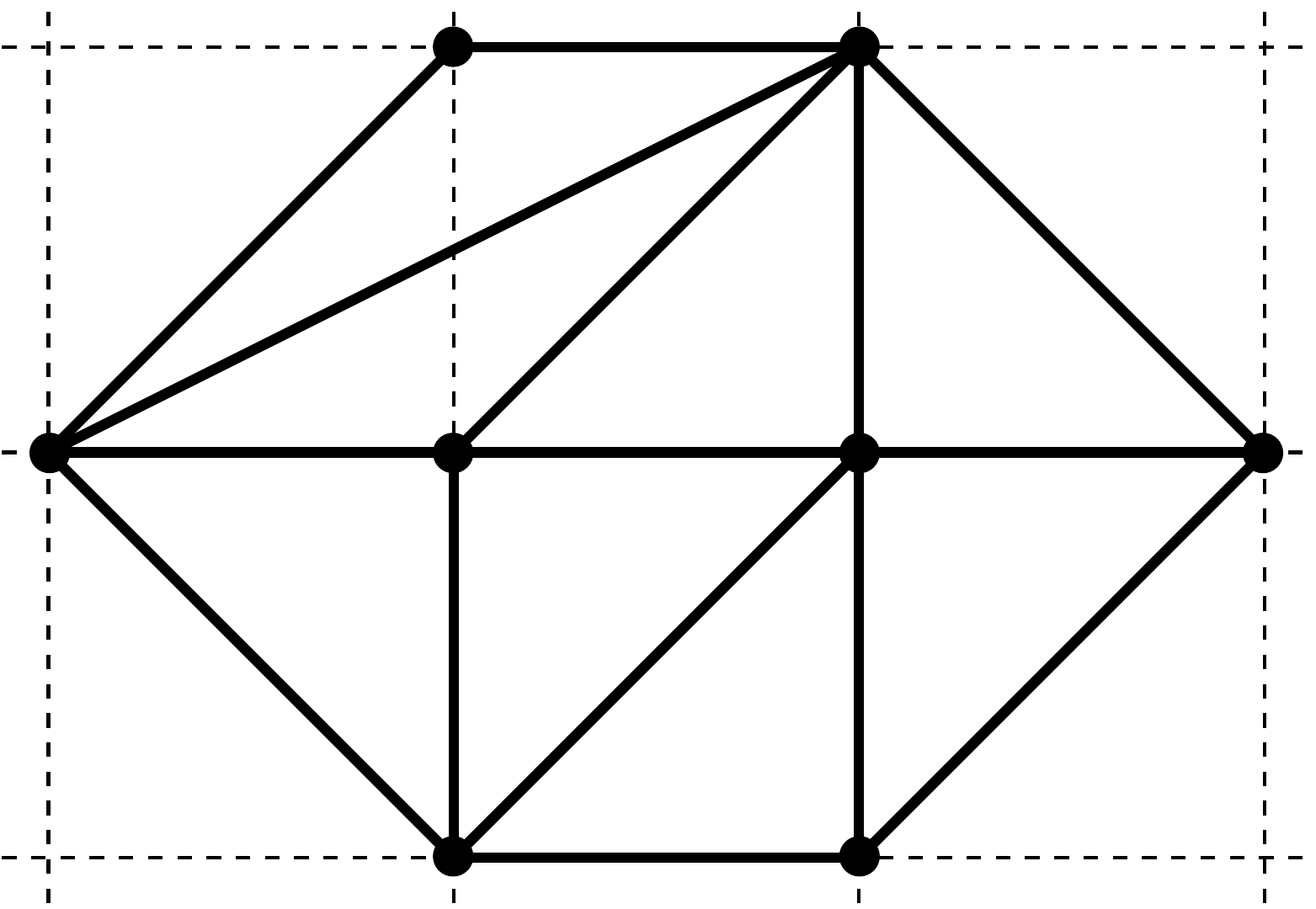}\label{fig:beetle res j}}\;
\subfigure[\small ]{\includegraphics[width=2.3cm]{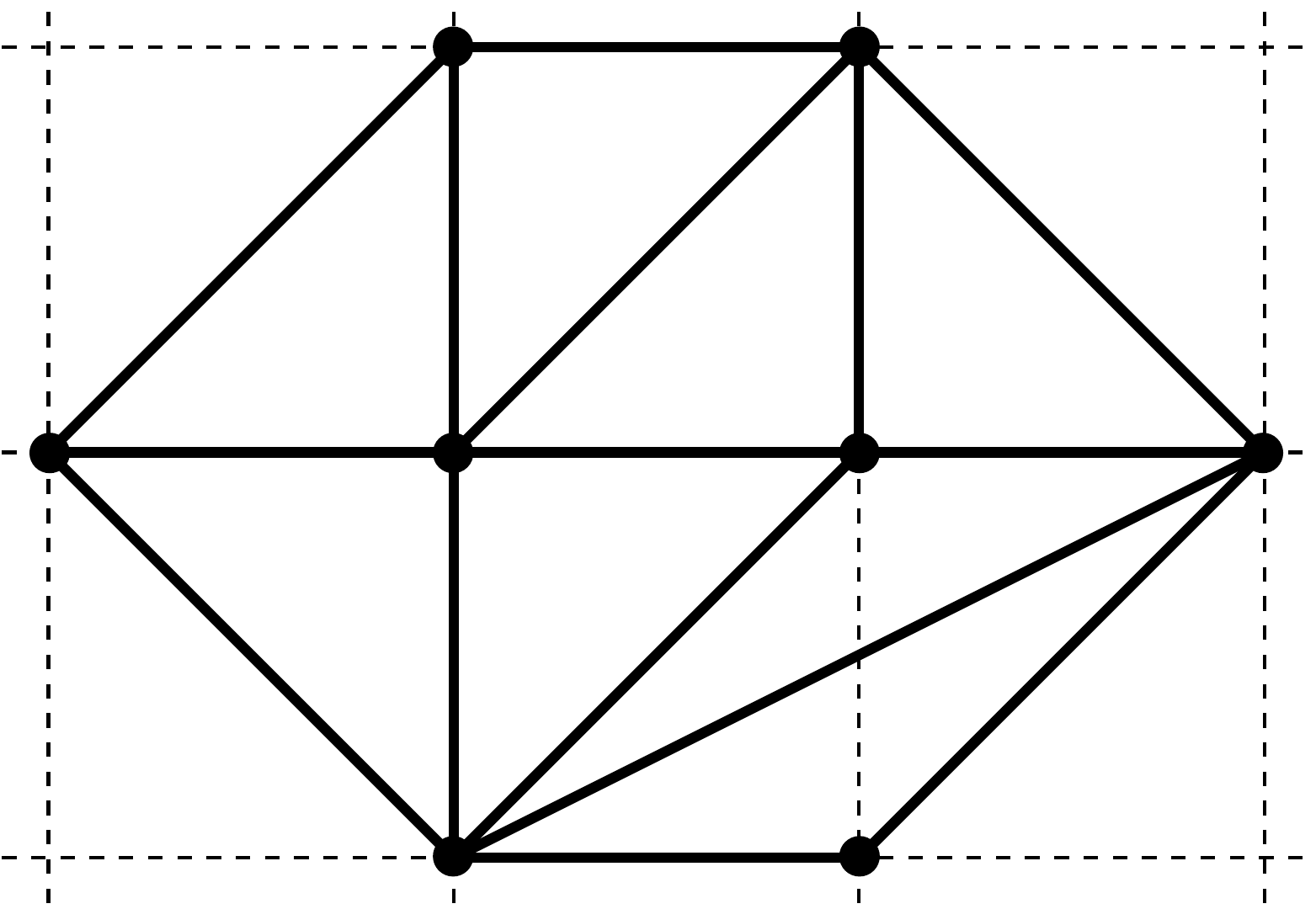}\label{fig:beetle res k}}\;
\subfigure[\small ]{\includegraphics[width=2.3cm]{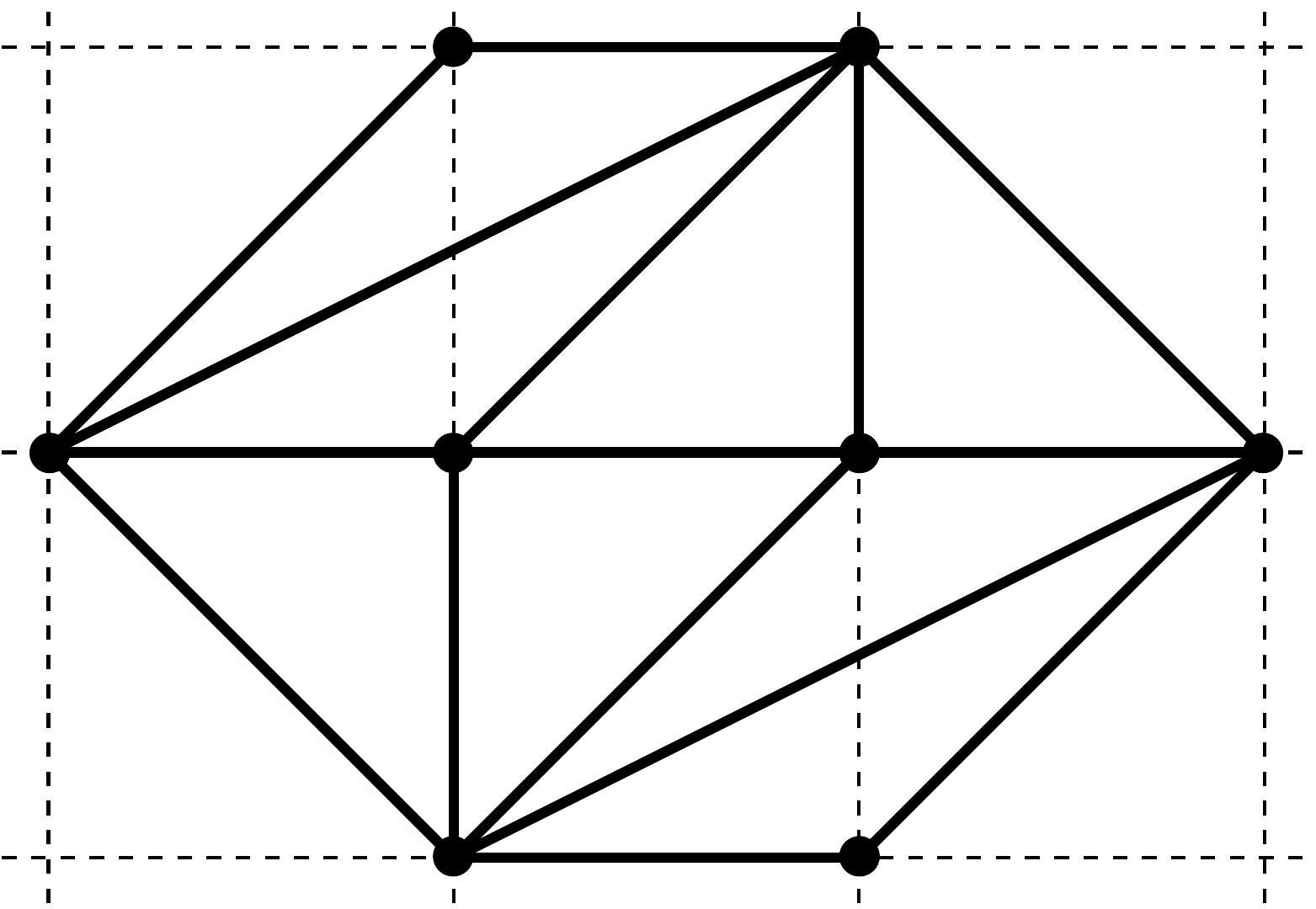}\label{fig:beetle res l}}\\
\subfigure[\small ]{\includegraphics[width=2.3cm]{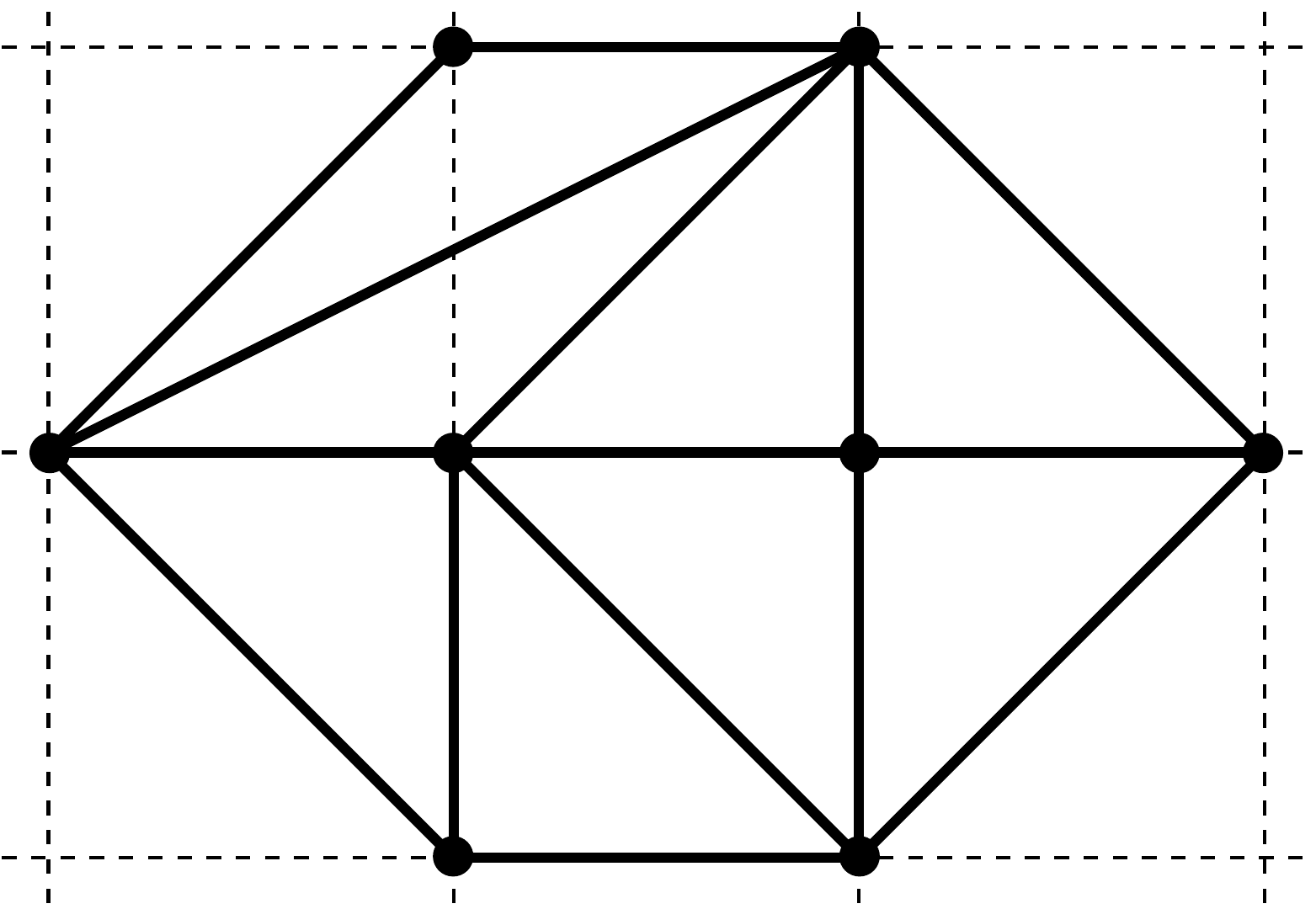}\label{fig:beetle res m}}\;
\subfigure[\small ]{\includegraphics[width=2.3cm]{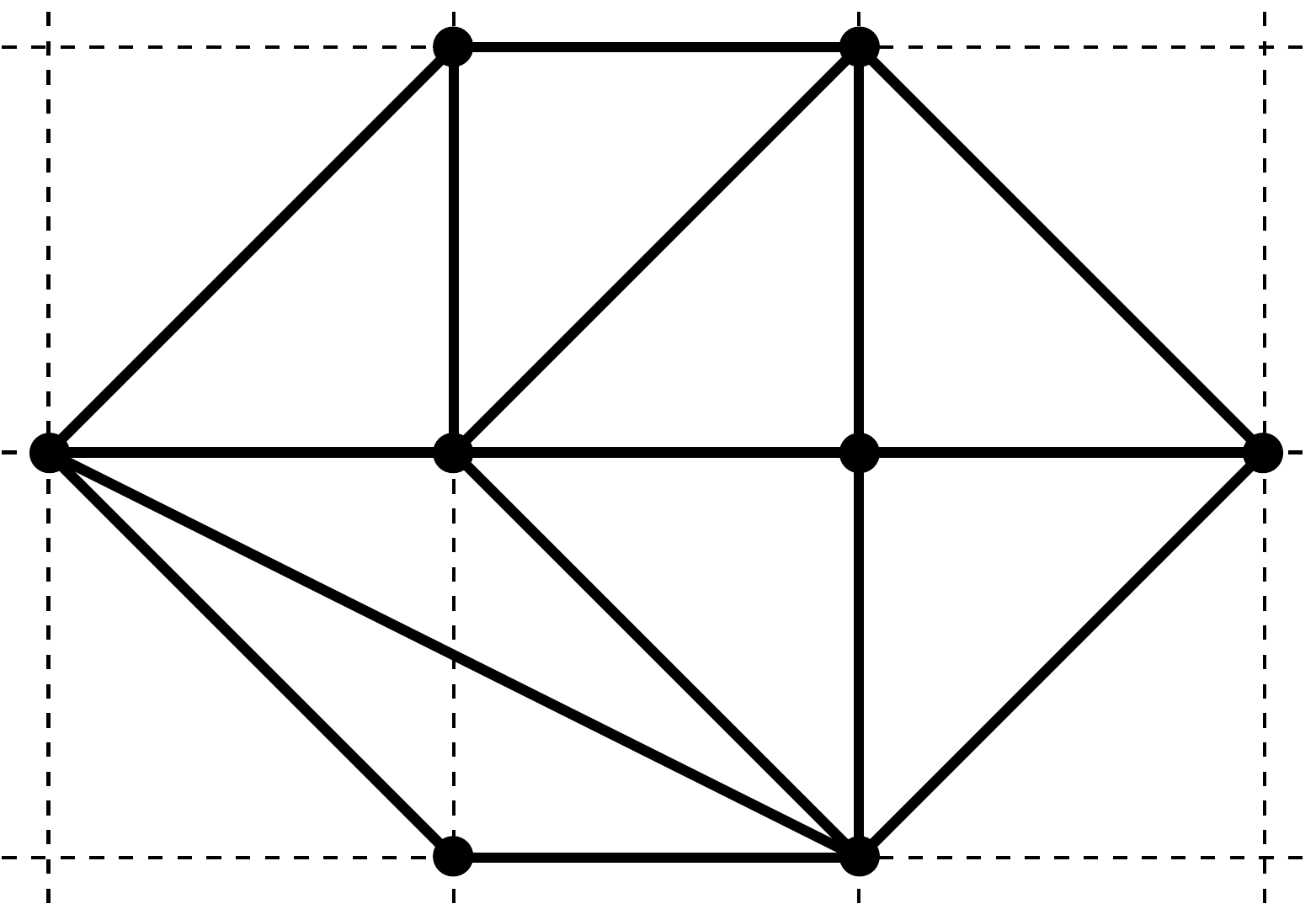}\label{fig:beetle res n}}\;
\subfigure[\small ]{\includegraphics[width=2.3cm]{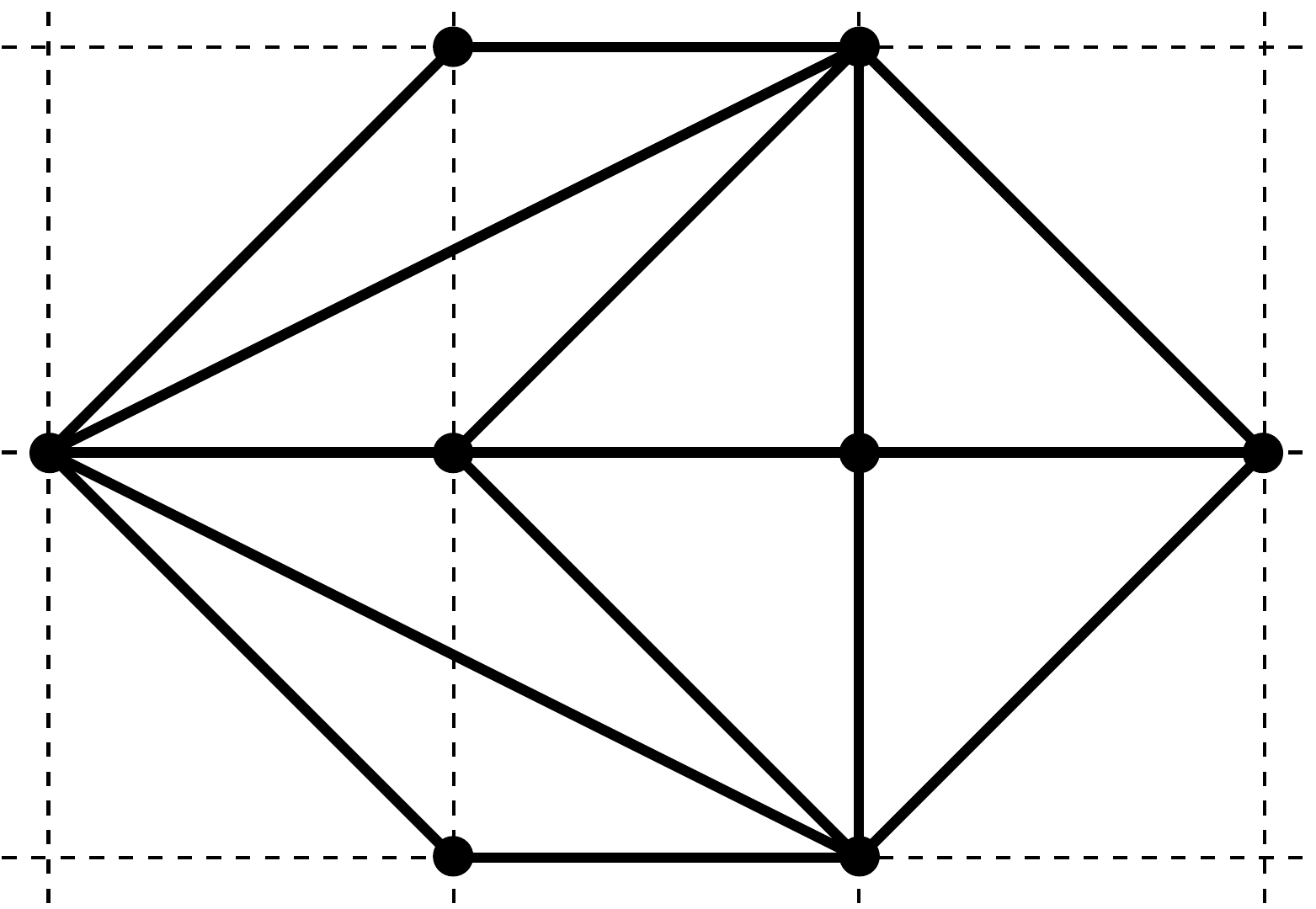}\label{fig:beetle res o}}\;
\subfigure[\small ]{\includegraphics[width=2.3cm]{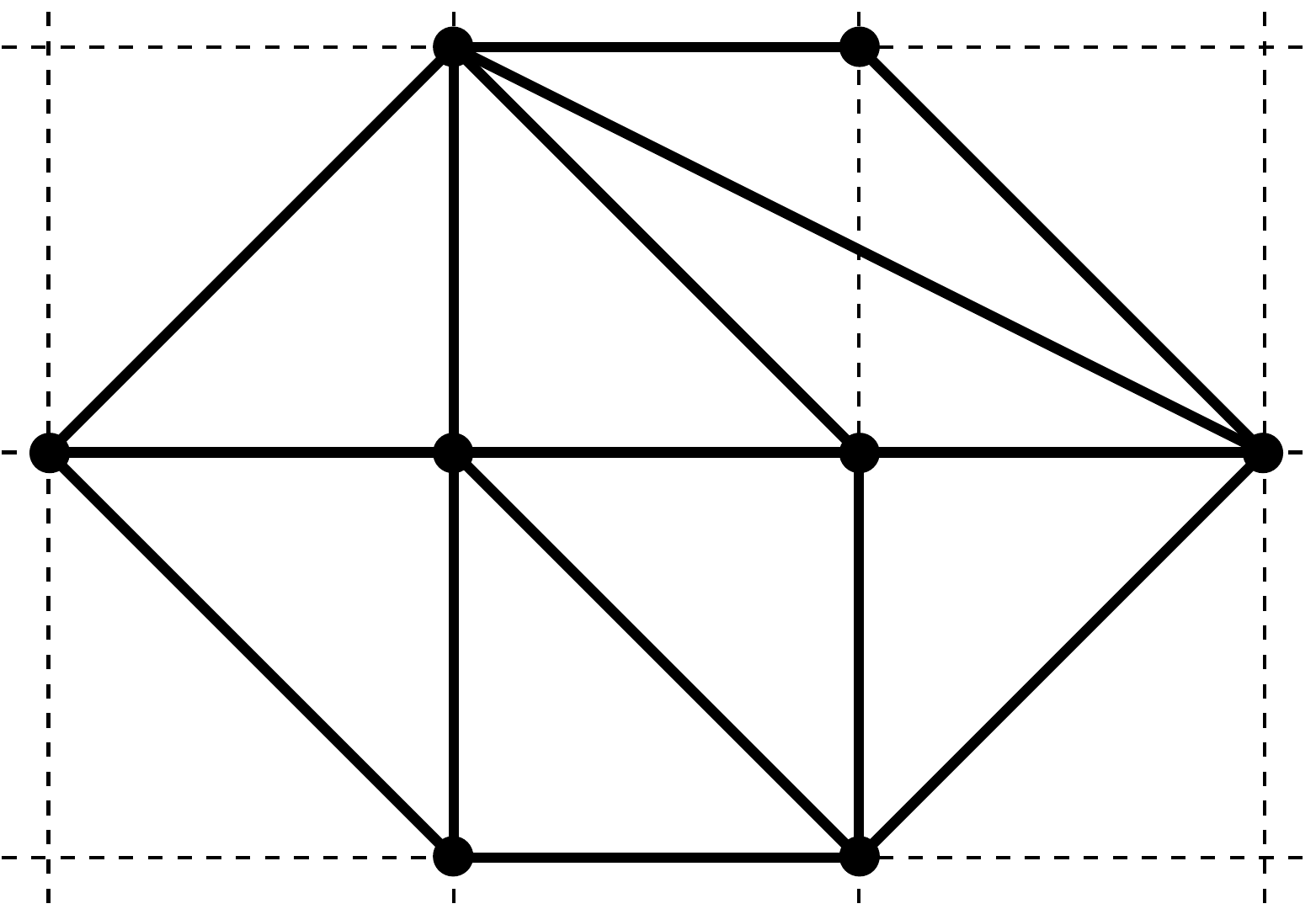}\label{fig:beetle res p}}\;
\subfigure[\small ]{\includegraphics[width=2.3cm]{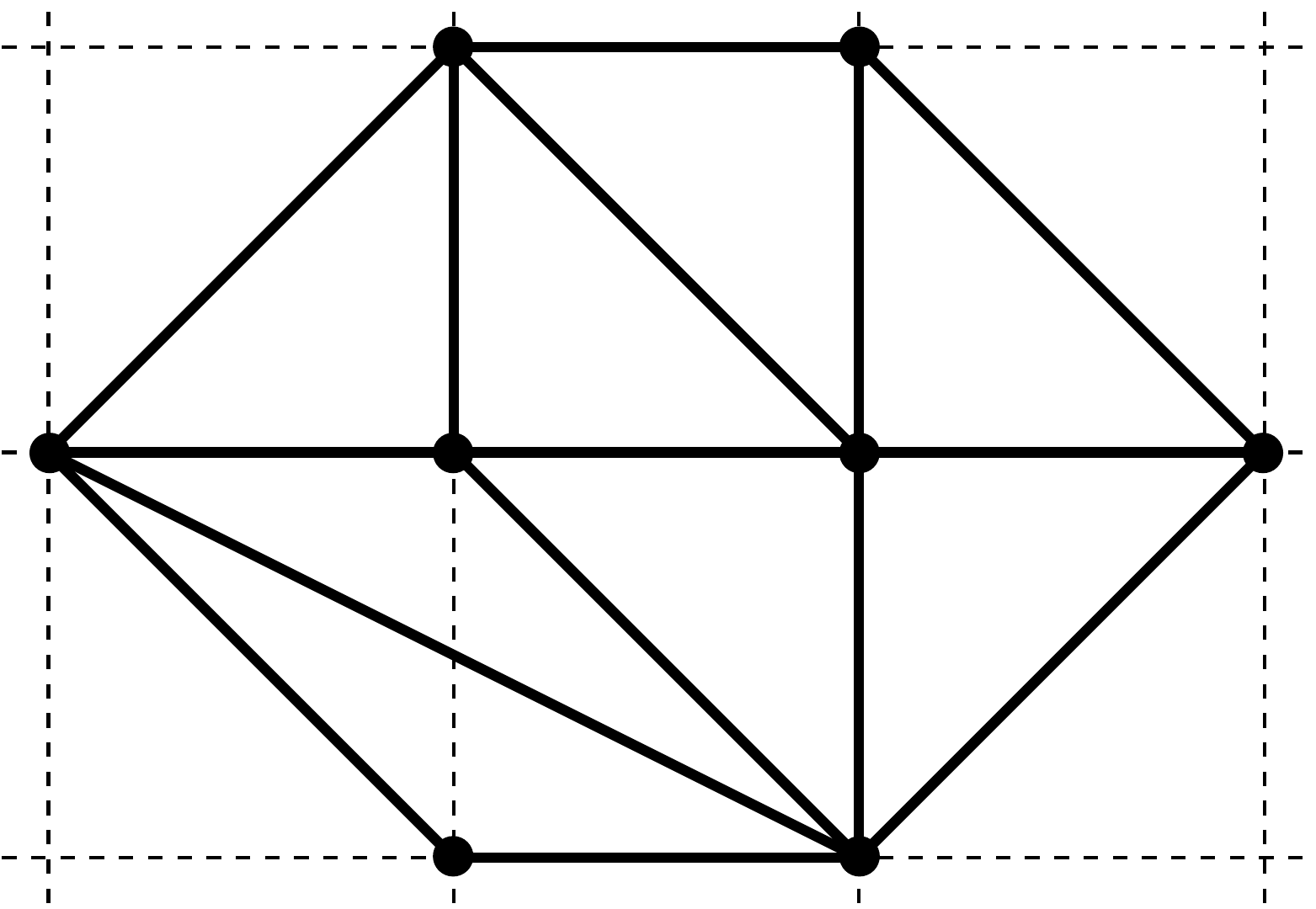}\label{fig:beetle res q}}\;
\subfigure[\small ]{\includegraphics[width=2.3cm]{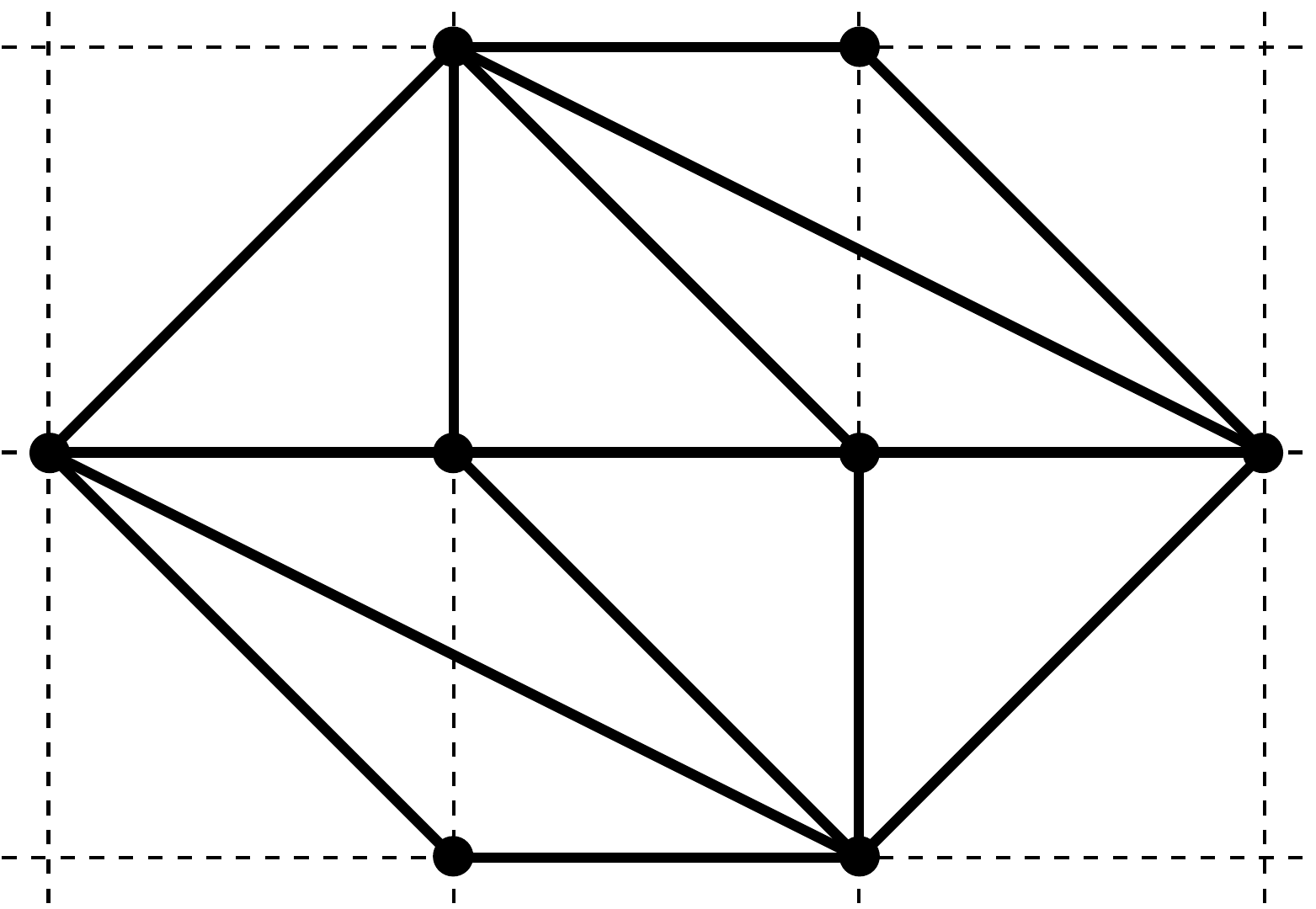}\label{fig:beetle res r}}
\\
%%%neither
\subfigure[\small ]{\includegraphics[width=2.3cm]{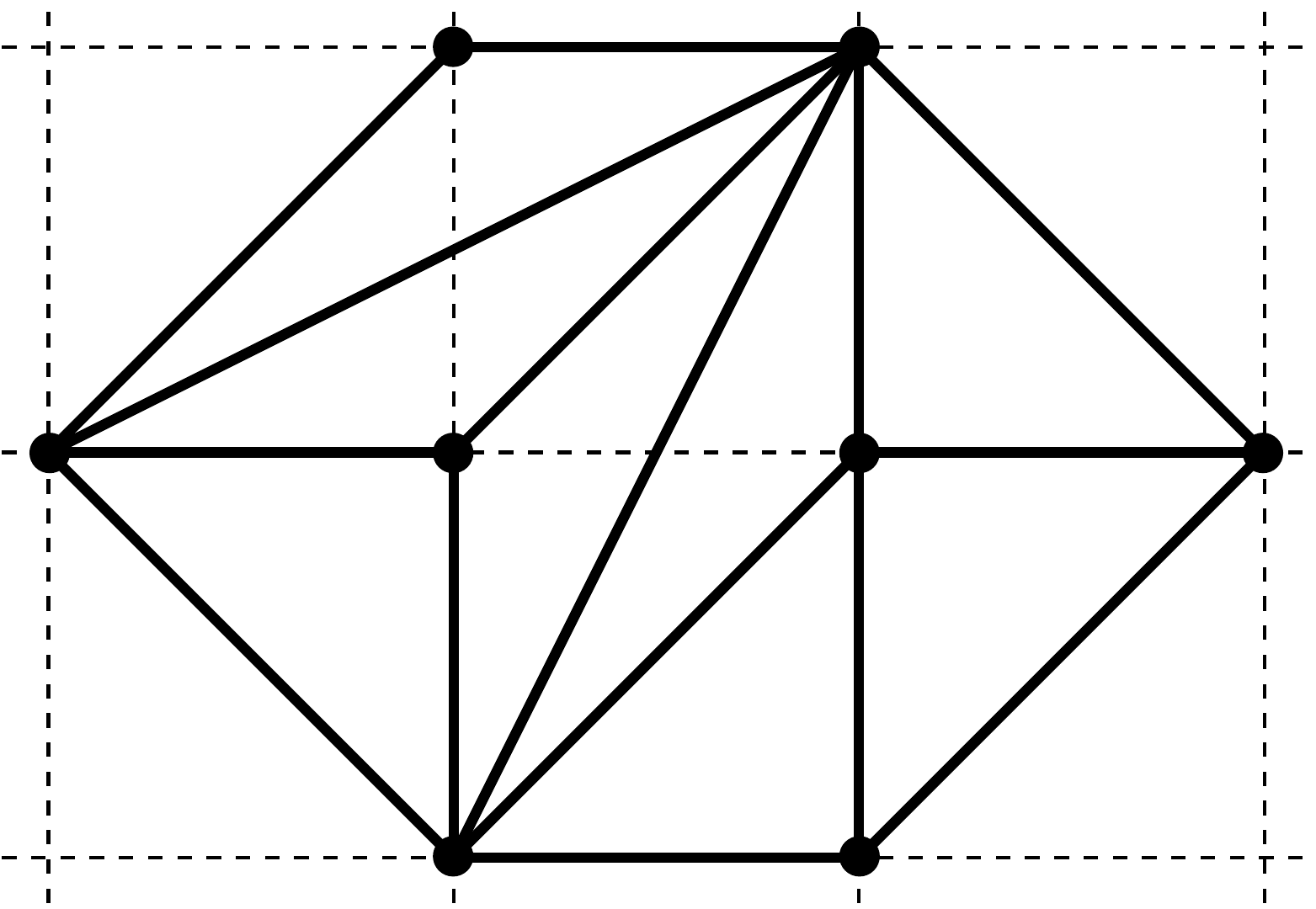}\label{fig:E2 res s}}\;
\subfigure[\small ]{\includegraphics[width=2.3cm]{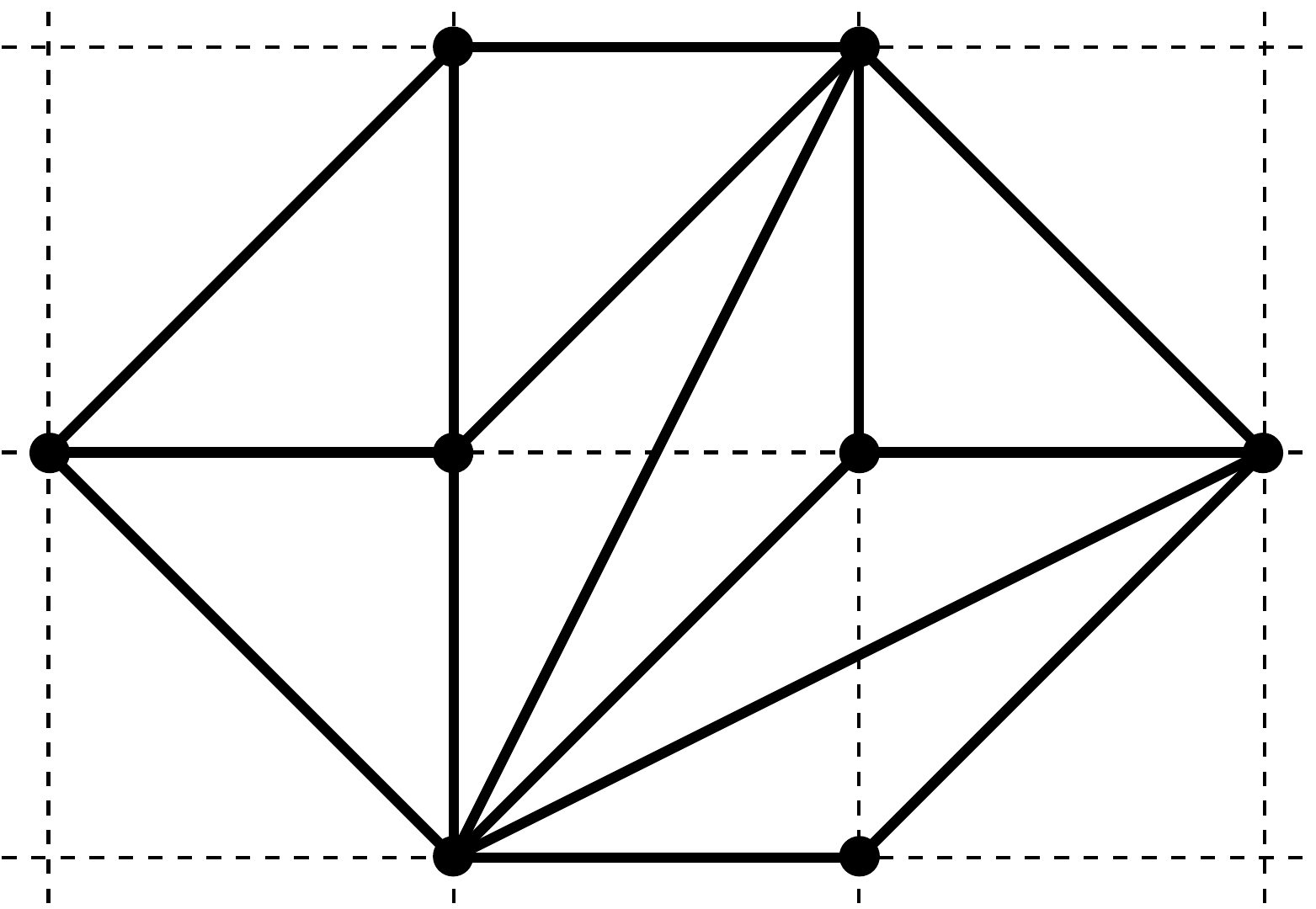}\label{fig:E2 res t}}\;
\subfigure[\small ]{\includegraphics[width=2.3cm]{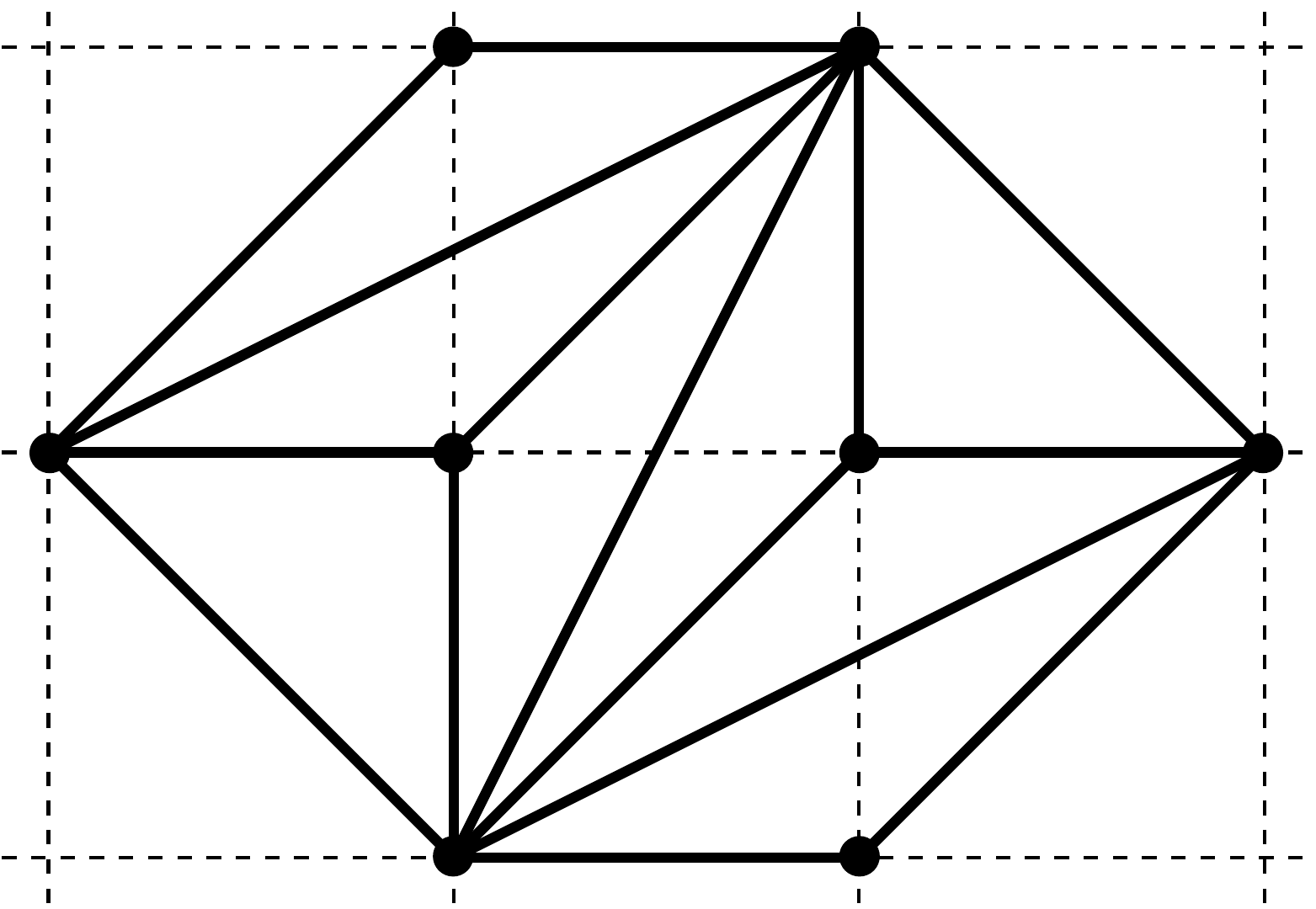}\label{fig:E2 res u}}\;
\subfigure[\small ]{\includegraphics[width=2.3cm]{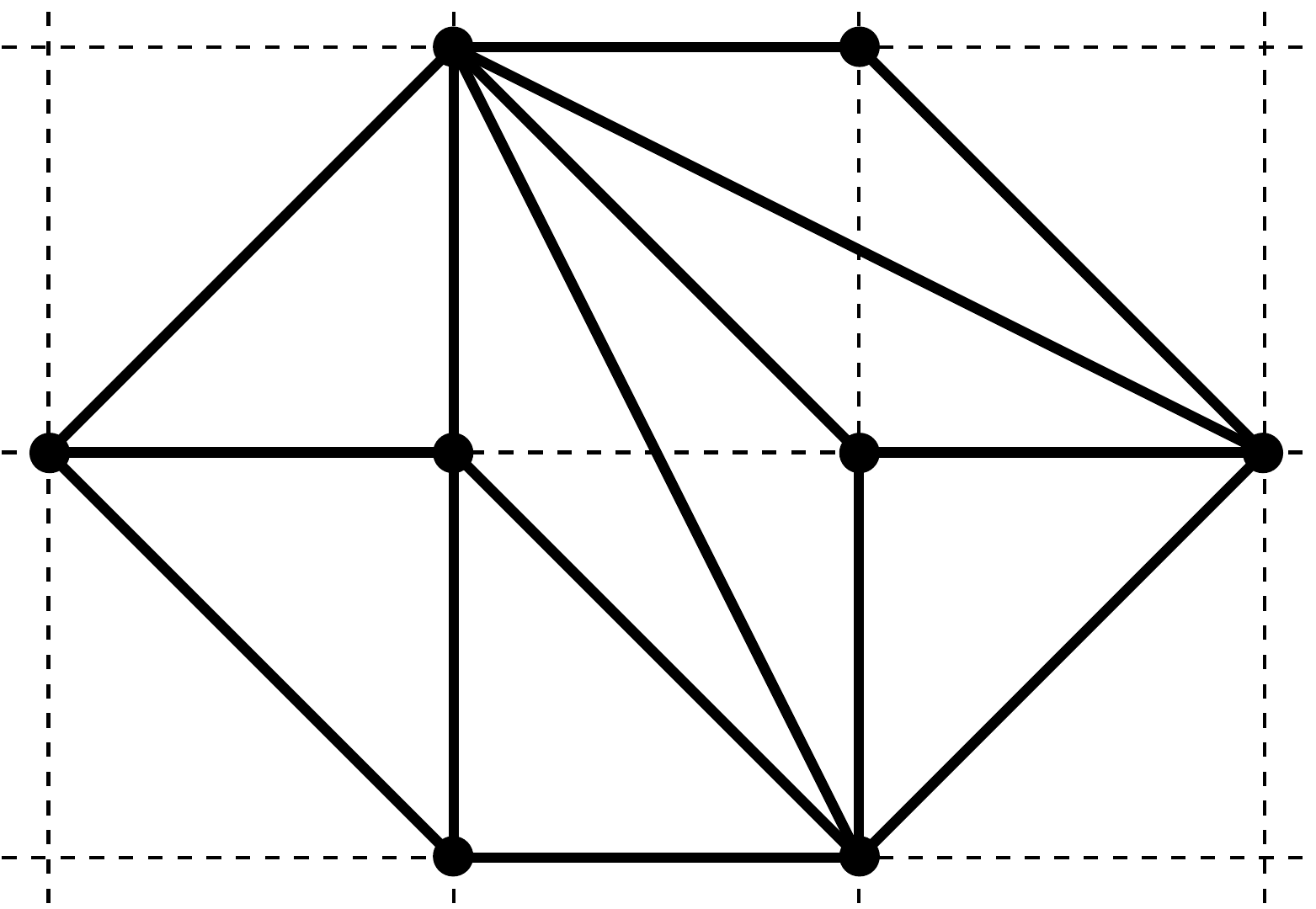}\label{fig:E2 res v}}\;
\subfigure[\small ]{\includegraphics[width=2.3cm]{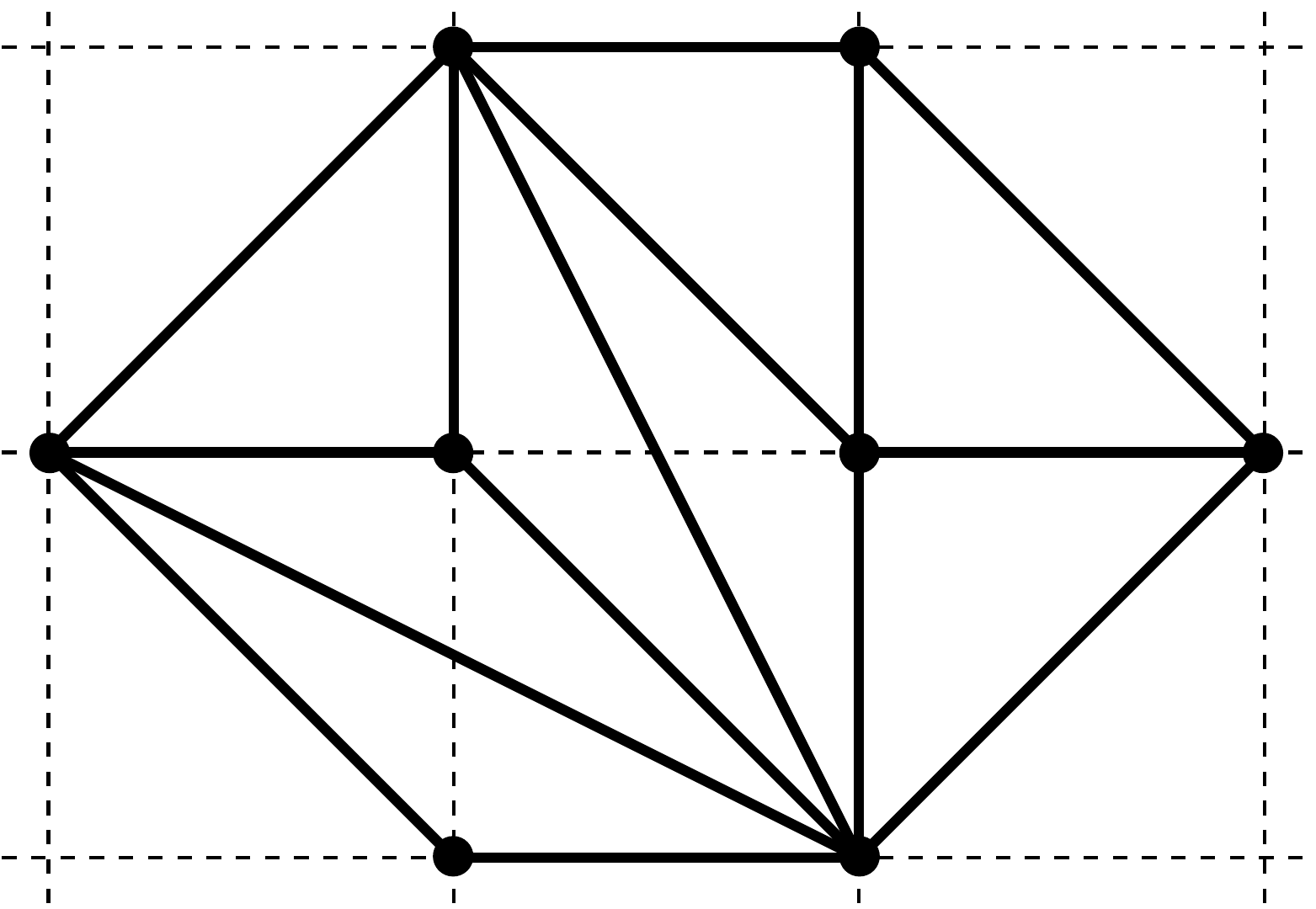}\label{fig:E2 res w}}\;
\subfigure[\small ]{\includegraphics[width=2.3cm]{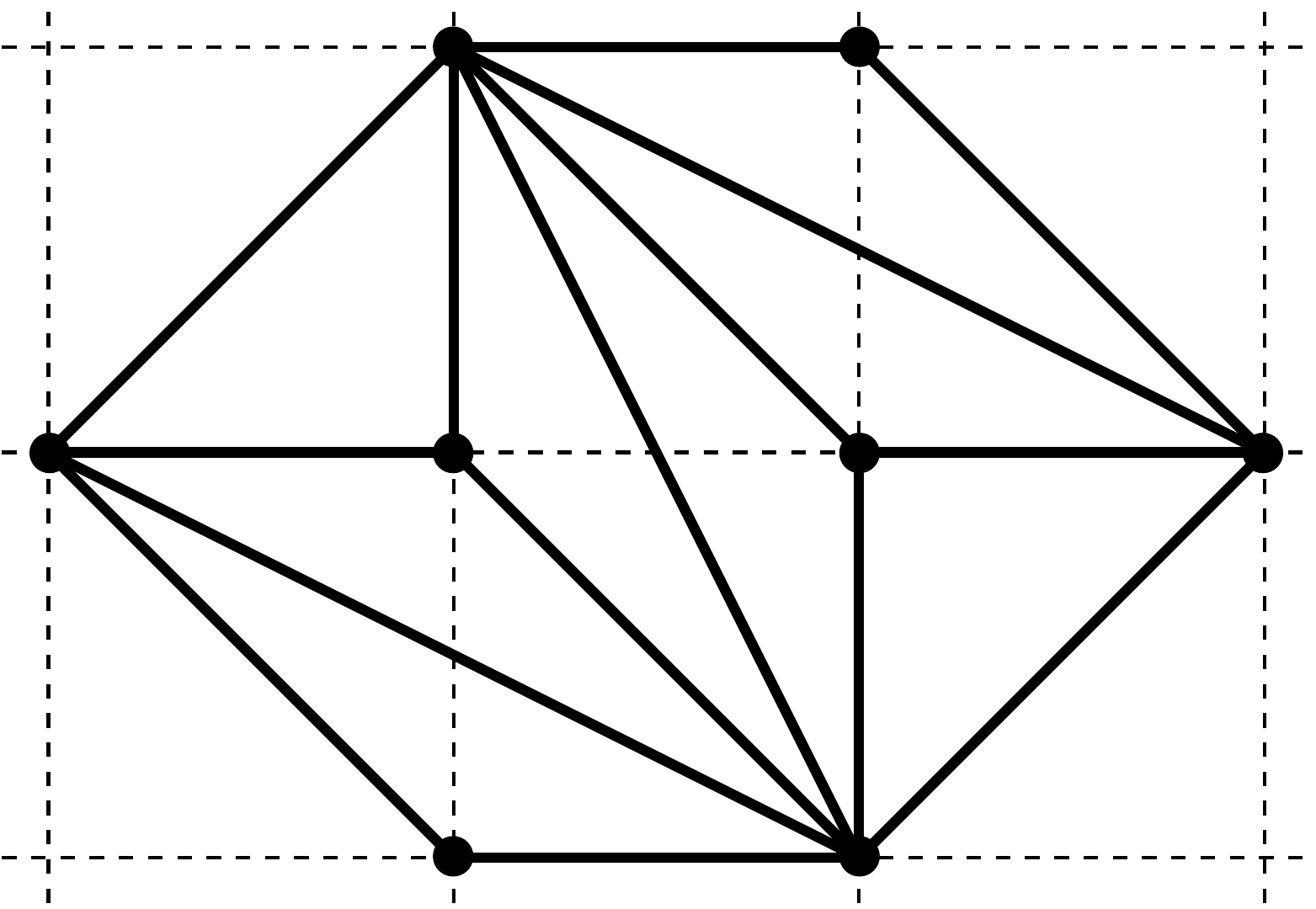}\label{fig:E2 res x}}
\caption{All 24 resolutions of the beetle geometry. The 6 resolutions (a)-(f) admit a vertical projection, the 16 resolutions (a), (b), (d), (e) and (g)-(r) admit a horizontal projection. Note that the 4 resolutions (a), (b), (d) and (e) admit both. The last 6 resolutions, (s)-(x), admit neither, hence do not have a gauge theory interpretation.  \label{fig:all beetles}}
 \end{center}
 \end{figure}

Let us now consider some more interesting ``rank-two'' geometry.
The toric singularity is shown in Figure~\ref{fig: beetle sing}---for lack of a better name, we call it ``the beetle.'' It has 24 distinct resolutions, shown in Figure~\ref{fig:all beetles}. We easily see that there are 6 allowed vertical projections and 16 allowed horizontal projections. The four resolutions (a), (b), (c) and (d) admit both a vertical and a horizontal reduction.

%%%%%%%%%
 %%%%%%%%%%%

It is also immediately clear that the vertical reduction leads to an $SU(2)\times SU(2)$ quiver gauge theory description, while the horizontal reduction leads to an $SU(3)$ gauge theory with $k_{\rm eff}=0$ and $N_f=2$ fundamental flavors. Thus, we have the S-duality relation:
\be
SU(2) \times SU(2) \qquad \longleftrightarrow \qquad SU(3)_0~, \, N_f=2~.
\ee
We will explore this ``UV duality'' relation in the following. Let us insist again on the fact that, for most resolutions in Fig.~\ref{fig:all beetles}, only one of the two gauge-theory descriptions is valid (or neither is valid, in the case of the resolutions (s) to (x)); yet, for the resolutions (a), (b), (d) and (e), both gauge-theory descriptions can be true {\it simultaneously.}  

Note also that the toric diagram of Figure~\ref{fig:all beetles} is parity invariant, in the sense of \eqref{C0 on w Parity}, therefore the SCFT is parity invariant. This will be confirmed by the IIA analysis. 

\subsubsection{M-theory geometry and prepotential}
The beetle singularity can be described by the following GLSM:
\be\label{glsm beetle gen}
\begin{tabular}{l|cccccccc|c}
 & $D_1$ &$D_2$& $D_3$&$D_4$&$D_5$&$D_6$&$\bE_1$&$\bE_2$& FI  \\
 \hline
  $\CC_2^{a}$  &$0$&$1$ &$0$ & $0$&$0$ &$0$&$-2$&$1$&$\xi_2$\\
   $\CC_3^{a}$  &$0$&$0$ &$-1$ & $1$&$0$ &$0$&$1$&$-1$&$\xi_3$\\
   $\CC_6^{a}$  &$1$&$0$ &$0$ & $0$&$1$ &$-1$&$0$&$-1$&$\xi_6$\\
   $\CC_7^{a}$  &$-1$&$0$ &$0$ & $0$&$0$ &$1$&$1$&$-1$&$\xi_7$\\
   $\CC_9^{a}$  &$1$&$0$ &$1$ & $0$&$0$ &$0$&$-2$&$0$&$\xi_9$
 \end{tabular}
\ee
Here, the rows are labelled by the curves in resolution (a), to be discussed below, in which case the FI terms $\xi_2, \xi_3, \xi_6, \xi_7$ and $\xi_9$ are all positive, but the same GLSM describes all 24 resolutions. We have the following relations amongst toric divisors:
\bea
&D_1+ D_6 \cong D_3+ D_4~, \cr
&\bE_1 \cong -D_1 -2 D_2 -D_3 +D_5~, \qquad
&\bE_2\cong D_2 -D_4 -2D_5 -D_6~,
\eea
which are independent of the particular resolution.
Let us consider:
\be\label{def S mu nu beetle}
S= \mu_1 D_1 + \mu_2 D_2 + \mu_6 D_6 + \nu_1 \bE_1+ \nu_2 \bE_2~.
\ee
The map between the FI parameters in \eqref{glsm beetle gen} and the parameters $\mu, \nu$ is:
\bea\label{mu to xi beetle}
&\mu_1= 4\xi_3 - 2 \xi_6 -2 \xi_7 + \xi_9~, \qquad\quad && \nu_1 = 2 \xi_3- \xi_6- \xi_7~, \\
& \mu_2 = \xi_2 +3 \xi_3 - \xi_6 - \xi_7~, \qquad&& \nu_2 = \xi_3 - \xi_6 - \xi_7~. \\
& \mu_6= 3 \xi_3 - 2 \xi_6 -\xi_7 +\xi_9~,
\eea
The geometric prepotential:
\be\label{F geom beetle}
\CF(\mu, \nu)= \CF(\xi)= -{1\ov 6} S^3
\ee
depends on the resolution. A complete list of the geometric prepotentials in the 24 K{\"a}hler chambers is given in Appendix~\ref{app: more on beetle}.

%%%%%%%%%
 %%%%%%%%%%%

\subsubsection{$SU(2)\times SU(2)$ phases}
Consider the quiver theory consisting of two $SU(2)$ gauge groups, $SU(2)_{(1)}\times SU(2)_{(2)}$, with one hypermultiplet $\CH$ in the bifundamental representation. We denote by $h_1$, $h_2$ and $\varphi_1$, $\varphi_2$ the two gauge couplings and the two Coulomb branch scalars, respectively, and by $m$ the mass of the bifundamental hypermultiplet. The prepotential reads:
\bea\label{CF su2 su2 quiver}
&\CF_{SU(2)\times SU(2)} &=&\; \; h_1 \varphi_1^2 +  h_2 \varphi_2^2 + {4\ov 3} \varphi_1^3 + {4\ov 3} \varphi_2^3  \\
&&&\;-{1\ov 6} \Theta(\varphi_1+\varphi_2+m)\, (\varphi_1+\varphi_2+m)^3\\
&&&\;-{1\ov 6} \Theta(\varphi_1-\varphi_2+m)\, (\varphi_1-\varphi_2+m)^3\\
&&&\;-{1\ov 6} \Theta(-\varphi_1+\varphi_2+m)\, (-\varphi_1+\varphi_2+m)^3\\
&&&\;-{1\ov 6} \Theta(-\varphi_1-\varphi_2+m)\, (-\varphi_1-\varphi_2+m)^3~.
\eea
By performing the vertical reduction, one can easily derive the precise relation between the geometry and field-theory parameters. In terms of the GLSM \eqref{glsm beetle gen}, we find:
\bea\label{xi to gauge su2su2}
&\xi_2 = h_1 + 2 \varphi_1 -\varphi_2 -m~,\quad\qquad &
&\xi_3 = -\varphi_1+ \varphi_2 -m~,\\
& \xi_6 = h_2 + \varphi_2 - 2m~, \qquad &
&\xi_7= - \varphi_1 + \varphi_2+ m~,\\
&\xi_9= 2 \varphi_1~.
\eea
Equivalently, plugging the above into \eqref{mu to xi beetle} we find:
\bea\label{mu to geom beetle}
&\mu_1= -2 h_2 -2m~, \qquad\quad && \nu_1 = -\varphi_1- h_2 -m~, \\
& \mu_2 = h_1-h_2 - 3m~, \qquad&& \nu_2 = -\varphi_2 -h_2~, \\
& \mu_6=-2 h_2~.
\eea
We will also find that the resolutions (a) to (f) from Figure~\ref{fig:all beetles} correspond to the following field-theory chambers:{\small
\bea\label{geom to gauge chamber su2su2}
&{\rm (a)} \quad \leftrightarrow\quad \begin{cases} \varphi_1 + \varphi_2 + m>0~, \;\; &  \varphi_1 - \varphi_2 + m<0~, \\
			     -\varphi_1 + \varphi_2 + m>0~, \;  \;& -\varphi_1-  \varphi_2 + m<0~,
			    \end{cases}  \\
&{\rm (b)}\quad \leftrightarrow\quad \begin{cases} \varphi_1 + \varphi_2 + m>0~, \;\; &  \varphi_1 - \varphi_2 + m>0~, \\
			     -\varphi_1 + \varphi_2 + m>0~, \;  \;& -\varphi_1-  \varphi_2 + m<0~,
			    \end{cases}  \\	   
&{\rm (c)} \quad \leftrightarrow\quad \begin{cases} \varphi_1 + \varphi_2 + m>0~, \;\; &  \varphi_1 - \varphi_2 + m>0~, \\
			     -\varphi_1 + \varphi_2 + m>0~, \;  \;& -\varphi_1-  \varphi_2 + m>0~,
			    \end{cases}  \\
&{\rm (d)} \quad \leftrightarrow\quad \begin{cases} \varphi_1 + \varphi_2 + m>0~, \;\; &  \varphi_1 - \varphi_2 + m>0~, \\
			     -\varphi_1 + \varphi_2 + m<0~, \;  \;& -\varphi_1-  \varphi_2 + m<0~,
			    \end{cases}  \\
&{\rm (e)}\quad \leftrightarrow\quad \begin{cases} \varphi_1 + \varphi_2 + m>0~, \;\; &  \varphi_1 - \varphi_2 + m<0~, \\
			     -\varphi_1 + \varphi_2 + m<0~, \;  \;& -\varphi_1-  \varphi_2 + m<0~,
			    \end{cases}  \\			    
&{\rm (f)} \quad \leftrightarrow\quad \begin{cases} \varphi_1 + \varphi_2 + m<0~, \;\; &  \varphi_1 - \varphi_2 + m<0~, \\
			     -\varphi_1 + \varphi_2 + m<0~, \;  \;& -\varphi_1-  \varphi_2 + m<0~.
			    \end{cases} 		 			      
\eea}
This is the expected match between the gauge-theory resolutions and the six distinct gauge-theory chambers. Note that we chose the fundamental $SU(2)\times SU(2)$ Weyl chamber, $\varphi_1 \geq 0$ and $\varphi_2\geq 0$, by convention, while the real mass $m$ can be of either sign. In Appendix~\ref{app: more on beetle}, we check that the M-theory and field-theory prepotentials precisely match, in those 6 chambers.

Note also that only resolution (c) is compatible with the field-theory decoupling limit $m \rightarrow \infty$, and similarly only resolution (f) is compatible with the limit $m \rightarrow -\infty$. In these limits, we integrate out the bifundamental hypermultiplet and obtain two decoupled $SU(2)_\pi$ gauge theories. This is compatible with the toric geometry shown in Fig.~\ref{fig:beetle res c} and Fig.~\ref{fig:beetle res f}, where $|m|\rightarrow \infty$ corresponds to blowing up a curve to infinite volume in such a way that the beetle singularity splits into two resolved $E_1$ singularities.

%%%%%%%%%
 \begin{figure}[t]
\begin{center}
\includegraphics[width=7.3cm]{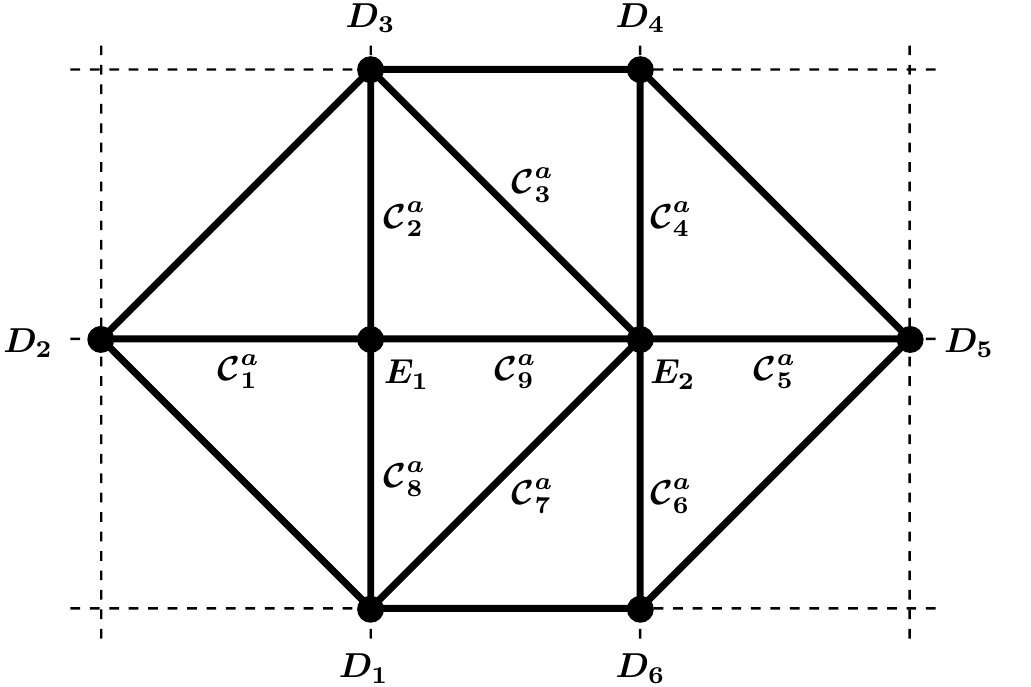}\;
\caption{Resolution (a) of the beetle geometry, with all the curves indicated.  \label{fig:IIA beetle res a curves}}
 \end{center}
 \end{figure} 
 %%%%%%

%%%%%%%%%
 \begin{figure}[t]
\begin{center}
\includegraphics[width=7.6cm]{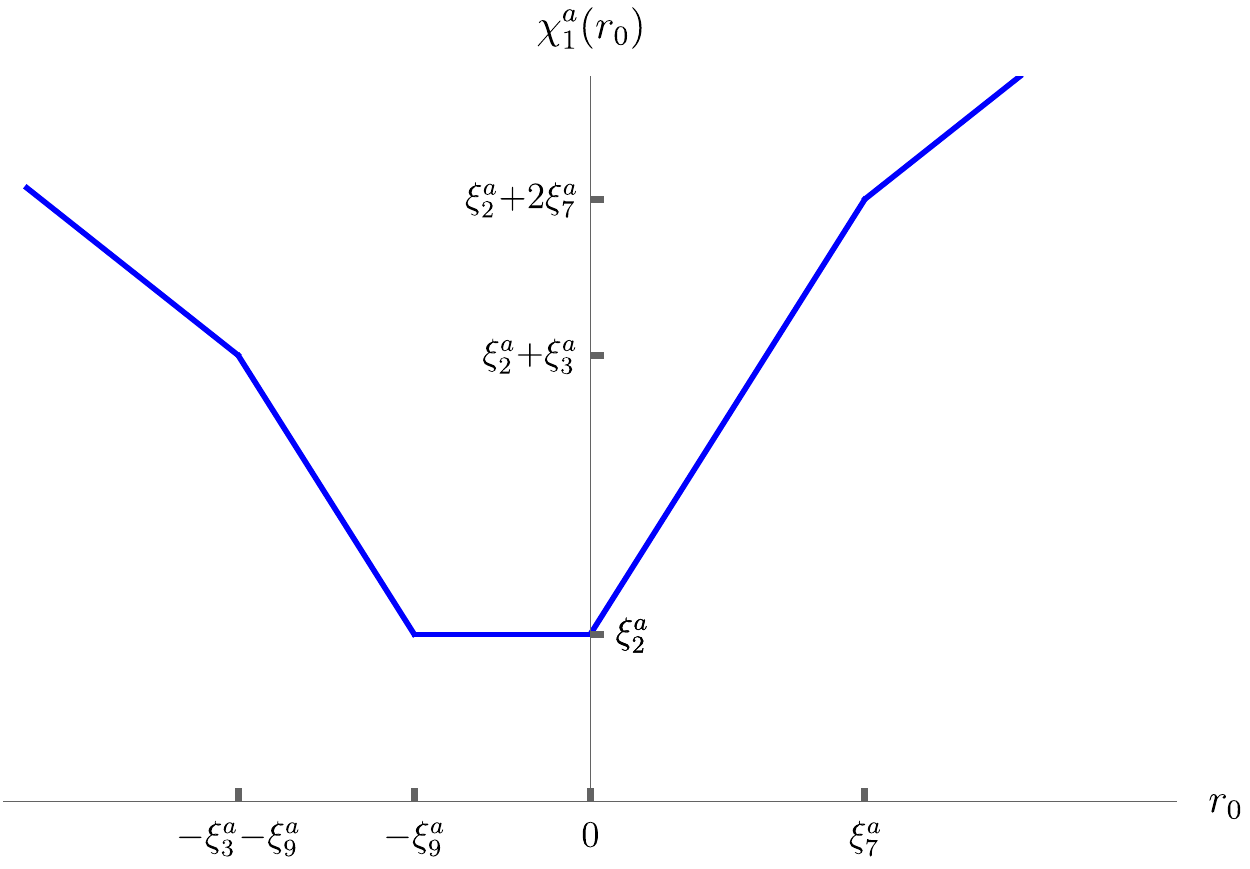}\;
\includegraphics[width=7.6cm]{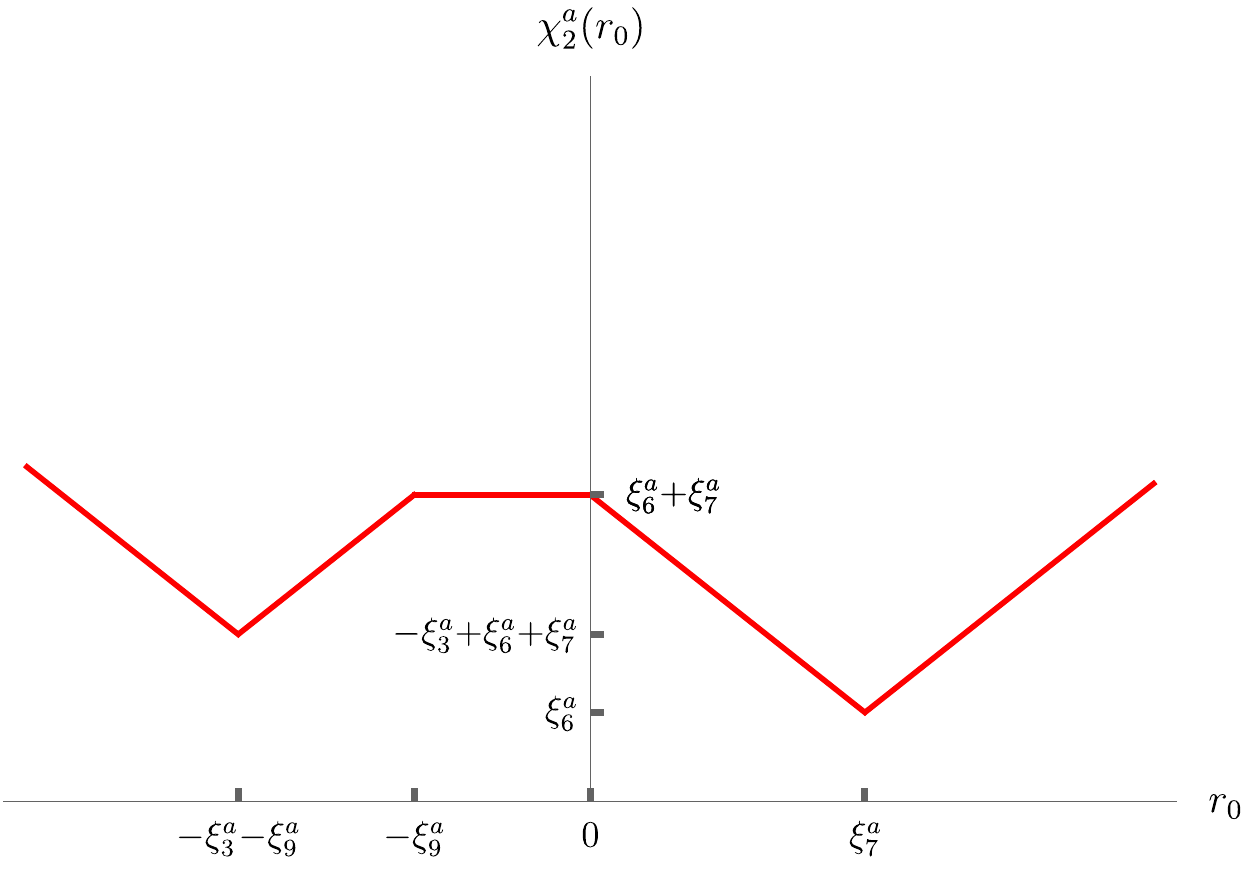}
\caption{IIA profiles for the vertical reduction of resolution (a). \label{fig:IIA profile a}}
 \end{center}
 \end{figure} 
 %%%%%%
\paragraph{Resolution (a).} Consider resolution (a) with the curves $\CC_1^a, \cdots, \CC_1^9$ as shown in Figure~\ref{fig:IIA beetle res a curves}. We have the linear relations:
\bea\label{rel Cs beetle a}
   \CC_1^{a} &\cong \CC_9^{a}, \qquad 
   \CC_8^{a} \cong \CC_2^{a}, \qquad
    \CC_3^{a} + \CC_4^{a} \cong \CC_6^a + \CC_7^a, \qquad
     \CC_5^{a} \cong \CC_3^{a} + \CC_7^{a}+ \CC_9^{a}~.
\eea
Let $\xi_i^a$ denote the K{\"a}hler volume of $\CC^a_i$. 
The vertical reduction is obtained from the GLSM:
\be\label{glsm beetle res a}
\begin{tabular}{l|cccccccc|c}
 & $D_1$ &$D_2$& $D_3$&$D_4$&$D_5$&$D_6$&$\bE_1$&$\bE_2$& FI  \\
 \hline
  $\CC_2^{a}$  &$0$&$1$ &$0$ & $0$&$0$ &$0$&$-2$&$1$&$\xi_2^a$\\
   $\CC_3^{a}$  &$0$&$0$ &$-1$ & $1$&$0$ &$0$&$1$&$-1$&$\xi_3^a$\\
   $\CC_6^{a}$  &$1$&$0$ &$0$ & $0$&$1$ &$-1$&$0$&$-1$&$\xi_6^a$\\
   $\CC_7^{a}$  &$-1$&$0$ &$0$ & $0$&$0$ &$1$&$1$&$-1$&$\xi_7^a$\\
   $\CC_9^{a}$  &$1$&$0$ &$1$ & $0$&$0$ &$0$&$-2$&$0$&$\xi_9^a$\\
   \hline\hline
   $U(1)_M$ & $-1$& $0$& $0$& $0$& $0$& $0$& $1$& $0$& $r_0$
 \end{tabular}
\ee
As anticipated, this is the same as in \eqref{glsm beetle gen}, with $\xi_i^a \equiv \xi_i \geq 0$, and we choose the curves $\{\CC_2^a, \CC_3^a, \CC_6^a, \CC_7^a, \CC_9^a\}$ to be the Mori cone generators, with the condition $-\xi_3^a+ \xi_6^a + \xi_7^a \geq 0$ understood in the definition of the  K{\"a}hler cone, due to the third relation in \eqref{rel Cs beetle a}.

Upon vertical reduction, the IIA background consists a (resolved) $A_2$ singularity with the exceptional curves $\P^1_s$ ($s=1,2$), with two D6-branes wrapping each $\P^1$, thus realizing the $SU(2) \times SU(2)$ gauge group, with the single bifundamental arising at the intersection of the $\P^1$'s. The $A_2$ singularity is non-trivially fibered over $\R\cong \{r_0\}$, with:
\be
\chi^a_1(r_0) = \left\{
                \begin{array}{ll}
                     \xi_2 + \xi_7 + r_0 &\text{ if }\quad r_0 \geq \xi_7\\
	        \xi_2 + 2 r_0 &\text{ if }\quad 0 \leq r_0 \leq \xi_7\\
	        	    \xi_2 &\text{ if }\quad  -\xi_9 \leq r_0 \leq 0\\
		     \xi_2 - 2 \xi_9 - 2 r_0 & \text{ if }\quad  -\xi_3-\xi_9 \leq r_0 \leq -\xi_9\\
                	    \xi_2 + \xi_3 - \xi_9 - r_0 \qquad&  \text{ if }\quad  r_0 \leq -\xi_3-\xi_9\\
                \end{array}
              \right.
 \ee
 and:
 \be
\chi^a_2(r_0) = \left\{
                \begin{array}{ll}
             \xi_6 - \xi_7 + r_0 & \text{ if }\quad   r_0 \geq \xi_7\\
               \xi_6 + \xi_7 - r_0 &\text{ if }\quad  0 \leq r_0 \leq \xi_7\\
	      \xi_6 + \xi_7 & \text{ if }\quad  -\xi_9 \leq r_0 \leq 0\\
         \xi_6 + \xi_7 + \xi_9 + r_0 &\text{ if }\quad  -\xi_3-\xi_9 \leq r_0 \leq -\xi_9\\
                	    -2\xi_3 + \xi_6 + \xi_7 - \xi_9 - r_0\qquad & \text{ if }\quad  r_0 \leq -\xi_3-\xi_9\\
                \end{array}
              \right.	
\ee
for  $\P^1_1$ and $\P^1_2$, respectively. This IIA profile is shown in Figure~\ref{fig:IIA profile a}. Note that $k_{s, {\rm eff}}=0$ here (as well as for all the other gauge-theory chambers), so the theory is indeed parity-invariant.

From this IIA description, we immediately read off the perturbative particles. The W-bosons for both gauge groups have masses:
\be
M(W_{(1)}) = 2\varphi_1 = \xi_9~, \qquad M(W_{(2)}) = 2\varphi_2 =\xi_3+ \xi_7+ \xi_9~,
\ee
and the masses of the four hypermultiplets modes are:
\bea
&M(\CH_{++}) = \varphi_1 + \varphi_2+m = \xi_7+ \xi_9~, \;\;\;
&& M(\CH_{+-}) = -\varphi_1 + \varphi_2-m = \xi_3~,\\
&M(\CH_{-+}) =- \varphi_1 + \varphi_2+m = \xi_7~, \;
&&M(\CH_{--}) = \varphi_1 + \varphi_2-m = \xi_3+ \xi_9~.
\eea
Finally, one can compute the instanton masses:
\bea\label{SU22chamberAmasses}
& M(\CI_{(1), 1})=M(\CI_{(1), 2})= h_1 +2 \varphi_1- \varphi_2 -m = \xi_2~,\\
&M(\CI_{(2), 1}) = h_2+ \varphi_2 - 2m = \xi_6~,
\qquad  M(\CI_{(2), 2}) = h_2+ \varphi_2  =-\xi_3 +  \xi_6+ \xi_7~.
\eea
This completes the derivation of the geometry-to-gauge-theory map \eqref{xi to gauge su2su2}.
One can also easily check that the string tensions computed from the IIA geometry agree with the field theory result:
\be
T_{1}^{(a)}= \d_{\varphi_1} \CF^{(a)} = \int_{-\xi_9}^0 \chi_1(r_0) dr_0~, \qquad
T_{2}^{(a)}= \d_{\varphi_2} \CF^{(a)} = \int_{-\xi_3{-}\xi_9}^{\xi_7} \chi_2(r_0) dr_0~, \qquad
\ee
Here, $\CF^{(a)}$ denotes the prepotential \eqref{CF su2 su2 quiver} in the gauge-theory chamber (a), as defined in \eqref{geom to gauge chamber su2su2}.

%%%%%%%%%
 \begin{figure}[t]
\begin{center}
\subfigure[\small Resolution (b).]{
\includegraphics[width=7.3cm]{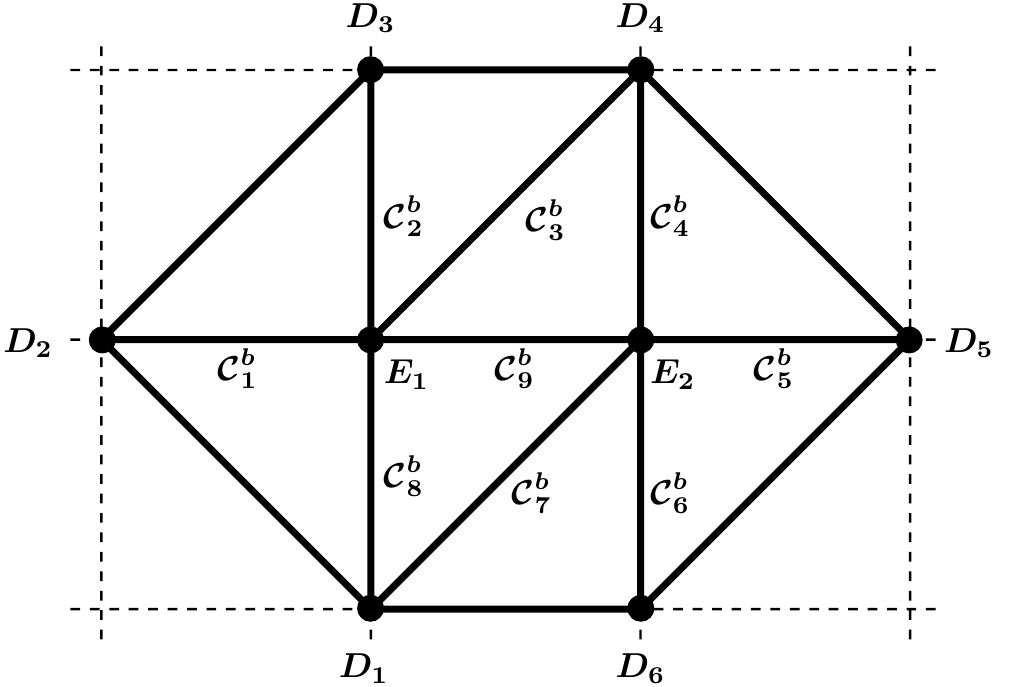}\;
\label{fig:IIA beetle res b curves}}\,
\subfigure[\small Resolution (c).]{
\includegraphics[width=7.3cm]{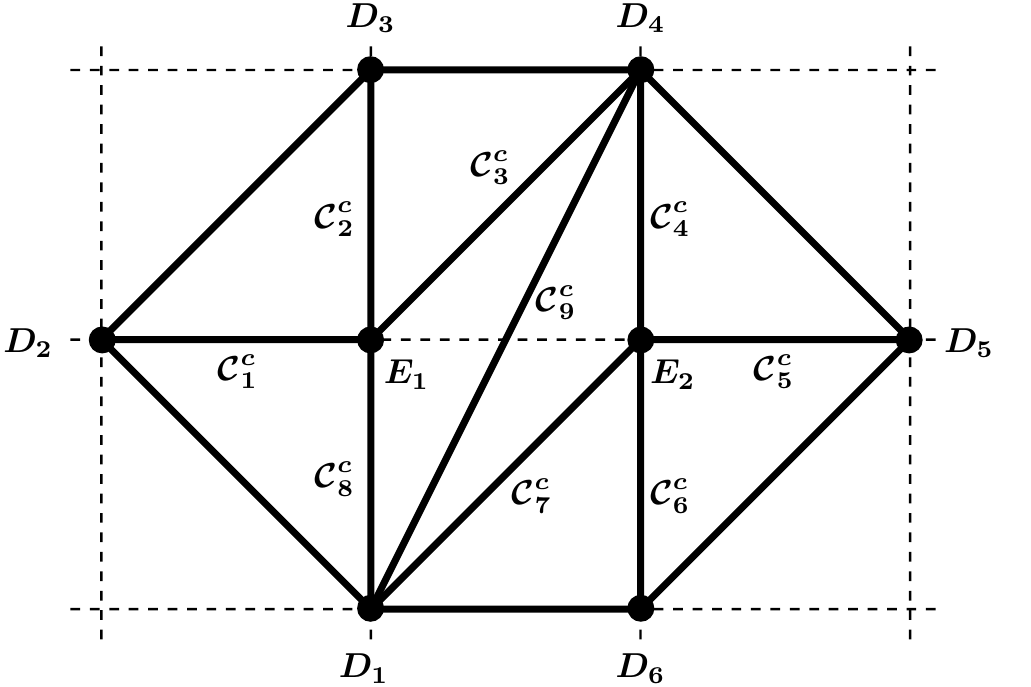}
\label{fig:IIA beetle res c curves}}
\caption{Resolutions (b) and (c), with all the curves indicated. \label{fig:beetle IIA vert ab}}
 \end{center}
 \end{figure} 
 %%%%%%
%%%%%%%%%
 \begin{figure}[t]
\begin{center}
\includegraphics[width=7.6cm]{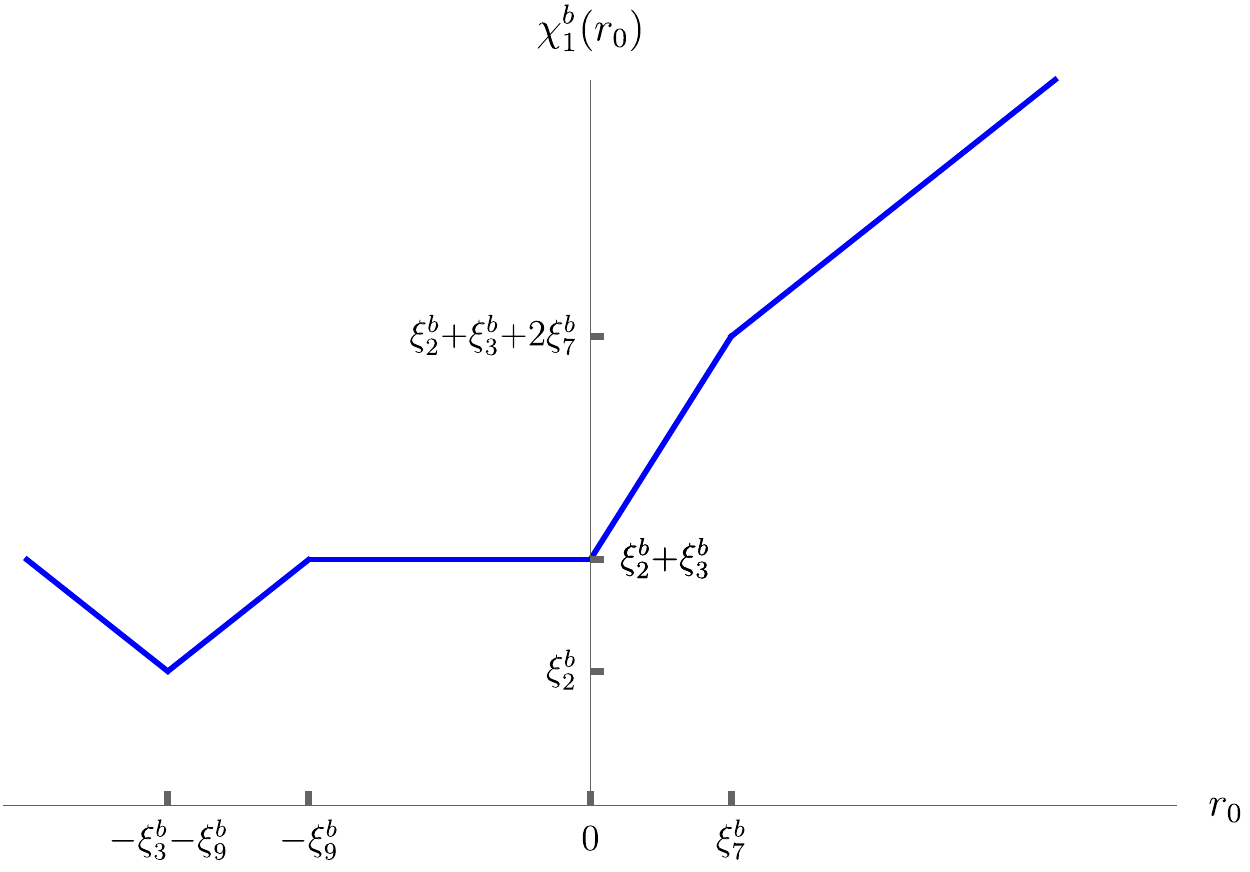}\;
\includegraphics[width=7.6cm]{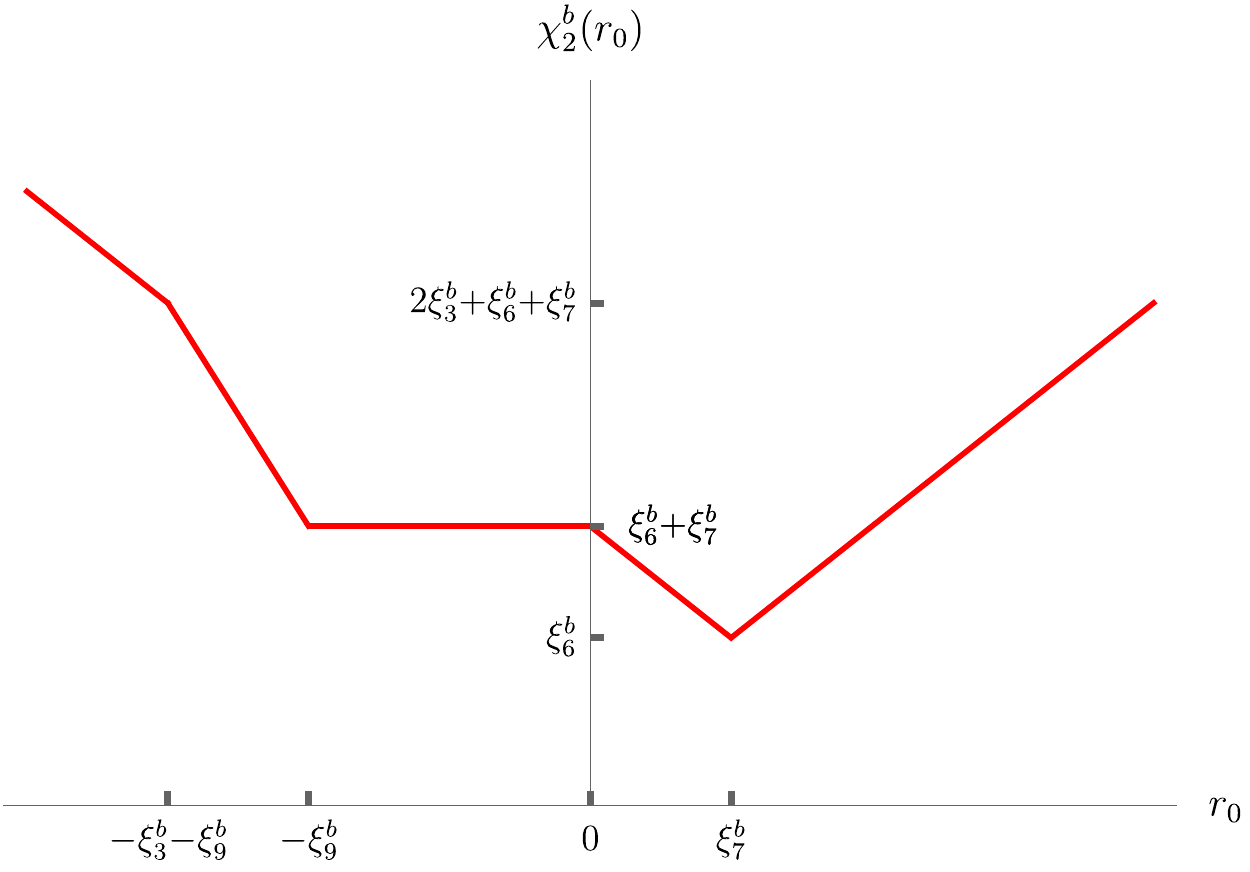}
\caption{IIA profiles for the vertical reduction of resolution (b). \label{fig:IIA profile b}}
 \end{center}
 \end{figure} 
 %%%%%%
\paragraph{Resolution (b).} Consider resolution (b), shown in Figure~\ref{fig:IIA beetle res b curves}. This is obtained from resolution (a) by flopping the curve $\CC_3^a$.
The relations amongst the curves are:
\be
 \CC_2^{b} + \CC_3^{b} \cong \CC_8^{b}, \qquad \CC_6^{b} + \CC_7^{b} \cong \CC_4^{b}, \qquad \CC_1^{b} \cong \CC_3^{b}+\CC_9^{b}, \qquad \CC_5^{b} = \CC_7^{b} + \CC_9^{b}~.
 \ee
 We can take $\{\CC_2^b, \CC_3^b, \CC_6^b, \CC_7^b, \CC_9^b\}$ to generate the Mori cone.
 The K{\"a}hler volumes of the curves $\CC_i^b$, denoted by $\xi_i^b$, are related to the FI terms $\xi_i$ of the GLSM \eqref{glsm beetle gen} by: 
\be\label{xib to xi}
\xi_2^b = \xi_2 + \xi_3~, \qquad
\xi_3^b= -\xi_3~, \qquad
\xi_6^b= \xi_6~, \qquad
\xi_7^b = \xi_7~, \qquad
\xi_9^b=\xi_9 + \xi_3~,
\ee
The IIA background is easily obtained from the GLSM:
\be\label{glsm beetle res b}
\begin{tabular}{l|cccccccc|c}
 & $D_1$ &$D_2$& $D_3$&$D_4$&$D_5$&$D_6$&$\bE_1$&$\bE_2$& FI  \\
 \hline
  $\CC_2^{b}$  &$0$&$1$ &$-1$ & $1$&$0$ &$0$&$-1$&$0$&$\xi_2^b$\\
   $\CC_3^{b}$  &$0$&$0$ &$1$ & $-1$&$0$ &$0$&$-1$&$1$&$\xi_3^b$\\
   $\CC_6^{b}$  &$1$&$0$ &$0$ & $0$&$1$ &$-1$&$0$&$-1$&$\xi_6^b$\\
   $\CC_7^{b}$  &$-1$&$0$ &$0$ & $0$&$0$ &$1$&$1$&$-1$&$\xi_7^b$\\
   $\CC_9^{b}$  &$1$&$0$ &$0$ & $1$&$0$ &$0$&$-1$&$-1$&$\xi_9^b$\\
   \hline\hline
   $U(1)_M$ & $-1$& $0$& $0$& $0$& $0$& $0$& $1$& $0$& $r_0$
 \end{tabular}
\ee
This is equivalent to \eqref{glsm beetle res a}, with the rows corresponding to the Mori cone generators. We find:
\be
\chi_1^b(r_0) = \left\{ \begin{array}{ll}
  	   \xi_2^b + \xi_3^b + \xi_7^b + r_0 \qquad& \text{if} \quad r_0 \geq \xi_7^b\\
	     	   \xi_2^b + \xi_3^b + 2r_0 & \text{if} \quad 0 \leq r_0 \leq \xi_7^b\\
		     	   \xi_2^b + \xi_3^b & \text{if} \quad -\xi_9^b\leq r_0 \leq 0\\
			     	   \xi_2^b + \xi_3^b + \xi_9^b + r_0 & \text{if} \quad-\xi_3^b-\xi_9^b\leq r_0 \leq -\xi_9^b\\
  	   \xi_2^b - \xi_3^b - \xi_9^b - r_0 & \text{if} \quad r_0 \leq -\xi_3^b-\xi_9^b
	    \end{array}
  \right.	
\ee
\be
  \chi_2^b(r_0) = \left\{ \begin{array}{ll}
    	   \xi_6^b - \xi_7^b + r_0 & \text{if} \quad r_0 \geq \xi_7^b\\
	     	   \xi_6^b + \xi_7^b - r_0 & \text{if} \quad 0 \leq r_0 \leq \xi_7^b\\
  	   \xi_6^b + \xi_7^b & \text{if} \quad -\xi_3^b\leq r_0 \leq 0\\	
	     	   \xi_6^b + \xi_7^b - 2\xi_9^b - 2 r_0 & \text{if} \quad -\xi_3^b-\xi_9^b\leq r_0 \leq -\xi_9^b\\   
  	   \xi_3^b + \xi_6^b + \xi_7^b - \xi_9^b - r_0 \qquad& \text{if} \quad r_0 \leq -\xi_3^b-\xi_9^b
 \end{array}
  \right.	
\ee
This is pictured in Fig.~\ref{fig:IIA profile b}.
From this profile, we read off the W-bosons and hypermultiplet masses:
\bea
 &M(W_{(1)}) = 2\varphi_1 = \xi_3^b+ \xi_9^b~, \quad  
 &&M(W_{(2)}) = 2\varphi_2 =\xi_7^b+ \xi_9^b~,\\
&M(\CH_{++}) = \varphi_1 + \varphi_2+m =\xi_3^b+ \xi_7^b+ \xi_9^b~, \;
&& M(\CH_{+-}) = \varphi_1 - \varphi_2+m = \xi_3^b~,\\
&M(\CH_{-+}) =- \varphi_1 + \varphi_2+m = \xi_7^b~, \;
&&M(\CH_{--}) = \varphi_1 + \varphi_2-m = \xi_9^b~.
\eea
We also have the instanton masses:
\bea
& M(\CI_{(1), 1})= h_1 + \varphi_1-2m = \xi_2^b~,\;\;
&& M(\CI_{(1), 2})= h_1 +2 \varphi_1-\varphi_2-m = \xi_2^b+ \xi_3^b~,\\
&M(\CI_{(2), 1}) = h_2+ \varphi_2 - 2m = \xi_6^b~,
&& M(\CI_{(2), 2}) = h_2-\varphi_1+2 \varphi_2- m  =  \xi_6^b+ \xi_7^b~.
\eea
Of course, this reproduces \eqref{xi to gauge su2su2}. The string tensions similarly match.

\paragraph{Resolution (c).} Consider the resolution (c), shown in Figure~\ref{fig:IIA beetle res c curves}. It can be obtained from resolution (b) by flopping the curve $\CC^b_9$ to $\CC^c_9 \cong - \CC^b_9$. Here, we have the following linear relations amongst curves:
\be
\CC_1^c \cong \CC_3^c~, \qquad 
\CC_4^c \cong \CC_6^c + \CC_7^c~, \qquad
\CC_5^c \cong \CC_7^c~, \qquad
\CC_8^c \cong \CC_2^c +\CC_3^c~.
\ee
Thus we can choose $\{\CC_2^c, \CC_3^c, \CC_6^c, \CC_7^c, \CC_9^c \}$ as generators of the Mori cone.
The K{\"a}hler volumes $\xi_i^c$ are related to the FI terms $\xi_i$ in  \eqref{glsm beetle gen}  by: 
\be
\xi_2^c = \xi_2 + \xi_3~, \qquad 
\xi_3^c = \xi_9~, \qquad
\xi_6^c= \xi_6~, \qquad
\xi_7^c = \xi_3+\xi_7+\xi_9~, \qquad
\xi_9^c =- \xi_3-\xi_9~.
\ee
We derive the IIA background from the GLSM:
\be\label{glsm beetle res e}
\begin{tabular}{l|cccccccc|c}
 & $D_1$ &$D_2$& $D_3$&$D_4$&$D_5$&$D_6$&$\bE_1$&$\bE_2$& FI  \\
 \hline
  $\CC_2^{c}$  &$0$&$1$ &$-1$ & $1$&$0$ &$0$&$-1$&$0$&$\xi_2^c$\\
   $\CC_3^{c}$  &$1$&$0$ &$1$ & $0$&$0$ &$0$&$-2$&$0$&$\xi_3^c$\\
   $\CC_6^{c}$  &$1$&$0$ &$0$ & $0$&$1$ &$-1$&$0$&$-1$&$\xi_6^c$\\
   $\CC_7^{c}$  &$0$&$0$ &$0$ & $1$&$0$ &$1$&$0$&$-2$&$\xi_7^c$\\
   $\CC_9^{c}$  &$-1$&$0$ &$0$ & $-1$&$0$ &$0$&$1$&$1$&$\xi_9^c$\\
   \hline\hline
   $U(1)_M$ & $-1$& $0$& $0$& $0$& $0$& $0$& $1$& $0$& $r_0$
 \end{tabular}
\ee
%%%%%%%%%
 \begin{figure}[t]
\begin{center}
\includegraphics[width=7.6cm]{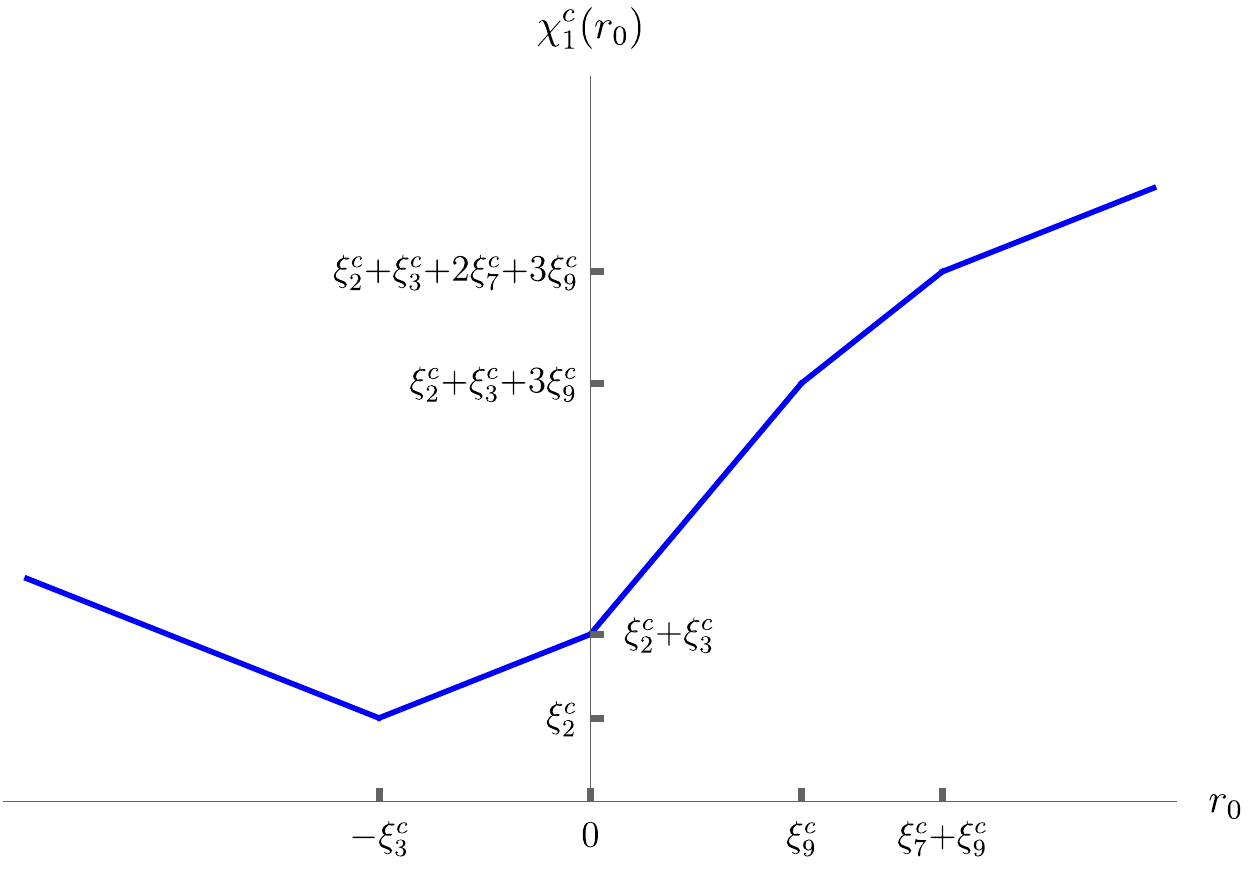}\;
\includegraphics[width=7.6cm]{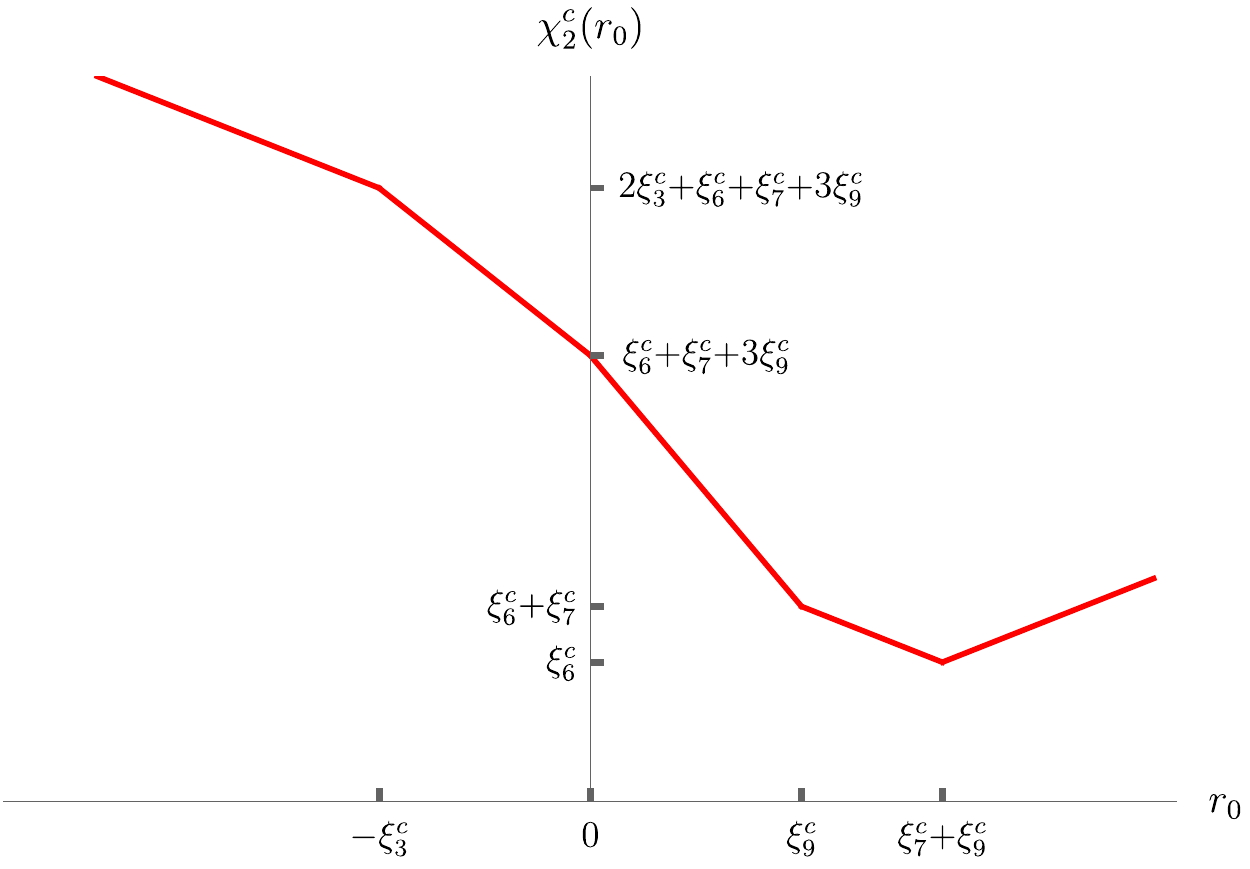}
\caption{IIA profiles for the vertical reduction of resolution (c). \label{fig:IIA profile c}}
 \end{center}
 \end{figure} 
 %%%%%%
We find the following profiles for $\chi_I(r_0)$:
\be
\chi^c_1(r_0) = \left\{
                \begin{array}{ll}
                                	    \xi^c_2+ \xi^c_3  + \xi^c_7 + 2\xi^c_9 + r_0 \quad & \text{if}\quad r_0 \geq \xi^c_7 + \xi^c_9\\
	                    	    \xi^c_2+ \xi^c_3+ \xi^c_9 + 2 r_0 & \text{if}\quad \xi^c_9 \leq r_0 \leq \xi^c_7 + \xi^c_9\\
		                	  \xi^c_2 +  \xi^c_3 + 3 r_0 & \text{if}\quad 0 \leq r_0 \leq \xi^c_9\\
                	     \xi^c_2+\xi^c_3  + r_0 & \text{if}\quad -\xi^c_3 \leq r_0 \leq 0\\
                	    \xi^c_2 -\xi^c_3  - r_0 & \text{if}\quad r_0 \leq -\xi^c_3
                \end{array}
              \right.	
\ee
\be
\chi^c_2(r_0) = \left\{
                \begin{array}{ll}
                      	    \xi^c_6 -\xi^c_7  - \xi^c_9 + r_0 & \text{if}\quad r_0 \geq \xi^c_7 + \xi^c_9\\
                	  \xi^c_6 +  \xi^c_7 + \xi^c_9 - r_0 & \text{if}\quad \xi^c_9 \leq r_0 \leq \xi^c_7 + \xi^c_9\\
             	    \xi^c_6+ \xi^c_7  + 3\xi^c_9 - 3 r_0 & \text{if}\quad 0 \leq r_0 \leq \xi^c_9\\
                	    \xi^c_6+ \xi^c_7  + 3 \xi^c_9 - 2 r_0 & \text{if}\quad -\xi^c_3 \leq r_0 \leq 0\\	    	    	    
                	    \xi^c_3 +\xi^c_6+ \xi^c_7  + 3 \xi^c_9 - r_0\quad & \text{if}\quad r_0 \leq -\xi^c_3
                \end{array}
              \right.	
\ee
This is shown in Figure~\ref{fig:IIA profile c}.
The masses of the perturbative particles are:
\bea
 &M(W_{(1)}) = 2\varphi_1 = \xi_3^c~, \quad  
 &&M(W_{(2)}) = 2\varphi_2 =\xi_7^c~,\\
&M(\CH_{++}) = \varphi_1 + \varphi_2+m =\xi_3^c+ \xi_7^c+ \xi_9^c~, \;
&& M(\CH_{+-}) = \varphi_1 - \varphi_2+m = \xi_3^c+\xi_9^c~,\\
&M(\CH_{-+}) =- \varphi_1 + \varphi_2+m = \xi_7^c+\xi_9^c~, \;
&&M(\CH_{--}) =- \varphi_1 - \varphi_2+m = \xi_9^c~.
\eea
and the instanton masses are:
\bea
& M(\CI_{(1), 1})= h_1 + \varphi_1-2m = \xi_2^c~,\;\;
&& M(\CI_{(1), 2})= h_1 +3 \varphi_1-2m = \xi_2^c+ \xi_3^c~,\\
&M(\CI_{(2), 1}) = h_2+ \varphi_2 - 2m = \xi_6^c~,
&& M(\CI_{(2), 2}) = h_2+3\varphi_2- 2m  =  \xi_6^c+ \xi_7^c~.
\eea
As mentioned above, this resolution is compatible with the limit $m \rightarrow \infty$, corresponding to $\xi_9^c\rightarrow \infty$. That this limit gives rise to two $SU(2)_\pi$ gauge groups is readily apparent from Figure~\ref{fig:IIA profile c}.

\paragraph{Resolution (d), (e) and (f).} The three other resolutions admitting a vertical reduction are completely similar, and can be obtained from the cases (a), (b) and (c), respectively, by a reflection along the vertical axis, thus exchanging the role of the two SU(2) gauge groups.

%%%%%%%%%%%%%

\subsubsection{$SU(3)$ phases}
The horizontal reduction of the beetle geometry, when allowed, leads to an $SU(3)$ gauge theory coupled to two hypermultiplets in the fundamental representation, with $k_{\rm eff=0}$. The prepotential of this gauge theory reads:
\bea\label{F SU3 nf2 phase}
&\CF_{SU(3)}^{N_f=2}= \t h_0 (\t \varphi_1^2 + \t \varphi_2^2 -\t\varphi_1 \t\varphi_2)+ \half (\t\varphi_1^2\t\varphi_2-\t\varphi_1\t\varphi_2^2)\\
&\quad\qquad+ {4\ov 3}(\t\varphi_1^3+\t\varphi_2^3) -\half (\t\varphi_1^2\t\varphi_2+\t\varphi_1\t\varphi_2^2) 
 -{1\ov 6} \sum_{i=1}^2\Big( \Theta(\t\varphi_1+ \t m_i) (\t\varphi_1+ \t m_i)^3\\
&\quad\qquad+ \Theta(-\t\varphi_1+\t\varphi_2+ \t m_i)(-\t\varphi_1+\t\varphi_2+ \t m_i)^3
 + \Theta(-\t\varphi_2+ \t m_i) (-\t\varphi_2+ \t m_i)^3
\Big)~.
\eea
 Here and in the rest of this section, we dress the $SU(3)$ variables with tildes, to distinguish them from the $SU(2) \times SU(2)$ gauge-theory variables. Note that the last term on the first line of \eqref{F SU3 nf2 phase} is a bare $SU(3)$ CS level $k=1$, which is necessary so that $k_{\rm eff}=0$, given our choice of ``$\kappa=-\half$'' quantization for the two fundamental hypermultiplets.

We have chosen the fundamental Weyl chamber:
\be\label{SU3 weyl chamber}
2\t\varphi_1- \t\varphi_2 \geq 0~, \qquad 
-\t\varphi_1+2\t\varphi_2 \geq 0~.
\ee
 There are 16 distinct gauge-theory chambers that can be obtained by varying $\t\varphi_1, \t\varphi_2$ and the real masses $\t m_1, \t m_2 \in \R$. The chambers correspond to all the possible choices of signs for the masses of the 6 hypermultiplets modes $\t\varphi_1 + \t m_i$, $-\t\varphi_1+\t\varphi_2+ \t m_i$, and $-\t\varphi_2 + \t m_i$ ($i=1,2)$, compatible with the Weyl-chamber condition \eqref{SU3 weyl chamber}.

We will now derive the following map between the GLSM parameters and the $SU(3)$ variables:
\bea\label{xi to gauge SU3Nf2}
&\xi_2 =  2 \t\varphi_1 -\t\varphi_2~,\quad\qquad\qquad &
&\xi_3 = -\t\varphi_1+\t \varphi_2 +\t m_1~,\\
& \xi_6 = \t\varphi_2 - \t m_2~, \qquad &
&\xi_7= - \t\varphi_1 + \t\varphi_2+ \t m_2~,\\
&\xi_9= \t h_0 + 2\t\varphi_1-\t m_1 - \t m_2~.\qquad
\eea
Equivalently, we find:
\bea\label{mu to geom beetle SU3}
&\mu_1= \t h_0 + 3 \t m_1 - \t m_2~, \qquad\qquad && \nu_1 = -\t\varphi_1+2\t m_1~, \\
& \mu_2 = 3\t m_1~, \qquad&& \nu_2 = -\varphi_2 + \t m_1~, \\
& \mu_6=\t h_0 +2 \t m_1~.
\eea
By comparing to \eqref{mu to geom beetle}, this implies an ``S-duality map'' which we discuss in subsection~\ref{subsec: S duality map beetle} below.

\paragraph{Resolution (a).} Consider the horizontal reduction of resolution (a) in Fig.~\ref{fig:IIA beetle res a curves}. We use the same GLSM as in \eqref{glsm beetle res a}, but with a different $U(1)_M$ charge:
\be\label{u1M beetle hor res a}
\begin{tabular}{l|cccccccc|c}
 & $D_1$ &$D_2$& $D_3$&$D_4$&$D_5$&$D_6$&$\bE_1$&$\bE_2$& FI  \\
   \hline\hline
   $U(1)_{M'}$ & $0$& $0$& $0$& $0$& $0$& $0$& $-1$& $1$& $r_0$
 \end{tabular}
\ee
The resulting IIA background consists of a resolved $A_1$ singularity fibered over $\R$, with three D6-branes wrapped over the exceptional $\P^1$, and with two D6-branes wrapped over non-compact divisors $\cong \C$, thus giving rise to an $SU(3)$ theory with two flavors. The profile $\chi(r_0)$ reads:
\be\label{chi beetle a su3}
\chi^a(r_0) = \left\{
                \begin{array}{ll}
                                	    -2\xi_2 + \xi_9 + 2 r_0 & \text{if} \quad  r_0 \geq \xi_2\\
                	    \xi_9 & \text{if} \quad  0 \leq r_0 \leq \xi_2\\
                	    \xi_9 - 2 r_0 & \text{if} \quad  -\xi_3 \leq r_0 \leq 0\\
                	    \xi_3 + \xi_9 - r_0 & \text{if} \quad  -\xi_7 \leq r_0 \leq -\xi_3\\
                	    \xi_3 + \xi_7 + \xi_9 & \text{if} \quad  -\xi_6-\xi_7 \leq r_0 \leq -\xi_7\\    	    	    	    
                	    \xi_3 - 2\xi_6 - \xi_7 + \xi_9 - 2r_0 \qquad&  \text{if} \quad  r_0 \leq -\xi_6-\xi_7
                \end{array}
              \right.	
 \ee
Recall that   $\xi_i = \xi_i^a$.  
%%%%%%%%%
 \begin{figure}[t]
\begin{center}
\includegraphics[width=8.5cm]{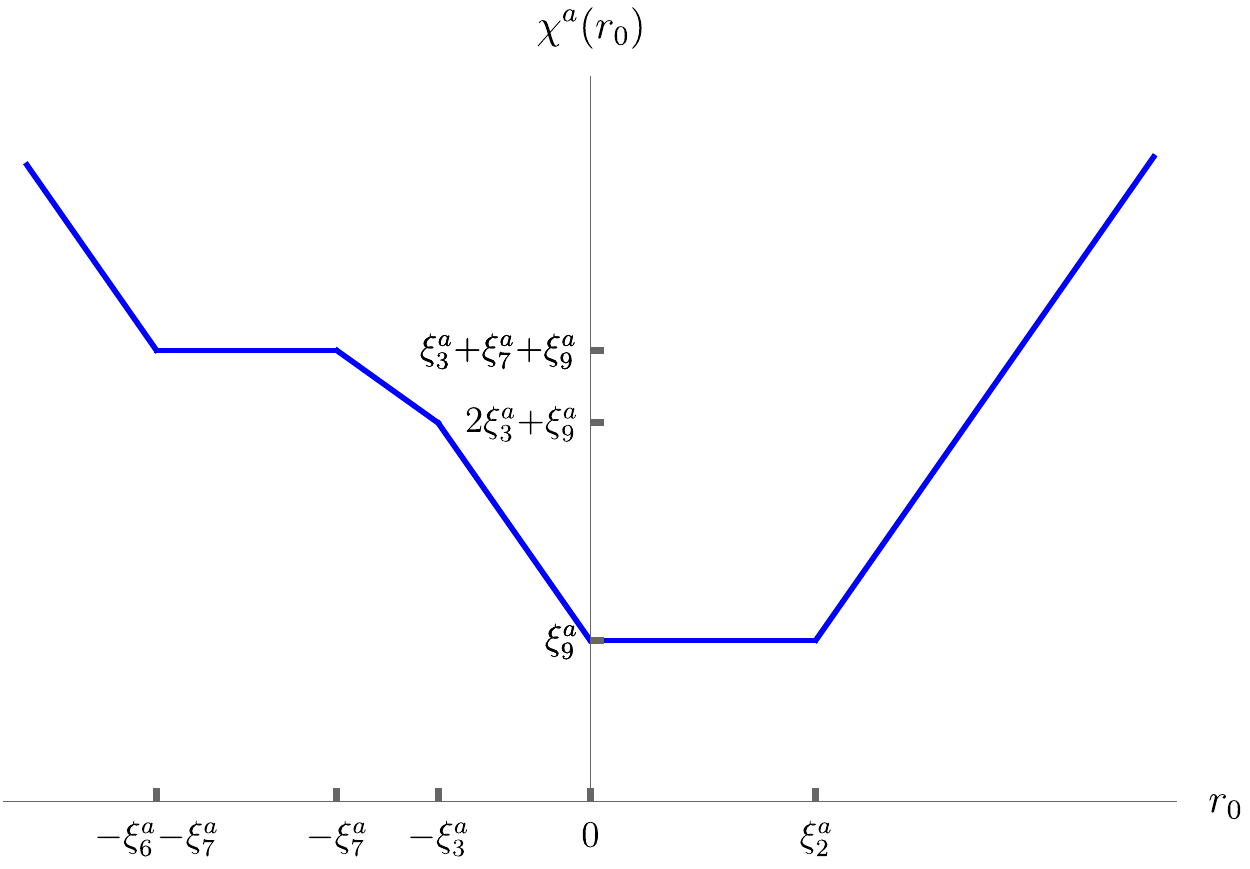}
\caption{IIA profiles for the horizontal reduction of resolution (a), giving rise an an $SU(3)$ theory with two flavors. \label{fig:IIA profile a hor}}
 \end{center}
 \end{figure} 
 %%%%%%
 From the IIA profile \eqref{chi beetle a su3}, shown in Figure~\ref{fig:IIA profile a hor}, we directly read off 
the effective CS level $k_{\rm eff}=0$, 
 the masses of the $SU(3)$ W-bosons:
\be
M(W_1) =2 \t \varphi_1- \t\varphi_2= \xi_2~, \qquad
M(W_2) = -\t \varphi_1+2 \t\varphi_2=\xi_6+\xi_7~,
\ee 
and of the hypermultiplet modes:
\bea\label{su3 res a hypers}
&M(\CH_{1,1})=\t\varphi_1+\t m_1 = \xi_2+\xi_3~,  && M(\CH_{1,2})=\t\varphi_1+\t m_2 = \xi_2+\xi_7~, \\
&M(\CH_{2,1})=-\t\varphi_1+\t\varphi_2+\t m_1 =\xi_3~, && M(\CH_{2,2})=-\t\varphi_1+\t\varphi_2+\t m_2=\xi_7  \\
&M(\CH_{3,1})=\t\varphi_2-\t m_1 = -\xi_3+\xi_6+\xi_7~,\quad&&M(\CH_{3,2})=\t\varphi_2-\t m_2= \xi_6
\eea
The instanton particle masses are given by: 
 \be
 M(\CI_1) = M(\CI_2)= \t h_0 +2 \t\varphi_1 -\t m_1- \t m_2 = \xi_9~, \qquad
 M(\CI_3) = \t h_0 +2\t \varphi_2  = \xi_3 + \xi_7 +\xi_9~.
 \ee           
This establishes the map \eqref{xi to gauge SU3Nf2} between the geometry and the $SU(3)$ gauge-theory parameters. As usual, one can also match the magnetic string tensions.

From this analysis, we find that this resolution corresponds to the field-theory chamber:
\be
\t\varphi_1+\t m_i >0~, \qquad -\t\varphi_1+\t\varphi_2+\t m_i>0~, \qquad
-\t\varphi_2+\t m_i <0~, \qquad i=1,2~.
\ee
By successive flops, it is straightforward to map the 16 different gauge-theory chambers to the 16 resolutions with an allowed horizontal reduction. For instance, by flopping the curve $\CC_4^a \cong - \CC_3^a + \CC_6^a+\CC^a_7$ we obtain resolution (g)---Figure~\ref{fig:beetle res g}. From \eqref{su3 res a hypers}, we see that this corresponds to changing the sign of the mass of the hypermultiplet mode $\CH_{3,1}$. If, in addition, we also flop the curve $\CC_6^a$, we obtain resolution (i) in Figure~\ref{fig:beetle res i}, corresponding to also changing the sign of $M(\CH_{3,2})$. In that resolution, we have $\rho(\varphi)+ m_1 >0$ and $\rho(\varphi)+ m_2 >0$ for the fundamental hypermultiplets, and therefore we can consider the simultaneous limit $m_1 \rightarrow \infty$ and $m_2 \rightarrow \infty$. In the field theory, this gives us an $SU(3)_{-1}$ gauge theory without matter. The toric geometry in that limit is obtained by removing $D_4$ and $D_6$ from the toric diagram. This is in perfect agreement with our discussion of $SU(N)_k$ gauge theories in section~\ref{sec: SUNk}.

\begin{figure}[t]
\begin{center}
\includegraphics[width=8.5cm]{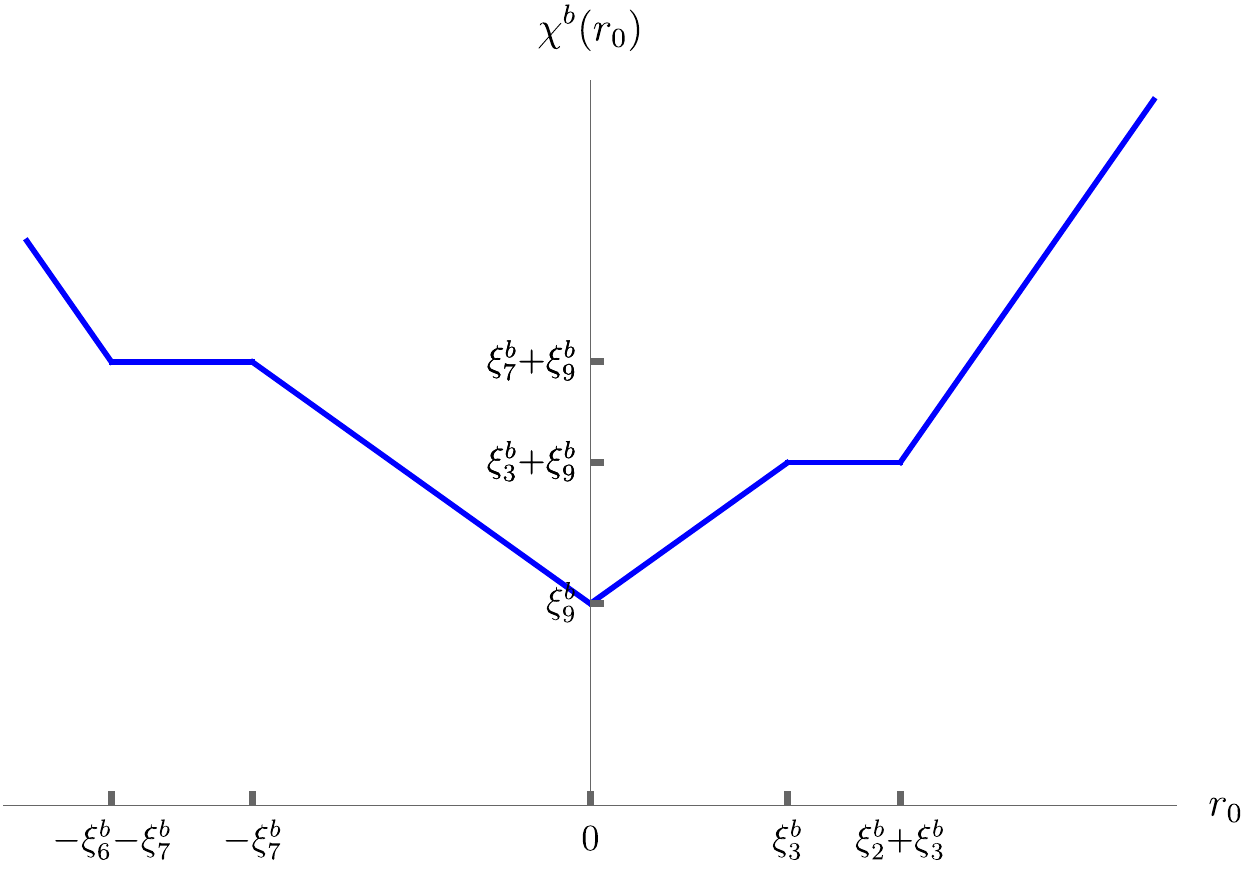}
\caption{IIA profiles for the horizontal reduction of resolution (b). \label{fig:IIA profile b hor}}
 \end{center}
 \end{figure}

\paragraph{Resolution (b).} As another example, consider the horizontal reduction of resolution (b), which is obtained from resolution (a) by a single flop on $\CC^a_3$---recall the relation \eqref{xib to xi} between the K{\"a}hler parameters. We find:
\be
\chi^b(r_0) = \left\{
                \begin{array}{ll}
                              	    -2\xi^b_2-\xi^b_3+\xi^b_9+2r_0\qquad & \text{if}\quad  r_0 \geq \xi^b_2 + \xi^b_3\\
                	    \xi^b_3+\xi^b_9 & \text{if}\quad  \xi^b_3 \leq r_0 \leq \xi^b_2+\xi^b_3\\
	           	    \xi^b_9+r_0 & \text{if}\quad  0 \leq r_0 \leq \xi^b_3\\
                	    \xi^b_9-r_0 & \text{if}\quad  -\xi^b_7 \leq r_0 \leq 0\\
	    	 \xi^b_7 + \xi^b_9 & \text{if}\quad  -\xi^b_6-\xi^b_7 \leq r_0 \leq -\xi^b_7\\	        	    
                	    -2\xi^b_6 - \xi^b_7 + \xi^b_9 - 2 r_0 &  \text{if}\quad  r_0 \leq -\xi^b_6-\xi^b_7
                \end{array}
              \right.
\ee
This is shown in Figure~\ref{fig:IIA profile b hor}.
%%%%%%%%%
 %%%%%%
 We have the perturbative particles:
\bea
&M(W_1) =2 \t \varphi_1- \t\varphi_2= \xi_2^b+ \xi_3^b~, \qquad
&& M(W_2) = -\t \varphi_1+2 \t\varphi_2=\xi_6^b+\xi_7^b~,\\
&M(\CH_{1,1})=\t\varphi_1+\t m_1 = \xi_2^b~,  && M(\CH_{1,2})=\t\varphi_1+\t m_2 = \xi_2^b+\xi_3^b+\xi_7^b~, \\
&M(\CH_{2,1})=\t\varphi_1-\t\varphi_2-\t m_1 =\xi_3^b~, && M(\CH_{2,2})=-\t\varphi_1+\t\varphi_2+\t m_2=\xi_7^b  \\
&M(\CH_{3,1})=\t\varphi_2-\t m_1 = \xi_3^b+\xi_6^b+\xi_7^b~,\quad&&M(\CH_{3,2})=\t\varphi_2-\t m_2= \xi_6^b
\eea
and the instanton particles:
 \bea
& M(\CI_1) =\t h_0 +2 \t\varphi_1 -\t m_1- \t m_2 =\xi_3^b+ \xi_9^b~,&& \\
& M(\CI_2) =\t h_0 + \t\varphi_1 +\t \varphi_2- \t m_2 = \xi_9^b~, \qquad
&& M(\CI_3) = \t h_0 +2\t \varphi_2  = \xi_7^b +\xi_9^b~.
 \eea
Of course,  using \eqref{xib to xi}, these relations also reproduce \eqref{xi to gauge SU3Nf2}.

 \paragraph{Resolutions (d), (e) and (g)-(r).}  All the other horizontal reductions can be performed similarly. The precise map between the resolutions and the 16 field-theory chambers is further discussed in Appendix~\ref{app: more on beetle}.

\subsubsection{The S-duality map}\label{subsec: S duality map beetle}
By comparing the geometry-to-gauge-theory maps for the vertical and horizontal reductions, \eqref{xi to gauge su2su2} and \eqref{xi to gauge SU3Nf2}, respectively, we obtain an ``S-duality map'' between the $SU(3)_0, N_f=2$ and the $SU(2)\times SU(2)$ gauge-theory parameters:
\bea\label{Sdual beetle map}
&\t m_1= {1\ov 3}(h_1-h_2) -m  \qquad\qquad\qquad && \t\varphi_1 = \varphi_1 + {1\ov 3}(2 h_1+h_2) - m~, \\
&\t m_2= {1\ov 3}(h_1-h_2) +m  \qquad && \t\varphi_2 = \varphi_2 + {1\ov 3} (h_1 +2 h_2) - m~, \\
& \t h_0= -{2\ov 3}(h_1+2 h_2)+ 2m~.
\eea
This same map was recently derived in \cite{Assel:2018rcw} by studying Wilson loops.~\footnote{Our result agrees with the ones of \protect\cite{Assel:2018rcw} up to some shifts of the gauge couplings by the real masses---for instance, our $\t h_0$ is equal to $t-\half(m_1+m_2)$ in their notation (their $m_i$ is our $-\t m_i$). These shifts arise from our different treatment of the parity anomaly.}

\paragraph{S-duality for the resolution (a).} In resolution (a), we have the following dictionary between particle states in the two descriptions:
\be\nn
\begin{tabular}{l|cccccccc}
{\rm Geometry:} & $ \CC_1^a\cong \CC_9^a$ &$\CC_5^a$& $\CC_2^a\cong \CC_8^a$& $\CC_6^a+\CC_7^a$ &$\CC_3^a$ &$\CC_4^a$ & $\CC_6^a$& $\CC_7^a$\\
\hline\hline
{\small $SU(2)\times SU(2)$}     &$ W_{(1)}$&$  W_{(2)}$ & $ \CI_{(1),1}$& $ \CI_{(2),1}{\oplus}\CH_{-+}$ & $\CH_{+-}$& $\CI_{(2),2}$ &$\CI_{(2),1}$ & $\CH_{-+}$  \\
{\small $SU(3), N_f=2$}    &$\CI_1$&$ \CI_3$  &$W_1$ &$W_2$ & $\CH_{2,1}$ &$\CH_{3,1}$  &$\CH_{3,2}$ & $\CH_{2,2}$ 
 \end{tabular}
\ee
The curves are as indicated in Fig.~\ref{fig:IIA beetle res a curves}, with the relations \eqref{rel Cs beetle a}.
As expected from the $(p,q)$-web picture \cite{Aharony:1997bh, Bergman:2013aca}, instanton particles are mapped to perturbatives particles in the S-dual description.

\paragraph{S-duality for the resolution (b).}
In resolution (b), we find:
\be\nn
\begin{tabular}{l|ccccccccc}
{\rm Geom:} & $ \CC_1^b$ &$\CC_5^b$& $\CC_8^b$& $\CC_4^b$ &$\CC_2^b$ &$\CC_3^b$ & $\CC_6^b$& $\CC_7^b$ &$\CC_9^b$\\
\hline\hline
{\small $SU(2)^2$}    &$ W_{(1)}$&$  W_{(2)}$ & $ \CI_{(1),1}{\oplus}\CH_{+-}$& $ \CI_{(2),1}{\oplus}\CH_{-+}$&$\CI_{(1),1}$ & $\CH_{+-}$ &$\CI_{(2),1}$ & $\CH_{-+}$&$\CH_{--}$  \\
{\small $SU(3)$}    &$\CI_1$&$ \CI_3$  &$W_1$ &$W_2$ & $\CH_{1,1}$&$\CH_{2,1}$   &$\CH_{3,2}$ & $\CH_{2,2}$  &$\CI_2$
 \end{tabular}
\ee
We obtain resolution (b) from resolution (a) by flopping $\CC_3^a$ to $\CC_3^b \cong - \CC_3^a$. Since $\CC_3^a$ corresponds to an hypermultiplet mode, $\CH_{+-}$ or $\CH_{2,1}$, in either gauge-theory theory description, we can cross this K{\"a}hler wall ``perturbatively'' in both descriptions, since this simply amounts to changing the sign of that hypermultiplet mass. This is in agreement with the fact that resolutions (a) and (b) both admit the two gauge-theory descriptions simultaneously. 

\paragraph{Some examples of phase transitions.} In general, more interesting phenomena can happen. Consider the flop of $\CC_9^b$ in resolution (b), which gives rise to resolution (c). As discussed above, this simply corresponds to changing the sign of  mass of $\CH_{--}$ in the $SU(2)\times SU(2)$ description. On the other hand, in the $SU(3), N_f=2$ description, this would correspond to changing the sign of $M(\CI_2)$, the mass of an instanton particle, which cannot be done consistently in the low-energy $SU(3)$ gauge-theory language. 

Moreover, in both resolutions (a) and (b), M2-branes wrapped over curves that cannot be flopped---namely, $\CC_1^a$, $\CC_2^a$ and $\CC_5^a$ in resolution (a), and $\CC_1^b$, $\CC_4^b$, $\CC_5^b$, $\CC_8^b$ in resolution (b)---give rise to W-bosons in one of the two gauge-theory descriptions. When such a curve is blown down, we reach a non-traversable K{\"a}hler wall associated with the appearance of an $SU(2)$ non-abelian gauge symmetry \cite{Witten:1996qb}. Note that not all such walls are apparent from the point of view of a single low-energy gauge-theory description, since some of those walls also correspond to instanton particles becoming massless. 

Interestingly, from the point of view of a given gauge theory, the walls corresponding to massless instanton particles may either be traversable  or not; that determination requires to keep into account the string tensions. Let us consider resolution (a) in the $SU(2) \times SU(2)$ frame. In that chamber the string tensions are:
\be
\begin{aligned}
&T_1^{(a)} = 2 \varphi_1 (2\varphi_1 + h_1 - \varphi_2  - m)~,\\
&T_2^{(a)} = -m^2 - \varphi_1^2 + \varphi_2(3 \varphi_2  - 2 m + 2h_2)~.
\end{aligned}
\ee 
The consistency of the effective field theory description requires that $T_1>0$ and $T_2 >0$, which is an extra condition to be imposed on the Coulomb phase, and gives rise to hard walls, along $T_1=0$ or $T_2 =0$. In this case, the masses of the various particles were computed in \eqref{SU22chamberAmasses}. We see that the instantons $\CI_{(1),1}$ and $\CI_{(1),2}$ have masses:
\be
M(\CI_{(1),a}) = h_1 +2 \varphi_1- \varphi_2 -m~,  \quad a=1, 2~,
\ee
and become massless precisely along the hard wall $T_1=0$, hence the corresponding curves cannot be flopped. The instanton $\CI_{(2),2}$ has mass $\varphi_2 + h_2$ and cannot become massless within this chamber by definition. Finally, the instanton $\CI_{(2), 1}$ has mass:
\be
M(\CI_{(2), 1}) = h_2+ \varphi_2 - 2m = \xi_6~.
\ee
We see that, while $T_1$ and $T_2$ remain positive throughout this chamber, we hit a wall at $\xi_6=0$, where:
\be
\varphi_2 = 2 m - h_2~.
\ee
At this locus, the $SU(2)\times SU(2)$ description is no longer valid and the prepotential needs to be modified to account for the flop transition. Exploiting our map, we see that we are flopping curve $\CC_6^a$ in figure \ref{fig:IIA beetle res a curves}. This maps to resolution (h), which has an $SU(3)$ description but no $SU(2)\times SU(2)$ interpretation. The particle that arise on the other side of the wall is an $SU(3)$ hypermultiplet with real mass $- \t\varphi_2 + \t m_2>0$.

Similarly, in the $SU(3), N_f=2$ description of resolution (b), the instantons $\CI_1$ and $\CI_3$ are associated to non-traversable walls, while the instanton $\CI_2$ is associated to a traversable wall, corresponding to flopping the curve $\CC_9^b$. This is only apparent in the S-dual $SU(2)\times SU(2)$ description, where the two types of walls are associated to the W-bosons and to the hypermultiplet $\CH_{--}$, respectively.

In summary, we see that distinct inequivalent gauge theory phases are interconnected in a non-trivial way along the Coulomb branch of the deformed beetle SCFT. 

%%%%%%%%%
 \begin{figure}[t]
\begin{center}
\includegraphics[width=8.5cm]{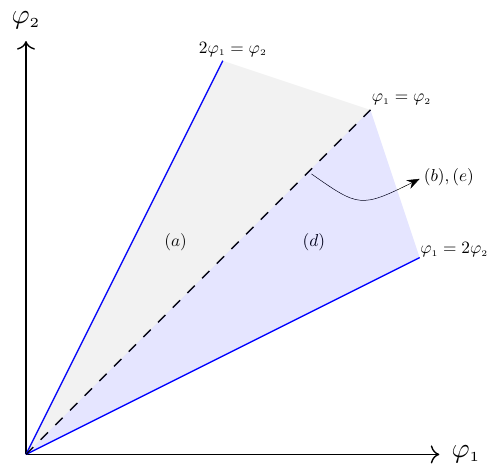}
\caption{The Coulomb branch of the rank-two beetle SCFT is shown in the shaded area. It consists of two chambers, corresponding to resolutions (a) and (d). The middle wall corresponds to a degeneration of both resolutions (b) and (e). \label{fig:beetle SCFT CB}}
 \end{center}
 \end{figure} 
 %%%%%%

\subsection{Probing the Coulomb branch of the 5d SCFT}
So far in this section, we considered $\mu$ and $\nu$ generic. Then, we often have a low-energy gauge-theory interpretation at energies much lower than the mass scale set by $\mu$. In the rest of this section, we would like to explore the opposite limit, $\mu \rightarrow 0$ with $\nu$ finite. This corresponds to a ``strong coupling limit'' where the weakly-coupled gauge-theory approximation breaks down entirely. Nonetheless, it is instructive to use the gauge-theory language in order to gain some intuition about the SCFT Coulomb branch, $\CM_C$, itself.

In rank-one cases, the $\mu=0$ limit is somewhat uninteresting: the Coulomb branch has the form $\CM_C\cong \R_+$---in the gauge-theory language, it is spanned by $\varphi >0$, with $\varphi=0$ the 5d fixed point.

 Let us then consider the beetle geometry, which corresponds to a rank-two SCFT. Setting $\mu=0$, we are left with two parameters:
 \be
 \nu_1=-\varphi_1~,\qquad \nu_2=-\varphi_2~,
 \ee
  which can be thought of as living in the Cartan of either $SU(2) \times SU(2)$ or $SU(3)$. Indeed, in the limit when all the mass parameters vanish, the ``S-duality map'' \eqref{Sdual beetle map} trivializes to $\varphi_1= \t\varphi_1$, $\varphi_2= \t \varphi_2$.

Moreover, out of the 24 K\"ahler chambers of Figure~\ref{fig:all beetles}, only 4 survive in the massless limit. They are:
\bea
&\text{chamber (a)}&& \; : \quad -\varphi_1+ \varphi_2 \geq 0~, \qquad 2 \varphi_1 - \varphi_2 \geq 0~, \cr
&\text{chamber (b)}&& \; : \quad \varphi_1= \varphi_2 \geq 0~, \cr
&\text{chamber (d)}&&\; : \quad \varphi_1- \varphi_2 \geq 0~, \qquad - \varphi_1 +2 \varphi_2 \geq 0~,\\
&\text{chamber (e)}&& \; : \quad \varphi_1= \varphi_2 \geq 0~,
\eea
as depicted in Figure~\ref{fig:beetle SCFT CB}. Not coincidentally, these are the four resolutions that have both an $SU(2)\times SU(2)$ and an $SU(3)$ $N_f=2$ field-theory interpretation. The prepotential in the $\mu=0$ limit reads:
\be
\CF_a= {4\ov 3} \varphi_1^3 -\varphi_1^2 \varphi_2 + \varphi_2^3~, \qquad
\CF_d= {4\ov 3} \varphi_1^3 -\varphi_1^2 \varphi_2 + \varphi_2^3~, \qquad
\CF_b=\CF_e=  {4\ov 3} \varphi_1^3~,
\ee
in the respective chambers. Note that the K\"ahler chambers (b) and (e) collapse to the line $\varphi_1= \varphi_2$, at the interface between chamber (a) and chamber (d). That wall corresponds to a simultaneous flop of the curves $\CC_3^a$ and $\CC_7^a$ in Figure~\ref{fig:IIA beetle res a curves}. In the gauge-theory language, the K\"ahler walls have the following interpretations:

\paragraph{$SU(2)\times SU(2)$ interpretation.} The outer wall of chamber (a), at $2\varphi_1-\varphi_2=0$ corresponds to an instanton particle for the first $SU(2)$ factor, $\CI_{(1)}$, going massless. At the same time, a magnetic string for that gauge group becomes tensionless at the wall, since:
\be
T_1=\d_{\varphi_1} \CF_a = 2\varphi_1(2\varphi_1-\varphi_2)~.
\ee
 Similarly, at the outer wall of chamber (d), at $-\varphi_1+2\varphi_2=0$, the instanton particle $\CI_{(2)}$ and the magnetic string of the second $SU(2)$ gauge group become massless. On the other hand, the middle wall at $\varphi_1= \varphi_2$ corresponds to the bifundamental hypermultiplet of the $SU(2) \times SU(2)$ quiver going massless. (All magnetic strings are tensionfull there.)

Note that, naively, the $SU(2) \times SU(2)$ Coulomb branch spans the whole quadrant $\varphi_1 \geq 0$, $\varphi_2 \geq 0$. However, as pointed out in \cite{Jefferson:2017ahm}, one should exclude the regions beyond the walls at which some non-perturbative states (instantons or monopole string) become massless.

\paragraph{$SU(3)$ $N_f=2$ interpretation.}  In the $SU(3)$ language, the outer walls are clearly hard walls, at which an $SU(3)$ W-boson goes massless. On the other hand, the middle wall at $\varphi_1=\varphi_2$ simply corresponds to an hypermultiplet mode going massless.

%%%%%%%%%%%%%%%%%%%%%%%%%%%%%%%%%%%%%%%%%%%%%%%%%%%%%%
\section{Non-isolated toric singularities and the 5d $T_N$ theory}\label{sec:nonTO}
In our discussion so far, we restricted ourselves to the study of {\it isolated} toric CY threefold singularities. This corresponds to toric diagrams $\Gamma$ which are strictly convex: no external point $w\in \Gamma$ lies inside an external line. More general toric diagrams correspond to non-isolated singularities---see Figure~\ref{fig:SU2 geoms singular toric} for some examples. 

There are some well-known difficulties with non-isolated toric singularities, which are dual to $(p,q)$-webs with parallel external legs \cite{Aharony:1997bh}; in particular, it is not clear that the low-energy theory at the singularity is really five-dimensional. In the $(p,q)$-web language, such difficulties can be alleviated by introducing 7-branes on which the external five-branes legs can end \cite{DeWolfe:1999hj}. That approach, however, takes us outside of the realm of toric geometry on the M-theory side.

In this section, we limit ourselves to making some general comments about non-isolated toric singularities, including the toric realization of the five-dimensional $T_N$ theory \cite{Benini:2009gi}, from the point of view of the M-theory/type IIA duality. We leave a more systematic study for future work.

\subsection{Rank-one toric singularities and $T_2$ gauging}
It is interesting to consider all the toric CY$_3$ singularities with a single exceptional divisor---that is, the toric diagrams with a single internal point. There are only 16 of them, up to an $SL(2,\Z)$ transformation, corresponding to the 16 two-dimensional toric Fano varieties \cite{cox2011toric}. Out of those 16, only 5 are isolated singularties, corresponding to the smooth del Pezzo surfaces $dP_{n}$, with $n\leq 3$, which we studied in section~\ref{sec:SU2 expls}---see Figure~\ref{fig:SU2 geoms}.

 \begin{figure}[t]
\begin{center}
\subfigure[\small $E_1^{(0,0)}$,  (1)]{
\includegraphics[height=2.4cm]{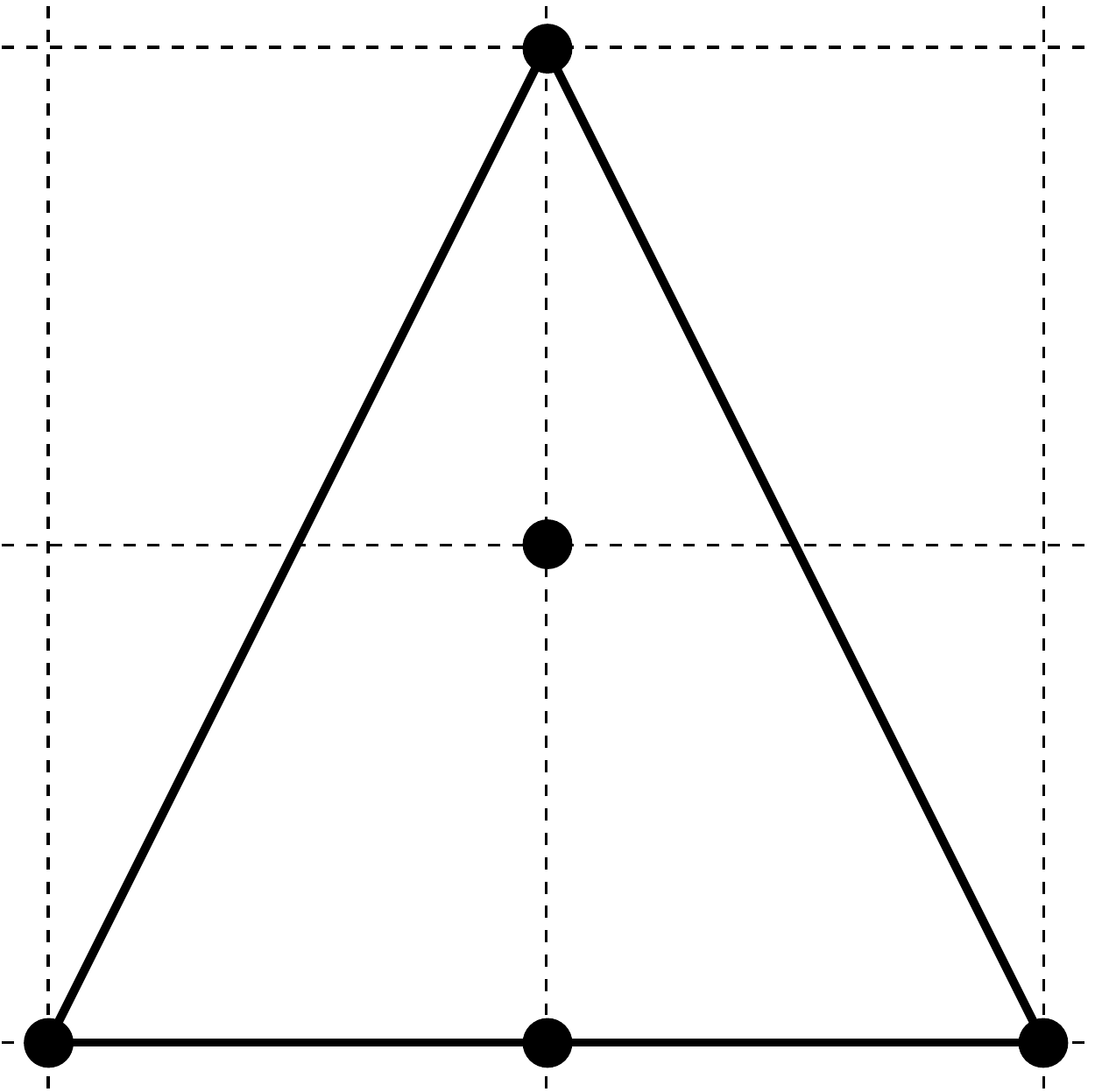}\label{fig:rk1 00}}\;
\subfigure[\small $E_2^{(0,1)}$, (3)]{
\includegraphics[height=2.4cm]{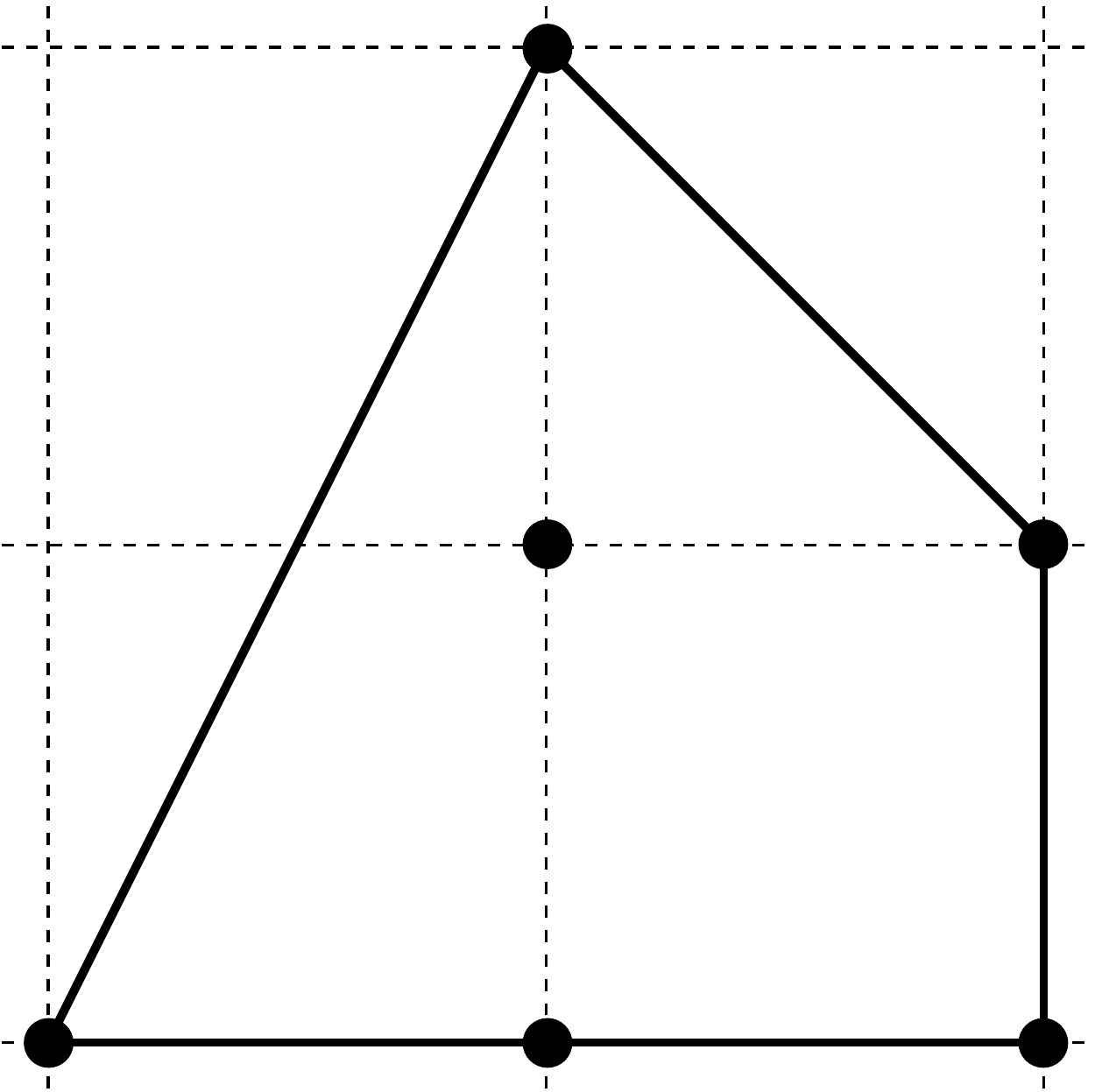}\label{fig:rk1 01}}\;
\subfigure[\small $E_3^{(1,1)}$, (9)]{
\includegraphics[height=2.4cm]{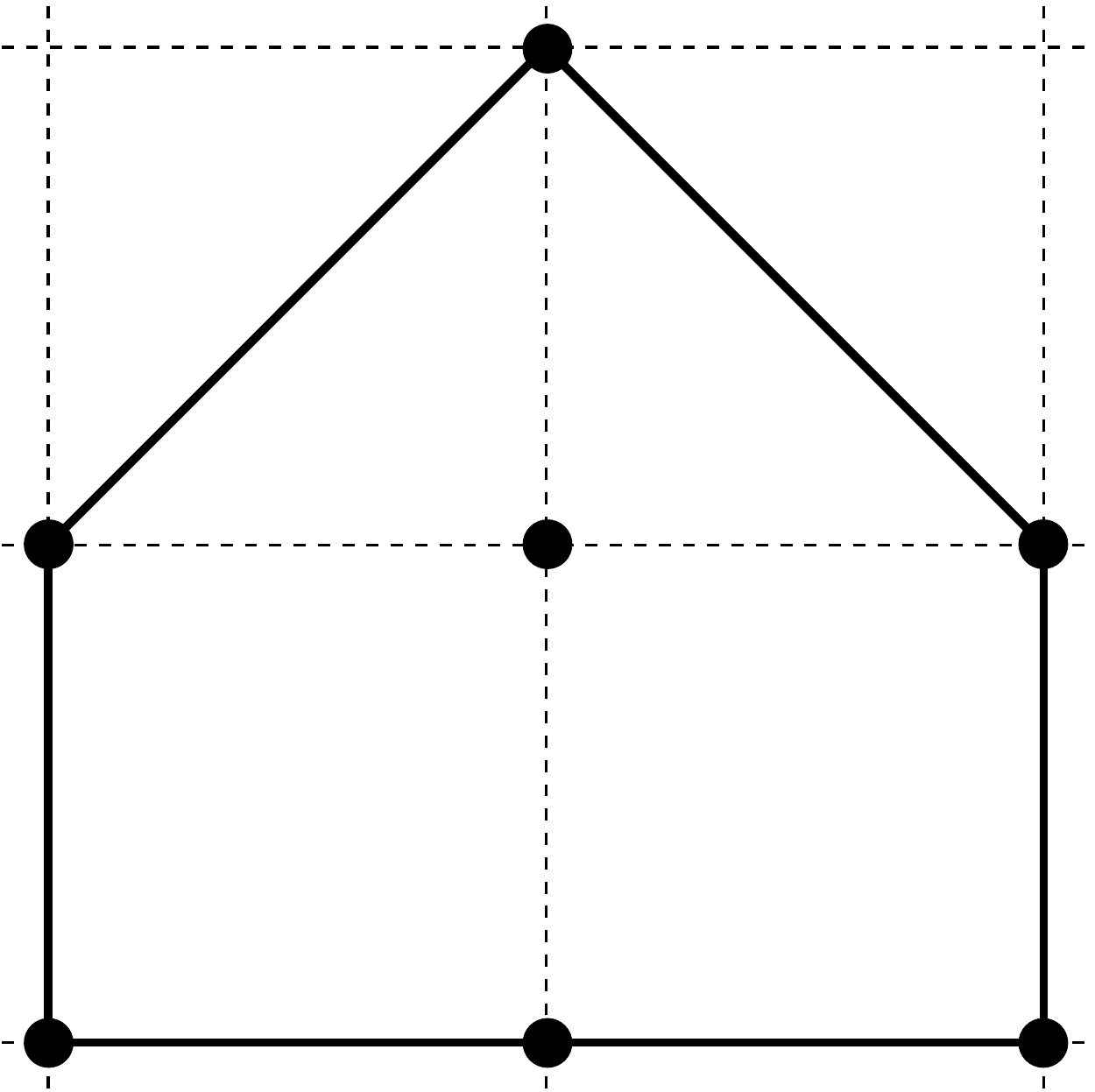}\label{fig:rk1 11}}\;
\subfigure[\small $E_3^{(0,2)}$, (6)]{
\includegraphics[height=2.4cm]{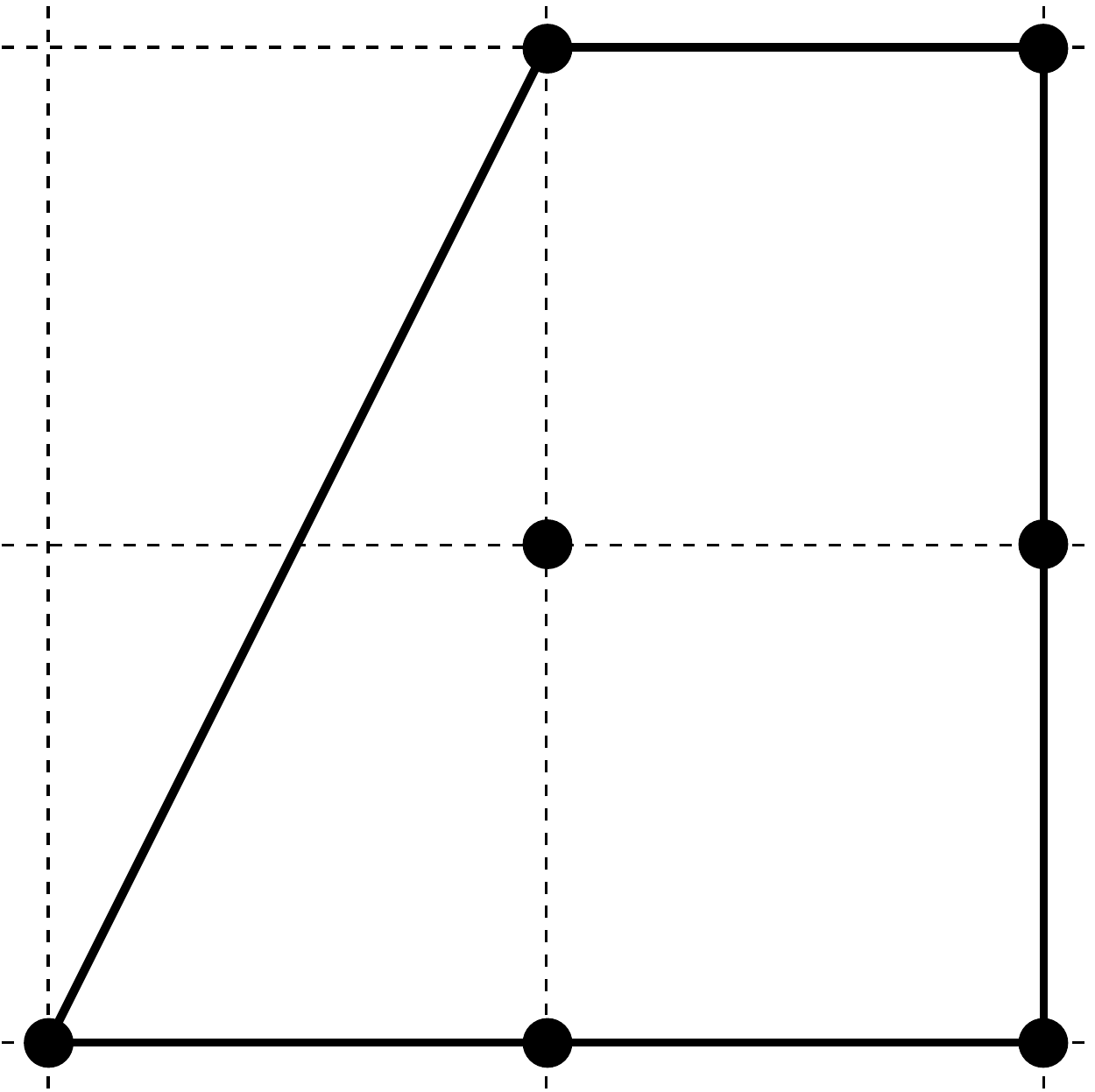}\label{fig:rk1 02}}\;
\subfigure[\small $E_3^{(0,T_2)}$, (3)]{
\includegraphics[height=2.4cm]{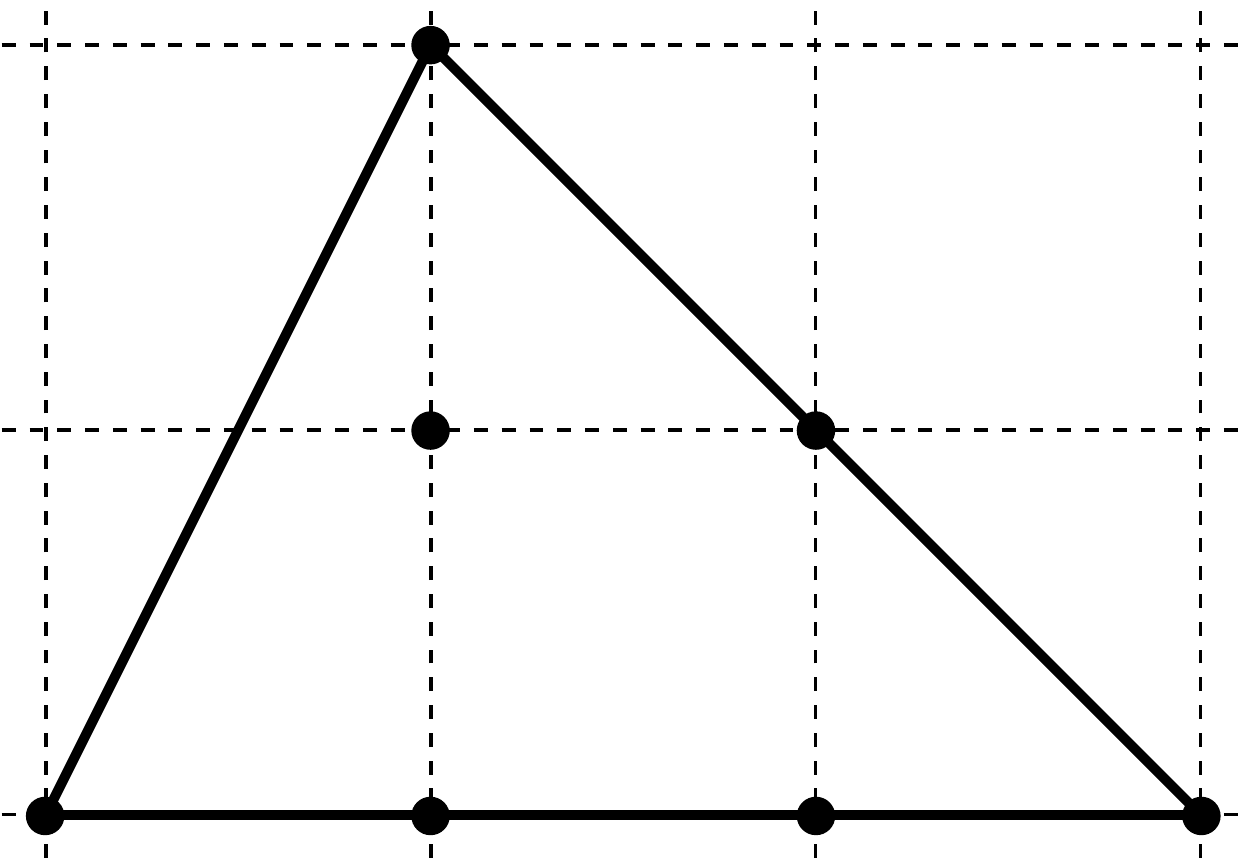}\label{fig:rk1 odd}}\\
\subfigure[\small $E_4^{(1,2)}$, (18)]{
\includegraphics[height=2.4cm]{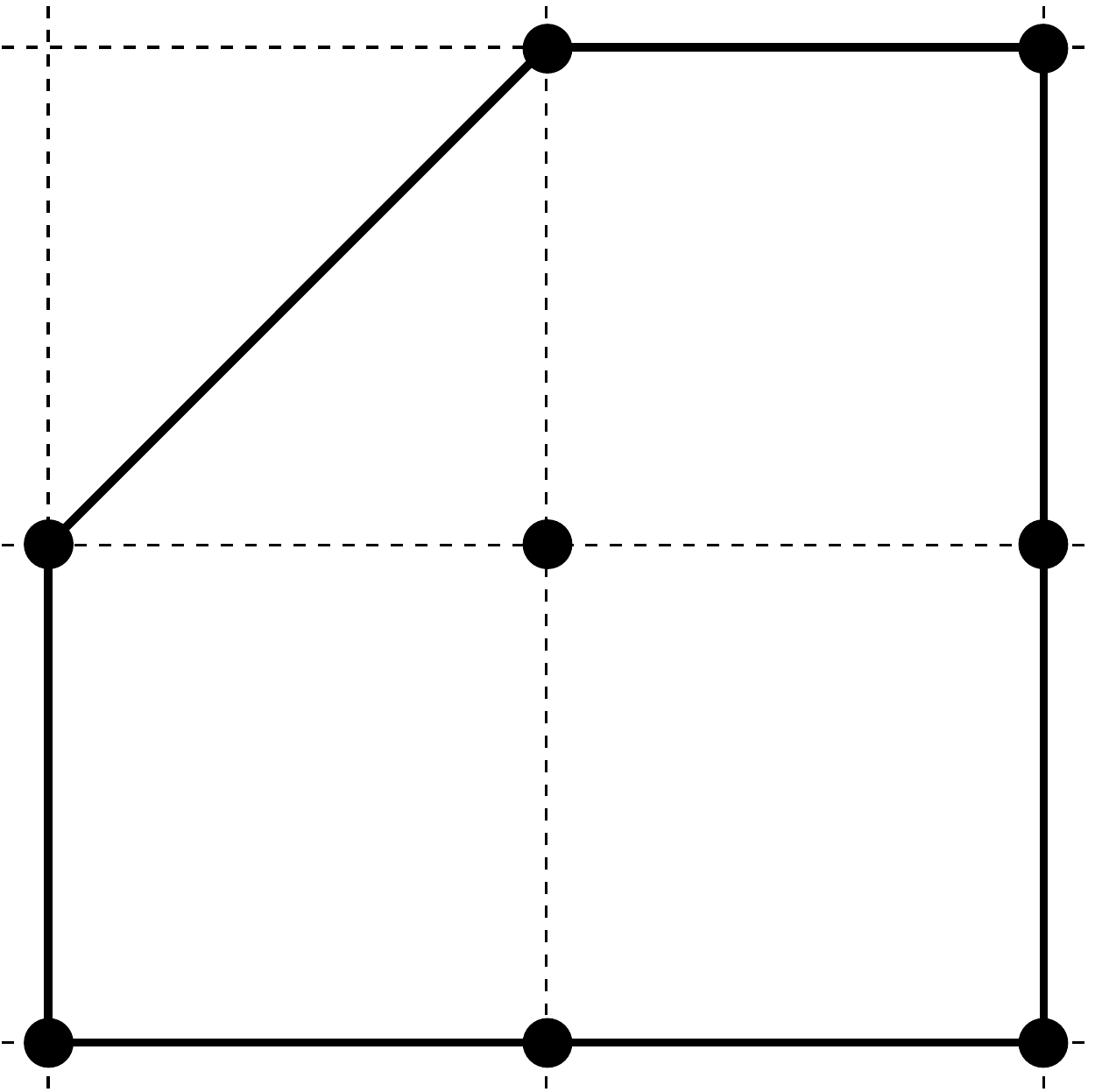}\label{fig:rk1 12}}\;
\subfigure[\small $E_4^{(0,3)}$, (10)]{
\includegraphics[height=3.4cm]{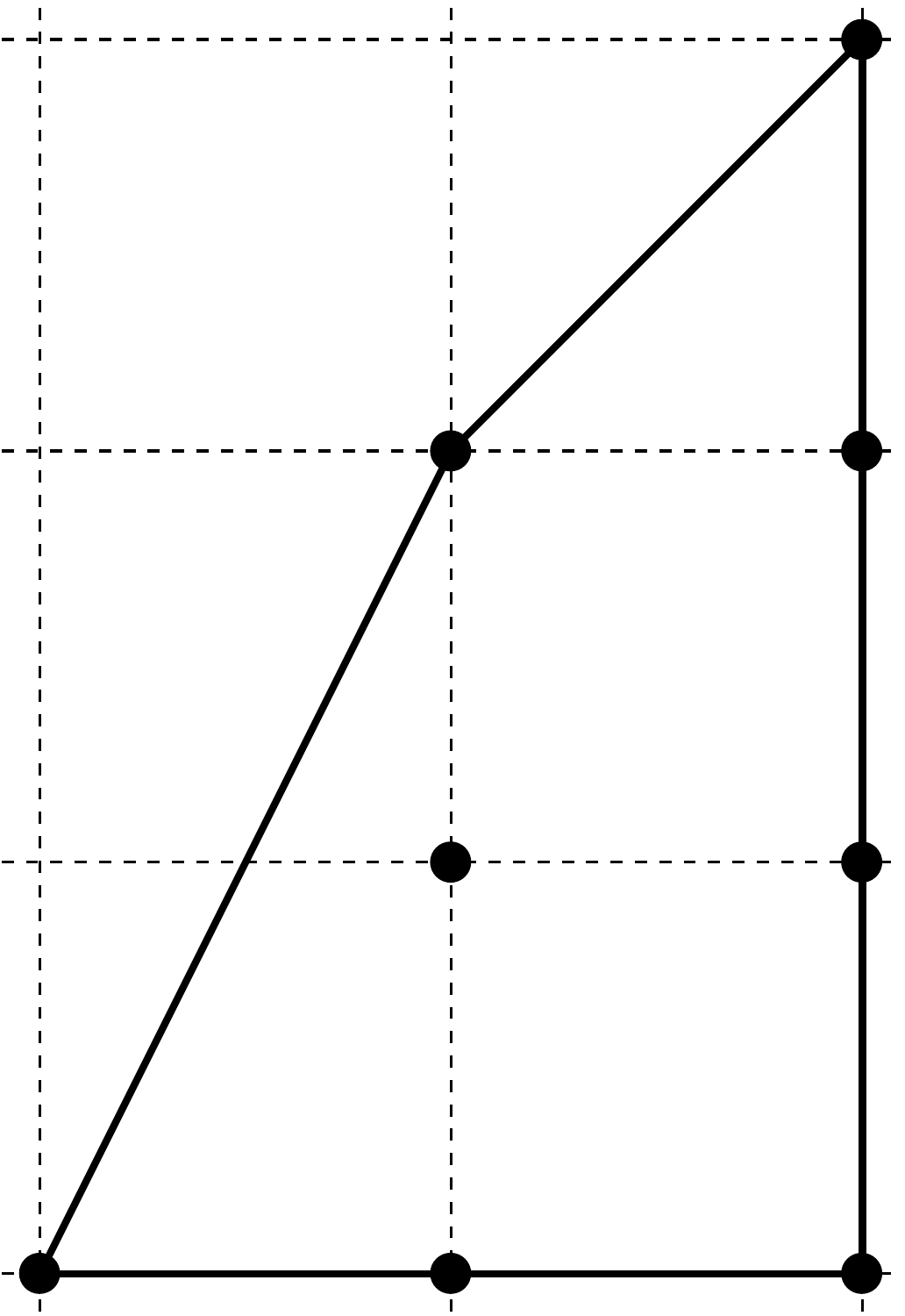}\label{fig:rk1 03}}\;
\subfigure[\small $E_5^{(2,2)}$, (36)]{
\includegraphics[height=2.4cm]{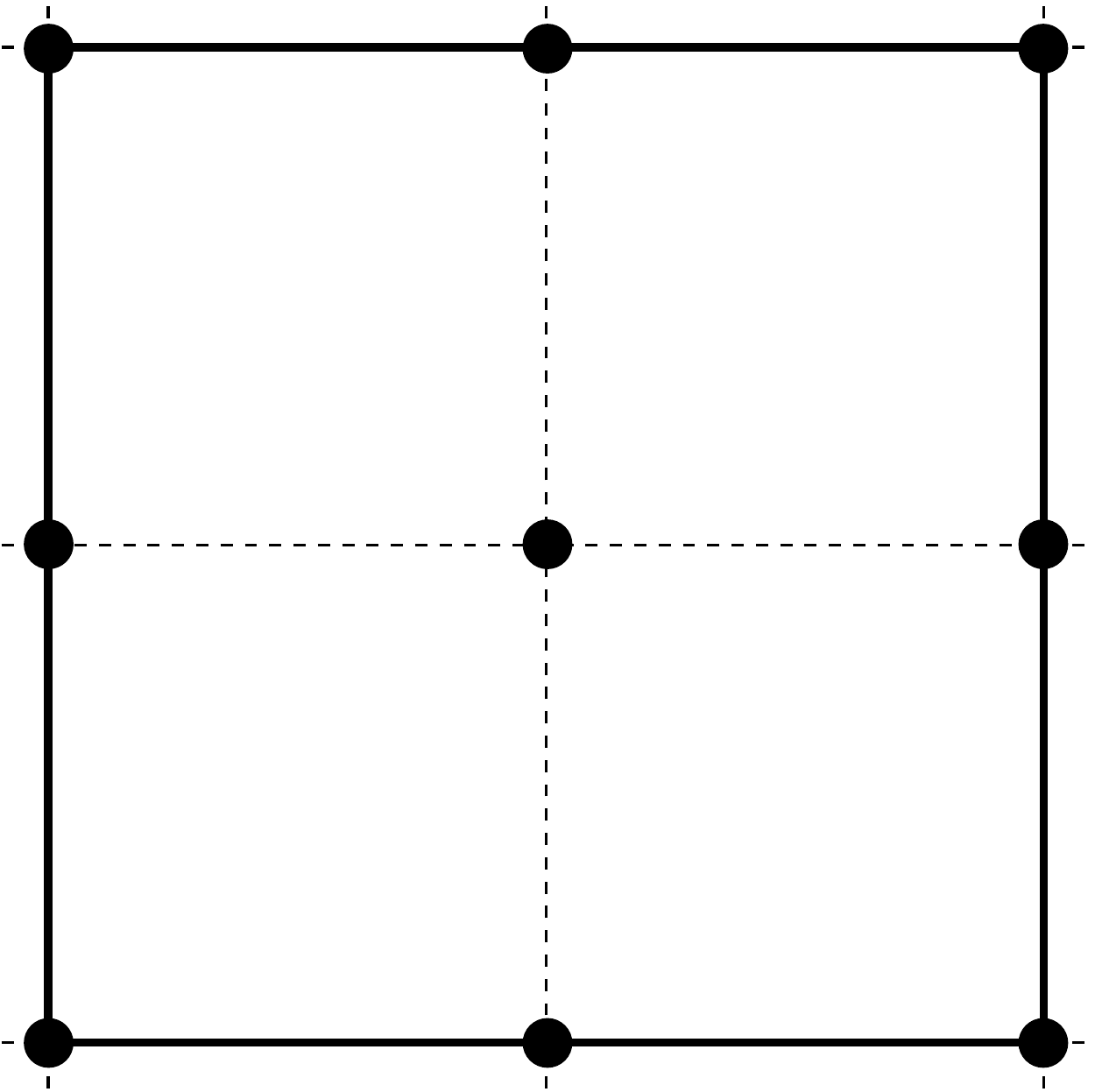}\label{fig:rk1 22}}\;
\subfigure[\small $E_5^{(1,3)}$, (30)]{
\includegraphics[height=3.4cm]{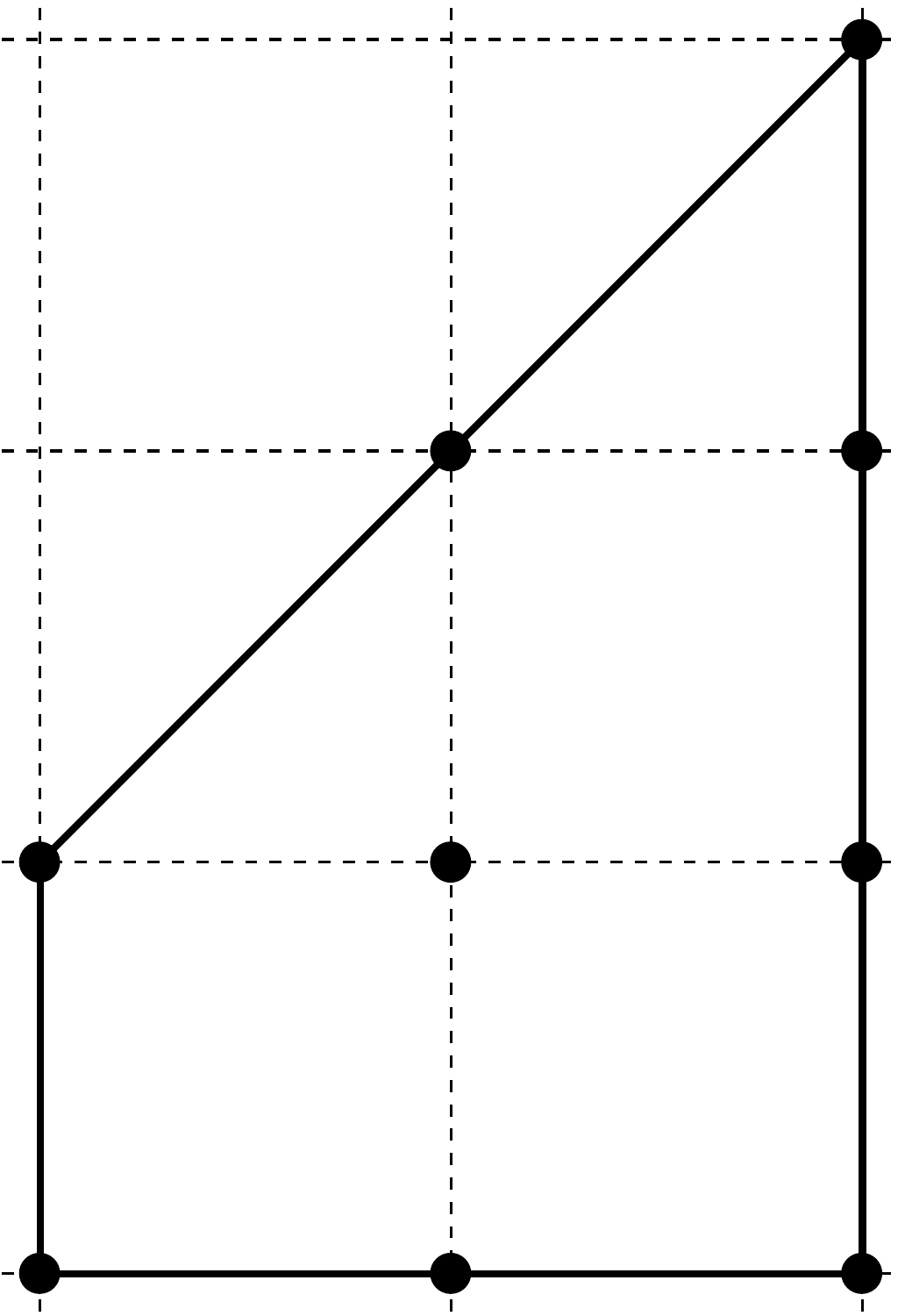}\label{fig:rk1 13}}\;
\subfigure[\small $E_5^{(0,4)}$, (15)]{
\includegraphics[height=4.4cm]{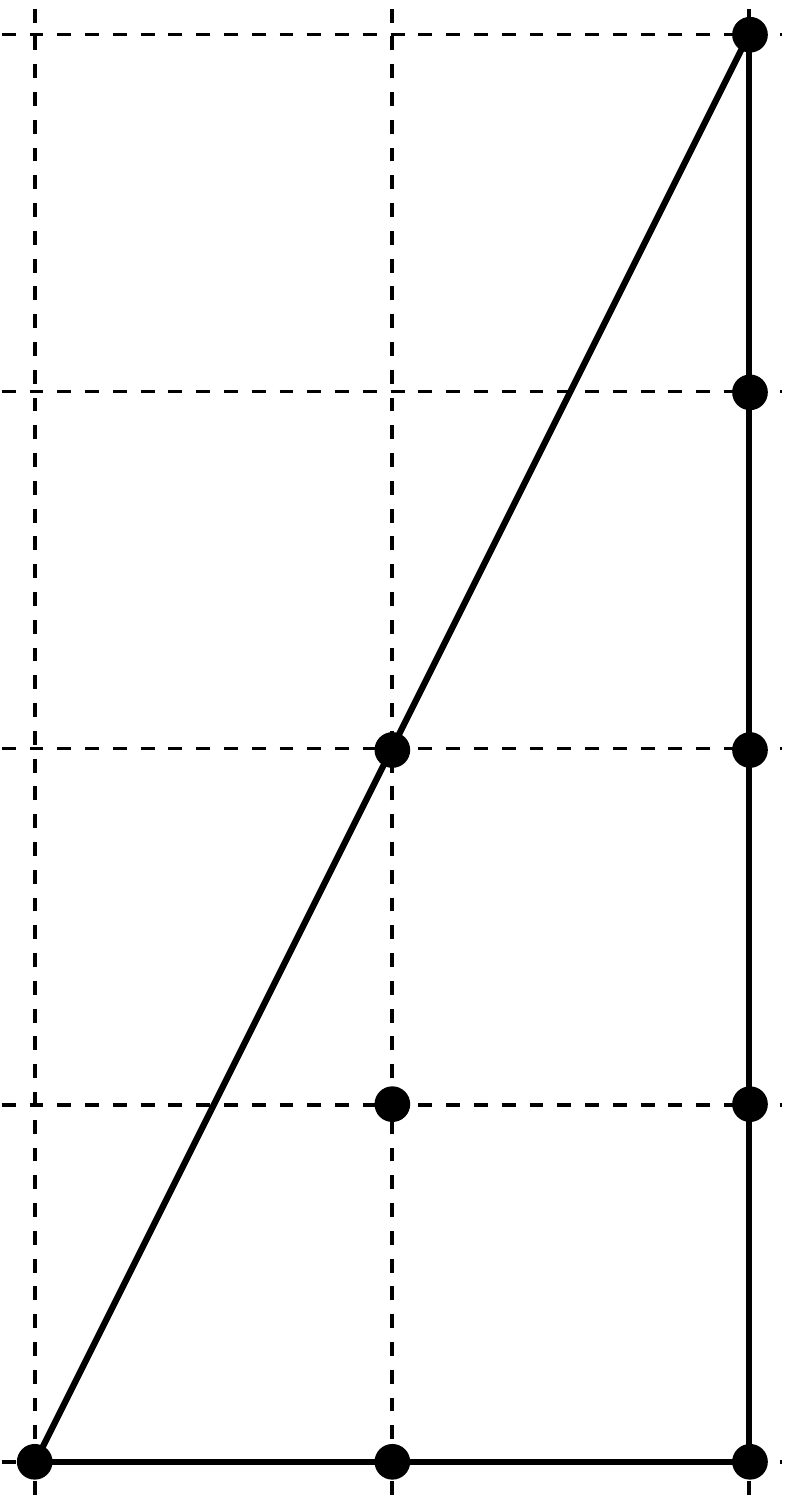}\label{fig:rk1 04}}\qquad\qquad
\caption{10 of the 11 non-isolated ``rank-one'' toric singularities. The number in parenthesis is the number of triangulations with an allowed vertical reduction. \label{fig:SU2 geoms singular toric}}
 \end{center}
 \end{figure} 
 %%%%%%%%%%%
We display the toric diagrams for 10 of the 11 non-isolated singularities in Figure~\ref{fig:SU2 geoms singular toric}, in an $SL(2, \Z)$ frame such that they all admit an obvious vertical reduction (for some of the possible triangulations). Conveniently, 9 of those 10 fit neatly in a family of singularities which we will denote by:
\be\label{En hLR name}
E_n^{(h_L, h_R)}~, \qquad n = h_L + h_R+1 \leq 5~. 
\ee 
The toric diagram contains the following $n+4$ points:
\bea
E_n^{(h_L, h_R)}\; : \qquad \Gamma = \begin{cases}  w_{i}^L=(-1,i)~, \quad & i=0, \cdots, h_L~, \\
 w_j=(0,j)~, \quad & j=0,1,2~, \\
w^R_k =(1, k)~, \quad & k=0, \cdots, h_R~,
\end{cases}
\eea
including the internal point $w_1=(0,1)$.
Given a triangulation of $\Gamma$ with an allowed vertical reduction, we can easily read off the low-energy field theory one obtains in type IIA. We simply have an $SU(2)$ theory with $N_f= n-1$ flavor, hence the name \eqref{En hLR name} for those non-isolated singularities (in a particular S-duality frame).

\begin{figure}[t]
\begin{center}
\subfigure[\small $T_2$.]{
\includegraphics[height=2.4cm]{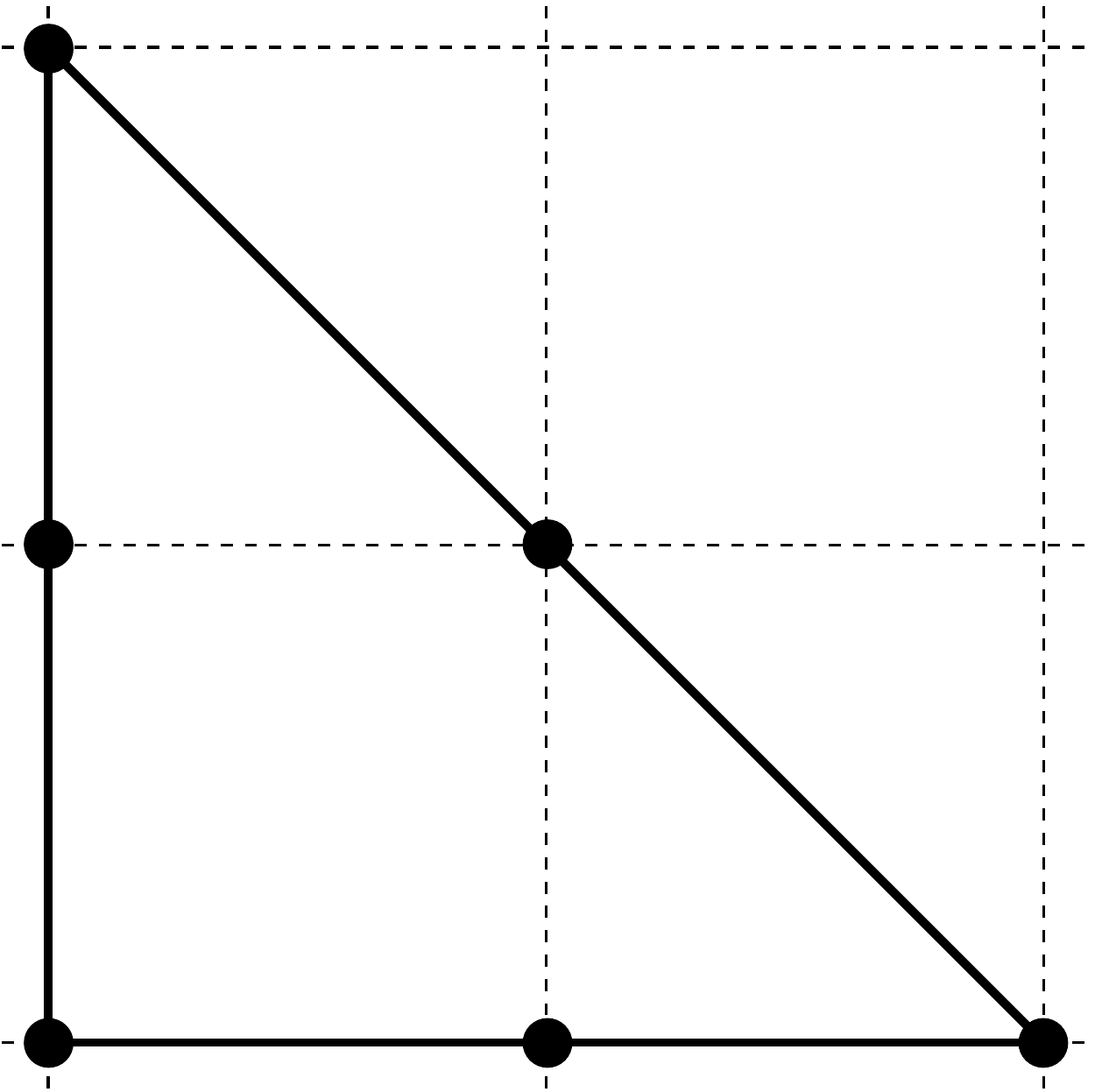}\label{fig:rk0 T2}}\qquad\qquad
\subfigure[\small $T_3 \cong E_6^{(3,T_2)}$.]{
\includegraphics[height=3.4cm]{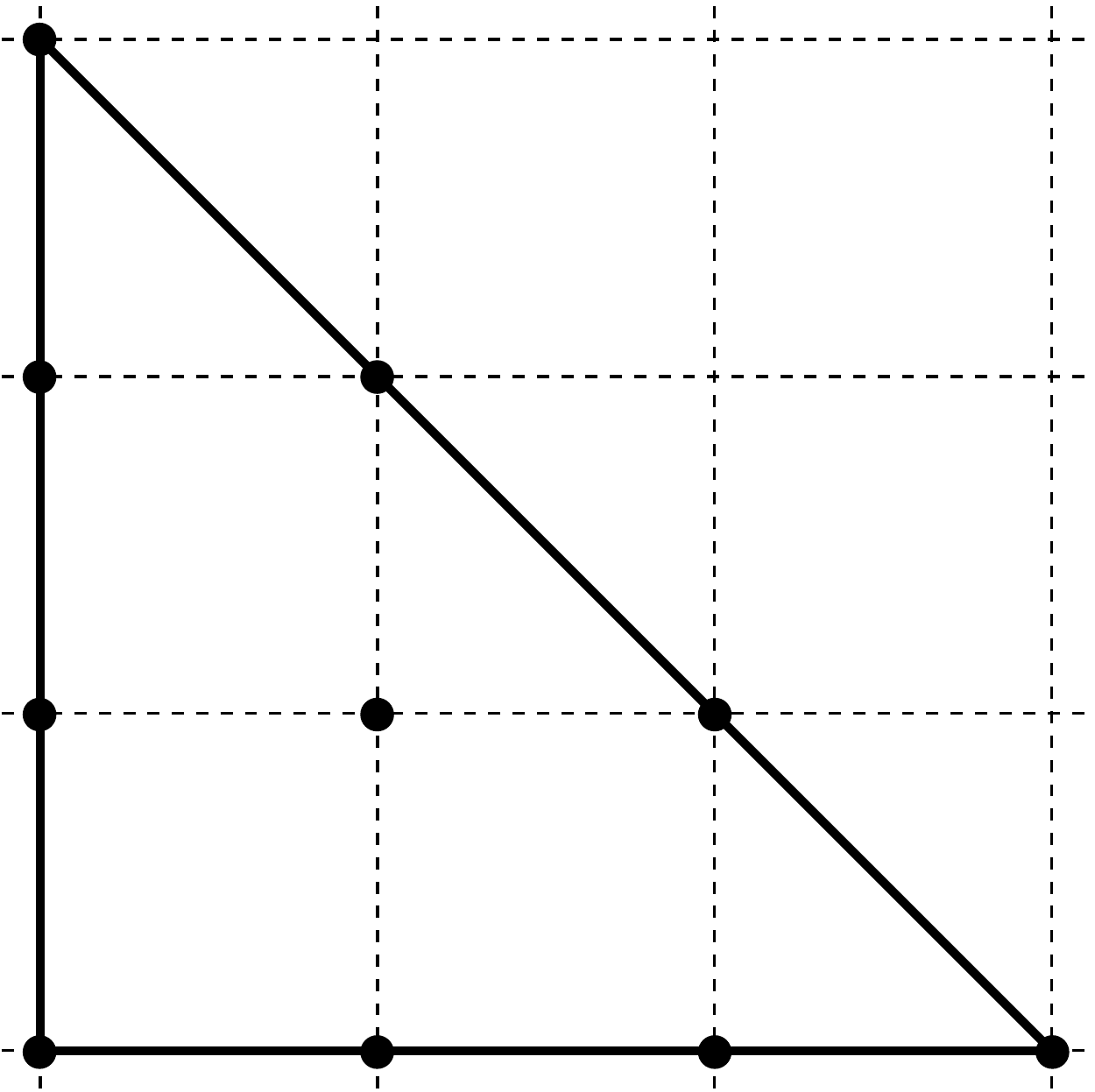}\label{fig:rk1 T3}}
\caption{The $T_2$ ``rank-zero'' toric geometry, and the $T_3$ toric geometry. They have 3 and 30 allowed vertical resolutions, respectively. \label{fig:T2 and T3}}
 \end{center}
 \end{figure}

Amongst the toric diagrams in Figure~\ref{fig:SU2 geoms singular toric}, the singularity shown in Fig.~\ref{fig:rk1 odd} is more peculiar. We claim that it also corresponds to an $SU(2)$ theory with $N_f=2$, but in a subtler way. As we will explain, one of the two flavors appears non-perturbatively in type IIA. We can think of the toric diagram of Fig.~\ref{fig:rk1 odd} as a ``gauging'' of the so-called $T_2$ toric singularity---a.k.a. the $\C^2/(\Z_2\times \Z_2)$ orbifold---shown in Figure~\ref{fig:rk0 T2}. The $T_2$ singularity realizes the 5d $T_2$ theory, the lowest member of the 5d $T_N$ family of 5d SCFTs. Moreover, an appropriate massive deformation of $T_2$ gives rise to free hypermultiplets filling two doublets of $SU(2)$ \cite{Hayashi:2014hfa}.  We will come back to the $T_N$ family at the end of this section. Since the singularity of Fig.~\ref{fig:rk1 odd} realizes an $SU(2)$ gauging of the $T_2$ SCFT ``on the right'' of the toric diagram, we denote it by $E_3^{(0,T_2)}$.

%%%%%%%%%%%%%%%%
%%%%%%%%%%%%%
%%%%%%%%%%%%%%%%
 \begin{figure}[t]
\begin{center}
\subfigure[\small $E_4^{(T_2, 1)} \cong E_4^{(0, 3)}$.]{
\includegraphics[height=2.4cm]{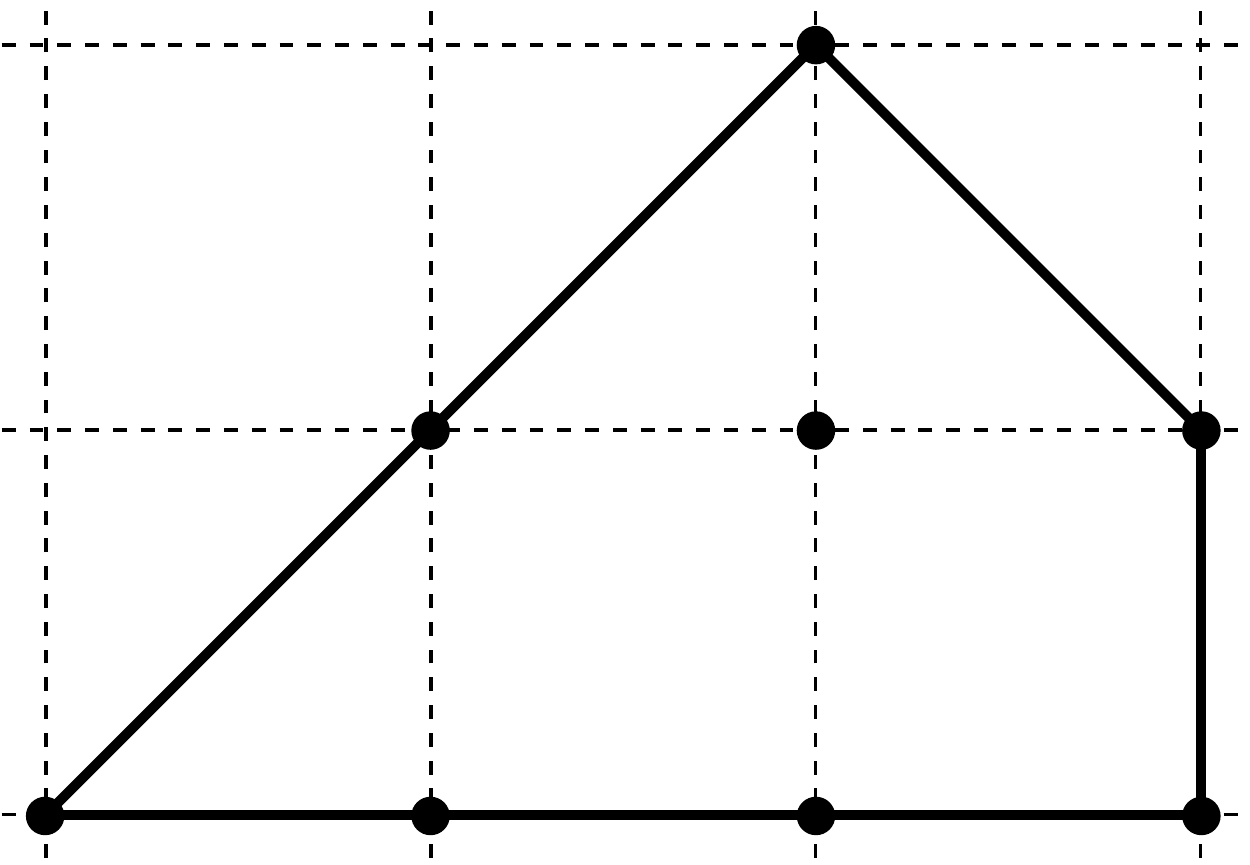}\label{fig:rk1 T2 1}}\qquad
\subfigure[\small $E_5^{(2,T_2)} \cong E_5^{(1, 3)}$.]{
\includegraphics[height=2.4cm]{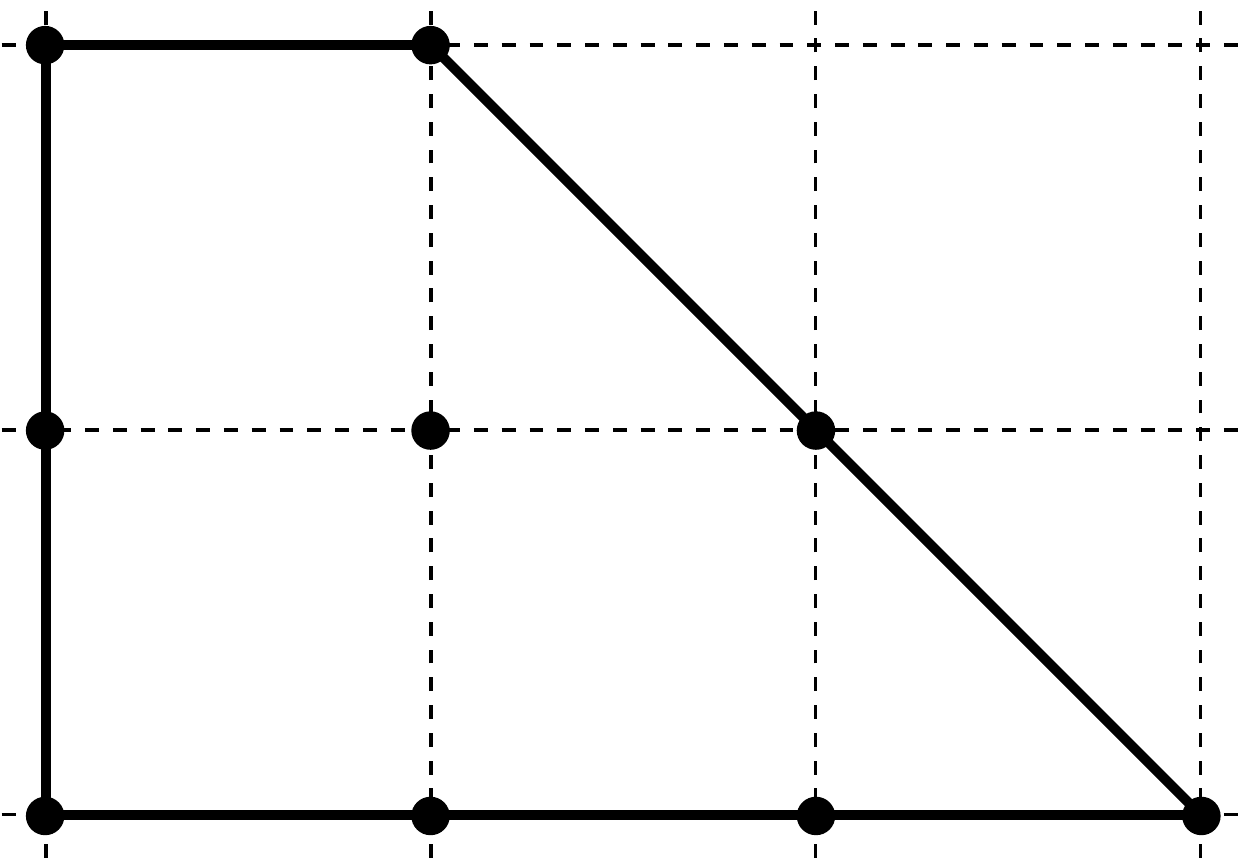}\label{fig:rk1 2 T2}}\qquad
\subfigure[\small $E_5^{(T_2,T_2)} \cong E_5^{(0, 4)}$.]{
\includegraphics[height=2.4cm]{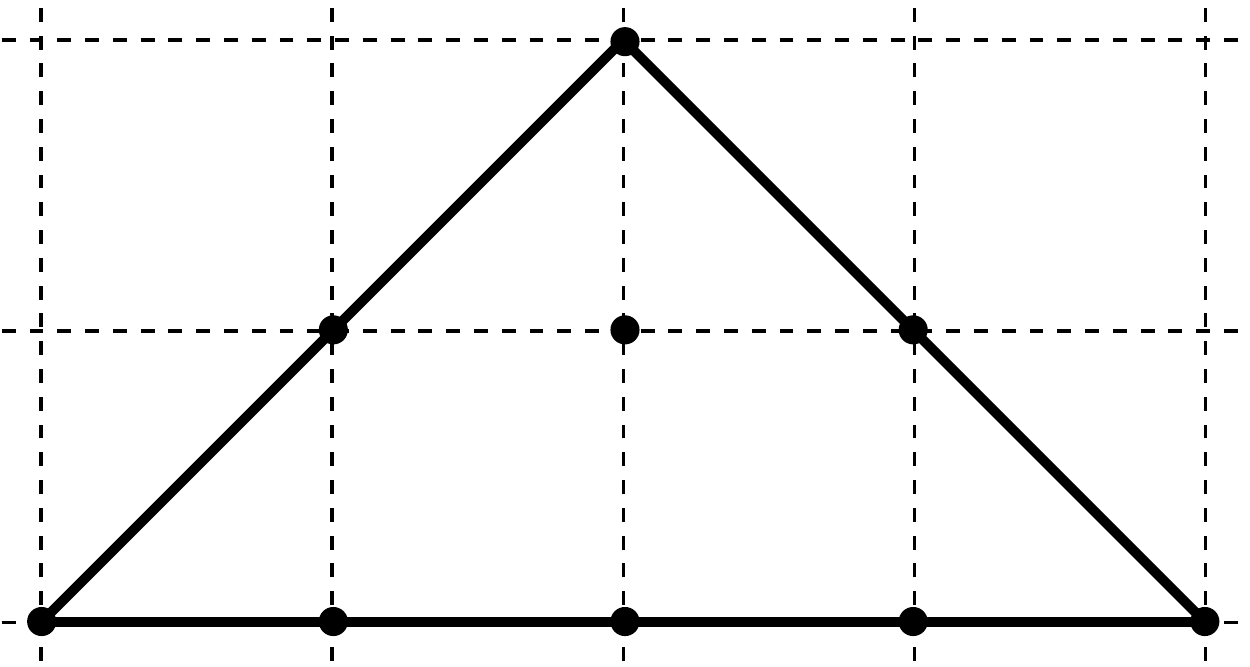}\label{fig:rk1 T2T2}}
\caption{$SL(2, \Z)$-transformed diagrams whose vertical reduction involves ``$T_2$ gauging.'' They have 9, 18 and 9 allowed vertical reductions, respectively. \label{fig:Sduals with T2}}
 \end{center}
 \end{figure} 
%%%%%%%%%%%%%
By including the $T_2$ gauging in our toolbox, we directly find several S-dual descriptions for the geometries in Figure~\ref{fig:SU2 geoms singular toric}, such as shown in Figure~\ref{fig:Sduals with T2}.

Finally, the last of the 16 ``rank-one'' toric diagrams is shown in Figure~\ref{fig:rk1 T3}. It consists of a $T_2$ gauging on the left, while the right-hand-side of the toric diagram corresponds to three additional flavors of $SU(2)$ upon vertical reduction to type IIA. Therefore, this geometry realizes the $SU(2)$ gauge theory with $N_f=5$ flavors. It is also known as the $T_3$ SCFT.

\subsubsection{Allowed vertical reductions and missing field-theory chambers}
It is interesting to count the triangulations of the toric diagrams $E_n^{(h_L, h_R)}$ (where $h_L, h_R$ could be $T_2$) in Figure~\ref{fig:SU2 geoms singular toric}, or of the S-dual diagrams in Figure~\ref{fig:Sduals with T2}. We focus on the number of triangulations {\it with an allowed vertical reduction.} One finds:
\be\label{Nallowed nis}
N_{\text{allowed vertical reductions}} =  n_{h_L} n_{h_R}~, \quad  \text{with:}\quad
{\small
\begin{tabular}{l|cccccc}
$h$ & $0$ &$1$& $2$ &$3$& $4$& $T_2$  \\
 \hline
$n_h$   & $1$  & $3$ & $6$ & $10$ & $15$ &$3$
 \end{tabular}
 }~.
\ee
We displayed these numbers in Figure~\ref{fig:SU2 geoms singular toric}. This should be compared to the number of distinct field-theory chambers of an $SU(2)$ gauge theory with $N_f$ flavors (varying both the Coulomb VEV and the mass parameters).  We have:
\be\label{Nf ft chamber counting}
N_{\text{FT chambers for } \text{SU(2)},  N_f}= 3^{N_f}~.
\ee
For the isolated toric singuarities, we found a one-to-one match between the number of allowed vertical reductions and the field-theory chambers. This is not true for the non-isolated singularities, since \eqref{Nallowed nis} is generally smaller than \eqref{Nf ft chamber counting} (for $N_f= h_L + h_R$). For instance, the geometry $E_5^{(0,4)}$ of Fig.~\ref{fig:rk1 04}  has 15 allowed vertical reductions, which only span 15 out of the 81 field theory chambers of $SU(2)$ with $N_f=4$.

 The missing chambers are related to the presence of non-isolated singularities: one can check in examples that exploring those additional field theory chambers correspond to flopping curves that cannot be flopped in those particular toric geometries. In the IIA setup, that would correspond to D6-branes crossings which are disallowed because some segments of the profiles $\chi_s(r_0)$ would become negative.
 
Those pathologies are symptoms of the fact that non-isolated singularities, by themselves, do not define five-dimensional SCFTs. In the present case, of course, every known rank-one SCFT can be realized at an isolated singularity, the complex cone over a smooth del Pezzo surface, $dP_{N_f+1}$, which happens to be non-toric for $N_f>2$. The non-isolated toric singularities considered here are complex cones over singular ``pseudo-del Pezzo'' toric varieties obtained by blowing up $\mathbb{P}^2$ at non-generic points \cite{Leung:1997tw}. They are still interesting, in particular because one can easily engineer 5d $\CN=1$ gauge theories by resolving them, but one has to keep the above caveats in mind.

\subsubsection{The $E_3^{(0,T_2)}$ geometry and the $SU(2)$ $N_f=2$ gauge theory}\label{subsec:E3T2}
\begin{figure}[t]
\begin{center}
\subfigure[\small $\CF_{(a)}$]{
\includegraphics[height=1.9cm]{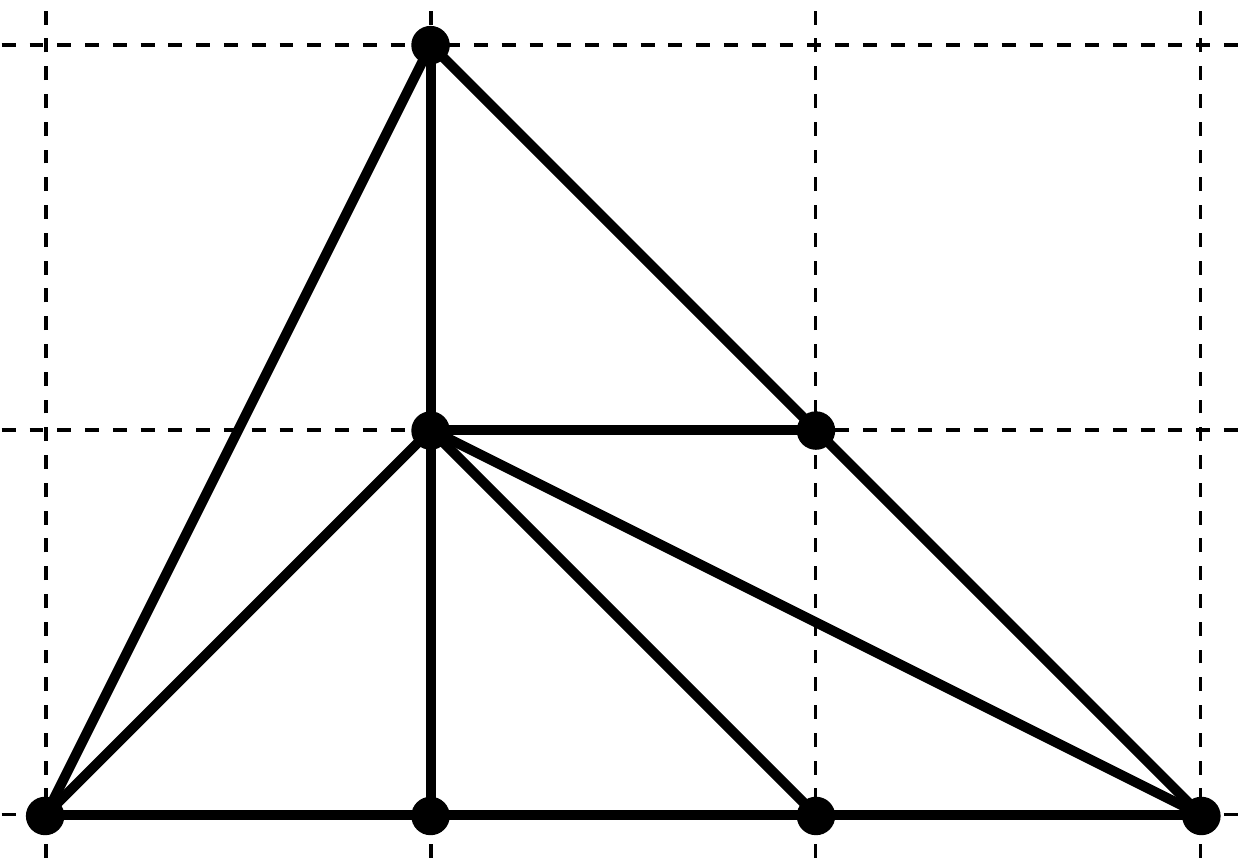}\label{fig:T2SU2 a}}
\subfigure[\small $\CF_{(c)}$]{
\includegraphics[height=1.9cm]{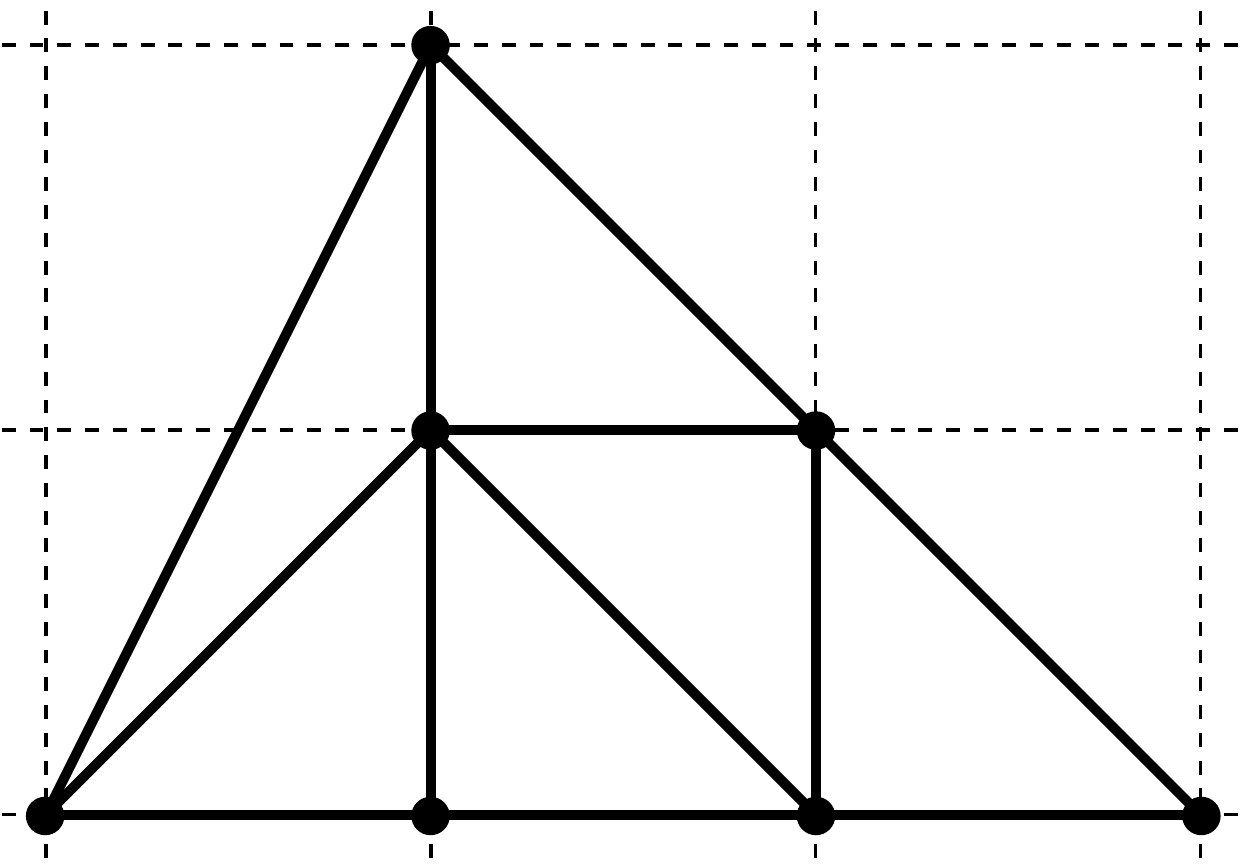}\label{fig:T2SU2 c}}
\subfigure[\small  $\CF_{(g)}$]{
\includegraphics[height=1.9cm]{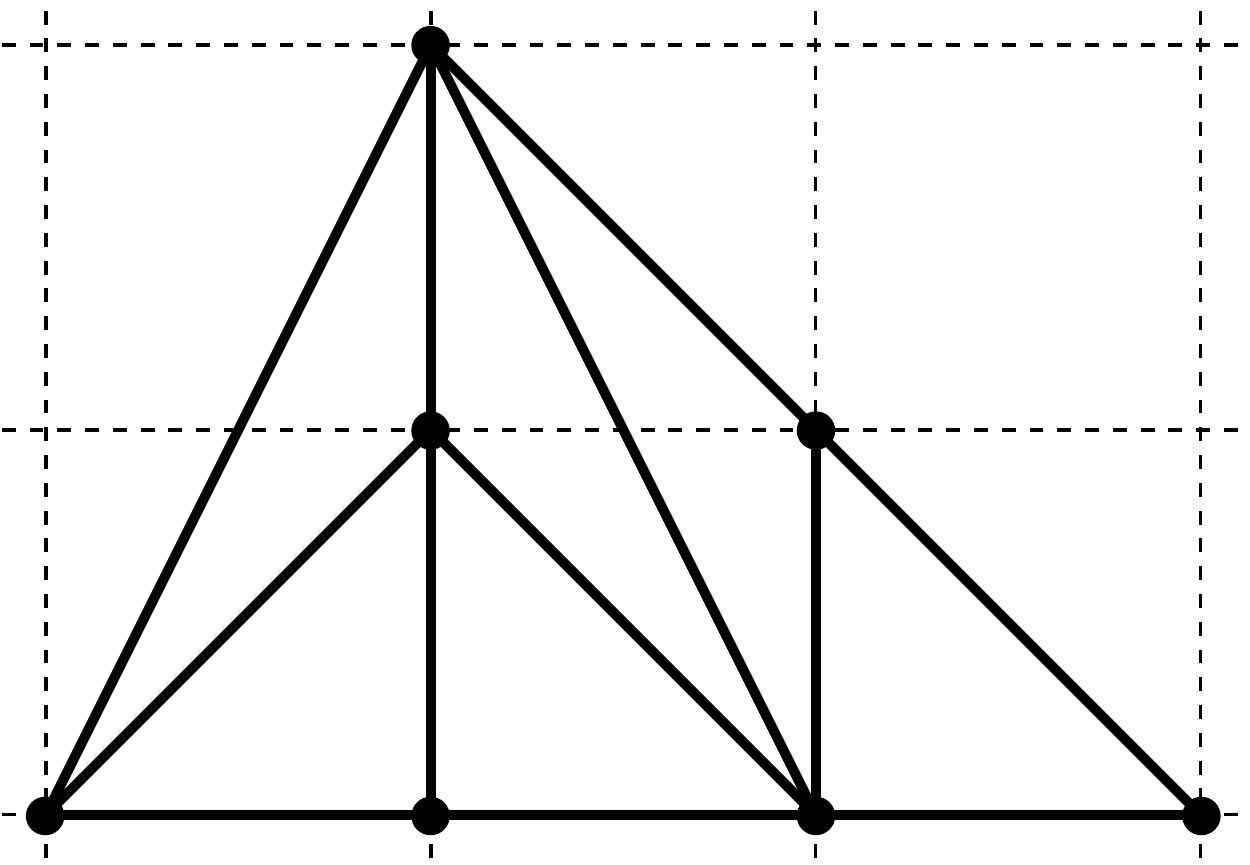}\label{fig:T2SU2 g}}
\subfigure[\small  $\CF_{(i)}$]{
\includegraphics[height=1.9cm]{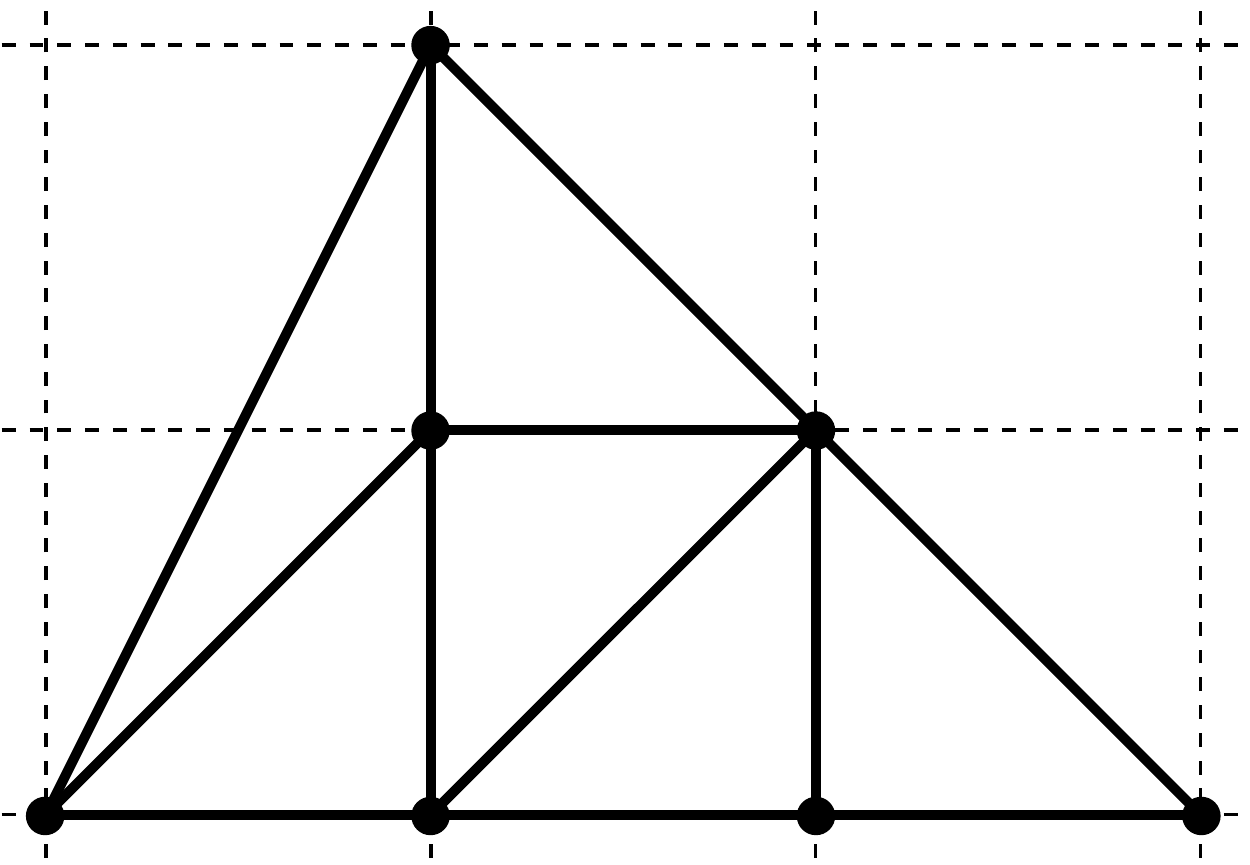}\label{fig:T2SU2 i}}
\subfigure[\small ]{
\includegraphics[height=1.9cm]{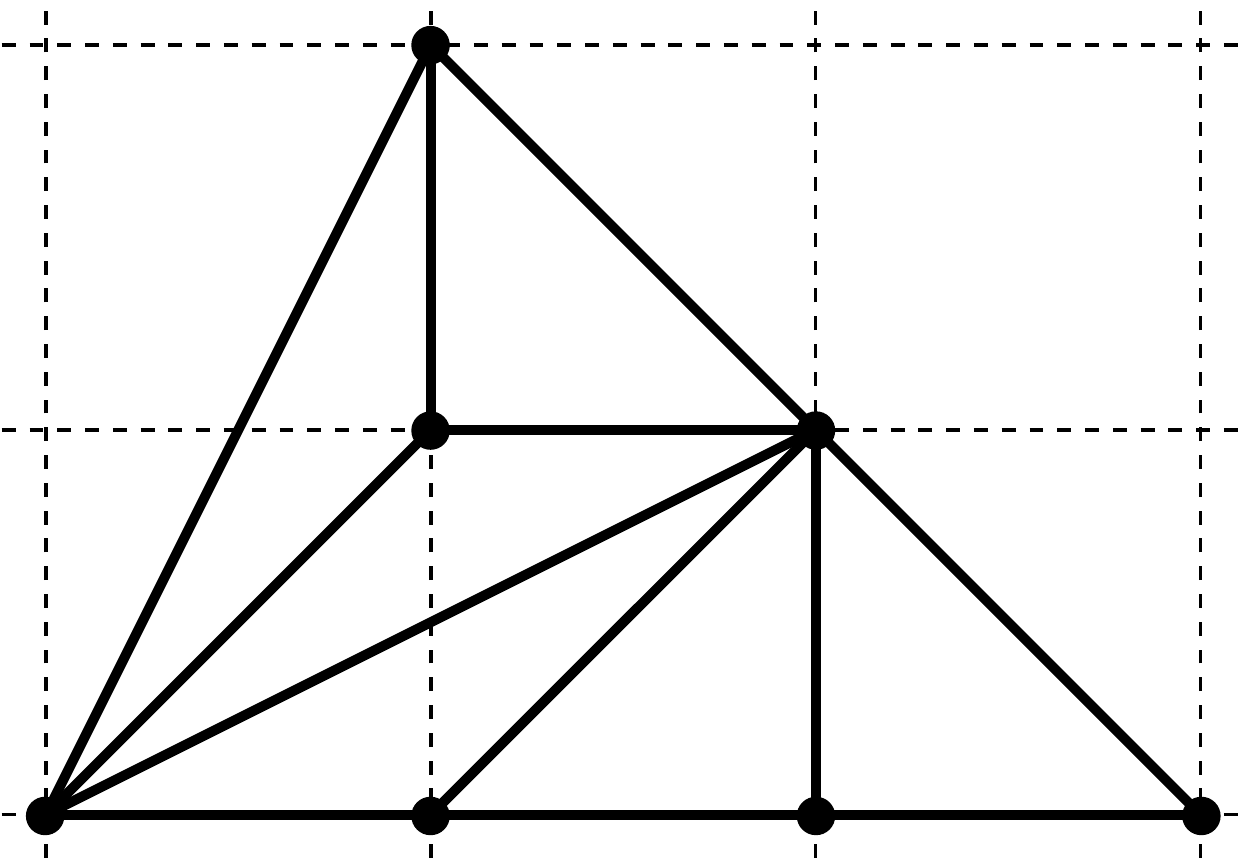}\label{fig:T2SU2 r}}
\caption{The 5 triangulations of the $E_3^{(0,T_2)}$ toric diagram. For the first 4 of them, we indicated the $SU(2)$ $N_f=2$ field-theory prepotentials they correspond to, as we explain in the text. \label{fig:triangulations T2 SU2}}
 \end{center}
 \end{figure}

We would like to obtain a better understanding of the ``$T_2$ gauging'' alluded to above.
For that purpose, it is sufficient to focus on the $E_3^{(0,T_2)}$ singularity of Figure~\ref{fig:rk1 odd}. It has five toric resolutions, as shown in Figure~\ref{fig:triangulations T2 SU2}. The three resolutions of Fig.~\ref{fig:T2SU2 c}, Fig.~\ref{fig:T2SU2 g} and Fig.~\ref{fig:T2SU2 i} obviously have an allowed vertical reduction, which  gives a quiver of the form:
\be\label{SU2SU1 quiver}
SU(2)  \; \rule[2pt]{0.8cm}{.6pt}\;\, SU(1) \qquad \cong\qquad  SU(2)  \; \rule[2pt]{0.8cm}{.6pt}\;\, T_2~.
\ee
The naive ``$SU(1)$'' factor is realized by a single wrapped D6-brane in type IIA. We will show that it can also be understood as two fundamentals of $SU(2)$, one of which arises non-perturbatively. This is denoted by the $T_2$ factor in \eqref{SU2SU1 quiver}.

%%%%%%%%%%%%%%%%
\begin{figure}[t]
\begin{center}
\subfigure[\small Toric divisors.]{
\includegraphics[height=4cm]{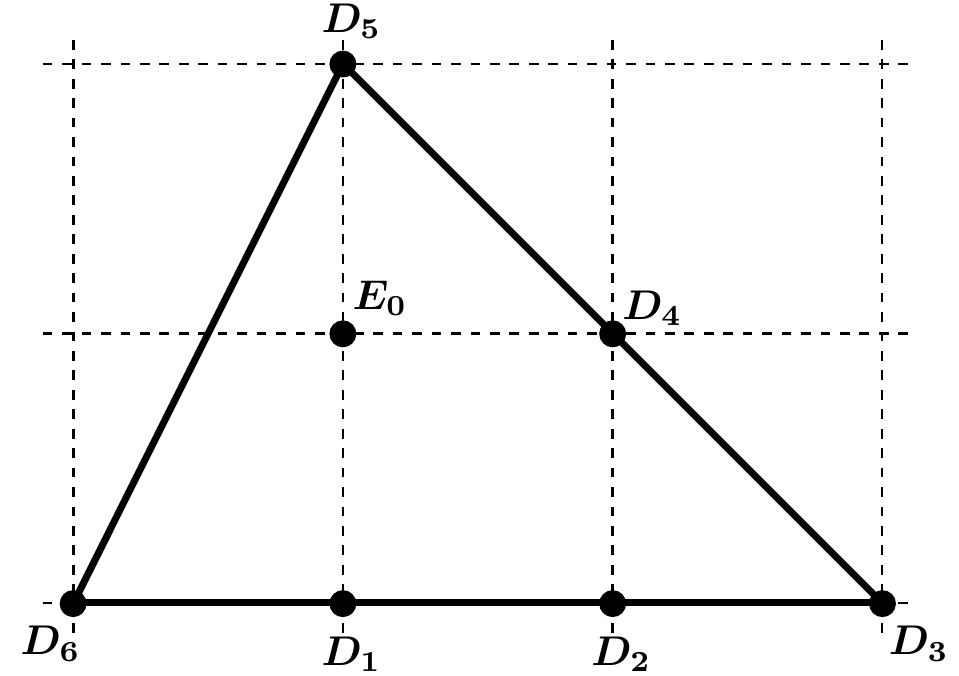}\label{fig:T2SU2 div labels}}\qquad
\subfigure[\small Resolution ``$\CF_{(c)}$.'']{
\includegraphics[height=4cm]{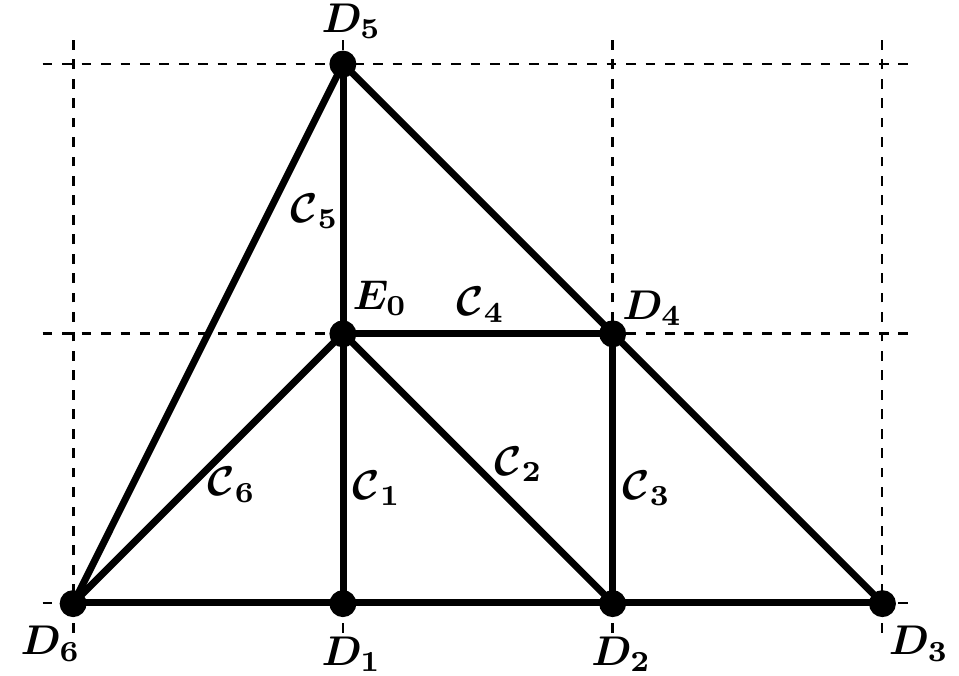}\label{fig:T2SU2 c labels}}
\caption{The $E_3^{(0,T_2)}$ toric geometry, resolution ``(c)'' and its curves. \label{fig:T2 SU2 labs}}
 \end{center}
 \end{figure}

%%%%%%%%%%%%%
\paragraph{Geometric prepotential and gauge theory interpretation.}
Consider the toric divisors of $E_3^{(0,T_2)}$ as indicated in Figure~\ref{fig:T2SU2 div labels}.   
We may take:
\be\label{S def E3T2}
S= \mu_3 D_3 + \mu_5 D_5 + \mu_6 D_6+ \nu \bE_0~.
\ee
The geometric prepotential $\CF=-{1\ov 6} S^3$ for the first four resolutions in Figure~\ref{fig:triangulations T2 SU2} reads:
\bea\label{Facgi SU2T2}
&\CF_{(a)}= -\nu^3 + \half (\mu_3+ 3\mu_5 + 2 \mu_6)\nu^2 +\half (\mu_3^2 -\mu_5^2 - 2 \mu_5\mu_6)\nu~,\cr
&\CF_{(c)}= -{7\ov 6}\nu^3 +\half (3\mu_5 + 2 \mu_6) \nu^2 -\half(\mu_5^2 +2 \mu_5 \mu_6)\nu~,\cr
&\CF_{(g)}=  -{4\ov 3}\nu^3 + (2\mu_5 + \mu_6) \nu^2 -(\mu_5^2 + \mu_5 \mu_6)\nu~,\cr
&\CF_{(i)}= -{4\ov 3}\nu^3 +\half (3\mu_5 + 2 \mu_6) \nu^2 -\half(\mu_5^2 +2 \mu_5 \mu_6)\nu~,
\eea
as one can check by direct computation of the intersection numbers. We claim that these four geometric resolutions corresponds to the field theory chambers denoted by (a), (c), (g) and (i) of the $SU(2)$ $N_f=2$ gauge theory, as studied in section~\ref{subsec:E3 SCFT}. The geometry-to-field-theory map is given by:
\be\label{mu to FT SU2T2}
\mu_3= m_1-m_2~, \qquad
\mu_5= -2m_2~, \qquad
\mu_6= h_0-m_1~, \qquad
\nu= -\varphi-m_1~.
\ee
Indeed, plugging \eqref{mu to FT SU2T2} into \eqref{Facgi SU2T2}, one reproduces the corresponding field-theory prepotentials in \eqref{Fabc FT dP3}-\eqref{Fabc FT dP3 bis}, modulo the constant term.
Interestingly, the resolution of Fig.~\ref{fig:T2SU2 a} also has a gauge-theory interpretation, even though it doesn't have an allowed IIA reduction, strictly speaking. To understand this point better, let us study the resolution of Fig.~\ref{fig:T2SU2 c} and its IIA reduction in more detail.

%%%%%%%%%%%%%%%%
%%%%%%%%%%%%%%%%
 \begin{figure}[t]
\begin{center}
\subfigure[\small $\chi_1(r_0)$]{
\includegraphics[height=5cm]{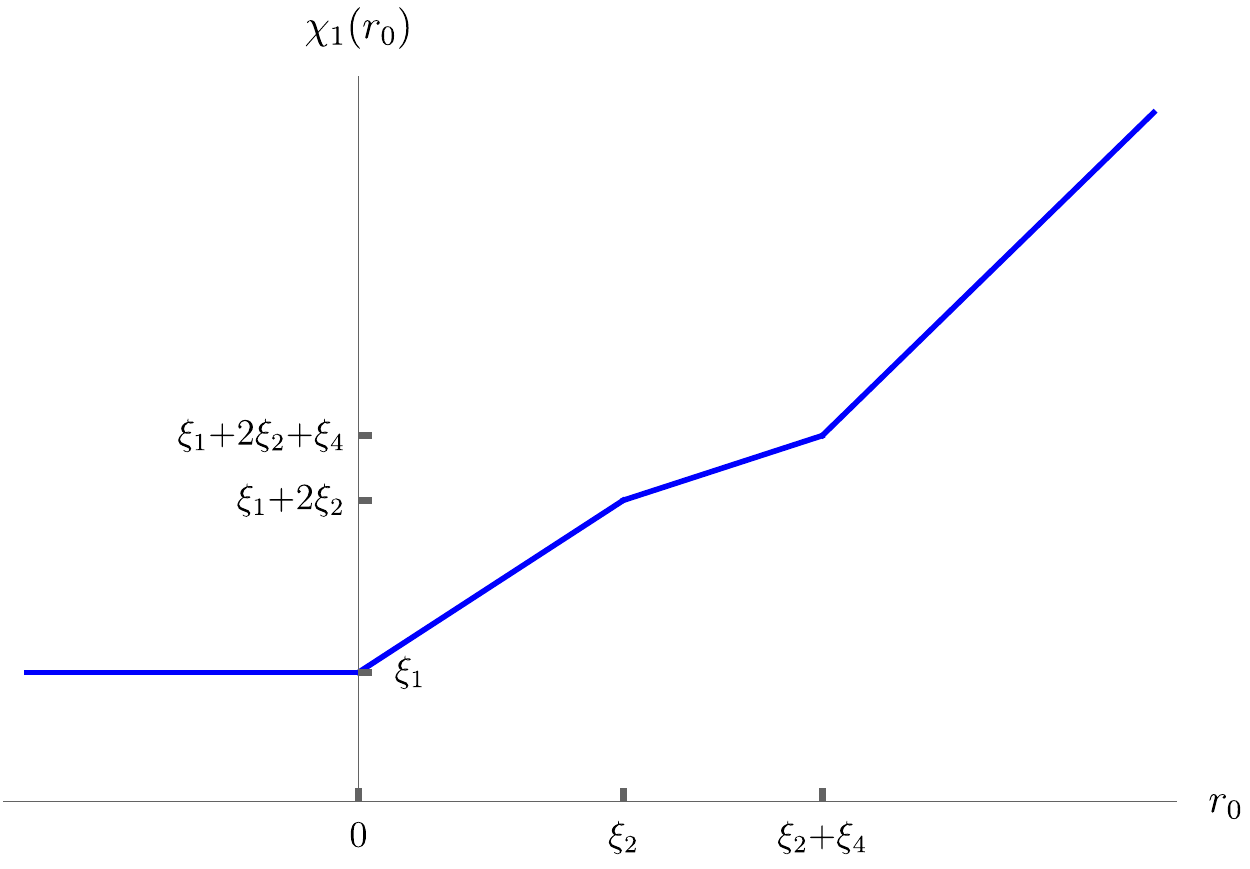}\label{fig:T2SU2 IIA chi1}}\;\;
\subfigure[\small $\chi_2(r_0)$]{
\includegraphics[height=5cm]{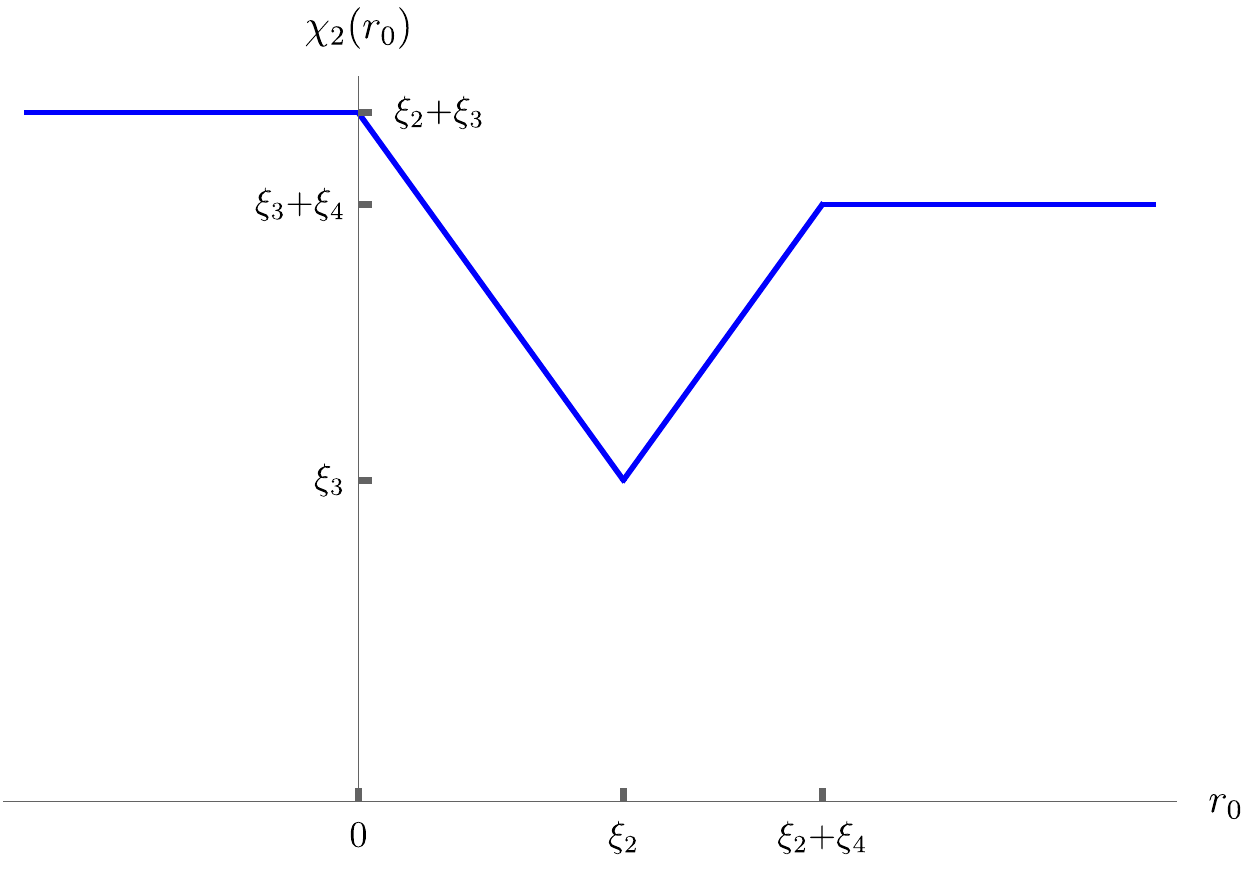}\label{fig:T2SU2 IIA chi2}}
\caption{IIA profile for the resolution ``(c)'' of the $E_3^{(0,T_2)}$ geometry.   \label{fig:T2 SU2 IIA}}
 \end{center}
 \end{figure} 
%%%%%%%%%%%%%
\paragraph{Vertical resolution and field theory chamber (c).}  Consider the resolution of Figure~\ref{fig:T2SU2 c}, with its curves labelled as in Figure~\ref{fig:T2SU2 c labels}. 
We have the relations:
\be
\CC_5\cong \CC_1 +2 \CC_2+ \CC_4~, \qquad \CC_6\cong \CC_2+ \CC_4~.
\ee
The GLSM reads:
\be\label{glsm SU2T2}
\begin{tabular}{l|ccccccc|c}
 & $D_1$ &$D_2$& $D_3$&$D_4$& $D_5$ & $D_6$ & $\bE_0$& \\
 \hline
$\CC_1$   & $-2$ & $1$ & $0$ & $0$ & $0$ & $1$   & $0$ & $\xi_1$\\
$\CC_2$   & $1$  & $-1$ & $0$ & $1$ & $0$ & $0$  & $-1$ & $\xi_2$\\
$\CC_3$   & $0$  & $-1$ & $1$ & $-1$ & $0$ & $0$  & $1$ & $\xi_3$\\
$\CC_4$   & $0$  & $1$ & $0$ & $-1$ & $1$ & $0$ & $-1$ & $\xi_4$\\
\hline\hline
$U(1)_M$ & $1$ &$0$ &$0$ & $0$ & $0$ & $0$ & $-1$ & $r_0$
 \end{tabular}
\ee
The relation between the $\mu, \nu$ parameters in \eqref{S def E3T2} and the FI parameters is:
\be\label{mu to xi SU2T2}
\mu_3= \xi_2 + \xi_3~, \qquad \mu_5= -\xi_2 +\xi_4~, \qquad \mu_6= \xi_1~, \qquad
\nu= -\xi_2~.
\ee
A vertical reduction of this geometry gives an $A_2$ singularity in type IIA, with two D6-branes wrapped on $\mathbb{P}^1_1$, and a single D6-brane wrapped over $\mathbb{P}^1_2$. The IIA profiles for the two exceptional curves are:
\be
\chi_1(r_0) = \begin{cases} 3r_0 + \xi_1-\xi_2 -2\xi_4& {\rm if}\; r_0 \geq \xi_2+ \xi_4\\
r_0 + \xi_1 +\xi_2 & {\rm if}\;  \xi_2 \leq r_0 \leq \xi_2+ \xi_4\\
2 r_0 +\xi_1  \qquad & {\rm if}\;0 \leq r_0 \leq \xi_2 \\
\xi_1   \qquad & {\rm if}\; r_0 \leq 0\end{cases}~, 
\ee
\be
\chi_2(r_0) = \begin{cases}  \xi_3+ \xi_4& {\rm if}\; r_0 \geq \xi_2+ \xi_4\\
r_0- \xi_2 +\xi_3& {\rm if}\;  \xi_2 \leq r_0 \leq \xi_2+ \xi_4\\
- r_0 +\xi_2+\xi_3  \qquad & {\rm if}\; 0 \leq r_0 \leq \xi_2\\
\xi_2+\xi_3 \qquad & {\rm if}\; r_0 \leq 0 \end{cases}~.
\ee
This is displayed in Figure~\ref{fig:T2 SU2 IIA}. The profile for $\chi_1(r_0)$ shows the two gauge D6-branes wrapped over $\mathbb{P}^1_1$, realizing the $SU(2)$ gauge group, and one ``flavor'' brane between them, corresponding to the single D6-brane wrapped over $\mathbb{P}^1_2$. We then naively find a single flavor of $SU(2)$ realized by open strings, with:
\be
M(\CH_{1,1}) =\varphi+ m_1=\xi_2~, \quad
M(\CH_{2,1}) =\varphi- m_1= \xi_4~, \quad M(W_\alpha)= 2\varphi = \xi_2 + \xi_4~.
\ee
Indeed, in the limit $\xi_3 \rightarrow \infty$, one can decouple the divisor $D_3$ and the toric diagram in Fig.~\ref{fig:T2SU2 c labels} goes over to the toric diagram of $E_2^{(0,1)}$ in Fig.~\ref{fig:rk1 01}, corresponding to $SU(2)$ with $N_f=1$. Thus, we expect that, for finite values of $\xi_3$, M2-branes wrapped over $\CC_3$ realize one fundamental hypermultiplet of the gauged $SU(2)$.

To understand this better, we look at the instanton particles. From the IIA setup, the naive unitary quiver is:
\be
U(2)_{-{3\ov 2}}  \; \rule[2pt]{0.8cm}{.6pt}\;\,  U(1)_0~,
\ee
where the subscripts are the effective CS levels, which are read off from the IIA profile.
In our conventions, that means that the bare CS levels for the $U(2)$ and $U(1)$ gauge groups are $k_{U(2)}=-1$ and $k_{U(1)}=1$, respectively. The corresponding prepotential \eqref{F UN} reads:
\bea\label{U2U1 quiver F}
&\CF^{U(2){-}U(1)} &=&\; \sum_{i=1}^2 \left( \half h_0 \phi_i^2 -{1\ov 6} \phi_i^3 -{1\ov 6} \Theta(\phi_i + \phi_0 + m_1) (\phi_i + \phi_0 + m_1)^3 \right) \cr
&&&\; + {1\ov 6} (\phi_1-\phi_2)^3 + \half \t h_0 \phi_0^2 + {1\ov 6} \phi_0^3~.
\eea
Here, $\t h_0$ is the $U(1)$ gauge coupling, $\phi_0$ is the $U(1)$ Coulomb branch parameter, and $m_1$ is the mass of the bifundamental hypermultiplet coupling the $U(2)$ and $U(1)$ gauge groups. Using \eqref{I mass heuristic}, we find:
\be
M(\CI_{1, SU(2)})= h_0 - m_1 = \xi_1~, \qquad 
M(\CI_{1, SU(2)})= h_0 + 3\varphi= \xi_1+2\xi_2 + \xi_4~,
\ee
for the $SU(2)$ instanton masses, where the identification with the FI terms follows from the IIA profile in Fig.~\ref{fig:T2SU2 IIA chi1}. We also have a more mysterious ``$SU(1)$ instanton,'' with mass:
\be
M(\CI_{SU(1)})= \t h_0 - m_1 -\varphi = \xi_3~.
\ee
The identification with $\xi_3$ is clear from Fig.~\ref{fig:T2SU2 IIA chi2}. The fact that there can be non-trivial contributions to the low-energy physics from ``stringy instantons'' at a quiver node with trivial gauge group---here, $SU(1)$---, due to wrapped D-branes, is familiar in string theory (see {\it e.g.} \cite{Aharony:2007pr}). We claim that this particular ``stringy instanton'' particle is equivalent to an hypermultiplet in the fundamental of $SU(2)$. Indeed, if we identify the ``$SU(1)$ gauge coupling'' with the mass difference:
\be
\t h_0 = m_1- m_2~,
\ee
we obtain the relation:
\be
\xi_3= - \varphi - m_2~,
\ee 
which is the positive mass of the second hypermultiplet mode in the correct field theory chamber. We can then identify:
\be\label{rel xi3 to phim2 SU2T2}
M(\CI_{SU(1)})= M(\CH_{2,1})= \xi_3=  - \varphi - m_2~, \quad
M(\CH_{2,2})= \xi_2+\xi_3+\xi_4 = \varphi - m_2~,
\ee
for the two modes of this ``non-perturbative'' hypermultiplet; it corresponds to M2-branes wrapped over $\CC_3$ and $\CC_3+ \CC_6$, respectively. Some additional discussion of this ``$SU(1)$'' mode can be found in Appendix~\ref{app: conifold}.

In summary, we have the following relation between the FI terms in the GLSM \eqref{glsm SU2T2} and the  parameters of an $SU(2)$ $N_f=2$ gauge theory:
\be\label{xi to FT SU2T2} 
\xi_1=h_0 - m_1~, \qquad
\xi_2=\varphi+ m_1~, \qquad
\xi_4=\varphi-m_1~, \qquad
\xi_3= - \varphi - m_2~.
\ee
Plugging this into \eqref{mu to xi SU2T2}, we obtain the relations \eqref{mu to FT SU2T2}, as anticipated.

\paragraph{Flowing to the $T_2$ theory.} We can obtain the $T_2$ geometry of Fig.~\ref{fig:rk0 T2} as a decoupling limit from the geometry in Fig.~\ref{fig:T2SU2 c labels}, simply by taking the limit $\xi_1\rightarrow \infty$ (and thus $\xi_5 \rightarrow \infty$). This has the effect of decoupling the toric divisor $D_6$. It corresponds to the limit of vanishing gauge coupling, $h_0 \rightarrow \infty$, so that the gauged $SU(2)$ becomes a global symmetry.

\paragraph{Other resolutions of the $E_3^{(0,T_2)}$ singularity.} The other two resolutions that admit a vertical reduction, Figures~\ref{fig:T2SU2 g} and \ref{fig:T2SU2 i}, correspond to the field-theory chambers (g) and (i), respectively, of the $SU(2)$ $N_f=2$ gauge theory in \eqref{res to FT chambers dP3} (see the Figures~\ref{fig:dp3-g} and \ref{fig:dp3-i} for the corresponding $E_3$ resolutions). The field theory chambers (c), (g), (i) are the only three chambers such that $\pm \varphi_2+m_2 <0$, as we can see from \eqref{res to FT chambers dP3}. Correspondingly, the second hypermultiplet does not contribute at all to the prepotential (due to the step function). For this reason, these are the only 3 field theory chambers (out of 9) in which the $SU(2)$ $N_f=2$ prepotential is compatible with the prepotential \eqref{U2U1 quiver F}, which corresponds to the unitary quiver seen by open strings in the type-IIA picture. 

The vertical reduction can be performed for the field theory chambers (g) and (i) exactly as for chamber (c), and one confirms the map \eqref{xi to FT SU2T2} between geometry and field theory. What is perhaps more surprising is that the resolution of Figure \ref{fig:T2SU2 a}, which has no allowed vertical reduction, is nonetheless described by the field-theory chamber (a). This is necessary for consistency of the whole picture: one can go from ``chamber (c)'' to ``chamber (a)'' by flopping the curve $\CC_3$ in Fig.~\ref{fig:T2SU2 c labels}; due to the indentification \eqref{rel xi3 to phim2 SU2T2}, this corresponds to flipping the sign of the mass of the hypermultiplet $\CH_{2,1}$.

We note also that the toric singularity $E_3^{(0,T_2)}$ has only five K\"ahler chambers, compared to the 24 K\"ahler chambers for the $E_3$ singularity that we studied in section~\ref{subsec:E3 SCFT}. 
Moreover, in the 3 chambers where both singularities have the same gauge-theory description, the spectrum of instanton operators is nonetheless different between the two geometries. We again interpret these observations as an indication that non-isolated singularities do not give rise to well-defined 5d SCFTs.

\subsection{General $T_N$ quiver}
%%%%%%
 \begin{figure}[t]
\begin{center}
\subfigure[\small $\chi_1(r_0)$]{
\includegraphics[height=5cm]{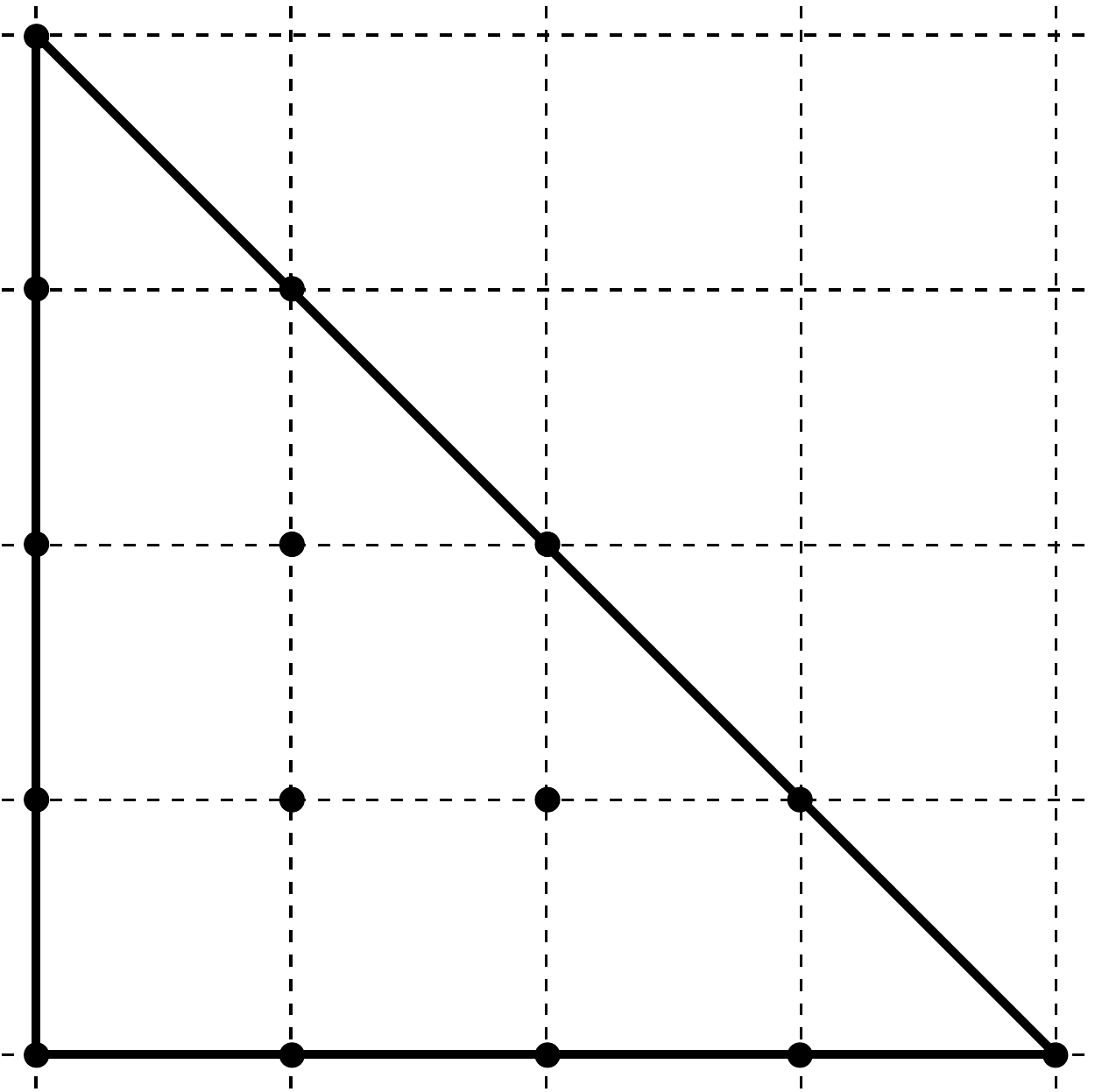}\label{fig:T4 sing}}\qquad
\subfigure[\small $\chi_1(r_0)$]{
\includegraphics[height=5cm]{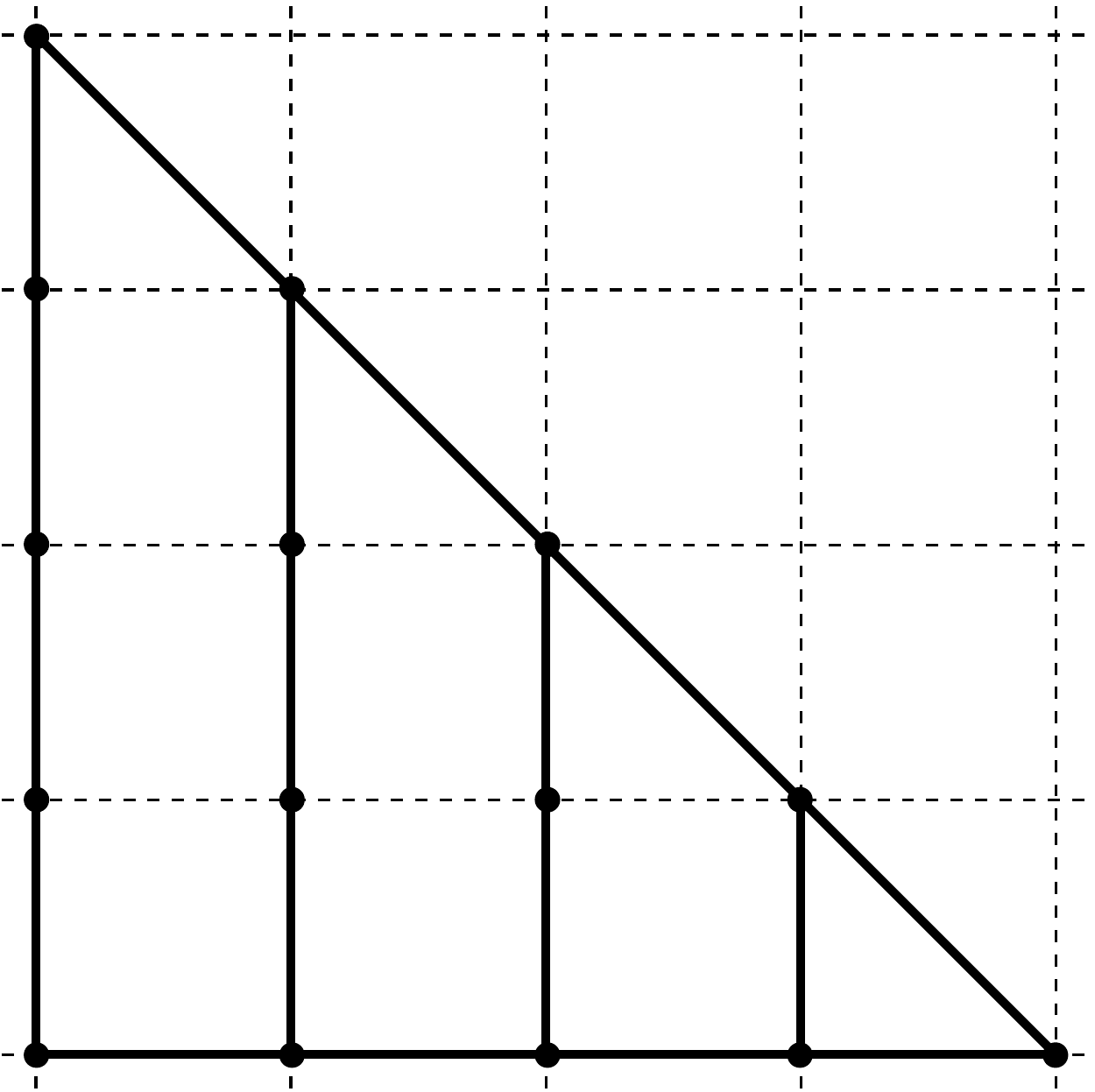}\label{fig:T4 part res}}
\caption{The $T_N$ toric diagram (a.k.a. $\C^3/(\Z_N \times \Z_N)$ and a partial resolution with an allowed vertical reduction, for $N=4$. \label{fig:TN toric diag}}
 \end{center}
 \end{figure} 
 %%%%%%%%
The four-dimensional 4d $\CN=2$ supersymmetric $T_N$ theory is a 4d SCFT with (at least) $SU(N)^3$ global symmetry, corresponding to $N$ M5-branes on a sphere with three full punctures \cite{Gaiotto:2009we}. Its five-dimensional uplift was proposed in \cite{Benini:2009gi}, as the 5D SCFT arising at the intersection of $N$ D5-branes, $N$ NS5-branes and $N$ $(1,1)$-fivebranes in type IIB. While all the 5-branes should end on 7-branes, it has been argued that one can send the 7-branes to infinity without changing the low-energy SCFT---possibly up to some ``decoupled factors.'' Then, one has a simple $(p,q)$-web, dual to a $\C^3/(\Z_N \times \Z_N)$ toric orbifold singularity, as shown in Figure~\ref{fig:T4 sing}.

For any given partial resolution of the singularity with an allowed vertical reduction, as shown in Figure~\ref{fig:T4 part res}, one can directly read off the quiver description \eqref{SUn quiver} from type IIA:
\be\label{TN quiver}
[U(N)]\; \rule[2pt]{0.8cm}{.6pt}\; SU(N-1) \; \rule[2pt]{0.8cm}{.6pt}\; \quad\cdots\quad \; \rule[2pt]{0.8cm}{.6pt}\; SU(3)  \; \rule[2pt]{0.8cm}{.6pt}\;   SU(2) \; \rule[2pt]{0.8cm}{.6pt}\; T_2
\ee
Here, all the effective CS levels vanish, as one can check by direct computation.  The ``$T_2$ tail'' corresponds to two flavors of $SU(2)$, as described above. 
In this way, our IIA perspective nicely reproduces the known quiver description of the mass-deformed $T_N$ theory  \cite{Bergman:2014kza, Hayashi:2014hfa}. 

In the rank-one case, we know that the $T_3$ theory is better described by the isolated singularity $\MG=E_6$, with resolution $\wh \MG={\rm Tot}(\CK \rightarrow dP_6)$, which is non-toric. It is tempting to conjecture that, for any $N$, there exists some (non-toric) isolated singularity $\MG$ whose crepant resolutions capture the full deformation space of the five-dimensional $\FT=T_N$ SCFT. We hope to come back to this question in future work.

%%%%%%%%%
\section*{Acknowledgements} We would like to thank 
Fabio Apruzzi, Benjamin Assel, Francesco Benini, Sergio Benvenuti, Daniel Butter, Thomas Dumitrescu, Marco Fazzi, Amihay Hanany, Kenneth Intriligator, Heeyeon Kim, Martin Ro\v{c}ek, Antonio Sciarappa and Brian Willett for interesting discussions and correspondence.
CC is a Royal Society University Research Fellow and a Research Fellow at St John's College, Oxford. 
VS is supported in part by NSF grant PHY-1620628.

%%%%%%%%%%%%%%%%%%%%%%%%%%%%%%%%%%%%%%%%%%%%%%%%%%%%%%%%%%%%%%%%%%%%%%%%%%%%%%%%%%%%%%%%%%%%%%%%%%%%%%%%%%%%%%%%%%%%%%%%%%%%%%%%%%%%%%%%%%%%%%%%%%%%%%%%%%%%%%%%%%%%%%%%
%%%%%%%%%%%%%%%%%%%%%%%%%%%%
%%%%%%%%%%%%%%%%%%%%%%%%%%%%
\appendix
%%%%%%%%%%%%%%%%%%%%%%%%%%%%%%%%%%%%%%%%%%%%%%%%%%%%%%%%%%%%%%%%%%%%%%%%%%%%%%%%%%%%

\section{The 5d $\CN=1$ gauge-theory prepotential, revisited}\label{app prepot}
In this Appendix, we revisit the derivation of the well-known one-loop prepotential $\CF(\varphi)$, which determines the low-energy theory on the Coulomb branch of a 5d $\CN=1$ gauge theory \cite{Seiberg:1996bd, Nekrasov:1996cz, Intriligator:1997pq}. 
The prepotential can be derived in a slightly roundabout way, by compactifying the theory on a circle, thus obtaining a 4d $\CN=2$ KK theory, and then taking the 5d limit on the 4d $\CN=2$ prepotential \cite{Nekrasov:1996cz}.

Let us denote by $a$ the 4d Coulomb branch complex scalar. For a four-dimensional gauge theory, the one-loop 4d $\CN=2$ prepotential is given by:
\bea
&\CF_{\rm 4d}(a) &=&\; {\tau\ov 2} \Tr(a^2) - {1\ov 8 \pi i}\sum_{\alpha\in \Delta} \alpha(a)^2 \left[\log\left({{\alpha(a)^2\ov \Lambda^2}}\right)- 3\right]\\
&&&\;+ {1\ov 8\pi i} \sum_{\rho, \omega} (\rho(a) +\omega(\mu))^2 \left[\log\left({{ (\rho(a) +\omega(\mu))^2\ov \Lambda^2}}\right)-3\right]+ \cdots~.
\eea
The ellipsis denotes the instanton corrections, which can be safely ignored because they will be suppressed in the 5d limit \cite{Nekrasov:1996cz}. The prepotential of the 5d $\CN=1$ theory on a circle is obtained by resumming the contributions of the KK modes. In particular, an hypermultiplet of charge $1$ under some $U(1)$ will contribute:
\be\label{CFH inter}
\CF_{\rm 4d}^\CH(a)= {1\ov 8\pi i} \sum_{n\in \Z} (a+n)^2  \left[\log{(a+n)^2}- 3\right]~,
\ee
formally. Here, we choose $a$ a dimensionless parameter, soaking up dimensions with the $S^1$ radius $\beta$:
\be
a= a_0 - i \beta \varphi~, \qquad a_0 = {1\ov 2\pi } \int_{S^1} A~,
\ee
with $\varphi$ and $A_\mu$ the 5d scalar and gauge field, respectively.
The diverging sum \eqref{CFH inter} can be regularized to:~\footnote{One quick way to obtain this result is to note that, formally, the fourth derivative of \protect\eqref{CFH inter} is a convergent sum, which gives:
\be\nn
\d_a^4 \CF_{\rm 4d}^\CH(a)= {\pi i \ov 2 \sin^2(\pi a)}~.
\ee
This determines $\CF_{\rm 4d}^\CH(a)$ up to a cubic polynomial in $a$, which we fix by requiring consistency with the decoupling limits $\varphi \rightarrow \pm \infty$.
}
\be\label{CF H 4d oneloop}
\CF_{\rm 4d}^\CH(a) = -{1\ov (2\pi i)^3} \trilog(e^{2\pi i a})~.
\ee
This, we claim, is the correct result for a single hypermultiplet in the $U(1)_{-\half}$ quantization, as discussed in section~\ref{subsec: parity anomaly}. Indeed, the limit $\varphi\rightarrow -\infty$ gives us:
\be
\lim_{{\rm Im}(a)\rightarrow - \infty}\CF_{\rm 4d}^\CH(a) =0~,\qquad
\lim_{{\rm Im}(a)\rightarrow  \infty}\CF_{\rm 4d}^\CH(a) ={1\ov 6} a^3+ {1\ov 4} a^2+ \half a~,
\ee
in agreement with \eqref{limit m large}. The cubic polynomial on the right-hand-side corresponds to the 5d CS action at level $k=-1$.

\paragraph{5d limit.} The 5d prepotential can be obtained from the 4d prepotential of the theory on a circle in the large-$\beta$ limit, with:
\be
\CF= \lim_{\beta\rightarrow \infty} {i\ov \beta^3} \CF_{\rm 4d}~.
\ee
The 4d gauge coupling is related to the 5d gauge coupling $g^2$ by:
\be
\tau={\theta\ov 2\pi} +{4 \pi i \ov g^2_{\rm 4d}}={\theta\ov 2\pi}+ i \beta h_0~, \qquad\qquad h_0\equiv {8\pi^2 \ov g^2}~.
\ee
We then obtain:
\be
\CF^{\rm classical} = \half h_0 \varphi^2 + {k \ov6} \varphi^3
\ee
for the YM and CS terms, for $\GG=U(1)$; the generalization to any $\GG$ is straightforward. The one-loop contribution \eqref{CF H 4d oneloop} from a single hypermultiplet gives us:
\be
\CF^\CH =  \begin{cases} 0 & \text{if} \;\varphi<0 \\ -{1\ov 6} \varphi^3 &  \text{if} \;\varphi>0  \end{cases}\;\;=  -{1\ov 6} \Theta(\varphi) \, \varphi^3~.
\ee
The analysis of the W-boson contribution is completely similar. We thus obtain the five-dimensional one-loop prepotential given in \eqref{F 5d full}, namely:
\bea\label{F 5d full app}
&\CF(\varphi, \mu)\;=\;  \half h_{0} \Tr(\varphi^2) +{k^{abc}\ov 6} \varphi_a \varphi_b \varphi_c   \,+{1\ov 6} \sum_{\alpha\in \Delta} \Theta\big(\alpha(\varphi)\big) \big(\alpha(\varphi)\big)^3 \cr
&\qquad\qquad\;\; \; -{1\ov 6}\sum_\omega \sum_{\rho \in \FR} \Theta\big(\rho(\varphi)+ \omega(m)\big) \big(\rho(\varphi)+ \omega(m)\big)^3~.
\eea

\paragraph{Comparing with IMS.}  At first sight, the result \eqref{F 5d full app} differs from the well-known IMS prepotential \cite{Intriligator:1997pq}, which reads:
\bea\label{F 5d full IMS app}
&\CF_{\rm IMS}(\varphi, \mu)\;=\;  \half h_{0} \Tr(\varphi^2) +{k_{\rm eff}^{abc}\ov 6} \varphi_a \varphi_b \varphi_c   \, +{1\ov 12}\sum_{\alpha \in \Delta} |\alpha(\sigma)|^3 \cr
&\qquad\qquad\;\; \;- {1\ov 12} \sum_{\rho, \omega} |\rho(\sigma) + \omega(m)|^3~.
\eea
Since the roots come in pairs, $\alpha$ and $-\alpha$, it is easy to see that the W-boson contribution in \eqref{F 5d full app} is the same as in \eqref{F 5d full IMS app}, namely:
\be
\CF^{\rm vec}(\varphi) = {1\ov 6} \sum_{\alpha\in \Delta} \Theta\big(\alpha(\varphi)\big) \big(\alpha(\varphi)\big)^3 = {1\ov 12} \sum_{\alpha\in \Delta} |\alpha(\varphi))\big|^3~.
\ee
On the other hand, our choice of a gauge-invariant quantization for the hypermultiplets leads to a small discrepancy between the IMS prepotential \eqref{F 5d full IMS app} and \eqref{F 5d full app}. 

Let us explain this point in more detail. At some level, it is only a matter of notation. For instance, consider a $U(1)$ theory with bare CS level $k\in \Z$ and one hypermultiplet of charge $Q$. We have:
\be
\CF= \half h_0 \varphi^2 + {k\ov 6} \varphi^3 - {1\ov 6} \Theta(Q\varphi) (Q\varphi)^3~. 
\ee
This is equal to the IMS prepotential:
\be
\CF_{\rm IMS}= \half h_0 \varphi^2 + {k_{\rm eff}\ov 6} \varphi^3 - {1\ov 12} |Q\varphi|^3~,
\ee
with:
\be
k_{\rm eff}= k- \half Q^3~.
\ee
This theory would generally be called ``$U(1)_{k_{\rm eff}}$ coupled to an hypermultiplet;'' we are being slightly pedantic in distinguishing between the integer-quantized bare CS level $k$ and the contribution $-\half Q^3$ from the hypermultiplet itself. More generally, the cubic terms in $\varphi$ agree between \eqref{F 5d full IMS app} and \eqref{F 5d full app} if we identify:
\be\label{k eff app}
k_{\rm eff}^{abc}= k^{abc}- \half \sum_{\rho, \omega} \rho^a \rho^b\rho^c~.
\ee
On the other hand, the lower-order terms (of order $\varphi^2$ and $\varphi$) are, in general, slightly different between the two expressions. In our conventions, all the IR contact terms $\kappa$, for both gauge and flavor symmetries (or mixed gauge-flavor), are integer-quantized at a generic point on the Coulomb branch. 

%%%%%%%%%%%
\section{Intersection numbers of toric CY threefolds}\label{app: intersect}
In this appendix, we discuss the computation of triple-interesection numbers in smooth local Calabi-Yau threefolds. We consider many examples, collecting results that are useful in the main text.

\subsection{Computing the M-theory prepotential}
Consider a toric CY$_3$ geometry $\wh\MG$, which we present as a GLSM  \eqref{CY3 glsm bis}. It will be convenient to write it as:
\be\label{app glsm gen}
\begin{tabular}{l|cc|c}
 & $D_l$ & $\bE_a$ & FI \\
 \hline
    $\CC^{\bf a}$&$Q^{\bf a}_l$   &$Q^{\bf a}_a$ &  $\xi^{\bf a}$\\
\end{tabular}~, 
\ee
where we distinguish between the non-compact toric divisors $D_i$, $l=1, \cdots, n_E$, and the compact ones, $\bE_a$, $a=1, \cdots, r$. For definiteness, we can choose the rows in \eqref{app glsm gen} to correspond to curves in $\wh\MG$.
The K\"ahler class of the threefold is parameterized by:
\be\label{S def app}
S= \mu^i D_i + \nu^a \bE_a~,
\ee
where the $D_i$'s appearing here are a chosen subset of $n_E-3$ elements amongst the $n_E$ non-compact toric divisors:
\be
\{ D_i\}_{i=1}^{n_E-3} \subset \{ D_l \}_{l=1}^{n_E}~.
\ee
We choose the $D_i$'s such that $(D_i, \bE_a)$ form a dual basis to the curves $\CC^{\bf a}$. In other words, we define the following square matrix of intersection numbers:
\be
({{\bf Q}^{\bf a}}_{\bf b}) = \left(Q^{\bf a}_i~, \, Q^{\bf a}_b\right)= \left(\CC^{\bf a} \cdot D_i~, \, \CC^{\bf a} \cdot \bE_b \right)~, \qquad\qquad {\bf b}= (i, b)~,
\ee
and we choose the $D_i$'s such that  $\det({\bf Q})\neq 0$.~\footnote{Whenever possible, we choose a basis such that $|\det({\bf Q})|=1$.} Then, the parameters $\mu, \nu$ in \eqref{S def app} are related to the K\"ahler parameters $\xi^{\bf a}$ (the FI parameters) as:
\be
\xi^{\bf a}= Q^{\bf a}_i \mu^i + Q^{\bf a}_b \nu^b \qquad \Leftrightarrow \qquad \mat{\mu \cr \nu}^{\bf a} = {({\bf Q}^{-1})^{\bf a}}_{\bf b}\, \xi^{\bf b}~.
\ee

\paragraph{The prepotential from M-theory.} The geometric prepotential of $\wh\MG$ is  given by:
\be
\CF= -{1\ov 6} S\cdot S \cdot S~.
\ee
It is convenient to expand it as:
\be
\CF= -{1\ov 6} \bE_a \bE_b \bE_c \, \nu^a \nu^b \nu^c-{1\ov 2} \bE_a \bE_b D_i \, \nu^a \nu^b \mu^i -{1\ov 2} \bE_a D_i D_j \nu^a \mu^i \mu^j - {1\ov 6}D_i D_j D_k \mu^i \mu^j \mu^k~.
\ee
For most purposes, we can discard the last term:
\be\label{CF 0 def}
\CF_0 \equiv - {1\ov 6}D_i D_j D_k \mu^i \mu^j \mu^k~,
\ee
 which is ill-defined on a non-compact Calabi-Yau. Whenever there is a gauge-theory interpretation, $\CF_0$ corresponds to a $\varphi$-independent term---namely, a constant term that does not affect any flat-space observable of the five-dimensional theory. Nonetheless, it is sometimes possible to give a useful prescription for \eqref{CF 0 def}, as we shall discuss momentarily. Let us also note that $\CF_0$ actually introduces local terms for background vector multiplet for the flavor symmetry, and therefore some ambiguity is indeed expected.
 
Neglecting $\CF_0$ for now, we must simply compute the intersection numbers that involve at least one compact divisor:
\be\label{all inter numbers to compute}
 \bE_a\cdot \bE_b\cdot \bE_c~,\quad \qquad \bE_a\cdot \bE_b\cdot D_i~, \qquad \quad
  \bE_a\cdot D_i\cdot D_j~.
\ee
This can be done systematically using toric methods, as we reviewed in section~\ref{subsec: triang and inter numb}.

\subsection{Triple intersection numbers and JK residue} 
One can also compute the intersection numbers using a residue formula, which allows us to consistently define the ``non-compact'' intersection numbers in \eqref{CF 0 def}, by introducing some equivariant parameters as regulators \cite{Szenes2004, Closset:2015rna, Honma:2018qcr}.

This computation is based on the topological A-model. Namely, even though we are considering the threefold $\wh \MG$ in M-theory, it is still useful to view the GLSM description of the toric geometry as 2d $\CN=(2,2)$ gauge theory, and to consider the corresponding topological A-twist. Then, the intersection numbers are simply the zero-instanton contribution to the genus-zero correlators. A useful localization formula for the latter can be given \cite{Szenes2004, Closset:2015rna} in terms of the so-called Jeffrey-Kirwan (JK) residue~\cite{JK1995}. 

\paragraph{The residue formula.}
Let us introduce the dummy variables $\sigma = (\sigma_{\bf a})$, and let us assign the following (formal) linear functions of $\sigma$ to the toric divisors:
\be
D_i \equiv  Q_i(\sigma) \equiv Q_i^{\bf a} \sigma_{\bf a}~, \qquad\qquad 
\bE_a \equiv Q_a(\sigma) = Q_a^{\bf a} \sigma_{\bf a}~.
\ee
We define the ``correlator'' of any polynomial $P$ of the $\sigma$'s as:
\be\label{Pcorr gen}
\langle P(\sigma) \rangle_0 \equiv \oint_{{\rm JK}(\eta=\xi)} {d\sigma_1\ov 2\pi i}  \wedge \cdots \wedge {d\sigma_{n-3}\ov 2\pi i} 
{P(\sigma) \ov  \prod_{l=1}^{n_E}  (Q_l(\sigma)+\lambda_l) \, \prod_{a=1}^r Q_a(\sigma)}~.
\ee
Here, note that the denominator is a product over {\it all} the $n= n_E+r$ toric divisors from \eqref{app glsm gen}.
The ``equivariant parameters'' $\lambda_l\in \C$ are regulators, which we choose to be generic.
The JK residue is a simple operation on the integrand. For $\lambda_l$ generic, it is essentially a sum of iterated residues at so-called ``regular singularities,'' where $n-3$ hyperplanes  $\{Q_l(\sigma)+ \lambda_l=0\} \subset \C^{n-3}$ or $\{Q_a(\sigma)=0\} \subset \C^{n-3}$  intersect at a point. Importantly, the singularities that contributes depend on the auxiliary vector:
\be
\eta^{\bf a} = \xi^{\bf a}~,
\ee
which coincides with the FI parameters of the GLSM. In this way, the JK residue gives a different answer for $\wh \MG$ in different K\"ahler chambers. We refer to \cite{Closset:2015rna} (and references therein) for a complete definition of the JK residue.

The triple intersection numbers \eqref{all inter numbers to compute} can be computed as ``correlators'' of the form:
\be
\langle \bE_a \bE_b \bE_c\rangle_0~, \quad\qquad \langle \bE_a \bE_b D_i\rangle_0~, \quad\qquad \langle
  \bE_a D_i D_j\rangle_0~,
\ee
respectively. These numbers are independent of the regulators $\lambda_l$ (assuming $\lambda_l$ is generic), and only depend on the FI parameters $\xi$ through the K\"ahler chamber in which $\xi$ sits. (We assume that $\xi$ is not on a K\"ahler wall, so that the JK residue is well-defined.)

The advantage of the residue formula is that it allows us to {\it define} the triple intersection numbers of three non-compact divisors, as:
\be\label{DDD nc def}
D_i\cdot D_j\cdot D_k \equiv \langle
 D_i D_j D_k\rangle_0~.
\ee
This approach was recently discussed in \cite{Honma:2018qcr}.
The result \eqref{DDD nc def} does depend on the regulator $\lambda_l$ in a non-trivial way. However, in simple-enough cases at least, one can always take some convenient limit on the $\lambda_l$'s to simplify the final answer. We will see some examples in the next subsection, where we can find an answer for \eqref{DDD nc def} that allows us to reproduce the constant term of the gauge-theory prepotential.

\subsection{Intersection numbers in examples}
Let us now consider various toric geometries studied in this paper, and compute the intersection numbers in each case..

\subsubsection{Resolution of the $E_1$ singularity (local $\mathbb{F}_0$)}\label{app: internumb E1}
Consider the $E_1$ singularity, studied in section~\ref{subsec: E1 gauge theory}, with the GLSM \eqref{CY2 F0}. The toric divisors are shown in Figure~\ref{fig:E1 res}. The intersection numbers involving the compact divisor $\bE_0$ are easily computed:
\be\label{inter nbr with E0 for E1}
\bE_0^3= 8~, \quad D_1\bE_0^2=D_2\bE_0^2 =-2~, \quad D_1^2 \bE_0=D_2^2 \bE_0=0~, \quad
D_1 D_2 \bE_0= 1~.
\ee
One can also check this using the JK residue. In this case, we have a unique K\"ahler chamber, corresponding to $\eta=(1,1)$, say, in the JK residue \eqref{Pcorr gen}. We thus have:
\be\nn
\langle P(\sigma) \rangle_0= \oint_{\substack{{\rm JK}\\ \eta=(1,1)}} {d\sigma_1 d\sigma_2\ov (2\pi i)^2} {P(\sigma)\ov (\sigma_1+\lambda_1)(\sigma_2+\lambda_2)(\sigma_1+\lambda_3)(\sigma_2+\lambda_4)(-2\sigma_1+-2\sigma_2)}~.
\ee
This reproduces the intersection numbers \eqref{inter nbr with E0 for E1}---for instance:
\be
\bE_0^3= \langle (-2\sigma_1 -2\sigma_2)^3 \rangle_0 =8~,  \qquad
D_1\bE_0^2= \langle \sigma_1 (-2\sigma_1 -2\sigma_2)^2 \rangle_0 =-2~, \quad {\rm etc.}
\ee
We can also use the residue formula to define the ``non-compact'' intersection numbers, as explained above. Let us choose:
\be\label{def DDD for E1 adhoc}
D_i D_j D_k = \lim_{(\lambda_1, \lambda_2, \lambda_3, \lambda_4) \rightarrow (-3, 1, -3, 1)\lambda} \langle D_i(\sigma) D_j(\sigma) D_j(\sigma) \rangle_0~.
\ee
This gives:
\be
D_1^3=0~, \quad D_1^2 D_2= 0~, \quad D_1 D_2^2 = -\half~, \quad D_2^3=\half~.
\ee
This is the result we quoted in \eqref{DDD E1}; of course, upon using the linear relation $\bE_0\cong -2 D_1-2D_2$, these interesection numbers also reproduce \eqref{inter nbr with E0 for E1}. We chose an {\it ad-hoc} limit \eqref{def DDD for E1 adhoc}  on the $\lambda$ regulator in order to reproduce the field-theory result for pure $SU(2)$ upon vertical reduction.

\subsubsection{Resolutions of the $\t E_1$ singularity (local dP$_1$)}\label{app: t E1}
Consider the $\t E_1$ singularity, studied in section~\ref{subsec: E1 sing}, with the GLSM \eqref{CY2 dP1}. It has two resolutions, shown in Figure~\ref{fig: resdP1}.  

\paragraph{Resolution (a).} Consider the resolution shown in Figure~\ref{fig:E1t res a}. One can easily compute the intersection numbers:
\bea
&\bE_0^3= 8~, \quad && D_1\bE_0^2=-2~, \quad &&D_2\bE_0^2 =-3~, \cr
& D_1^2 \bE_0=0~, \quad&& D_2^2 \bE_0=1~, \quad&&
D_1 D_2 \bE_0= 1~.
\eea
Using the residue formula, one can also define the ``non-compact'' intersection numbers. We again choose a convenient limit:
\be\label{DDD E1t adhoc}
D_i D_j D_k = \lim_{(\lambda_1, \lambda_2, \lambda_3, \lambda_4) \rightarrow (0, 1, 0, 1)\lambda} \langle D_i(\sigma) D_j(\sigma) D_j(\sigma) \rangle_0~.
\ee
Here, the JK residue is taken with $\eta=(1,1)$, since the K\"ahler chamber is such that $\xi_2>0$, $\xi_3>0$.
This gives:
\be
D_1^3=0~, \quad D_1^2 D_2= 0~, \quad D_1 D_2^2 = -\half~, \quad D_2^3=-{1\ov 4}~.
\ee

\paragraph{Resolution (b).} For completeness, let us give the intersection numbers in the second resolution, shown in Figure~\ref{fig:E1t res b}. One finds:
\bea
&\bE_0^3= 9~, \quad&& D_1\bE_0^2=-3~, \quad&& D_2\bE_0^2 =-3~, \cr
& D_1^2 \bE_0=1~, \quad&& D_2^2 \bE_0=1~, \quad&&
D_1 D_2 \bE_0= 1~.
\eea
Using the same limit \eqref{DDD E1t adhoc} with $\eta= (2,-1)$,  since the K\"ahler chamber is such that $\xi_2+\xi_3>0$, $\xi_3<0$. We find:
\be
D_1^3=-1~, \quad D_1^2 D_2= 0~, \quad D_1 D_2^2 = -\half~, \quad D_2^3=-{1\ov 4}~.
\ee

\subsubsection{Resolutions of the $E_2$ singularity (local dP$_2$)}\label{app: E2 sing}
Consider the $E_2$ singularity, studied in section~\ref{subsec: E2 sing}. It admits 5 distinct resolutions, shown in Figure~\ref{fig:dP2 res}. The toric divisors $D_1, \cdots, D_5$ and $\bE_0$ are as shown in Figure~\ref{fig:E2 res}, with the linear equivalences:
\bea
 D_1 &\sim D_3 + D_5, \quad D_2 \sim D_4 + D_5~, \quad \bE_0 \sim -2D_1-2D_2 + D_5~.
\eea
Let us focus on the three resolutions (a), (b), (c) with an allowed vertical reduction---one can similarly consider the resolutions (d) and (e).  
The redundant GLSM, showing the intersections between toric divisors and curves, are as follows:
{\small \be\label{intersect dP2a}
{\bf  (a)}\quad :\quad
\left\{\quad\begin{tabular}{l|cccccc|c}
 & $D_1$ &$D_2$& $D_3$&$D_4$& $D_5$ & $\bE_0$& \\
 \hline
$\CC_1$   & $1$  & $0$  & $1$  & $0$  & $0$   & $-2$ & $\xi_1$\\
$\CC_2$   & $0$  & $1$  & $0$  & $1$  & $0$   & $-2$ & $\xi_2$\\
$\CC_3$   & $1$  & $0$  & $0$  & $-1$ & $1$   & $-1$ & $\xi_3$\\
$\CC_4$   & $0$  & $1$  & $-1$ & $0$  & $1$   & $-1$ & $\xi_4$\\
$\CC_5$   & $0$  & $0$  & $1$  & $1$  & $-1$  & $-1$ & $\xi_5$\\
 \end{tabular}\right.
\ee
\be\label{intersect dP2b}
{\bf  (b)}\quad :\quad
\left\{\quad
\begin{tabular}{l|cccccc|c}
 & $D_1$ &$D_2$& $D_3$&$D_4$& $D_5$ & $\bE_0$& \\
 \hline
$\CC_1$   & $1$  & $0$  & $1$  & $0$   & $0$   & $-2$ & $\xi_1$\\
$\CC_2$   & $0$  & $1$  & $0$  & $1$   & $0$   & $-2$ & $\xi_2$\\
$\CC_3'$   & $1$  & $0$  & $1$  & $0$   & $0$   & $-2$ & $\xi_3+\xi_5$\\
$\CC_4'$   & $0$  & $1$  & $0$  & $1$   & $0$   & $-2$ & $\xi_4+\xi_5$\\
$\b\CC_5$   & $0$  & $0$  & $-1$ & $-1$  & $1$   & $1$  & $-\xi_5$\\
 \end{tabular}\right.
\ee
 \be\label{intersect dP2c}
 {\bf  (c)}\quad :\quad
\left\{\quad
\begin{tabular}{l|cccccc|c}
 & $D_1$ &$D_2$& $D_3$&$D_4$& $D_5$ & $\bE_0$& \\
 \hline
$\CC_1'$   & $1$  & $1$   & $0$  & $0$   & $1$   & $-3$ & $\xi_1+\xi_4$\\
$\CC_2$   & $0$  & $1$   & $0$  & $1$   & $0$   & $-2$ & $\xi_2$\\
$\CC_3$   & $1$  & $0$   & $0$  & $-1$  & $1$   & $-1$ & $\xi_3$\\
$\b\CC_4$   & $0$  & $-1$  & $1$  & $0$   & $-1$  & $1$  & $-\xi_4$\\
$\CC_5'$   & $0$  & $1$   & $0$  & $1$   & $0$   & $-2$ & $\xi_4+\xi_5$\\
 \end{tabular}\right.
\ee}

\paragraph{Intersection numbers.} To compute the geometric prepotential using $S$ in \eqref{xi to mu nu E2}, the relevant intersection numbers involving $\bE_0$ are:
\be\label{intersections dP2}
\begin{tabular}{l|c|c|c|c|c}
 & (a)&  (b)& (c)& (d) & (e)\\ \hline
$\bE_0^3$     & $7$  & $8$  & $8$  & $8$  & $9$   \\ \hline
$\bE_0^2 D_1$ & $-2$ & $-2$ & $-2$ & $-3$ & $-3$  \\ \hline
$\bE_0^2 D_5$ & $-1$ & $0$  & $-2$ & $-2$ & $-3$  \\ \hline
$\bE_0 D_1^2$ & $0$  & $0$  & $0$  & $1$  & $1$   \\ \hline
$\bE_0 D_5^2$ & $-1$ & $0$  & $0$  & $0$  & $1$   \\ \hline
$\bE_0 D_1 D_5$ & $0$ & $0$ & $0$  & $1$  & $1$  
 \end{tabular}~,
\ee
in the five resolutions.
Moreover, one can again define the ``non-compact'' intersection numbers using the JK residue. We choose:~\footnote{One can choose $\eta$ in the JK residue to be $\eta_a=(2,2,1)$,  $\eta_b=(1,1,-1)$, $\eta_c=(3,1,2)$, $\eta_d=(1,3,2)$, and $\eta_e=(2,2,3)$, respectively, for the five resolutions.}
\be
D_i D_j D_k = \lim_{(\lambda_1, \lambda_2, \lambda_3, \lambda_4, \lambda_5) \rightarrow (0, 1, 0, 0,0)\lambda} \langle D_i(\sigma) D_j(\sigma) D_j(\sigma) \rangle_0~.
\ee
For the relevant intersections amongst $D_1$ and $D_5$, this gives:
\be\label{intersections NC dP2}
\begin{tabular}{l|c|c|c|c|c}
 & (a)&  (b)& (c)& (d) & (e)\\ \hline
$D_1^3$            & $0$  & $0$  & $0$  & $-1$  & $-1$   \\ \hline
$D_1^2 D_5$ & $0$ & $0$ & $0$ & $-1$ & $-1$  \\ \hline
$D_1 D_5^2$ & $0$ & $0$  & $0$ & $-1$ & $-1$  \\ \hline
$D_5^3$          & $1$& $2$  & $0$  & $0$  & $-1$   
 \end{tabular}
\ee
For the resolutions (a), (b) and (c), which admit a vertical reduction, plugging the intersection numbers \eqref{intersections dP2} and \eqref{intersections NC dP2} into the M-theory prepotential reproduces precisely the field-theory prepotential, including the constant term.

\subsubsection{Resolutions of the $E_3$ singularity (local dP$_3$)}\label{app: E3 sing}
Consider the $E_3$ singularity, studied in section~\ref{subsec:E3 SCFT}. It admits 18 distinct resolutions,  shown in Figure~\ref{fig:dP3 res}. In the following, we list the triple intersection numbers that involve at least one compact divisor, in all 9 resolutions with an allowed vertical reduction. The toric divisors $D_1, \cdots, D_6$ and $\bE_0$ are as indicated in Figure~\ref{fig:E3 res}, with the linear equivalences:
{\small\bea
\hspace{-0.1in}  D_4 &\sim D_1 + D_2 - D_5~, \quad D_6 \sim D_2 + D_3 - D_5~, \quad \bE_0 \sim -2 D_1 - 3 D_2 - 2 D_3 + D_5~.
\eea}
We find:
{\footnotesize\begin{align}
{\rm \bf (a)} \quad:\quad
\left\{\begin{array}{c@{~,\quad}c@{~,\quad}c@{~,\quad}c@{~,}}
 \bE_0^3=6 & D_1 \bE_0^2=-1 & D_2 \bE_0^2=-1 & D_3 \bE_0^2=-1 \\
 D_4 \bE_0^2=-1 & D_5 \bE_0^2=-1 & D_6 \bE_0^2=-1 & D_1 D_2 \bE_0=1 \\
 D_1 D_3 \bE_0=0 & D_1 D_4 \bE_0=0 & D_1 D_5 \bE_0=0 & D_1 D_6 \bE_0=1 \\
 D_2 D_3 \bE_0=1 & D_2 D_4 \bE_0=0 & D_2 D_5 \bE_0=0 & D_2 D_6 \bE_0=0 \\
 D_3 D_4 \bE_0=1 & D_3 D_5 \bE_0=0 & D_3 D_6 \bE_0=0 & D_4 D_5 \bE_0=1 \\
 D_4 D_6 \bE_0=0 & D_5 D_6 \bE_0=1 & D_1^2 \bE_0=-1 & D_2^2 \bE_0=-1 \\
 D_3^2 \bE_0=-1 & D_4^2 \bE_0=-1 & D_5^2 \bE_0=-1 & D_6^2 \bE_0=-1 \\
\end{array}\right.	
\end{align}}
\vspace{-0.1cm}
{\footnotesize\begin{align}
{\rm \bf (b)}\quad  : \quad
\left\{\begin{array}{c@{~,\quad}c@{~,\quad}c@{~,\quad}c@{~,}}
 \bE_0^3=7 & D_1 \bE_0^2=0 & D_2 \bE_0^2=-2 & D_3 \bE_0^2=-1 \\
 D_4 \bE_0^2=-1 & D_5 \bE_0^2=-1 & D_6 \bE_0^2=-2 & D_1 D_2 \bE_0=0 \\
 D_1 D_3 \bE_0=0 & D_1 D_4 \bE_0=0 & D_1 D_5 \bE_0=0 & D_1 D_6 \bE_0=0 \\
 D_2 D_3 \bE_0=1 & D_2 D_4 \bE_0=0 & D_2 D_5 \bE_0=0 & D_2 D_6 \bE_0=1 \\
 D_3 D_4 \bE_0=1 & D_3 D_5 \bE_0=0 & D_3 D_6 \bE_0=0 & D_4 D_5 \bE_0=1 \\
 D_4 D_6 \bE_0=0 & D_5 D_6 \bE_0=1 & D_1^2 \bE_0=0 & D_2^2 \bE_0=0 \\
 D_3^2 \bE_0=-1 & D_4^2 \bE_0=-1 & D_5^2 \bE_0=-1 & D_6^2 \bE_0=0 \\
\end{array}\right.
\end{align}}
\vspace{-0.1cm}
{\footnotesize\begin{align}
{\rm \bf (c)} \quad :\quad
\left\{\begin{array}{c@{~,\quad}c@{~,\quad}c@{~,\quad}c@{~,}}
 \bE_0^3=7 & D_1 \bE_0^2=-2 & D_2 \bE_0^2=0 & D_3 \bE_0^2=-2 \\
 D_4 \bE_0^2=-1 & D_5 \bE_0^2=-1 & D_6 \bE_0^2=-1 & D_1 D_2 \bE_0=0 \\
 D_1 D_3 \bE_0=1 & D_1 D_4 \bE_0=0 & D_1 D_5 \bE_0=0 & D_1 D_6 \bE_0=1 \\
 D_2 D_3 \bE_0=0 & D_2 D_4 \bE_0=0 & D_2 D_5 \bE_0=0 & D_2 D_6 \bE_0=0 \\
 D_3 D_4 \bE_0=1 & D_3 D_5 \bE_0=0 & D_3 D_6 \bE_0=0 & D_4 D_5 \bE_0=1 \\
 D_4 D_6 \bE_0=0 & D_5 D_6 \bE_0=1 & D_1^2 \bE_0=0 & D_2^2 \bE_0=0 \\
 D_3^2 \bE_0=0 & D_4^2 \bE_0=-1 & D_5^2 \bE_0=-1 & D_6^2 \bE_0=-1 \\
\end{array}\right.
\end{align}}
\vspace{-0.1cm}
{\footnotesize\begin{align}
{\rm \bf (d)} \quad :\quad
\left\{\begin{array}{c@{~,\quad}c@{~,\quad}c@{~,\quad}c@{~,}}
 \bE_0^3=7 & D_1 \bE_0^2=-1 & D_2 \bE_0^2=-1 & D_3 \bE_0^2=-2 \\
 D_4 \bE_0^2=0 & D_5 \bE_0^2=-2 & D_6 \bE_0^2=-1 & D_1 D_2 \bE_0=1 \\
 D_1 D_3 \bE_0=0 & D_1 D_4 \bE_0=0 & D_1 D_5 \bE_0=0 & D_1 D_6 \bE_0=1 \\
 D_2 D_3 \bE_0=1 & D_2 D_4 \bE_0=0 & D_2 D_5 \bE_0=0 & D_2 D_6 \bE_0=0 \\
 D_3 D_4 \bE_0=0 & D_3 D_5 \bE_0=1 & D_3 D_6 \bE_0=0 & D_4 D_5 \bE_0=0 \\
 D_4 D_6 \bE_0=0 & D_5 D_6 \bE_0=1 & D_1^2 \bE_0=-1 & D_2^2 \bE_0=-1 \\
 D_3^2 \bE_0=0 & D_4^2 \bE_0=0 & D_5^2 \bE_0=0 & D_6^2 \bE_0=-1 \\
\end{array}\right.
\end{align}}
\vspace{-0.1cm}
{\footnotesize\begin{align}
{\rm \bf (e)} \quad :\quad
\left\{\begin{array}{c@{~,\quad}c@{~,\quad}c@{~,\quad}c@{~,}}
 \bE_0^3=7 & D_1 \bE_0^2=-1 & D_2 \bE_0^2=-1 & D_3 \bE_0^2=-1 \\
 D_4 \bE_0^2=-2 & D_5 \bE_0^2=0 & D_6 \bE_0^2=-2 & D_1 D_2 \bE_0=1 \\
 D_1 D_3 \bE_0=0 & D_1 D_4 \bE_0=0 & D_1 D_5 \bE_0=0 & D_1 D_6 \bE_0=1 \\
 D_2 D_3 \bE_0=1 & D_2 D_4 \bE_0=0 & D_2 D_5 \bE_0=0 & D_2 D_6 \bE_0=0 \\
 D_3 D_4 \bE_0=1 & D_3 D_5 \bE_0=0 & D_3 D_6 \bE_0=0 & D_4 D_5 \bE_0=0 \\
 D_4 D_6 \bE_0=1 & D_5 D_6 \bE_0=0 & D_1^2 \bE_0=-1 & D_2^2 \bE_0=-1 \\
 D_3^2 \bE_0=-1 & D_4^2 \bE_0=0 & D_5^2 \bE_0=0 & D_6^2 \bE_0=0 \\
\end{array}\right.
\end{align}}
\vspace{-0.1cm}
{\footnotesize\begin{align}
{\rm \bf (f)} \quad :\quad
\left\{\begin{array}{c@{~,\quad}c@{~,\quad}c@{~,\quad}c@{~,}}
 \bE_0^3=8 & D_1 \bE_0^2=0 & D_2 \bE_0^2=-2 & D_3 \bE_0^2=-2 \\
 D_4 \bE_0^2=0 & D_5 \bE_0^2=-2 & D_6 \bE_0^2=-2 & D_1 D_2 \bE_0=0 \\
 D_1 D_3 \bE_0=0 & D_1 D_4 \bE_0=0 & D_1 D_5 \bE_0=0 & D_1 D_6 \bE_0=0 \\
 D_2 D_3 \bE_0=1 & D_2 D_4 \bE_0=0 & D_2 D_5 \bE_0=0 & D_2 D_6 \bE_0=1 \\
 D_3 D_4 \bE_0=0 & D_3 D_5 \bE_0=1 & D_3 D_6 \bE_0=0 & D_4 D_5 \bE_0=0 \\
 D_4 D_6 \bE_0=0 & D_5 D_6 \bE_0=1 & D_1^2 \bE_0=0 & D_2^2 \bE_0=0 \\
 D_3^2 \bE_0=0 & D_4^2 \bE_0=0 & D_5^2 \bE_0=0 & D_6^2 \bE_0=0 \\
\end{array}\right.
\end{align}}
\vspace{-0.1cm}
{\footnotesize\begin{align}
{\rm \bf (g)} \quad :\quad
\left\{\begin{array}{c@{~,\quad}c@{~,\quad}c@{~,\quad}c@{~,}}
 \bE_0^3=8 & D_1 \bE_0^2=-2 & D_2 \bE_0^2=0 & D_3 \bE_0^2=-2 \\
 D_4 \bE_0^2=-2 & D_5 \bE_0^2=0 & D_6 \bE_0^2=-2 & D_1 D_2 \bE_0=0 \\
 D_1 D_3 \bE_0=1 & D_1 D_4 \bE_0=0 & D_1 D_5 \bE_0=0 & D_1 D_6 \bE_0=1 \\
 D_2 D_3 \bE_0=0 & D_2 D_4 \bE_0=0 & D_2 D_5 \bE_0=0 & D_2 D_6 \bE_0=0 \\
 D_3 D_4 \bE_0=1 & D_3 D_5 \bE_0=0 & D_3 D_6 \bE_0=0 & D_4 D_5 \bE_0=0 \\
 D_4 D_6 \bE_0=1 & D_5 D_6 \bE_0=0 & D_1^2 \bE_0=0 & D_2^2 \bE_0=0 \\
 D_3^2 \bE_0=0 & D_4^2 \bE_0=0 & D_5^2 \bE_0=0 & D_6^2 \bE_0=0 \\
\end{array}\right.
\end{align}}
\vspace{-0.1cm}
{\footnotesize\begin{align}
{\rm \bf (h)} \quad :\quad
\left\{\begin{array}{c@{~,\quad}c@{~,\quad}c@{~,\quad}c@{~,}}
 \bE_0^3=8 & D_1 \bE_0^2=0 & D_2 \bE_0^2=-2 & D_3 \bE_0^2=-1 \\
 D_4 \bE_0^2=-2 & D_5 \bE_0^2=0 & D_6 \bE_0^2=-3 & D_1 D_2 \bE_0=0 \\
 D_1 D_3 \bE_0=0 & D_1 D_4 \bE_0=0 & D_1 D_5 \bE_0=0 & D_1 D_6 \bE_0=0 \\
 D_2 D_3 \bE_0=1 & D_2 D_4 \bE_0=0 & D_2 D_5 \bE_0=0 & D_2 D_6 \bE_0=1 \\
 D_3 D_4 \bE_0=1 & D_3 D_5 \bE_0=0 & D_3 D_6 \bE_0=0 & D_4 D_5 \bE_0=0 \\
 D_4 D_6 \bE_0=1 & D_5 D_6 \bE_0=0 & D_1^2 \bE_0=0 & D_2^2 \bE_0=0 \\
 D_3^2 \bE_0=-1 & D_4^2 \bE_0=0 & D_5^2 \bE_0=0 & D_6^2 \bE_0=1 \\
\end{array}\right.
\end{align}}
\vspace{-0.1cm}
{\footnotesize\begin{align}
{\rm \bf (i)} \quad :\quad
\left\{\begin{array}{c@{~,\quad}c@{~,\quad}c@{~,\quad}c}
 \bE_0^3=8 & D_1 \bE_0^2=-2 & D_2 \bE_0^2=0 & D_3 \bE_0^2=-3~, \\
 D_4 \bE_0^2=0 & D_5 \bE_0^2=-2 & D_6 \bE_0^2=-1 & D_1 D_2 \bE_0=0~, \\
 D_1 D_3 \bE_0=1 & D_1 D_4 \bE_0=0 & D_1 D_5 \bE_0=0 & D_1 D_6 \bE_0=1~, \\
 D_2 D_3 \bE_0=0 & D_2 D_4 \bE_0=0 & D_2 D_5 \bE_0=0 & D_2 D_6 \bE_0=0~, \\
 D_3 D_4 \bE_0=0 & D_3 D_5 \bE_0=1 & D_3 D_6 \bE_0=0 & D_4 D_5 \bE_0=0~, \\
 D_4 D_6 \bE_0=0 & D_5 D_6 \bE_0=1 & D_1^2 \bE_0=0 & D_2^2 \bE_0=0~, \\
 D_3^2 \bE_0=1 & D_4^2 \bE_0=0 & D_5^2 \bE_0=0 & D_6^2 \bE_0=-1~. \\
\end{array}\right.
\end{align}}
 
 %%%%%%%%%%%%%%%%%%%
\subsubsection{Resolutions of the beetle geometry}\label{app:all intersection numbers beetle}
Consider the ``beetle singularity'' that we studied in section~\ref{sec: beetle}. It admits 24 distinct resolutions, shown in Figure~\ref{fig:all beetles}. The toric divisors $D_1, \cdots, D_6$ and $\bE_1, \bE_2$ are as shown in Figure~\ref{fig: beetle sing}, with the linear equivalences:
{\small\bea
 & D_4 \sim D_1 - D_3 + D_6~, \quad\qquad& \bE_1 \sim -2D_1-2D_2+D_4+D_5-D_6~, \cr
&  \bE_2 \sim D_2 - D_4 - 2D_5 - D_6~.
\eea}
In the following, we compute the intersection numbers for every resolution, and we verify that the geometric prepotential matches precisely with the $SU(2)\times SU(2)$ and/or with the $SU(3)$, $N_f=2$ prepotential, whenever a gauge-theory interpretation exists. 

\paragraph{Resolutions with a vertical reduction.} The resolutions (a) to (f) have an allowed vertical reduction. Their intersection numbers are:
{\footnotesize\begin{align}
  {\rm \bf (a)} :
\left\{\begin{array}{c@{~,\quad}c@{~,\quad}c@{~,\quad}c@{~,\quad}c@{~,}}
 \bm{E_1}^3=8 & \bm{E_2}^3=6 & D_1^2 \bm{E_2}=-1 & D_1 D_2 \bm{E_2}=0 & D_2^2 \bm{E_2}=0 \\
 D_1 D_6 \bm{E_2}=1 & D_2 D_6 \bm{E_2}=0 & D_6^2 \bm{E_2}=-1 & D_1 \bm{E_1}^2=-2 & D_2 \bm{E_1}^2=-2 \\
 D_6 \bm{E_1}^2=0 & \bm{E_1}^2 \bm{E_2}=-2 & D_1 \bm{E_2}^2=-1 & D_2 \bm{E_2}^2=0 & D_6 \bm{E_2}^2=-1 \\
 D_1^2 \bm{E_1}=0 & D_1 D_2 \bm{E_1}=1 & D_2^2 \bm{E_1}=0 & D_1 D_6 \bm{E_1}=0 & D_2 D_6 \bm{E_1}=0 \\
 D_6^2 \bm{E_1}=0 & \bm{E_1} \bm{E_2}^2=0 & D_1 \bm{E_1} \bm{E_2}=1 & D_2 \bm{E_1} \bm{E_2}=0 & D_6 \bm{E_1} \bm{E_2}=0 \\
\end{array}\right.
\end{align}}
\vspace{-0.1cm}
{\footnotesize\begin{align}
{\rm \bf (b)}  :
\left\{\begin{array}{c@{~,\quad}c@{~,\quad}c@{~,\quad}c@{~,\quad}c@{~,}}
 \bm{E_1}^3=7 & \bm{E_2}^3=7 & D_1^2 \bm{E_2}=-1 & D_1 D_2 \bm{E_2}=0 & D_2^2 \bm{E_2}=0 \\
 D_1 D_6 \bm{E_2}=1 & D_2 D_6 \bm{E_2}=0 & D_6^2 \bm{E_2}=-1 & D_1 \bm{E_1}^2=-2 & D_2
   \bm{E_1}^2=-2 \\
 D_6 \bm{E_1}^2=0 & \bm{E_1}^2 \bm{E_2}=-1 & D_1 \bm{E_2}^2=-1 & D_2 \bm{E_2}^2=0 & D_6
   \bm{E_2}^2=-1 \\
 D_1^2 \bm{E_1}=0 & D_1 D_2 \bm{E_1}=1 & D_2^2 \bm{E_1}=0 & D_1 D_6 \bm{E_1}=0 & D_2
   D_6 \bm{E_1}=0 \\
 D_6^2 \bm{E_1}=0 & \bm{E_1} \bm{E_2}^2=-1 & D_1 \bm{E_1} \bm{E_2}=1 & D_2 \bm{E_1} \bm{E_2}=0 & D_6
   \bm{E_1} \bm{E_2}=0 \\
\end{array}\right.
\end{align}}
\vspace{-0.1cm}
{\footnotesize\begin{align}
  {\rm \bf (c)} :
\left\{\begin{array}{c@{~,\quad}c@{~,\quad}c@{~,\quad}c@{~,\quad}c@{~,}}
 \bm{E_1}^3=8 & \bm{E_2}^3=8 & D_1^2 \bm{E_2}=0 & D_1 D_2 \bm{E_2}=0 & D_2^2 \bm{E_2}=0 \\
 D_1 D_6 \bm{E_2}=1 & D_2 D_6 \bm{E_2}=0 & D_6^2 \bm{E_2}=-1 & D_1 \bm{E_1}^2=-3 & D_2
   \bm{E_1}^2=-2 \\
 D_6 \bm{E_1}^2=0 & \bm{E_1}^2 \bm{E_2}=0 & D_1 \bm{E_2}^2=-2 & D_2 \bm{E_2}^2=0 & D_6 \bm{E_2}^2=-1
   \\
 D_1^2 \bm{E_1}=1 & D_1 D_2 \bm{E_1}=1 & D_2^2 \bm{E_1}=0 & D_1 D_6 \bm{E_1}=0 & D_2 D_6
   \bm{E_1}=0 \\
 D_6^2 \bm{E_1}=0 & \bm{E_1} \bm{E_2}^2=0 & D_1 \bm{E_1} \bm{E_2}=0 & D_2 \bm{E_1} \bm{E_2}=0 & D_6 \bm{E_1}
   \bm{E_2}=0 \\
\end{array}\right.
\end{align}}
\vspace{-0.1cm}
{\footnotesize\begin{align}
{\rm \bf (d)} :
\left\{\begin{array}{c@{~,\quad}c@{~,\quad}c@{~,\quad}c@{~,\quad}c@{~,}}
 \bm{E_1}^3=6 & \bm{E_2}^3=8 & D_1^2 \bm{E_2}=0 & D_1 D_2 \bm{E_2}=0 & D_2^2 \bm{E_2}=0 \\
 D_1 D_6 \bm{E_2}=0 & D_2 D_6 \bm{E_2}=0 & D_6^2 \bm{E_2}=0 & D_1 \bm{E_1}^2=-1 & D_2
   \bm{E_1}^2=-2 \\
 D_6 \bm{E_1}^2=-1 & \bm{E_1}^2 \bm{E_2}=0 & D_1 \bm{E_2}^2=0 & D_2 \bm{E_2}^2=0 & D_6
   \bm{E_2}^2=-2 \\
 D_1^2 \bm{E_1}=-1 & D_1 D_2 \bm{E_1}=1 & D_2^2 \bm{E_1}=0 & D_1 D_6 \bm{E_1}=1 & D_2
   D_6 \bm{E_1}=0 \\
 D_6^2 \bm{E_1}=-1 & \bm{E_1} \bm{E_2}^2=-2 & D_1 \bm{E_1} \bm{E_2}=0 & D_2 \bm{E_1} \bm{E_2}=0 & D_6
   \bm{E_1} \bm{E_2}=1 \\
\end{array}\right.
\end{align}}
\vspace{-0.1cm}
{\footnotesize\begin{align}
{\rm \bf (e)} :
\left\{\begin{array}{c@{~,\quad}c@{~,\quad}c@{~,\quad}c@{~,\quad}c@{~,}}
 \bm{E_1}^3=7 & \bm{E_2}^3=7 & D_1^2 \bm{E_2}=0 & D_1 D_2 \bm{E_2}=0 & D_2^2 \bm{E_2}=0 \\
 D_1 D_6 \bm{E_2}=0 & D_2 D_6 \bm{E_2}=0 & D_6^2 \bm{E_2}=0 & D_1 \bm{E_1}^2=-1 & D_2
   \bm{E_1}^2=-2 \\
 D_6 \bm{E_1}^2=-1 & \bm{E_1}^2 \bm{E_2}=-1 & D_1 \bm{E_2}^2=0 & D_2 \bm{E_2}^2=0 & D_6
   \bm{E_2}^2=-2 \\
 D_1^2 \bm{E_1}=-1 & D_1 D_2 \bm{E_1}=1 & D_2^2 \bm{E_1}=0 & D_1 D_6 \bm{E_1}=1 & D_2
   D_6 \bm{E_1}=0 \\
 D_6^2 \bm{E_1}=-1 & \bm{E_1} \bm{E_2}^2=-1 & D_1 \bm{E_1} \bm{E_2}=0 & D_2 \bm{E_1} \bm{E_2}=0 & D_6
   \bm{E_1} \bm{E_2}=1 \\
\end{array}\right.
\end{align}}
\vspace{-0.1cm}
{\footnotesize\begin{align}
{\rm \bf (f)} :
\left\{\begin{array}{c@{~,\quad}c@{~,\quad}c@{~,\quad}c@{~,\quad}c}
 \bm{E_1}^3=8 & \bm{E_2}^3=8 & D_1^2 \bm{E_2}=0 & D_1 D_2 \bm{E_2}=0 & D_2^2 \bm{E_2}=0~, \\
 D_1 D_6 \bm{E_2}=0 & D_2 D_6 \bm{E_2}=0 & D_6^2 \bm{E_2}=1 & D_1 \bm{E_1}^2=-1 & D_2
   \bm{E_1}^2=-2~, \\
 D_6 \bm{E_1}^2=-2 & \bm{E_1}^2 \bm{E_2}=0 & D_1 \bm{E_2}^2=0 & D_2 \bm{E_2}^2=0 & D_6
   \bm{E_2}^2=-3~, \\
 D_1^2 \bm{E_1}=-1 & D_1 D_2 \bm{E_1}=1 & D_2^2 \bm{E_1}=0 & D_1 D_6 \bm{E_1}=1 & D_2
   D_6 \bm{E_1}=0~, \\
 D_6^2 \bm{E_1}=0 & \bm{E_1} \bm{E_2}^2=0 & D_1 \bm{E_1} \bm{E_2}=0 & D_2 \bm{E_1} \bm{E_2}=0 & D_6 \bm{E_1}
   \bm{E_2}=0~. \\
\end{array}\right.
\end{align}}

\paragraph{Resolutions with an horizontal reduction.} The resolutions (g) to (r) have an horizontal reduction. So do the resolutions (a), (b), (d), (e) above. The intersection numbers are:
{\footnotesize\begin{align}
{\rm \bf (g)} :
\left\{\begin{array}{c@{~,\quad}c@{~,\quad}c@{~,\quad}c@{~,\quad}c@{~,}}
 \bm{E_1}^3=8 & \bm{E_2}^3=7 & D_1^2 \bm{E_2}=-1 & D_1 D_2 \bm{E_2}=0 & D_2^2 \bm{E_2}=0 \\
 D_1 D_6 \bm{E_2}=1 & D_2 D_6 \bm{E_2}=0 & D_6^2 \bm{E_2}=-1 & D_1 \bm{E_1}^2=-2 & D_2
   \bm{E_1}^2=-2 \\
 D_6 \bm{E_1}^2=0 & \bm{E_1}^2 \bm{E_2}=-2 & D_1 \bm{E_2}^2=-1 & D_2 \bm{E_2}^2=0 & D_6 \bm{E_2}^2=-1
   \\
 D_1^2 \bm{E_1}=0 & D_1 D_2 \bm{E_1}=1 & D_2^2 \bm{E_1}=0 & D_1 D_6 \bm{E_1}=0 & D_2 D_6
   \bm{E_1}=0 \\
 D_6^2 \bm{E_1}=0 & \bm{E_1} \bm{E_2}^2=0 & D_1 \bm{E_1} \bm{E_2}=1 & D_2 \bm{E_1} \bm{E_2}=0 & D_6 \bm{E_1}
   \bm{E_2}=0 \\
\end{array}\right.
\end{align}}
\vspace{-0.1cm}
{\footnotesize\begin{align}
{\rm \bf (h)} :
\left\{\begin{array}{c@{~,\quad}c@{~,\quad}c@{~,\quad}c@{~,\quad}c@{~,}}
 \bm{E_1}^3=8 & \bm{E_2}^3=7 & D_1^2 \bm{E_2}=0 & D_1 D_2 \bm{E_2}=0 & D_2^2 \bm{E_2}=0 \\
 D_1 D_6 \bm{E_2}=0 & D_2 D_6 \bm{E_2}=0 & D_6^2 \bm{E_2}=0 & D_1 \bm{E_1}^2=-2 & D_2
   \bm{E_1}^2=-2 \\
 D_6 \bm{E_1}^2=0 & \bm{E_1}^2 \bm{E_2}=-2 & D_1 \bm{E_2}^2=-2 & D_2 \bm{E_2}^2=0 & D_6 \bm{E_2}^2=0
   \\
 D_1^2 \bm{E_1}=0 & D_1 D_2 \bm{E_1}=1 & D_2^2 \bm{E_1}=0 & D_1 D_6 \bm{E_1}=0 & D_2 D_6
   \bm{E_1}=0 \\
 D_6^2 \bm{E_1}=0 & \bm{E_1} \bm{E_2}^2=0 & D_1 \bm{E_1} \bm{E_2}=1 & D_2 \bm{E_1} \bm{E_2}=0 & D_6 \bm{E_1}
   \bm{E_2}=0 \\
\end{array}\right.
\end{align}}
\vspace{-0.1cm}
{\footnotesize\begin{align}
{\rm \bf (i)} :
\left\{\begin{array}{c@{~,\quad}c@{~,\quad}c@{~,\quad}c@{~,\quad}c@{~,}}
 \bm{E_1}^3=8 & \bm{E_2}^3=8 & D_1^2 \bm{E_2}=0 & D_1 D_2 \bm{E_2}=0 & D_2^2 \bm{E_2}=0 \\
 D_1 D_6 \bm{E_2}=0 & D_2 D_6 \bm{E_2}=0 & D_6^2 \bm{E_2}=0 & D_1 \bm{E_1}^2=-2 & D_2
   \bm{E_1}^2=-2 \\
 D_6 \bm{E_1}^2=0 & \bm{E_1}^2 \bm{E_2}=-2 & D_1 \bm{E_2}^2=-2 & D_2 \bm{E_2}^2=0 & D_6 \bm{E_2}^2=0
   \\
 D_1^2 \bm{E_1}=0 & D_1 D_2 \bm{E_1}=1 & D_2^2 \bm{E_1}=0 & D_1 D_6 \bm{E_1}=0 & D_2 D_6
   \bm{E_1}=0 \\
 D_6^2 \bm{E_1}=0 & \bm{E_1} \bm{E_2}^2=0 & D_1 \bm{E_1} \bm{E_2}=1 & D_2 \bm{E_1} \bm{E_2}=0 & D_6 \bm{E_1}
   \bm{E_2}=0 \\
\end{array}\right.
\end{align}}
\vspace{-0.1cm}
{\footnotesize\begin{align}
{\rm \bf (j)} :
\left\{\begin{array}{c@{~,\quad}c@{~,\quad}c@{~,\quad}c@{~,\quad}c@{~,}}
 \bm{E_1}^3=8 & \bm{E_2}^3=7 & D_1^2 \bm{E_2}=-1 & D_1 D_2 \bm{E_2}=0 & D_2^2 \bm{E_2}=0 \\
 D_1 D_6 \bm{E_2}=1 & D_2 D_6 \bm{E_2}=0 & D_6^2 \bm{E_2}=-1 & D_1 \bm{E_1}^2=-2 & D_2 \bm{E_1}^2=-3 \\
 D_6 \bm{E_1}^2=0 & \bm{E_1}^2 \bm{E_2}=-1 & D_1 \bm{E_2}^2=-1 & D_2 \bm{E_2}^2=0 & D_6 \bm{E_2}^2=-1 \\
 D_1^2 \bm{E_1}=0 & D_1 D_2 \bm{E_1}=1 & D_2^2 \bm{E_1}=1 & D_1 D_6 \bm{E_1}=0 & D_2 D_6 \bm{E_1}=0 \\
 D_6^2 \bm{E_1}=0 & \bm{E_1} \bm{E_2}^2=-1 & D_1 \bm{E_1} \bm{E_2}=1 & D_2 \bm{E_1} \bm{E_2}=0 & D_6 \bm{E_1} \bm{E_2}=0 \\
\end{array}\right.
\end{align}}
\vspace{-0.1cm}
{\footnotesize\begin{align}
{\rm \bf (k)} :
\left\{\begin{array}{c@{~,\quad}c@{~,\quad}c@{~,\quad}c@{~,\quad}c@{~,}}
 \bm{E_1}^3=7 & \bm{E_2}^3=8 & D_1^2 \bm{E_2}=0 & D_1 D_2 \bm{E_2}=0 & D_2^2 \bm{E_2}=0 \\
 D_1 D_6 \bm{E_2}=0 & D_2 D_6 \bm{E_2}=0 & D_6^2 \bm{E_2}=0 & D_1 \bm{E_1}^2=-2 & D_2
   \bm{E_1}^2=-2 \\
 D_6 \bm{E_1}^2=0 & \bm{E_1}^2 \bm{E_2}=-1 & D_1 \bm{E_2}^2=-2 & D_2 \bm{E_2}^2=0 & D_6 \bm{E_2}^2=0
   \\
 D_1^2 \bm{E_1}=0 & D_1 D_2 \bm{E_1}=1 & D_2^2 \bm{E_1}=0 & D_1 D_6 \bm{E_1}=0 & D_2 D_6
   \bm{E_1}=0 \\
 D_6^2 \bm{E_1}=0 & \bm{E_1} \bm{E_2}^2=-1 & D_1 \bm{E_1} \bm{E_2}=1 & D_2 \bm{E_1} \bm{E_2}=0 & D_6 \bm{E_1}
   \bm{E_2}=0 \\
\end{array}\right.
\end{align}}
\vspace{-0.1cm}
{\footnotesize\begin{align}
{\rm \bf (l)} :
\left\{\begin{array}{c@{~,\quad}c@{~,\quad}c@{~,\quad}c@{~,\quad}c@{~,}}
 \bm{E_1}^3=8 & \bm{E_2}^3=8 & D_1^2 \bm{E_2}=0 & D_1 D_2 \bm{E_2}=0 & D_2^2 \bm{E_2}=0 \\
 D_1 D_6 \bm{E_2}=0 & D_2 D_6 \bm{E_2}=0 & D_6^2 \bm{E_2}=0 & D_1 \bm{E_1}^2=-2 & D_2 \bm{E_1}^2=-3 \\
 D_6 \bm{E_1}^2=0 & \bm{E_1}^2 \bm{E_2}=-1 & D_1 \bm{E_2}^2=-2 & D_2 \bm{E_2}^2=0 & D_6 \bm{E_2}^2=0 \\
 D_1^2 \bm{E_1}=0 & D_1 D_2 \bm{E_1}=1 & D_2^2 \bm{E_1}=1 & D_1 D_6 \bm{E_1}=0 & D_2 D_6 \bm{E_1}=0 \\
 D_6^2 \bm{E_1}=0 & \bm{E_1} \bm{E_2}^2=-1 & D_1 \bm{E_1} \bm{E_2}=1 & D_2 \bm{E_1} \bm{E_2}=0 & D_6 \bm{E_1} \bm{E_2}=0 \\
\end{array}\right.
\end{align}}
\vspace{-0.1cm}
{\footnotesize\begin{align}
{\rm \bf (m)} :
\left\{\begin{array}{c@{~,\quad}c@{~,\quad}c@{~,\quad}c@{~,\quad}c@{~,}}
 \bm{E_1}^3=7 & \bm{E_2}^3=8 & D_1^2 \bm{E_2}=0 & D_1 D_2 \bm{E_2}=0 & D_2^2 \bm{E_2}=0 \\
 D_1 D_6 \bm{E_2}=0 & D_2 D_6 \bm{E_2}=0 & D_6^2 \bm{E_2}=0 & D_1 \bm{E_1}^2=-1 & D_2
   \bm{E_1}^2=-3 \\
 D_6 \bm{E_1}^2=-1 & \bm{E_1}^2 \bm{E_2}=0 & D_1 \bm{E_2}^2=0 & D_2 \bm{E_2}^2=0 & D_6 \bm{E_2}^2=-2
   \\
 D_1^2 \bm{E_1}=-1 & D_1 D_2 \bm{E_1}=1 & D_2^2 \bm{E_1}=1 & D_1 D_6 \bm{E_1}=1 & D_2 D_6
   \bm{E_1}=0 \\
 D_6^2 \bm{E_1}=-1 & \bm{E_1} \bm{E_2}^2=-2 & D_1 \bm{E_1} \bm{E_2}=0 & D_2 \bm{E_1} \bm{E_2}=0 & D_6 \bm{E_1}
   \bm{E_2}=1 \\
\end{array}\right.
\end{align}}
\vspace{-0.1cm}
{\footnotesize\begin{align}
{\rm \bf (n)} :
\left\{\begin{array}{c@{~,\quad}c@{~,\quad}c@{~,\quad}c@{~,\quad}c@{~,}}
 \bm{E_1}^3=7 & \bm{E_2}^3=8 & D_1^2 \bm{E_2}=0 & D_1 D_2 \bm{E_2}=0 & D_2^2 \bm{E_2}=0 \\
 D_1 D_6 \bm{E_2}=0 & D_2 D_6 \bm{E_2}=0 & D_6^2 \bm{E_2}=0 & D_1 \bm{E_1}^2=0 & D_2
   \bm{E_1}^2=-3 \\
 D_6 \bm{E_1}^2=-2 & \bm{E_1}^2 \bm{E_2}=0 & D_1 \bm{E_2}^2=0 & D_2 \bm{E_2}^2=0 & D_6 \bm{E_2}^2=-2
   \\
 D_1^2 \bm{E_1}=0 & D_1 D_2 \bm{E_1}=0 & D_2^2 \bm{E_1}=1 & D_1 D_6 \bm{E_1}=0 & D_2 D_6
   \bm{E_1}=1 \\
 D_6^2 \bm{E_1}=0 & \bm{E_1} \bm{E_2}^2=-2 & D_1 \bm{E_1} \bm{E_2}=0 & D_2 \bm{E_1} \bm{E_2}=0 & D_6 \bm{E_1}
   \bm{E_2}=1 \\
\end{array}\right.
\end{align}}
\vspace{-0.1cm}
{\footnotesize\begin{align}
{\rm \bf (o)} :
\left\{\begin{array}{c@{~,\quad}c@{~,\quad}c@{~,\quad}c@{~,\quad}c@{~,}}
 \bm{E_1}^3=8 & \bm{E_2}^3=8 & D_1^2 \bm{E_2}=0 & D_1 D_2 \bm{E_2}=0 & D_2^2 \bm{E_2}=0 \\
 D_1 D_6 \bm{E_2}=0 & D_2 D_6 \bm{E_2}=0 & D_6^2 \bm{E_2}=0 & D_1 \bm{E_1}^2=0 & D_2
   \bm{E_1}^2=-4 \\
 D_6 \bm{E_1}^2=-2 & \bm{E_1}^2 \bm{E_2}=0 & D_1 \bm{E_2}^2=0 & D_2 \bm{E_2}^2=0 & D_6 \bm{E_2}^2=-2
   \\
 D_1^2 \bm{E_1}=0 & D_1 D_2 \bm{E_1}=0 & D_2^2 \bm{E_1}=2 & D_1 D_6 \bm{E_1}=0 & D_2 D_6
   \bm{E_1}=1 \\
 D_6^2 \bm{E_1}=0 & \bm{E_1} \bm{E_2}^2=-2 & D_1 \bm{E_1} \bm{E_2}=0 & D_2 \bm{E_1} \bm{E_2}=0 & D_6 \bm{E_1}
   \bm{E_2}=1 \\
\end{array}\right.
\end{align}}
\vspace{-0.1cm}
{\footnotesize\begin{align}
{\rm \bf (p)} :
\left\{\begin{array}{c@{~,\quad}c@{~,\quad}c@{~,\quad}c@{~,\quad}c@{~,}}
 \bm{E_1}^3=7 & \bm{E_2}^3=8 & D_1^2 \bm{E_2}=0 & D_1 D_2 \bm{E_2}=0 & D_2^2 \bm{E_2}=0 \\
 D_1 D_6 \bm{E_2}=0 & D_2 D_6 \bm{E_2}=0 & D_6^2 \bm{E_2}=0 & D_1 \bm{E_1}^2=-1 & D_2
   \bm{E_1}^2=-2 \\
 D_6 \bm{E_1}^2=-1 & \bm{E_1}^2 \bm{E_2}=-1 & D_1 \bm{E_2}^2=0 & D_2 \bm{E_2}^2=0 & D_6
   \bm{E_2}^2=-2 \\
 D_1^2 \bm{E_1}=-1 & D_1 D_2 \bm{E_1}=1 & D_2^2 \bm{E_1}=0 & D_1 D_6 \bm{E_1}=1 & D_2 D_6
   \bm{E_1}=0 \\
 D_6^2 \bm{E_1}=-1 & \bm{E_1} \bm{E_2}^2=-1 & D_1 \bm{E_1} \bm{E_2}=0 & D_2 \bm{E_1} \bm{E_2}=0 & D_6
   \bm{E_1} \bm{E_2}=1 \\
\end{array}\right.
\end{align}}
\vspace{-0.1cm}
{\footnotesize\begin{align}
{\rm \bf (q)} :
\left\{\begin{array}{c@{~,\quad}c@{~,\quad}c@{~,\quad}c@{~,\quad}c@{~,}}
 \bm{E_1}^3=8 & \bm{E_2}^3=7 & D_1^2 \bm{E_2}=0 & D_1 D_2 \bm{E_2}=0 & D_2^2 \bm{E_2}=0 \\
 D_1 D_6 \bm{E_2}=0 & D_2 D_6 \bm{E_2}=0 & D_6^2 \bm{E_2}=0 & D_1 \bm{E_1}^2=0 & D_2
   \bm{E_1}^2=-3 \\
 D_6 \bm{E_1}^2=-2 & \bm{E_1}^2 \bm{E_2}=-1 & D_1 \bm{E_2}^2=0 & D_2 \bm{E_2}^2=0 & D_6 \bm{E_2}^2=-2
   \\
 D_1^2 \bm{E_1}=0 & D_1 D_2 \bm{E_1}=0 & D_2^2 \bm{E_1}=1 & D_1 D_6 \bm{E_1}=0 & D_2 D_6
   \bm{E_1}=1 \\
 D_6^2 \bm{E_1}=0 & \bm{E_1} \bm{E_2}^2=-1 & D_1 \bm{E_1} \bm{E_2}=0 & D_2 \bm{E_1} \bm{E_2}=0 & D_6 \bm{E_1}
   \bm{E_2}=1 \\
\end{array}\right.
\end{align}}
\vspace{-0.1cm}
{\footnotesize\begin{align}
{\rm \bf (r)} :
\left\{\begin{array}{c@{~,\quad}c@{~,\quad}c@{~,\quad}c@{~,\quad}c}
 \bm{E_1}^3=8 & \bm{E_2}^3=8 & D_1^2 \bm{E_2}=0 & D_1 D_2 \bm{E_2}=0 & D_2^2 \bm{E_2}=0~, \\
 D_1 D_6 \bm{E_2}=0 & D_2 D_6 \bm{E_2}=0 & D_6^2 \bm{E_2}=0 & D_1 \bm{E_1}^2=0 & D_2
   \bm{E_1}^2=-3~, \\
 D_6 \bm{E_1}^2=-2 & \bm{E_1}^2 \bm{E_2}=-1 & D_1 \bm{E_2}^2=0 & D_2 \bm{E_2}^2=0 & D_6 \bm{E_2}^2=-2~,   \\
 D_1^2 \bm{E_1}=0 & D_1 D_2 \bm{E_1}=0 & D_2^2 \bm{E_1}=1 & D_1 D_6 \bm{E_1}=0 & D_2 D_6
   \bm{E_1}=1~, \\
 D_6^2 \bm{E_1}=0 & \bm{E_1} \bm{E_2}^2=-1 & D_1 \bm{E_1} \bm{E_2}=0 & D_2 \bm{E_1} \bm{E_2}=0 & D_6 \bm{E_1}
   \bm{E_2}=1~. \\
\end{array}\right.
\end{align}}
%%%%%%%%%%%
\paragraph{Resolutions without a gauge-theory phase.} For completeness, let us also give the intersection numbers for the resolutions (s)-(x), which do not admit a gauge-theory description.
%%%
{\footnotesize\begin{align}
{\rm \bf (s)} :
\left\{\begin{array}{c@{~,\quad}c@{~,\quad}c@{~,\quad}c@{~,\quad}c@{~,}}
 \bm{E_1}^3=9 & \bm{E_2}^3=8 & D_1^2 \bm{E_2}=0 & D_1 D_2 \bm{E_2}=0 & D_2^2 \bm{E_2}=0 \\
 D_1 D_6 \bm{E_2}=1 & D_2 D_6 \bm{E_2}=0 & D_6^2 \bm{E_2}=-1 & D_1 \bm{E_1}^2=-3 & D_2
   \bm{E_1}^2=-3 \\
 D_6 \bm{E_1}^2=0 & \bm{E_1}^2 \bm{E_2}=0 & D_1 \bm{E_2}^2=-2 & D_2 \bm{E_2}^2=0 & D_6 \bm{E_2}^2=-1
   \\
 D_1^2 \bm{E_1}=1 & D_1 D_2 \bm{E_1}=1 & D_2^2 \bm{E_1}=1 & D_1 D_6 \bm{E_1}=0 & D_2 D_6
   \bm{E_1}=0 \\
 D_6^2 \bm{E_1}=0 & \bm{E_1} \bm{E_2}^2=0 & D_1 \bm{E_1} \bm{E_2}=0 & D_2 \bm{E_1} \bm{E_2}=0 & D_6 \bm{E_1}
   \bm{E_2}=0 \\
\end{array}\right.
\end{align}}
\vspace{-0.1cm}
{\footnotesize\begin{align}
{\rm \bf (t)} :
\left\{\begin{array}{c@{~,\quad}c@{~,\quad}c@{~,\quad}c@{~,\quad}c@{~,}}
 \bm{E_1}^3=8 & \bm{E_2}^3=9 & D_1^2 \bm{E_2}=1 & D_1 D_2 \bm{E_2}=0 & D_2^2 \bm{E_2}=0 \\
 D_1 D_6 \bm{E_2}=0 & D_2 D_6 \bm{E_2}=0 & D_6^2 \bm{E_2}=0 & D_1 \bm{E_1}^2=-3 & D_2
   \bm{E_1}^2=-2 \\
 D_6 \bm{E_1}^2=0 & \bm{E_1}^2 \bm{E_2}=0 & D_1 \bm{E_2}^2=-3 & D_2 \bm{E_2}^2=0 & D_6 \bm{E_2}^2=0
   \\
 D_1^2 \bm{E_1}=1 & D_1 D_2 \bm{E_1}=1 & D_2^2 \bm{E_1}=0 & D_1 D_6 \bm{E_1}=0 & D_2 D_6
   \bm{E_1}=0 \\
 D_6^2 \bm{E_1}=0 & \bm{E_1} \bm{E_2}^2=0 & D_1 \bm{E_1} \bm{E_2}=0 & D_2 \bm{E_1} \bm{E_2}=0 & D_6 \bm{E_1}
   \bm{E_2}=0 \\
\end{array}\right.
\end{align}}
\vspace{-0.1cm}
{\footnotesize\begin{align}
{\rm \bf (u)} :
\left\{\begin{array}{c@{~,\quad}c@{~,\quad}c@{~,\quad}c@{~,\quad}c@{~,}}
 \bm{E_1}^3=9 & \bm{E_2}^3=9 & D_1^2 \bm{E_2}=1 & D_1 D_2 \bm{E_2}=0 & D_2^2 \bm{E_2}=0 \\
 D_1 D_6 \bm{E_2}=0 & D_2 D_6 \bm{E_2}=0 & D_6^2 \bm{E_2}=0 & D_1 \bm{E_1}^2=-3 & D_2
   \bm{E_1}^2=-3 \\
 D_6 \bm{E_1}^2=0 & \bm{E_1}^2 \bm{E_2}=0 & D_1 \bm{E_2}^2=-3 & D_2 \bm{E_2}^2=0 & D_6 \bm{E_2}^2=0
   \\
 D_1^2 \bm{E_1}=1 & D_1 D_2 \bm{E_1}=1 & D_2^2 \bm{E_1}=1 & D_1 D_6 \bm{E_1}=0 & D_2 D_6
   \bm{E_1}=0 \\
 D_6^2 \bm{E_1}=0 & \bm{E_1} \bm{E_2}^2=0 & D_1 \bm{E_1} \bm{E_2}=0 & D_2 \bm{E_1} \bm{E_2}=0 & D_6 \bm{E_1}
   \bm{E_2}=0 \\
\end{array}\right.
\end{align}}
\vspace{-0.1cm}
{\footnotesize\begin{align}
{\rm \bf (v)} :
\left\{\begin{array}{c@{~,\quad}c@{~,\quad}c@{~,\quad}c@{~,\quad}c@{~,}}
 \bm{E_1}^3=8 & \bm{E_2}^3=9 & D_1^2 \bm{E_2}=0 & D_1 D_2 \bm{E_2}=0 & D_2^2 \bm{E_2}=0 \\
 D_1 D_6 \bm{E_2}=0 & D_2 D_6 \bm{E_2}=0 & D_6^2 \bm{E_2}=1 & D_1 \bm{E_1}^2=-1 & D_2
   \bm{E_1}^2=-2 \\
 D_6 \bm{E_1}^2=-2 & \bm{E_1}^2 \bm{E_2}=0 & D_1 \bm{E_2}^2=0 & D_2 \bm{E_2}^2=0 & D_6 \bm{E_2}^2=-3
   \\
 D_1^2 \bm{E_1}=-1 & D_1 D_2 \bm{E_1}=1 & D_2^2 \bm{E_1}=0 & D_1 D_6 \bm{E_1}=1 & D_2 D_6
   \bm{E_1}=0 \\
 D_6^2 \bm{E_1}=0 & \bm{E_1} \bm{E_2}^2=0 & D_1 \bm{E_1} \bm{E_2}=0 & D_2 \bm{E_1} \bm{E_2}=0 & D_6 \bm{E_1}
   \bm{E_2}=0 \\
\end{array}\right.
\end{align}}
\vspace{-0.1cm}
{\footnotesize\begin{align}
{\rm \bf (w)} :
\left\{\begin{array}{c@{~,\quad}c@{~,\quad}c@{~,\quad}c@{~,\quad}c@{~,}}
 \bm{E_1}^3=9 & \bm{E_2}^3=8 & D_1^2 \bm{E_2}=0 & D_1 D_2 \bm{E_2}=0 & D_2^2 \bm{E_2}=0 \\
 D_1 D_6 \bm{E_2}=0 & D_2 D_6 \bm{E_2}=0 & D_6^2 \bm{E_2}=1 & D_1 \bm{E_1}^2=0 & D_2
   \bm{E_1}^2=-3 \\
 D_6 \bm{E_1}^2=-3 & \bm{E_1}^2 \bm{E_2}=0 & D_1 \bm{E_2}^2=0 & D_2 \bm{E_2}^2=0 & D_6 \bm{E_2}^2=-3
   \\
 D_1^2 \bm{E_1}=0 & D_1 D_2 \bm{E_1}=0 & D_2^2 \bm{E_1}=1 & D_1 D_6 \bm{E_1}=0 & D_2 D_6
   \bm{E_1}=1 \\
 D_6^2 \bm{E_1}=1 & \bm{E_1} \bm{E_2}^2=0 & D_1 \bm{E_1} \bm{E_2}=0 & D_2 \bm{E_1} \bm{E_2}=0 & D_6 \bm{E_1}
   \bm{E_2}=0 \\
\end{array}\right.
\end{align}}
\vspace{-0.1cm}
{\footnotesize\begin{align}
{\rm \bf (x)} :
\left\{\begin{array}{c@{~,\quad}c@{~,\quad}c@{~,\quad}c@{~,\quad}c}
 \bm{E_1}^3=9 & \bm{E_2}^3=9 & D_1^2 \bm{E_2}=0 & D_1 D_2 \bm{E_2}=0 & D_2^2 \bm{E_2}=0~, \\
 D_1 D_6 \bm{E_2}=0 & D_2 D_6 \bm{E_2}=0 & D_6^2 \bm{E_2}=1 & D_1 \bm{E_1}^2=0 & D_2
   \bm{E_1}^2=-3~, \\
 D_6 \bm{E_1}^2=-3 & \bm{E_1}^2 \bm{E_2}=0 & D_1 \bm{E_2}^2=0 & D_2 \bm{E_2}^2=0 & D_6 \bm{E_2}^2=-3~,   \\
 D_1^2 \bm{E_1}=0 & D_1 D_2 \bm{E_1}=0 & D_2^2 \bm{E_1}=1 & D_1 D_6 \bm{E_1}=0 & D_2 D_6
   \bm{E_1}=1~, \\
 D_6^2 \bm{E_1}=1 & \bm{E_1} \bm{E_2}^2=0 & D_1 \bm{E_1} \bm{E_2}=0 & D_2 \bm{E_1} \bm{E_2}=0 & D_6 \bm{E_1}
   \bm{E_2}=0~. \\
\end{array}\right.
\end{align}}
%%%%%%%%%%%%%

\section{Beetle geometry prepotential and gauge-theory phases}\label{app: more on beetle}
In this appendix, we consider more in detail the M-theory prepotential for the beetle geometry of section~\ref{sec: beetle}, and we verify that it matches precisely the prepotential of the two S-dual field theory descriptions, in the appropriate K\"ahler chambers.  

\subsection{M-theory prepotential}
We parameterize the K\"ahler class as in equation \eqref{def S mu nu beetle}, namely:
\be
S= \mu_1 D_1 + \mu_2 D_2 + \mu_6 D_6 + \nu_1 \bE_1+ \nu_2 \bE_2~.
\ee
The M-theory prepotential is then:
\be\label{CF Mth beetle gen}
\CF = -{1\ov 6} \bE_a \bE_b \bE_c\, \nu_a \nu_b\nu_c -\half D_i  \bE_a \bE_b  \, \mu_i \nu_a \nu_b
 -\half D_i D_j \bE_a\, \mu_i \mu_j \nu_a~,
\ee
where $a, b, c=1, 2$ and $i, j=1, 2, 6$, and repeated indices are summed over. Here and in the following, we neglect the constant ($\nu$-independent) term, for simplicity. Plugging the intersection numbers listed in Appendix~\ref{app:all intersection numbers beetle} into \eqref{CF Mth beetle gen}, we find the geometric prepotential in every chamber.

\paragraph{Resolutions with a vertical reduction:} For the first 6 resolutions of Figure~\ref{fig:all beetles}, we have:
{\footnotesize\bea\nn
 \mathcal{F}_{a} &= -\frac{4 \nu _1^3}{3}+\left(\mu _1+\mu _2\right) \nu _1^2+\nu _2 \nu _1^2-\mu _1 \mu _2 \nu _1-\mu _1 \nu _2 \nu _1-\nu _2^3+\left(\frac{\mu _1}{2}+\frac{\mu _6}{2}\right) \nu _2^2+\left(\frac{\mu _1^2}{2}-\mu _6 \mu _1+\frac{\mu _6^2}{2}\right) \nu _2~, \\
 \mathcal{F}_{b} &= -\frac{7 \nu _1^3}{6}+\left(\mu _1+\mu _2\right) \nu _1^2+\frac{1}{2} \nu _2 \nu _1^2+\frac{1}{2} \nu _2^2 \nu _1-\mu _1 \mu _2 \nu _1-\mu _1 \nu _2 \nu _1-\frac{7 \nu _2^3}{6}+\left(\frac{\mu _1}{2}+\frac{\mu _6}{2}\right) \nu _2^2+\left(\frac{\mu _1^2}{2}-\mu _6 \mu _1+\frac{\mu _6^2}{2}\right) \nu _2~, \\
 \mathcal{F}_{c} &= -\frac{4 \nu _1^3}{3}+\left(\frac{3 \mu _1}{2}+\mu _2\right) \nu _1^2+\left(-\frac{\mu _1^2}{2}-\mu _2 \mu _1\right) \nu _1-\frac{4 \nu _2^3}{3}+\left(\mu _1+\frac{\mu _6}{2}\right) \nu _2^2+\left(\frac{\mu _6^2}{2}-\mu _1 \mu _6\right) \nu _2~, \\
 \mathcal{F}_{d} &= -\nu _1^3+\left(\frac{\mu _1}{2}+\mu _2+\frac{\mu _6}{2}\right) \nu _1^2+\nu _2^2 \nu _1+\left(\frac{\mu _1^2}{2}-\mu _2 \mu _1-\mu _6 \mu _1+\frac{\mu _6^2}{2}\right) \nu _1-\mu _6 \nu _2 \nu _1-\frac{4 \nu _2^3}{3}+\mu _6 \nu _2^2~, \\
 \mathcal{F}_{e} &= -\frac{7 \nu _1^3}{6}+\left(\frac{\mu _1}{2}+\mu _2+\frac{\mu _6}{2}\right) \nu _1^2+\frac{1}{2} \nu _2 \nu _1^2+\frac{1}{2} \nu _2^2 \nu _1+\left(\frac{\mu _1^2}{2}-\mu _2 \mu _1-\mu _6 \mu _1+\frac{\mu _6^2}{2}\right) \nu _1-\mu _6 \nu _2 \nu _1-\frac{7 \nu _2^3}{6}+\mu _6 \nu _2^2~, \\
 \mathcal{F}_{f} &= -\frac{4 \nu _1^3}{3}+\left(\frac{\mu _1}{2}+\mu _2+\mu _6\right) \nu _1^2+\left(\frac{\mu _1^2}{2}-\mu _2 \mu _1-\mu _6 \mu _1\right) \nu _1-\frac{4 \nu _2^3}{3}+\frac{3}{2} \mu _6 \nu _2^2-\frac{1}{2} \mu _6^2 \nu _2~.
\eea}

\paragraph{Resolutions (g) to (r):}
{\footnotesize\bea\nn
 \mathcal{F}_{g} &= -\frac{4 \nu _1^3}{3}+\left(\mu _1+\mu _2\right) \nu _1^2+\nu _2 \nu _1^2-\mu _1 \mu _2 \nu _1-\mu _1 \nu _2 \nu _1-\frac{7 \nu _2^3}{6}+\left(\frac{\mu _1}{2}+\frac{\mu _6}{2}\right) \nu _2^2+\left(\frac{\mu _1^2}{2}-\mu _6 \mu _1+\frac{\mu _6^2}{2}\right) \nu _2~, \\
 \mathcal{F}_{h} &= -\frac{4 \nu _1^3}{3}+\left(\mu _1+\mu _2\right) \nu _1^2+\nu _2 \nu _1^2-\mu _1 \mu _2 \nu _1-\mu _1 \nu _2 \nu _1-\frac{7 \nu _2^3}{6}+\mu _1 \nu _2^2~, \\
 \mathcal{F}_{i} &= -\frac{4 \nu _1^3}{3}+\left(\mu _1+\mu _2\right) \nu _1^2+\nu _2 \nu _1^2-\mu _1 \mu _2 \nu _1-\mu _1 \nu _2 \nu _1-\frac{4 \nu _2^3}{3}+\mu _1 \nu _2^2~, \\
 \mathcal{F}_{j} &= -\frac{4 \nu _1^3}{3}+\left(\mu _1+\frac{3 \mu _2}{2}\right) \nu _1^2+\frac{1}{2} \nu _2 \nu _1^2+\frac{1}{2} \nu _2^2 \nu _1+\left(-\frac{\mu _2^2}{2}-\mu _1 \mu _2\right) \nu _1-\mu _1 \nu _2 \nu _1-\frac{7 \nu _2^3}{6}+\left(\frac{\mu _1}{2}+\frac{\mu _6}{2}\right) \nu _2^2\\ &\qquad+\left(\frac{\mu _1^2}{2}-\mu _6 \mu _1+\frac{\mu _6^2}{2}\right) \nu _2~, \\
 \mathcal{F}_{k} &= -\frac{7 \nu _1^3}{6}+\left(\mu _1+\mu _2\right) \nu _1^2+\frac{1}{2} \nu _2 \nu _1^2+\frac{1}{2} \nu _2^2 \nu _1-\mu _1 \mu _2 \nu _1-\mu _1 \nu _2 \nu _1-\frac{4 \nu _2^3}{3}+\mu _1 \nu _2^2~, \\
 \mathcal{F}_{l} &= -\frac{4 \nu _1^3}{3}+\left(\mu _1+\frac{3 \mu _2}{2}\right) \nu _1^2+\frac{1}{2} \nu _2 \nu _1^2+\frac{1}{2} \nu _2^2 \nu _1+\left(-\frac{\mu _2^2}{2}-\mu _1 \mu _2\right) \nu _1-\mu _1 \nu _2 \nu _1-\frac{4 \nu _2^3}{3}+\mu _1 \nu _2^2~,
\eea
\bea\nn
\mathcal{F}_{m} &= -\frac{7 \nu _1^3}{6}+\left(\frac{\mu _1}{2}+\frac{3 \mu _2}{2}+\frac{\mu _6}{2}\right) \nu _1^2+\nu _2^2 \nu _1+\left(\frac{\mu _1^2}{2}-\mu _2 \mu _1-\mu _6 \mu _1-\frac{\mu _2^2}{2}+\frac{\mu _6^2}{2}\right) \nu _1-\mu _6 \nu _2 \nu _1-\frac{4 \nu _2^3}{3}+\mu _6 \nu _2^2~, \\
 \mathcal{F}_{n} &= -\frac{7 \nu _1^3}{6}+\left(\frac{3 \mu _2}{2}+\mu _6\right) \nu _1^2+\nu _2^2 \nu _1+\left(-\frac{\mu _2^2}{2}-\mu _6 \mu _2\right) \nu _1-\mu _6 \nu _2 \nu _1-\frac{4 \nu _2^3}{3}+\mu _6 \nu _2^2~, \\
 \mathcal{F}_{o} &= -\frac{4 \nu _1^3}{3}+\left(2 \mu _2+\mu _6\right) \nu _1^2+\nu _2^2 \nu _1+\left(-\mu _2^2-\mu _6 \mu _2\right) \nu _1-\mu _6 \nu _2 \nu _1-\frac{4 \nu _2^3}{3}+\mu _6 \nu _2^2~, \\
 \mathcal{F}_{p} &= -\frac{7 \nu _1^3}{6}+\left(\frac{\mu _1}{2}+\mu _2+\frac{\mu _6}{2}\right) \nu _1^2+\frac{1}{2} \nu _2 \nu _1^2+\frac{1}{2} \nu _2^2 \nu _1+\left(\frac{\mu _1^2}{2}-\mu _2 \mu _1-\mu _6 \mu _1+\frac{\mu _6^2}{2}\right) \nu _1-\mu _6 \nu _2 \nu _1-\frac{4 \nu _2^3}{3}+\mu _6 \nu _2^2~, \\
 \mathcal{F}_{q} &= -\frac{4 \nu _1^3}{3}+\left(\frac{3 \mu _2}{2}+\mu _6\right) \nu _1^2+\frac{1}{2} \nu _2 \nu _1^2+\frac{1}{2} \nu _2^2 \nu _1+\left(-\frac{\mu _2^2}{2}-\mu _6 \mu _2\right) \nu _1-\mu _6 \nu _2 \nu _1-\frac{7 \nu _2^3}{6}+\mu _6 \nu _2^2~, \\
 \mathcal{F}_{r} &= -\frac{4 \nu _1^3}{3}+\left(\frac{3 \mu _2}{2}+\mu _6\right) \nu _1^2+\frac{1}{2} \nu _2 \nu _1^2+\frac{1}{2} \nu _2^2 \nu _1+\left(-\frac{\mu _2^2}{2}-\mu _6 \mu _2\right) \nu _1-\mu _6 \nu _2 \nu _1-\frac{4 \nu _2^3}{3}+\mu _6 \nu _2^2~. \\
\eea
}

\paragraph{Resolutions (s) to (x):}
{\footnotesize\bea\nn
 \mathcal{F}_{s} &= -\frac{3 \nu _1^3}{2}+\left(\frac{3 \mu _1}{2}+\frac{3 \mu _2}{2}\right) \nu _1^2+\left(-\frac{\mu _1^2}{2}-\mu _2 \mu _1-\frac{\mu _2^2}{2}\right) \nu _1-\frac{4 \nu _2^3}{3}+\left(\mu _1+\frac{\mu _6}{2}\right) \nu _2^2+\left(\frac{\mu _6^2}{2}-\mu _1 \mu _6\right) \nu _2~, \\
 \mathcal{F}_{t} &= -\frac{4 \nu _1^3}{3}+\left(\frac{3 \mu _1}{2}+\mu _2\right) \nu _1^2+\left(-\frac{\mu _1^2}{2}-\mu _2 \mu _1\right) \nu _1-\frac{3 \nu _2^3}{2}+\frac{3}{2} \mu _1 \nu _2^2-\frac{1}{2} \mu _1^2 \nu _2~, \\
 \mathcal{F}_{u} &= -\frac{3 \nu _1^3}{2}+\left(\frac{3 \mu _1}{2}+\frac{3 \mu _2}{2}\right) \nu _1^2+\left(-\frac{\mu _1^2}{2}-\mu _2 \mu _1-\frac{\mu _2^2}{2}\right) \nu _1-\frac{3 \nu _2^3}{2}+\frac{3}{2} \mu _1 \nu _2^2-\frac{1}{2} \mu _1^2 \nu _2~, \\
 \mathcal{F}_{v} &= -\frac{4 \nu _1^3}{3}+\left(\frac{\mu _1}{2}+\mu _2+\mu _6\right) \nu _1^2+\left(\frac{\mu _1^2}{2}-\mu _2 \mu _1-\mu _6 \mu _1\right) \nu _1-\frac{3 \nu _2^3}{2}+\frac{3}{2} \mu _6 \nu _2^2-\frac{1}{2} \mu _6^2 \nu _2~, \\
 \mathcal{F}_{w} &= -\frac{3 \nu _1^3}{2}+\left(\frac{3 \mu _2}{2}+\frac{3 \mu _6}{2}\right) \nu _1^2+\left(-\frac{\mu _2^2}{2}-\mu _6 \mu _2-\frac{\mu _6^2}{2}\right) \nu _1-\frac{4 \nu _2^3}{3}+\frac{3}{2} \mu _6 \nu _2^2-\frac{1}{2} \mu _6^2 \nu _2~, \\
 \mathcal{F}_{x} &= -\frac{3 \nu _1^3}{2}+\left(\frac{3 \mu _2}{2}+\frac{3 \mu _6}{2}\right) \nu _1^2+\left(-\frac{\mu _2^2}{2}-\mu _6 \mu _2-\frac{\mu _6^2}{2}\right) \nu _1-\frac{3 \nu _2^3}{2}+\frac{3}{2} \mu _6 \nu _2^2-\frac{1}{2} \mu _6^2 \nu _2~.
 \eea }

\subsection{Matching to the $SU(2)\times SU(2)$ chambers}
The resolutions (a) to (f) can be matched to the six field theory chambers of the $SU(2)\times SU(2)$ quiver. The chambers are shown in equation~\ref{geom to gauge chamber su2su2}. It is convenient to introduce the notation:
\bea\label{thetas su2su2}
&\vartheta_a = (1,0,1,0)~, \qquad\qquad &&  \vartheta_d = (1,1,0,0)~,\\
&\vartheta_b = (1,1,1,0)~, \qquad\qquad  &&\vartheta_e = (1,0,0,0)~,\\
&\vartheta_c = (1,1,1,1)~,\qquad \qquad  &&\vartheta_f = (0,0,0,0)~.
\eea
Here, the vectors $\vartheta$ denote the field-theory chambers~\eqref{geom to gauge chamber su2su2} in the obvious way; the entries in the vector are $1$ or $0$ depending on whether the hypermultiplet real masses:
 \be\label{Ms su2su2}
\CM \equiv \left(\varphi_1+ \varphi_2+m~,\; \varphi_1- \varphi_2+m~, \;-\varphi_1+ \varphi_2+m~, \;-\varphi_1- \varphi_2+m\right)~,
\ee
are positive or negative, respectively. In this notation, the $SU(2)\times SU(2)$ prepotential \eqref{CF su2 su2 quiver} takes the simple form:
\be
\CF_{\rm \h x}^{SU(2)\times SU(2)}=  h_1 \varphi_1^2 +  h_2 \varphi_2^2 + {4\ov 3} \varphi_1^3 + {4\ov 3} \varphi_2^3 -{1\ov 6} \sum_{\alpha=1}^4 {\theta_{\h x}}^\alpha (\CM_\alpha)^3~,
\ee
where $\h x= (a, \cdots , f)$ runs over the 6 chambers, and $\alpha$ runs over the components of the vectors \eqref{thetas su2su2} and \eqref{Ms su2su2}. Plugging in the relations \eqref{mu to geom beetle} into the M-theory prepotentials given above, we find perfect agreement with the field theory in all 6 chambers (modulo the constant terms, which we did not keep track of).

\subsection{Matching to the $SU(3)$, $N_f=2$ chambers}
The 16 resolutions (a), (b), (d), (e) and (g) to (r) can be matched to the field theory chambers of the $SU(3)$, $N_f=2$ field theory. Using the same notation as above, we find:
\bea\label{thetas su3}
&\vartheta_a = (1,1,0,1,1,0), \qquad\qquad &&  \vartheta_k = (1,0,0,1,1,1)~,\\
&\vartheta_b = (1,0,0,1,1,0), \qquad\qquad &&  \vartheta_l = (0,0,0,1,1,1)~,\\
&\vartheta_d = (1,0,0,1,0,0), \qquad\qquad &&  \vartheta_m = (0,0,0,1,0,0)~,\\
&\vartheta_e = (1,1,0,1,0,0), \qquad\qquad &&  \vartheta_n = (1,0,0,0,0,0)~,\\
&\vartheta_g = (1,1,1,1,1,0), \qquad\qquad &&  \vartheta_o = (0,0,0,0,0,0)~,\\
&\vartheta_h = (1,1,0,1,1,1), \qquad\qquad &&  \vartheta_p = (1,1,1,1,0,0)~,\\
&\vartheta_i = (1,1,1,1,1,1), \qquad\qquad &&  \vartheta_q = (1,1,0,0,0,0)~,\\
&\vartheta_j = (0,0,0,1,1,0), \qquad\qquad &&  \vartheta_r = (1,1,1,0,0,0)~.
\eea
Here, the 6 hypermultiplet modes are:
\be
\CM\equiv \left(\t\varphi_1 + \t m_1~, \, -\t\varphi_1+\t\varphi_2 + \t m_1~, \,-\t\varphi_2+ \t m_1~, \;
\t\varphi_1 + \t m_2~, \, -\t\varphi_1+\t\varphi_2 + \t m_2~, \,-\t\varphi_2+ \t m_2\right)
\ee
The gauge-theory prepotential \eqref{F SU3 nf2 phase} in chamber $\h x$ then reads:
\bea\label{F SU3 nf2 phase}
&\CF_{\h x}^{SU(3), N_f=2}&=&\quad \t h_0 (\t \varphi_1^2 + \t \varphi_2^2 -\t\varphi_1 \t\varphi_2)+ \half (\t\varphi_1^2\t\varphi_2-\t\varphi_1\t\varphi_2^2)\\
&&&\quad+ {4\ov 3}(\t\varphi_1^3+\t\varphi_2^3) -\half (\t\varphi_1^2\t\varphi_2+\t\varphi_1\t\varphi_2^2) 
 -{1\ov 6} \sum_{\alpha=1}^6  {\theta_{\h x}}^\alpha (\CM_\alpha)^3~.
\eea
Plugging in the relations \eqref{mu to geom beetle SU3} into the M-theory prepotentials given above, we again find perfect agreement with the field theory description, in all 16 chambers \eqref{thetas su3}.

\subsection{Selected chambers of the deformed beetle SCFT}
\begin{figure}[ht]
\begin{center}
\begin{tabular}{ccc}
\includegraphics[width=2.8in]{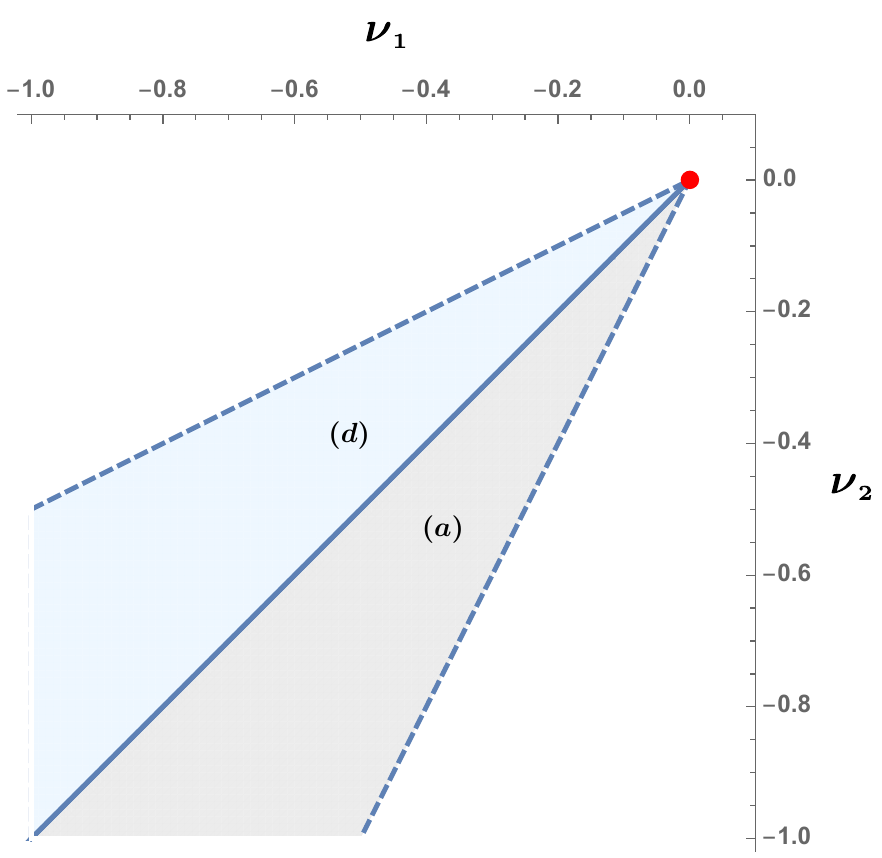}&$\qquad$&\includegraphics[width=2.8in]{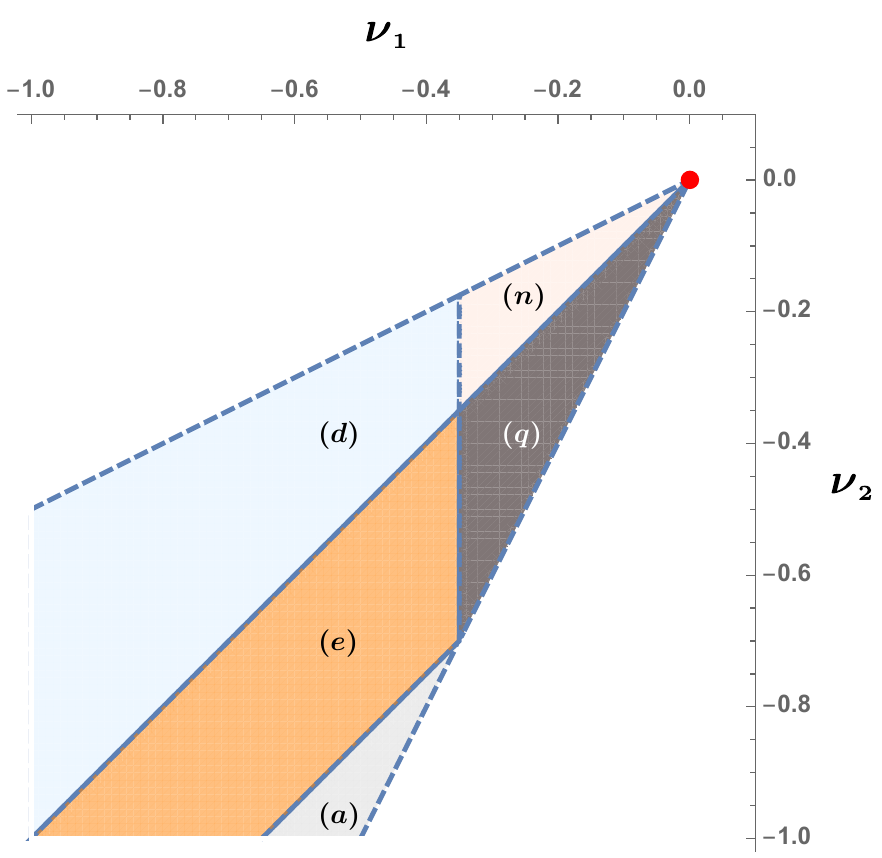}\\
{\small (i): phases: (a), (d)}& & {\small (ii): phases: (a), (d), (e), (q), (n)}\\
{\small  $\bm{\mu} = (0,0,0)$}&&{\small $\bm{\mu} = (0.35,0,0)$}\\
&&\\
\includegraphics[width=2.8in]{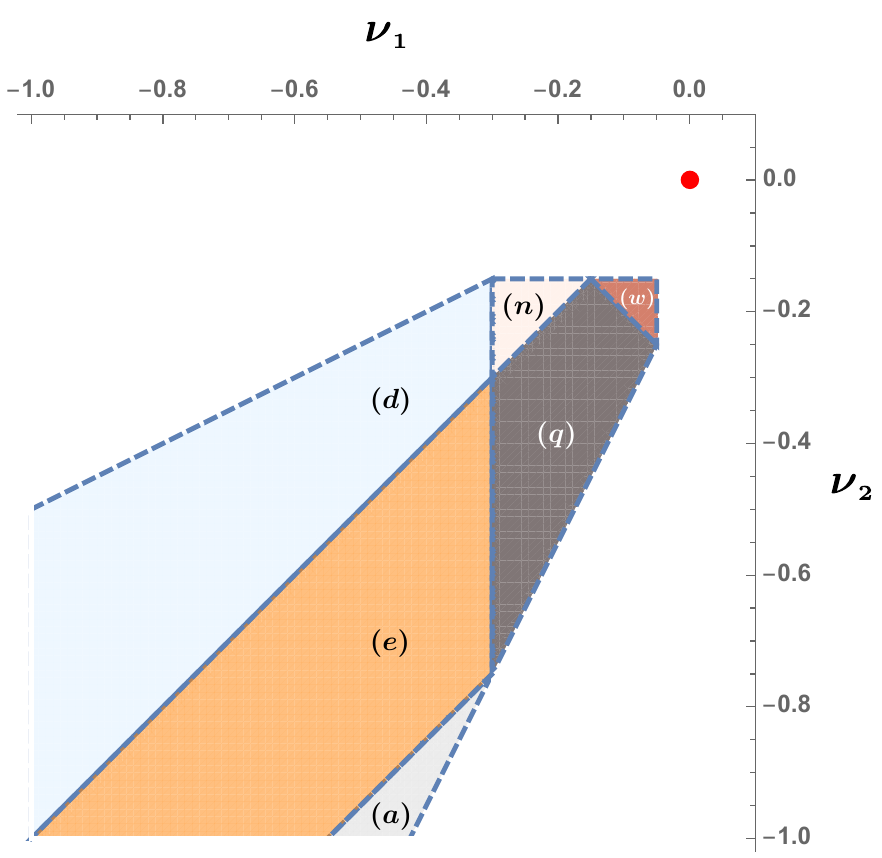}& &\includegraphics[width=2.8in]{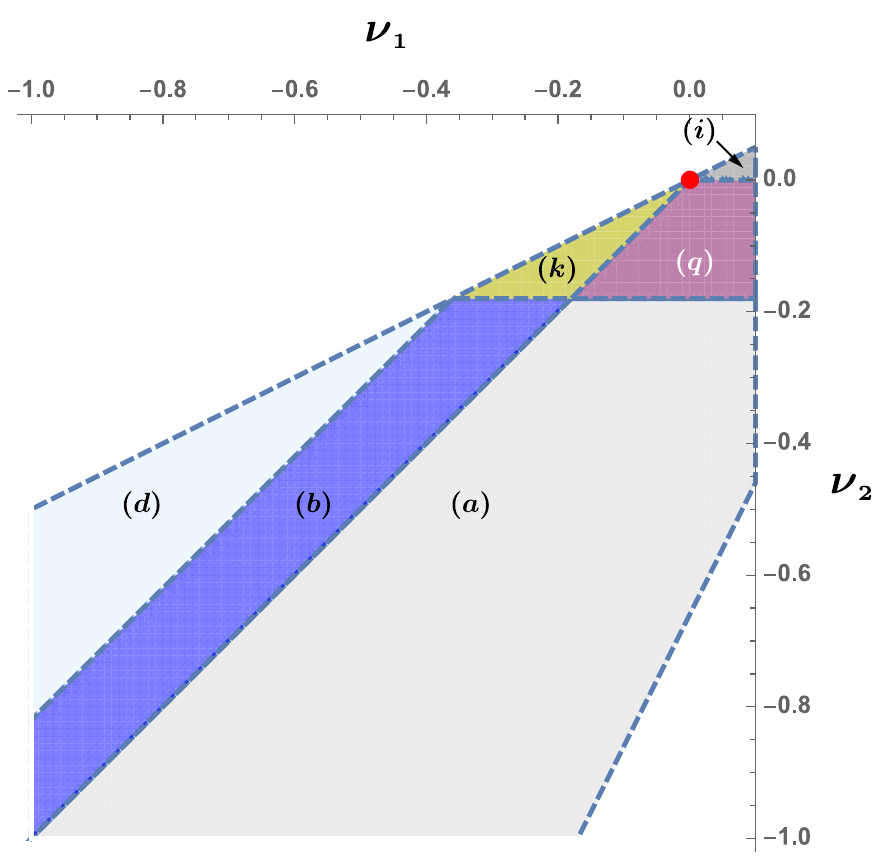}\\
{\small (iii): phases: (a), (d), (e), (q), (n), (w)} & &{\small (iv): phases: (a), (b), (d), (i), (k), (q)}\\
{\small  $\bm{\mu} =  (0.15,0.15,-0.30)$}&&{\small $\bm{\mu} =  (0.28,0.66,0.46)$}\\
\end{tabular}
\end{center}
\caption{Sample slices of the moduli space of the beetle geometry. In (i) we see the Coulomb Branch of the beetle SCFT, where chambers (b) and (e) degenerate along the wall. Turning on distinct mass deformations can open up distinct gauge theory phases, some of which may be connected to the origin as in (ii) and (iv), but also non-gauge theoretic phase such as (w) in (iii). \label{fig:beetle slices}}
\end{figure} 
%%%%%
\begin{figure}[ht]
\begin{center}
\begin{tabular}{ccc}
\includegraphics[width=2.8in]{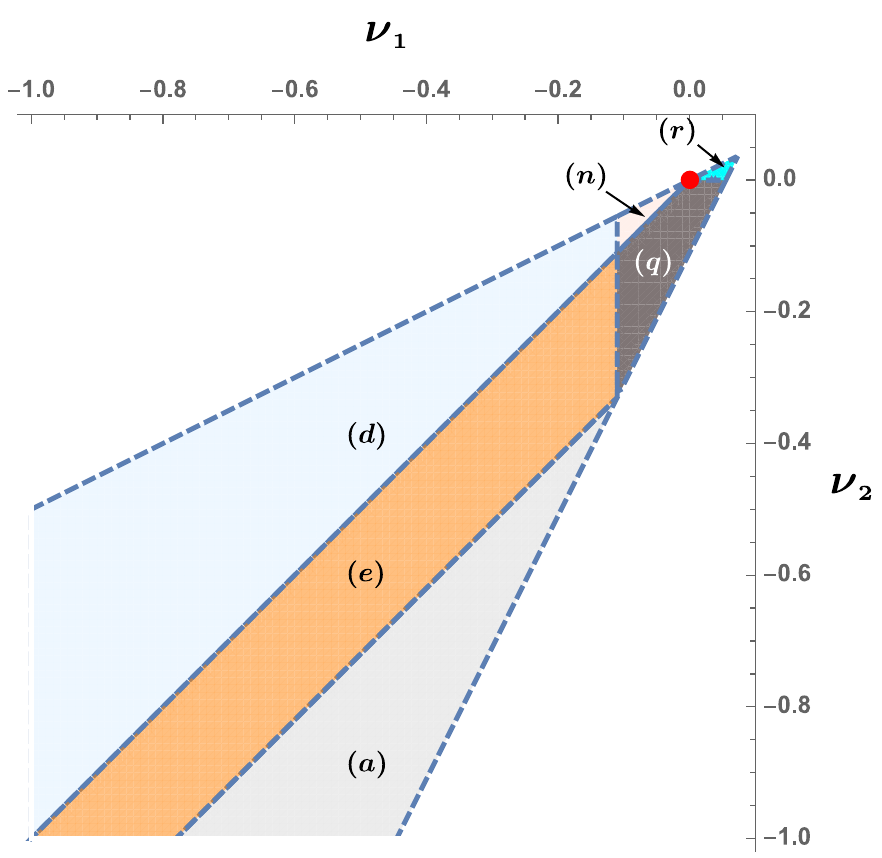}&&\includegraphics[width=2.8in]{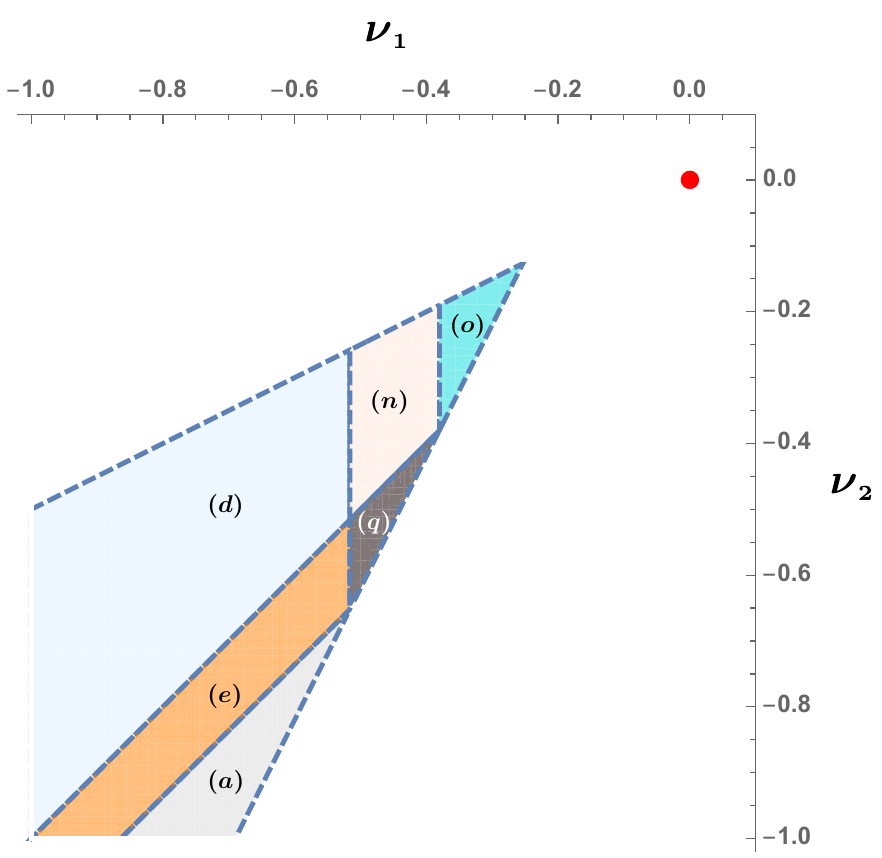}\\
{\small (v): phases: (a), (d), (e), (n), (q), (r)}& & {\small (vi): phases: (a), (d), (e), (n), (o), (q)}\\
{\small  $\bm{\mu} = (0.35,0.11,0.13)$}&&{\small $\bm{\mu} = (0.09,-0.38,-0.05)$}\\
&&\\
\includegraphics[width=2.8in]{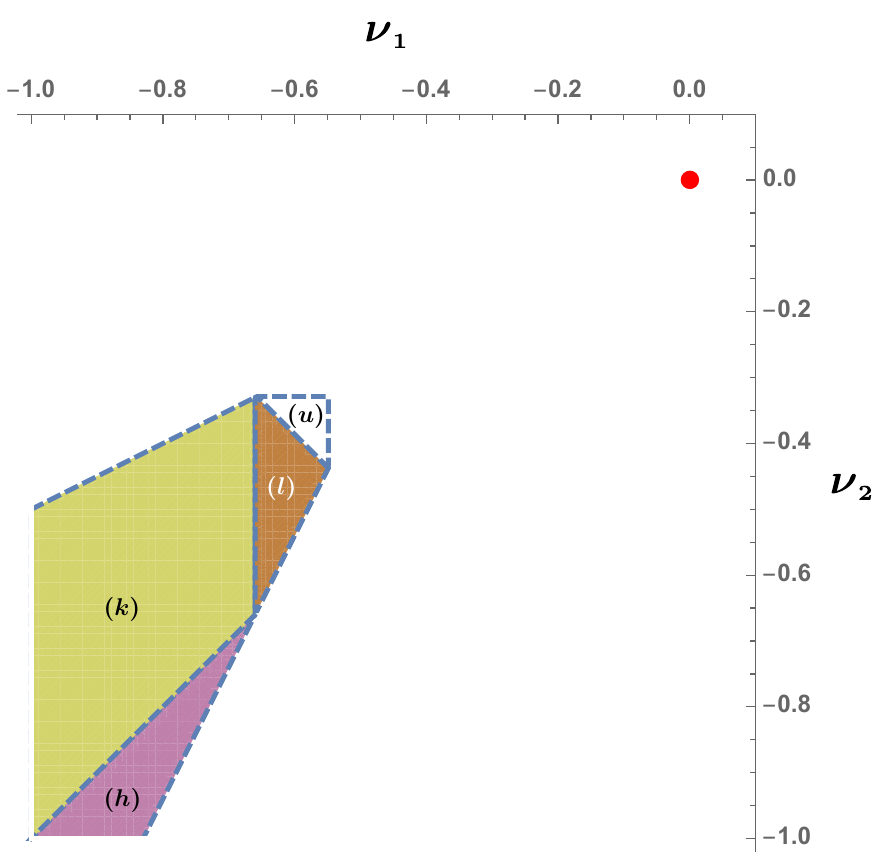}& &\includegraphics[width=2.8in]{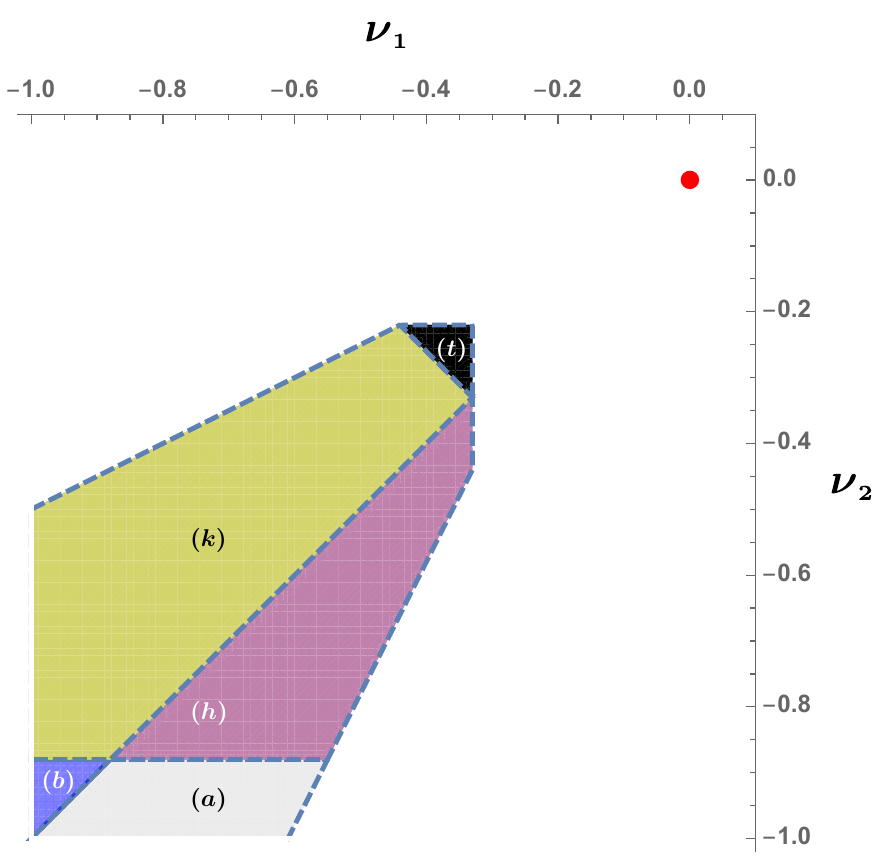}\\
{\small (vii): phases: (h), (k), (l), (u)} & &{\small (viii): phases: (a), (b), (h), (k), (t)}\\
{\small  $\bm{\mu} = (-0.99,-0.66,0.66)$}&&{\small $\bm{\mu} = (-0.66,-0.22,0.22)$}\\
\end{tabular}
\end{center}
\caption{Sample slices of the moduli space of the beetle geometry, (contd). Turning on different mass deformations reveals more $SU(3)$ $N_f=2$ phases such as (q), (n), (o) and (r) as in (i) and (ii), but also non-gauge theoretic phases such as (u) in (vii) and (t) in (viii).\label{fig:beetle slices contd}}
\end{figure} 
%%%%%%%
To conclude this appendix, we provide some representative examples of the phases of the beetle geometry, which illustrate the essential features of the phase structure captured by geometry---see Figures~\ref{fig:beetle slices} and \ref{fig:beetle slices contd}. The phase diagram of the beetle geometry, parametrized by $(\bm{\nu};\bm{\mu}) \equiv (\nu_1, \nu_2; \mu_1, \mu_2, \mu_6)$, is a five dimensional region, given by the (disjoint) union of the regions described by the defining inequalities of 24 K\"{a}hler chambers. We visualize it by taking slices at fixed values of $(\mu_1, \mu_2, \mu_6)$, revealing different chambers. Certain chambers vanish altogether for some ranges of $(\bm{\nu};\bm{\mu})$, whereas other ones collapse to real codimension-one boundaries in this parameter space (along which flops may occur).  On the various plots, the origin $(\nu_1,\nu_2) = (0,0)$ is denoted by a red dot. For the sake of clarity, in Figures~\ref{fig:beetle slices} and \ref{fig:beetle slices contd}, we only indicate chambers that have finite area in parameter space in the slices that we consider. Note also that, when $\bm{\mu}\neq 0$, the origin $\nu_1=\nu_2=0$ is not generally the origin of the gauge-theory Coulomb branch, when such a description is available, since the map \eqref{mu to geom beetle} between $\nu_1,\nu_2$ and $\varphi_1, \varphi_2$ for $SU(2)\times SU(2)$ is non-trivial, and similarly for the $SU(3)$ phases.

Flops can occur when two chambers are separated by interior walls. This is consistent with the allowed flop transitions between resolutions, as one can infer from Figure~\ref{fig:all beetles}. For instance, consider the plot (iii) in Figure~\ref{fig:beetle slices}. As we can see from the triangulations in Figure~\ref{fig:all beetles}, one can indeed start from resolution (w), then go to resolution (q) by flopping a single curve, then go to either resolution (n) or resolution (e) by a single flop, and so on and so forth.

%%%%%%%%%%%%%

\section{The conifold and the hypermultiplet}\label{app: conifold}
In this appendix, we consider a somewhat degenerate case: ``rank-zero'' isolated toric CY$_3$ singularities, whose toric diagrams have no internal points.  The only such singularity is the conifold, whose toric diagram is shown in Figure~\ref{fig:conifold 1}. It defines the simplest 5d SCFT, a free massless hypermultiplet. 

\subsection{Two ways of slicing a conifold}
\begin{figure}[t]
\begin{center}
\subfigure[\small  Conifold.]{
\includegraphics[height=2.5cm]{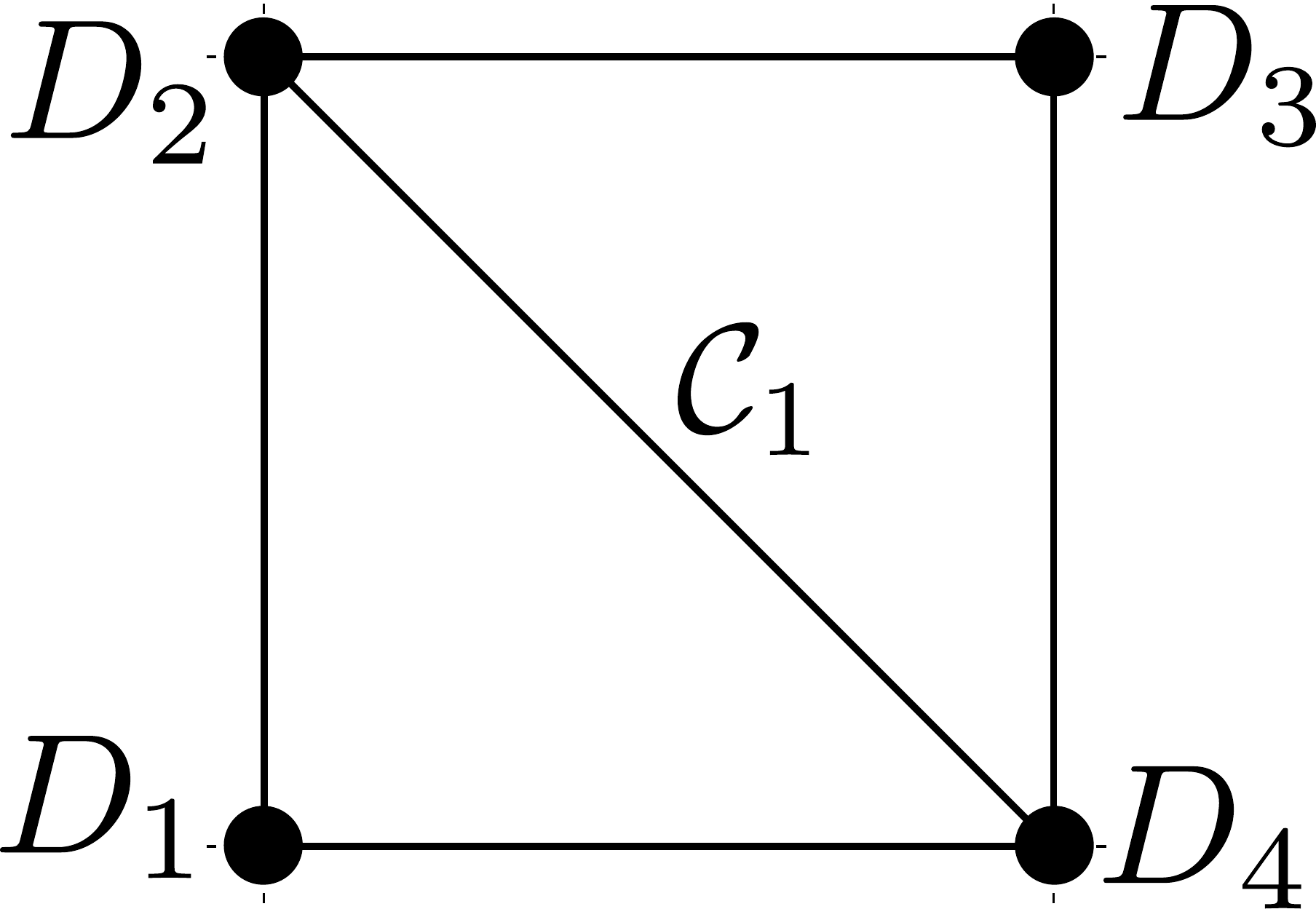}\label{fig:conifold 1}}\quad 
\subfigure[\small Conifold, bis.]{
\includegraphics[height=2.5cm]{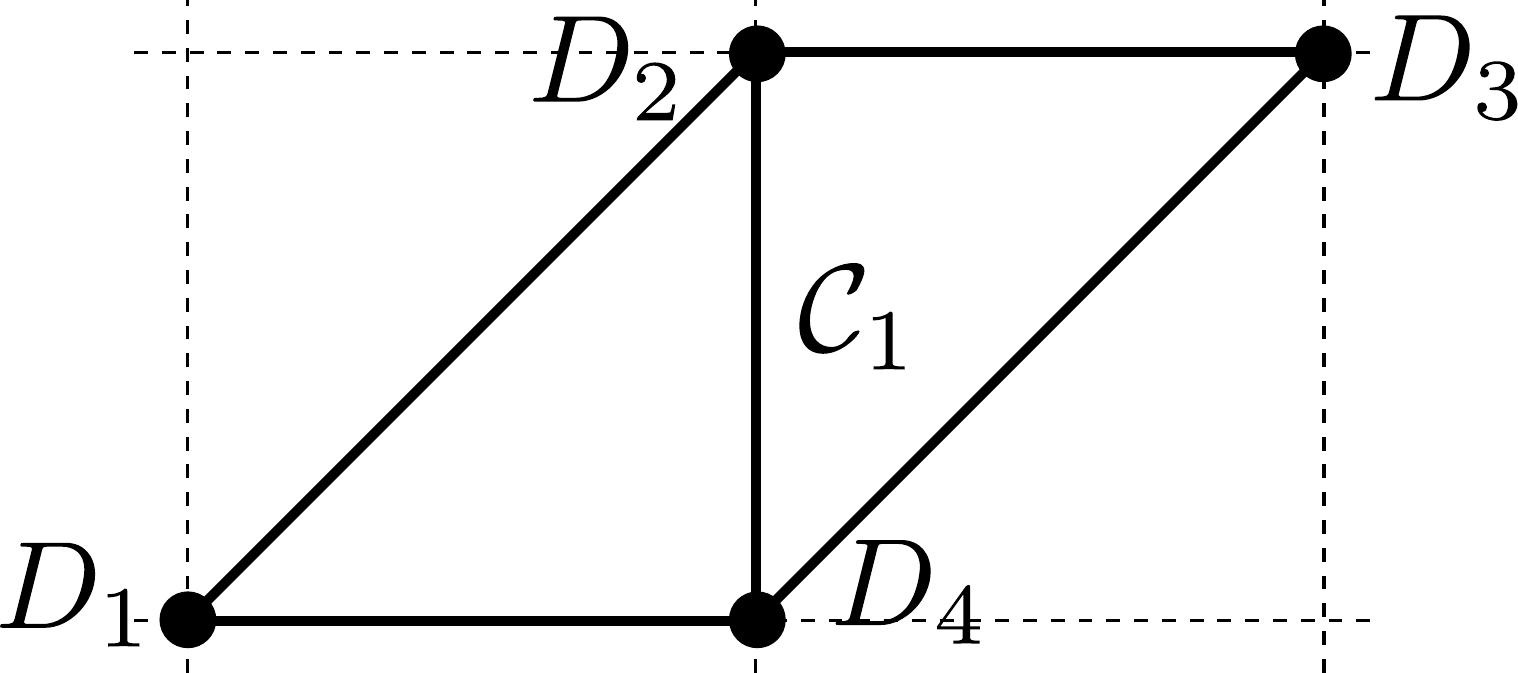}\label{fig:conifold 2}}\quad
\subfigure[\small  IIA profile.]{
\includegraphics[height=3cm]{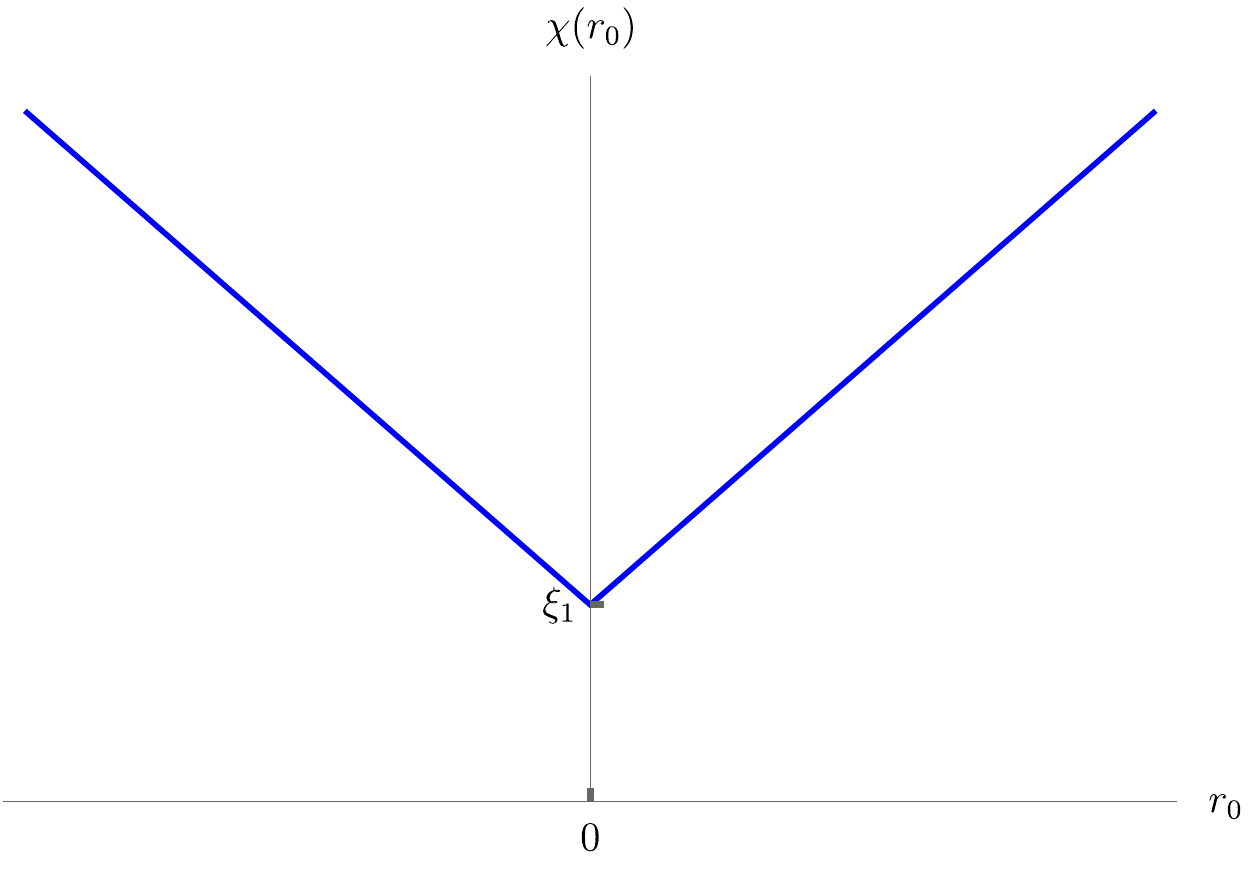}\label{fig:U10 coni}}
\caption{The two toric diagrams are related by an $SL(2, \Z)$ transformation. The vertical reduction of the second toric diagram gives the IIA profile displayed on the right. \label{fig:rankzero sings}}
 \end{center} 
 \end{figure}

The GLSM of the conifold reads:
\be\label{conifold GLSM}
\begin{tabular}{l|cccc|c}
 & $D_1$ &$D_2$& $D_3$&$D_4$& \\
 \hline
$\CC_1$    &$1$&$-1$ &$1$ & $-1$& $\xi_1$\\
\hline\hline
$U(1)_M$&$1$&$-1$ &$0$ &$0$ &$r_0$\\
\hline\hline
$U(1)_{M'}$&$0$&$-1$ &$0$ &$1$ &$r_0$
 \end{tabular}
\ee
Here we also introduced two distinct M-theory circles, to be discussed momentarily. 
The small resolution of the conifold has a single exceptional curve, $\CC_1 \cong \P^1$, with volume $\xi_1 >0$. The toric divisors are as indicated on Fig.~\ref{fig:conifold 1}.
%%%%%%%%%
 %%%%%%%%%%%

\paragraph{Conifold as hypermultiplet.}  If we perform the vertical reduction on the toric diagram of Fig.~\ref{fig:conifold 1}, it is clear that we obtain a IIA background $\CM_5 \cong \C^2 \times \R$, with two non-compact D6-branes along the two $\C$ factors in $\C^2$:
\be\label{D61D62}
\begin{tabular}{l|ccccc|cccc|c}
$x^\mu$& $x^0$ & $x^1$  & $x^2$ & $x^3$ & $x^4$ & $x^5$  & $x^6$ & $x^7$ & $x^8$ & $x^9=r_0$  \\
 \hline
 D6$_1$ &$\times$ &$\times$&$\times$&$\times$&$\times$&$\times$&$\times$&&&$r_0=0$\\
  D6$_2$ &$\times$ &$\times$&$\times$&$\times$&$\times$& & &$\times$&$\times$&$r_0=\xi_1$
 \end{tabular}
\ee
These D6-branes are located at $r_0=0$ and $r_0=\xi_1$, respectively. There is a single five-dimensional mode, the open string stretched between the two D6-branes, which gives rise to a single 5d $\CN=1$ hypermultiplet $\CH$, with a real mass $m=\xi_1$.

\paragraph{Conifold as ``$SU(1)$ gauge theory.''} 
The conifold admits another, inequivalent IIA reduction, as indicated by the $U(1)_{M'}$ charges in \eqref{fig:rankzero sings}, corresponding to a vertical reduction of the $SL(2,\Z)$-transformed toric diagram shown in Figure~\ref{fig:conifold 2}. The IIA description is in terms of a single D6-brane wrapping the exceptional $\P^1$ in a resolved $A_1$ singularity. This gives us a naive ``$SU(1)$ gauge theory'', which has no Coulomb branch parameter but still has an ``$SU(1)$ gauge coupling,'' $h_0$.  The corresponding IIA profile reads:
\be
\chi(r_0) = \begin{cases} r_0 + \xi_1 \quad &{\rm if}\;  r_0 >0~,\\
-r_0 + \xi_1 \quad &{\rm if}\;  r_0 <0~,\end{cases}
\ee
as shown in Figure~\ref{fig:U10 coni}. We thus find an effective CS level $k_{\rm eff}=0$ and the identification:
\be
\xi_1= h_0~,
\ee
between the K\"ahler parameter and the gauge coupling. The M2-brane wrapped on $\CC_1$ is an ``$SU(1)_0$ instanton,'' in this description, while it was giving rise to a single hypermultiplet in the previous description. This is the same ``duality'' that we invoked in our discussion of the (toric) $T_2$ theory in section~\ref{subsec:E3T2}.

%%%%%%%%%%%%%%%%%%%%%%%%%%%%%%%%%%%%%%
%%%%%%%%%%%%%%%%%%%%%%%%%%%%%%%%%%%%%%
%%%%%%%%%%%%%%%%%%%%%%%%%%%%%%%%%%%%%%
%%%%%%%%%%%%%%%%%%%%%%%%%%%%%%%%%%%%%%

%%%%%%%%%%%%%%%%%%%%%%%%%%%%%%%%%

\bibliographystyle{utphys}
\bibliography{bib5d}{}

\end{document}